\documentclass[8pt]{article}
\usepackage[a4paper]{geometry}
\usepackage[T1]{fontenc}
\usepackage[utf8x]{inputenc} 
\usepackage[affil-it]{authblk}
\usepackage{lmodern}
\usepackage[english]{babel}
\usepackage{amsfonts}
\usepackage{amssymb}
\usepackage{stmaryrd}
\usepackage{textcomp}
\usepackage{bbold}
\usepackage{amsmath}
\usepackage{authblk}
\usepackage{subfig}
\usepackage{caption}
\usepackage[toc,page]{appendix}
\usepackage{bbm} 
\usepackage{hyperref}
\usepackage{mathrsfs}
\usepackage[all]{xy} 
\usepackage{xcolor}
\usepackage{tikz} 
\usepackage{graphicx}
\usepackage{wrapfig}
\usepackage{lscape}
\usepackage{rotating}
\usepackage{epstopdf}
\usepackage{babel}
\usepackage{amsthm}
\usepackage{verbatim}
\usepackage{float}
\usepackage{setspace}
\usepackage{listings}
\PrerenderUnicode{é}
\usepackage{amsmath}

\date{September 7th 2017}

\title{Winning Investment Strategies Based on Financial Crisis Indicators}
\author[]{Antoine Kornprobst  \thanks{Electronic address: \texttt{antoinekor9042@gmail.com}}}
\affil[]{Université Paris 1 Panthéon-Sorbonne, Labex ReFi}

\begin{document}
\maketitle
\begin{abstract}
The aim of this work is to create systematic trading strategies built upon several financial crisis indicators based on the spectral properties of market dynamics. Within the limitations of our framework and data, we will demonstrate that our systematic trading strategies are able to make money, not as a result of pure luck but, in a reproducible way and while avoiding the pitfall of over fitting, as a result of the skill of the operators and their understanding and knowledge of the financial market. Using singular value decomposition (SVD) techniques in order to compute all spectra in an efficient way, we have built two kinds of financial crisis indicators with a demonstrable power of prediction. Firstly, there are those that compare at every date the distribution of the eigenvalues of a covariance or correlation matrix to a distribution of reference representing either a calm or agitated market reference. Secondly, we have those that merely compute at every date a chosen spectral property (trace, spectral radius or Frobenius norm) of a covariance or correlation matrix. Aggregating the signals provided by all the indicators in order to minimize false positive errors, we then build systematic trading strategies based on a discrete set of rules governing the investment decisions of the investor. Finally, we compare our active strategies to a passive reference as well as to random strategies in order to prove the usefulness of our approach and the added value provided by the out-of-sample predictive power of the financial crisis indicators upon which our systematic trading strategies are built.
\end{abstract}


 \section{Introduction}
Our goal in this paper is to build active trading strategies that are able to make money in a reproducible fashion, not as a result of luck but as a result of skill. While using only commercially available data \footnote{All the data in this study ha been obtained from Bloomberg}, we attempt to anticipate market movements and act on them beforehand by using the financial crisis indicators based on random matrix theory that were developed in the work of Douady and Kornprobst (2015). Simply speaking, our objective is to liquidate our positions before the prices start to drop and to acquire them back before the prices go back up again. The portfolios that we consider are constituted of a mix of ETF shares replicating an equity index and cash. The financial crisis indicators will work on the stock components of the equity index and  make a determination about whether it is prudent to convert the shares into cash because the probability of a crisis happening in the near future is getting higher, or on the contrary whether it is advisable to convert the cash into shares because the market is in a calm phase and the probability of a sudden price drop within a given forecast horizon is low. The cash, if present in a portfolio, earns the risk-free Libor rate.\\

We consider that our actions have no direct influence on the market and that we are always price takers. In other words, we consider that our trading operations are small enough to generate neither slippage nor feedback in the order book. This is of course a simplification, as even the smallest trades have some kind of impact on the order book. Our assumption however, is that this impact is negligible for our operations and that ignoring it does not change anything to our approach and the validity of the situations that we are attempting to describe. In order to focus entirely on demonstrating the power of predictions of our strategies and their potential for making money, we decided against modeling market frictions like transaction fees associated with our trading operations. While these market frictions do exist and represent a very real cost for market agents, our ambition at this stage is not to build a realistic simulation of market operation, which by the way would include many other variables than those we have considering here, but rather to provide a credible technical framework and methods that are ready to be applied by professionals.\\

There is a large literature, both applied and theoretical, on the topic of building successful systematic trading strategies, especially for the hedge fund industry as explained in Fung and Hsieh (1997). A first approach relies on what is commonly known in the financial world as \textit{technical analysis}. This mostly empirical approach aims at predicting market movements and stock returns by identifying certain well known patterns in stock prices and other market vital signs. Technical analysis shares a lot of similarities with our approach since our aim is also to empirically detect reproducible crisis announcing patterns inside the  market data and use that knowledge as a financial crisis indicator. The efficiency of optimized systematic trading strategies based on technical analysis has been demonstrated in the work of Gençay (1998). Indeed, clever technical analysis is usually able to do better than  passive buy-and-hold strategies. This predictability of stock returns as a way to establish successful investment strategies is also described in details in the work of Pesaran and Timmermann (1995), who discuss from a historical point of view how the predictability of U.S stock returns have been routinely exploited by investors since the 1960' and especially since the 1970's. Indeed, it is at times of higher market volatility that technical analysis strategies can reach their full potential. Technical analysis strategies of various kinds (momentum, mean reverting, moving average, etc...)  have also been successfully tested by Ratner and Leal (1999) where they discuss systematic trading strategies based on technical analysis and applied to the Latin American and Asian equity markets. The relevance of systematic trading strategies based on technical analysis, versus passive buy-and-hold strategies, has been also statistically demonstrated in the work of Kwon and Kish (2002) about the predictability of NYSE stocks.\\

Another approach successfully used in the financial industry in order to build systematic trading strategies is the dynamic time-series process and signal processing approach. This approach also shares a lot of similarities with our systematic trading strategies based on financial crisis indicators. Indeed we use a rolling window on market data time-series, which are constituted of the log-returns of the stock components of an equity index. The financial crisis indicators that we compute then produce a signal that is used in the decision-making process governing the active trading strategy. This time-series based approach to trading is discussed in the work of Farmer and Shareen (2002) about the price dynamics of trading strategies seen through the lenses of signal processing theory. While some strategies based on technical analysis techniques might increase undesirable noise inside a market and create volatility as well as instability, especially when a large proportions of investors is using them at the same time, their efficiency and their power to make money is usually proven beyond a reasonable doubt as demonstrated in the work of Brock, Lakonishok and LeBaron (1992).\\

Our own approach to systematic trading strategies based on financial crisis indicators combines elements from all of the approaches that we described previously: technical analysis, time-series, signal processing, statistical considerations. We thus create a new framework that is grounded in a solid theoretical background,  while still remaining flexible enough in order to be able to be applied in the real world and in order to be useful to traders and investors.

\section{The Data}

We work exclusively with daily data, because data with a daily frequency is more readily available, but there is nothing in our framework that would prevent the use of intraday data. Besides, working with daily data was easier because intraday datasets can become extremely large and require a lot of computing power to be properly exploited in a timely fashion. Our data is constituted of five global equity indices. We use the Standard \& Poor's 500 (SP500), the Bloomberg European 500 (BE500), the Shanghai-Shenzhen CSI 300 (SHSZ300), the NASDAQ and the CAC40. For all those equity indices, we also have at our disposal all of their respective components as well as a matching time-series of the U.S Government Bond of maturity 1 month (US0001M) that is going to be used as the riskless asset to compute what the cash earns, whenever it is present in one of our portfolios, at a given date. The data for the equity indices themselves is comprised of the closing price $P_{0}(t)$ at each date of the time period on which our study is conducted. The data for the $N$ the components is comprised, for all  $i \in  \llbracket 1,N  \rrbracket $ of the closing price $P_{i}(t)$, the daily volume traded $V_{i}(t)$, the closing market capitalization $C_{i}(t)$ and the closing financial leverage $L_{i}(t)$, which is the quotient of the total debt by the market capitalization. All prices have been adjusted for dividends and splits. At every time $t$, we compute the daily log-returns for both the index itself and its components. It is defined as the log-return between a given trading day and the previous trading day. For all $i \in \llbracket 0,N \rrbracket $:   $r_{i}(t) = log(\dfrac{P_{t}}{P_{t-1}})$.\\

The composition of the equity indices is dynamic over time, but our datasets need to have a stable composition. Indeed, if we consider a given index, companies regularly drop out and are replaced by new ones that better fit the membership criteria, which are defined by a committee and most often based on the listed companies' market capitalization. In order to conduct our study, we need to have for each of the five indices a stable set of index stock components over a sufficiently long period of time. Therefore it was not possible to keep all the current components of the indices. A compromise had to be found between keeping enough of the components of each index in order to have a sample of companies that is representative of the state of the financial market and the necessity to have a sufficient depth for our time-series in order to conduct a statistical study over a long enough period of time. All our datasets span between June 13th 2006 and March 15th 2016. The details about the five datasets that we are going to use in our study are the following. The exact composition of each dataset, given as a list of Bloomberg tickers, is given in appendix for each of the datasets. \\

\begin{itemize}
\item \textbf{Dataset-SP500} is comprised of 420 components of the Standard and Poor's 500 index, plus the index itself, which contains 500 of the largest companies ranked by market capitalization, having stock listed on the NYSE or NASDAQ stock exchanges.
\item \textbf{Dataset-BE500}  is made of  419 components of the Bloomberg European 500 index plus the index itself, which is the European counterpart to the SP500 and contains 500 of the largest companies listed on European stock exchanges.
\item \textbf{Dataset-SHSZ300}  is comprised of 147 components of the Shanghai-Shenzhen CSI 300 index, plus the index itself. The composition of this index, which has only been in existence since April 2005, is very much in a state of flux and its evolution reflects the profound transformations of the Chinese economy over the past decade. For that reason, less than half of the companies in the index today were already there at the time of its creation. We still decided to go forward and keep using this basket of 147 Chinese companies which are representative of the Chinese financial market since 2006 and which reflect its current state.
\item \textbf{Dataset-NASDAQ}  is comprised of 69 components of the NASDAQ Composite Index, plus the index itself which contains 100 of the largest companies, excluding financial companies, with respect to market capitalization, listed on the NASDAQ stock exchange.
\item \textbf{Dataset-CAC40}  is constituted of 37 components of the CAC40 index, plus the index itself, which contains 40 of the most important (selection by a committee) companies among the 100 companies with the highest market capitalization listed on the Euronext Paris stock exchange.
\end{itemize}

For all the datasets, we considered only the trading days, excluding week-ends and holidays. Moreover, when a specific stock wasn't traded on a given day or whenever an entry was missing in the data, we carried over the last valid available value to fill the gap in order to avoid neutralizing a large number of trading days in our study.\\

\section{The Financial Crisis Indicators}

The financial crisis indicators used as decision making tools in our systematic investment strategies are based on the work of Douady and Kornprobst (2017). For a given dataset constituted of the $N$ stock components of an equity index, we start by choosing the length $T$ of a rolling window. This choice has to be made carefully and is important for the quality of the financial crisis forecasts and therefore important for the success of the systematic trading strategies based upon those forecasts. $T$ represents the number of observations of the log-returns of the components of the equity index and it needs to be large enough for a given dataset to enable us to obtain a whole meaningful spectrum of the covariance or correlation matrix of the stock components, but it should also not be too large in order to preserve the responsiveness of the indicators and avoid giving them a too long memory.\\

To obtain a whole spectrum of the matrices, we must have $T > N$ otherwise the asset vectors are not long enough (there are not enough observations inside the rolling window) and thus can never be linearly independent and the covariance and correlation matrices will be degenerate. Although that would have been a sufficient size for $T$, we discard the possibility of choosing $T=N$. This is a technical requirement. Indeed, since we plan on using the singular value decomposition (SVD) of the rolling window to obtain the eigenvalues of the covariance and correlation matrices, the rolling window cannot be a square matrix, as explained in Horn and Johnson (2013) as well as in Golub and Van Loan (2013).\\

The requirement of having $T > N$ may however be in theory somewhat relaxed because the only part of the spectrum that really interests us is the larger eigenvalues. Larger eigenvalues are indicative of dynamical instability and our financial crisis indicators are based on that property. If $T< N$, we will not be able to obtain the whole meaningful spectrum of the matrices, but as long as $T$ is not too small, the eigenvalues that we will miss are going to be the smaller ones, which are the ones that are less interesting to us. That is why the indicators in Douady and Kornprobst (2017) do still work when applied on a dataset containing 226 components of the SP500, while using a rolling window of only 150 days. In practice however in this study, where we are going to aggregate the signals produced by many different indicators, truncating the spectrum of the covariance or correlation matrix in any way did create technical problems and tended to degrade the quality of the forecasts, especially for the indicators that are called the A-series in Douady and Kornprobst (2017) and which compare the distribution of the whole spectrum to chosen references that represents either a calm or an agitated market.\\

We therefore decided to choose from now on in this paper $T=1.1 \times N$ (1) for all the datasets. Even though it means that the rolling window becomes quite large for the datasets that contain the most assets, like Dataset-SP500 or Dataset-BE500, the benefit of working with a spectrum that has not been truncated outweighs the drawbacks in terms of loss of responsiveness of our indicators due to their longer memory for some of our datasets. We therefore choose for the whole study the following sizes for the rolling window of each of the datasets ; 462 days for Dataset-SP500, 461 days for Dataset-BE500, 162 days for Dataset-SHSZ300, 76 days for Dataset-NASDAQ and 41 days for Dataset-CAC40.\\

For a given dataset and at a given date $t_{0}$, we consider the matrix of the log-returns : 
$$A(t_{0}) = (a_{i,j}(t_{0}))_{i \in \llbracket 1,N \rrbracket ; j \in \llbracket t_{0}-T, t_{0}-1 \rrbracket}$$

The rows of $A(t_{0})$ represent time-series of individual stock component and the columns of $A(t_{0})$ represent the observation dates. The coefficients of this matrix are the scaled and centered daily log-returns of the stock components of one of the equity indices considered in this study : $$a_{i,j}(t_{0}) = \dfrac{1}{\sqrt{T}}\lbrace r_{i}(j)-\dfrac{\sum_{k=1}^{T}r_{i}(t_{0}-k)}{T}\rbrace \quad (2)$$

From the matrix  $A$, we derive five matrices of interest :

\begin{itemize}

\item The unmodified matrix $A(t_{0})$ itself enables us, through its singular value decomposition, which we will write in details below, to compute the spectrum of the covariance matrix of the $N$ stock components over the time period $T$. The covariance matrix contains both the correlation and volatility effects of the stock components of a given equity index and both of those signals are going to be useful.

\item The matrix $B_{0}(t_{0}) = (b^{0}_{i,j}(t_{0}))_{i \in \llbracket 1,N \rrbracket ; j \in \llbracket t_{0}-T, t_{0}-1\rrbracket} $ is obtained by normalizing each coefficient of $A$ with the standard deviation of the row to which it belongs. This normalization removes the volatility information that was contained in $A(t_{0})$ and  $B_{0}(t_{0})$ contains only the pure correlation effect between the stock components.

$$b^{0}_{i,j}(t_{0}) = \dfrac{a_{i,j}(t_{0})}{std(\llbracket a_{i,t_{0}-T} \ldots a_{i, t_{0}-1} \rrbracket)} \quad (3)$$

From the singular value decomposition of $B_{0}(t_{0})$ we obtain the spectrum of the correlation matrix of the $N$ stock components over the time period $T$.

\item The matrix $B_{1}(t_{0}) = (b^{1}_{i,j}(t_{0}))_{i \in \llbracket 1,N \rrbracket ; j \in \llbracket t_{0}-T, t_{0}-1\rrbracket} $ is obtained by weighting each of the coefficients of $B(t_{0})$ by the relative importance of the corresponding stock's daily traded volume at $t_{0}-1$, with respect to all the other components of the index. The idea behind $B_{1}$ is to build a different flavor of the correlation matrix, one which gives more importance to the stocks which are the most liquid and the most traded inside the index. Indeed, those liquid stocks are more susceptible to drive the movements of the market, especially the down movements during and preceding a crisis event.

$$b^{1}_{i,j}(t_{0}) = b^{0}_{i,j}(t_{0}) \frac{V_{i}(t_{0})}{\sum_{k=1}^{N} V_{k}(t_{0})} \quad (4)$$

From the singular value decomposition of $B_{1}(t_{0})$ we will obtain the spectrum of the correlation matrix of the $N$ stock components over the time period $T$ weighted by volume traded.

\item The matrix $B_{2}(t_{0}) = (b^{2}_{i,j}(t_{0}))_{i \in \llbracket 1,N \rrbracket ; j \in \llbracket t_{0}-T, t_{0}-1\rrbracket} $ is constructed by applying to each of the coefficients of  $B(t_{0})$ a weight proportional to the market capitalization of the corresponding company. The idea is to give more importance in the computation of the spectrum of the correlation matrix to the companies which have the largest market capitalization and which may therefore drive the movements, though their sheer size, of their entire industry sector or of the financial market as a whole. By computing the singular value decomposition of $B_{2}(t_{0})$ we will obtain the spectrum of another flavor of the correlation matrix, one which is weighted by market capitalization.

 $$b^{2}_{i,j}(t_{0}) = b^{0}_{i,j}(t_{0}) \frac{C_{i}(t_{0})}{\sum_{k=1}^{N} C_{k}(t_{0})} \quad (5)$$

\item Finally, the matrix $B_{3}(t_{0}) = (b^{3}_{i,j}(t_{0}))_{i \in \llbracket 1,N \rrbracket ; j \in \llbracket t_{0}-T, t_{0}-1\rrbracket} $ is another flavor of the weighted correlation matrix. This time, we apply to the each of the coefficients a weight proportional to the financial leverage of the corresponding firm. Since the financial leverage can be interpreted as a measure of the financial health of a company, it provides us with a valuable way of giving more importance in our computations to the firms which are, because of their higher financial leverage, more exposed to the risk of suffering major adverse effects during a crisis or to start a domino effect. Indeed, their higher proportion of debt with respect to their own assets, puts them in a more precarious financial situation, or at least in a situation where a sudden random downturn of the market would render them unable to service that debt and therefore prone to failure. Many global financial institutions fell prey to this vicious circle  during the 2007-2008 financial crisis for example.

 $$b^{3}_{i,j}(t_{0}) = b^{0}_{i,j}(t_{0}) \frac{L_{i}(t_{0})}{\sum_{k=1}^{N} L_{k}(t_{0})} \quad (6)$$

\end{itemize}

We then apply the singular value decomposition (SVD) technique, as detailed in Horn and Johnson (2013) as well as in Golub and Van Loan (2013), to our five rolling matrices ($A$, $B_{0}$, $B_{1}$, $B_{2}$ and $B_{3}$) in order to obtain at each date $t_{0}$ the spectrum of the corresponding covariance, correlation or weighted correlation matrix. The main advantage that we obtain from using SVD is that the algorithm computing the singular values does not require any matrix multiplication like in Douady and Kornprobst (2017), where obtaining the covariance and correlation matrices required the product of a matrix by its transpose. Therefore, there will not be any added numerical errors during execution of the code on a computer. Indeed, a classical spectrum determination approach would start by computing the product of the rolling matrix by its transpose before applying standard techniques, like numerically finding all the roots of the characteristic polynomial using Newton-Raphson's method or similar techniques as detailed in the work of Abbasbandy (2003). The SVD approach bypasses the need to use those multiplications of matrices.\\

Taking into consideration an $N \times T$, with  ($N<T$), matrix $M$ with $M=A$ or $M=B_{k}$ for $k \in \llbracket0,3 \rrbracket $ ), the singular value decomposition is written :
 $$M = U \ \Sigma  V^{T} ; \Sigma = 
\left[\begin{array}{cccc|ccc}
\sigma_1 & 0 & \ldots &  0    & 0 & \ldots & 0 \\
   0  & \sigma_2   & \ldots &  0    & 0 & \ldots & 0 \\
\vdots & \vdots & \ddots  & \vdots    & \vdots & \ddots & \vdots \\
   0   &   0 & \ldots  &\sigma_N     &  0 & \ldots & 0 \\
\end{array}  \right]  \quad (7)$$

The eigenvalues $\lambda_{i}$ of $MM^{T}$ are obtained from the singular values $\sigma_{i}$ found in $\Sigma$ : $$\forall i \in \llbracket 1,N \rrbracket, \lambda_{i} = \sigma_{i}^{2} \quad (8)$$

After computing the whole spectrum of the rolling matrices at each date $t_{0}$, we can then use the two kinds of financial crisis indicators described in the work of Kornprobst and Douady (2017).

\begin{itemize}
\item Firstly, there are the indicators that compare the whole spectrum of the covariance, correlation or weighted correlation matrices associated with our five rolling matrices, to a reference spectrum distribution. We call these financial crisis indicators the $\alpha$-series. The reference distribution may represent either an agitated or a calm market reference. Following the work of Douady and Kornprobst (2017), we consider three different references distributions. Two represent a calm market and one represents an agitated market. In the sense of the Hellinger distance, which is the metric adopted, the observed empirical distribution is expected to move away from a calm reference distribution and move closer to an agitated reference distribution when the risk of a financial crisis is increasing in the market. The main idea is to measure when the whole spectrum of the eigenvalues of the covariance matrix, correlation matrix or weighted correlation matrix shifts towards the higher eigenvalues, which is a situation indicative of instability in the market and therefore may indicate the possibility of an upcoming crisis. The first reference distribution modeling an ideal calm market, called $\mathscr{R}_{1}$, is the Marchenko-Pastur distribution, introduced in Marchenko and Pastur (1967). It is the distribution of the spectrum of a correlation matrix corresponding to a rolling matrix constituted of independent identically distributed normal Gaussian coefficients. The second reference distribution, called $\mathscr{R}_{2}$ represents a more realistic calm market. This numerically computed reference is the distribution of the eigenvalues of the covariance matrix corresponding to a rolling matrix constituted of Gaussian coefficients correlated to one another at the level of the mean of the long term correlation coefficients between all the assets of the whole sample contained in the chosen dataset (around 50\%), as explained in Douady and Kornprobst (2017). The distribution $\mathscr{R}_{3}$ representing an agitated market reference is the numerically computed distribution of the eigenvalues of a covariance matrix corresponding to a rolling matrix constituted of Student ($t=3$) coefficients which are correlated to one another by the same method as the one used in  $\mathscr{R}_{2}$. The resulting coefficients follow a fat tailed distribution and are correlated and therefore constitute an adequate representation of an agitated market, where the log-returns of the stock component of an equity index are highly volatile and prone to extreme losses. These market conditions simulate the kind of situation we expect to exist in days the days leading to a financial crisis. To summarize, we are considering five matrices ($A$, $B_{0}$, $B_{1}$, $B_{2}$ and $B_{3}$) and three references ($\mathscr{R}_{1}$, $\mathscr{R}_{2}$ and $\mathscr{R}_{3}$), which therefore gives us 15 financial crisis indicators of the $\alpha$-series.\\

\item Secondly, there are the financial crisis indicators that compute a specific spectral property of the covariance matrix, correlation matrix or weighted correlation matrices. We call these financial crisis indicators the $\beta$-series. We take into consideration three spectral properties: the spectral radius (the largest of the eigenvalues), the trace (the sum of the eigenvalues) and the Frobenius norm (the sum of the squared eigenvalues). The basic idea behind those indicators is similar than in the case of the indicators of the $\alpha$-series, which compare the whole distribution of the spectrum to a reference distribution. Indeed, a shift of the spectrum to the right, toward the larger eigenvalues is indicative of market instability. It means increased correlation and volatility in the market and therefore an increased risk of a crisis taking place. These effects are studied under different points of views corresponding to the five rolling matrices considered. We therefore take into consideration five matrices ($A$, $B_{0}$, $B_{1}$, $B_{2}$ and $B_{3}$) and three spectral properties (spectral radius, trace and Frobenius norm), which gives us 14 financial crisis indicators of the $\beta$-series. Indeed the trace of the covariance matrix is useless as an indicator because it is constant and equal to the number $N$ of stock components in the equity index considered. The traces of the weighted correlation matrices obtained from $B_{1}$ (volume traded), $B_{2}$ (market capitalization) and $B_{3}$ (financial leverage) are on the other hand perfectly valid indicators which bring their unique point of view to the study and are therefore very valuable. Indeed, as written in equations (4), (5) and (6), the weighting is applied to the coefficients of the rolling matrix $B_{0}$ and not to the coefficients of $A$. Therefore the normalization is applied before the weighting and the sum of the eigenvalues of the weighted correlation matrices, which are obtained par SVD decomposition of $B_{1}$, $B_{2}$ and $B_{3}$, has no reason to be constant.
\end{itemize}

When aggregating the indicators of the $\alpha$-series and of the $\beta$-series, we therefore obtain 29 financial crisis indicators overall. At a given date $t_{0}$, when an investment decision needs to be made, each of those financial crisis indicators is considered as a expert opinion. When a strategy has to be defined, the core of our approach is going to be to attempt to find a consensus among those 29 opinions.

\section{Calibration}

The daily results provided by our 29 financial crisis indicators are just numbers at the moment. We need to translate those numbers in terms of financial crisis forecasts. We must define an in-sample calibration period and recognize patterns in the values taken by the 29 indicators during that calibration period that are reproducible and that we are going to use later in the out-of-sample study to make previsions.\\

In order to quantitatively define a financial crisis, we introduce at each given date $t_0$ the notion of maximum draw down ($MDD(t_0)$) at a given time horizon $H$ in the future. We will choose $H$ at 100 days for the remainder of our study since it's a commonly used value, both in the literature and  in the industry. We define a \textit{financial crisis} or \textit{market event} as the crossing of a chosen threshold of maximum draw down, which can be for example down to 5\% if we want to consider mild market events, or up to 40\%, if our intention is to consider mostly very large crises. The order of magnitude of the maximum draw down never climbs much higher than 40\% for most indices, not even at the height of the 2008 crisis during the failure of Lehman Brothers. The choice of a maximum draw down threshold is an important part of the construction of a successful systematic trading strategy in the framework that we are building. For a given dataset, the reference asset for which the maximum draw down is computed is the index itself (or an ETF replicating the index) and the price considered is the last price of the day $P_{0}(t_0)$.

$$MDD(t_0) = \max_{t_{0} \leq t \leq \tau \leq t_{0}+H}  \lbrace 1 - \frac{P_{0}(\tau)}{P_{0}(t)}  \rbrace \quad (9)$$

Obviously, in the out-of-sample study, the value at a date $t_{0}$ of $MDD(t_{0})$ is not known at $t_{0}$, we would need knowledge from the future for that. Indeed, we want interpret the value of our indicators as a probability of crossing a chosen maximum draw down threshold over the course of the forecast horizon $H$ in the future. In the in-sample, training period of our indicators, we will however match the value of our indicators at a given date $t_{0}$ with the value of $MDD(t_{0})$, computed in advance by using data from the future from the point of view of an observer who exists at the date $t_{0}$. We will do this in order to precisely learn which are the values taken by our indicators which correspond to the crossing of a given MDD threshold in the future within the forecast horizon $H$. That process is at the heart of the calibration of our 29 indicators.\\

We must choose two things in the calibration process. Firstly, we decide the size $K$ of the in-sample calibration period during which our indicators learn how to recognize which are the values they take that match the largest MDD values. Obviously, the larger the calibration period, the better in theory, but since our data spans only around 10 years, we must preserve a large enough out-of-sample period to make previsions and validate the viability of our systematic trading strategies. It is also necessary that the 2007-2008 financial crisis be included in that calibration period because the indicators must be confronted at least once to a major crisis in order to learn its signature and hopefully be able to recognize similar events in advance in the out-of sample period. Besides all those considerations, the calibration period $K$ must at the very least fit the rolling window $T$ for a given dataset, which as we explained before in Equation (1) is equal to $1.1 \times N$, with $N$ the number of stock components present in the dataset. Using those notations, there are going to $K-T$ usable dates for the training of the indicators and the previsions will be able to start at the date $K+H$. Figure 1 summarizes the situation of the calibration period with the rolling window represented as an orange rectangle.

\begin{center}
\includegraphics[scale=0.70]{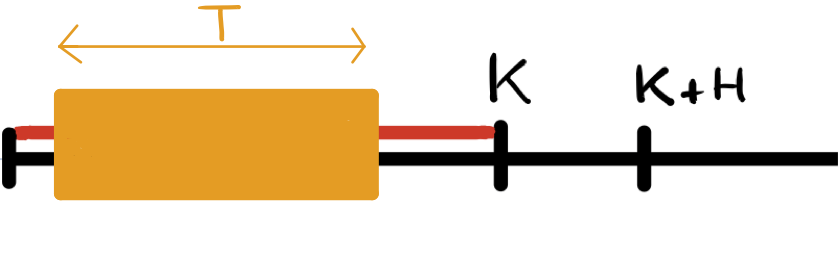}\\
Figure 1 : Calibration period (red) and rolling window (orange)
\end{center}

In order to give the indicators the chance to have enough usable calibration dates for every one of our five datasets and keeping in mind that we must try to treat them as equally as possible, even though they are of very different sizes in terms of the number of stock components that they contain, we establish the following rule :$$K = max(500, T+50) \quad (10)$$
This seems like a reasonable choice. Indeed, Formula (10) guarantees that the indicators will encounter the 2007-2008 financial crisis for all the datasets, even the smaller ones like the CAC40. It assures that the training period will not significantly be larger than 2 years (assuming around 250 trading days per year), even for the larger datasets like the SP500 and BE500. Finally, the indicators are guaranteed at least 50 training dates for all the datasets, even the larger ones, which is the minimum to guarantee their proper calibration. Using Formula (10), we obtain an in-sample calibration period of 512 days for Dataset-SP500, 511 days for Dataset-BE500 and 500 days for Dataset-SHSZ300, Dataset-NASDAQ and Dataset-CAC40.\\

For a given dataset and for each of our 29 financial crisis indicators, we draw the scatter plot of the maximum draw down at a given date versus the numerical value of the indicator at the same date. The plots for all the five datasets and the 29 financial crisis indicators of both the $\alpha$-series and the $\beta$-series are provided in appendix. The points corresponding to the dates of the in-sample calibration period are in red and the points corresponding to the dates of the out-of sample prediction period are in blue. The most notable feature that immediately emerges from observing all those plots is that they are all roughly bell-shaped and, most importantly, that bell-shaped structure is present both in the calibration period and the prediction period. The values of the indicator corresponding to the higher values of MDD are the same during the in-sample training period and the out-of-sample prediction period. That means that it makes sense to teach the indicators during their calibration period which ones are the values they take that actually correspond to the highest values of MDD (i.e the financial crises). Those structures are obviously easier to see when the number of usable dates inside the calibration period is larger (i.e for the smaller datasets in terms of the number of stock components inside the equity index), but it is almost always visible nonetheless.\\

We can interpret that bell-shaped structure from a financial point of view in the following manner. When the value of an indicator is very small (resp. very high for the indicators based on $\mathscr{R}_{3}$), it means that the probability of a crisis happening withing the chosen 100 days time horizon $H$ is very small. On the other hand, when the value of an indicator is very high (resp. very small for the indicators based on $\mathscr{R}_{3}$), then it means that the market is very likely already in the midst of a crisis and what the indicators are seeing at the $H=100$ days horizon is the post-crisis recovery. That leaves in the middle a \textit{danger zone} which corresponds to the values taken by an indicator which corresponds to the highest values of MDD at the 100 days horizon. In light of the previous discussion, "calibrating the indicators" therefore means "finding the danger zone for each indicator " and in order to do this in a quantitative and reproducible manner, we need to choose the first of the two parameters that define a systematic trading strategy in our framework: the MDD threshold $\mathscr{T}$. This choice is very important and will have profound implications for the eventual success of a systematic trading strategy.\\

By choosing an MDD threshold $\mathscr{T}$, we tell each of the 29 indicators to forget any point below $\mathscr{T}$ in the scatter plot (value vs. MDD) graph drawn during the calibration period and then to  place its danger zone (which is of a fixed width equal to 15\% of the total width of the graph, after having discarded possible outliers) such that the number of points inside it, is maximum. The choice of $\mathscr{T}$ determines whether we decide to design our strategies to attempt to detect in advance a large number of small crises (we choose in that case a low $\mathscr{T}$ around 5\% or 10\%) or whether we prefer to bet on the successful forecasts of a small number of large crises (we choose in that case a $\mathscr{T}$ of 15\% or more). We must find the good balance because both cases, large or small $\mathscr{T}$, present advantages and drawbacks.\\

If we choose $\mathscr{T}$ to be small, then according to the work of Douady and Kornprobst (2017), there is going to be some false positives (i.e. an indicator forecasts a crisis within its 100 days horizon of previsions, but nothing actually happens) and some false negatives as well (i.e. an indicator fails to predict a market event). The relatively low number of false positives for an individual indicator is clearly an advantage and will give more focus to our systematic trading strategies. The presence of some false negatives is potentially more dangerous but this may not be a disaster however, because the market events that we are betting on forecasting in order to build our strategy are typically small in that case, so even if a few of them are missed, we could still obtain acceptable results in terms of performance. If on the other hand we choose $\mathscr{T}$ quite high, then there is going to be much more false positives, further increasing the noise in the signal of an individual indicator, but there is going to be only a small risk that the indicators will miss a large market event. If we bet on accurately forecasting large financial crises, there is a lower risk of missing one, however if the indicators still do fail to correctly forecast a large market event, then in that case we would be instantly ruined.\\

As an illustration of the process of calibration of the indicators, we consider Dataset-CAC40 and the financial crisis indicator defined by the Hellinger distance between the empirical distribution of the spectrum of the correlation matrix weighted by financial leverage and the reference distribution $\mathscr{R}_{2}$ (calm market reference). In Figure 2a, we draw the scatter plot (indicator value vs. MDD), with the MDD represented on the y-axis. The bell-shaped structure is clearly visible, both for the red dots, which represent the calibration period and the blue dots, which represent the out-of-sample forecast period. In Figure 2b, we have chosen $\mathscr{T}=10\%$ and all the red points below $\mathscr{T}$ are forgotten. The danger zone is represented as the area between the two vertical red lines and it is chosen automatically by the computer code, which maximizes the number of remaining red points inside a vertical stripe of width 15\% of the total width of the scatter plot. In that example, the danger zone is defined by values of the indicator roughly between 224 and 238 for the Hellinger distance.

\begin{center}
\includegraphics[scale=0.25]{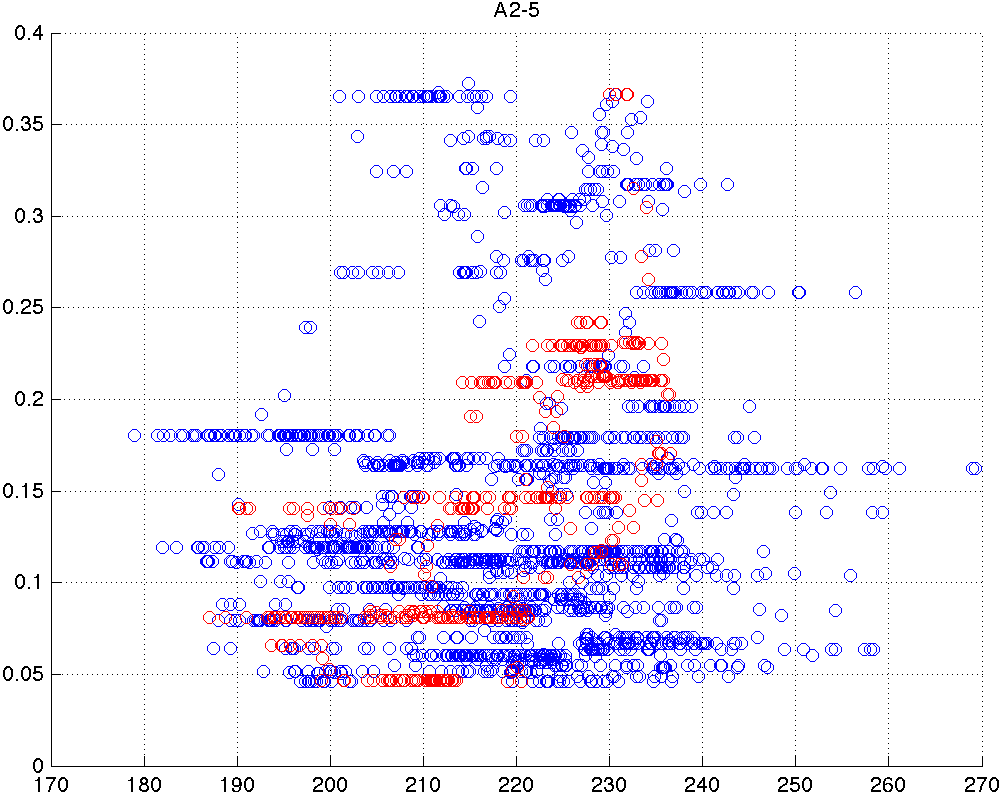}\\
Figure 2a: MDD vs. Hellinger distance between $\mathscr{R}_{2}$ and the distribution of the eigenvalues of the correlation matrix weighted by financial leverage for Dataset-CAC40
\end{center}
\begin{center}
\includegraphics[scale=0.15]{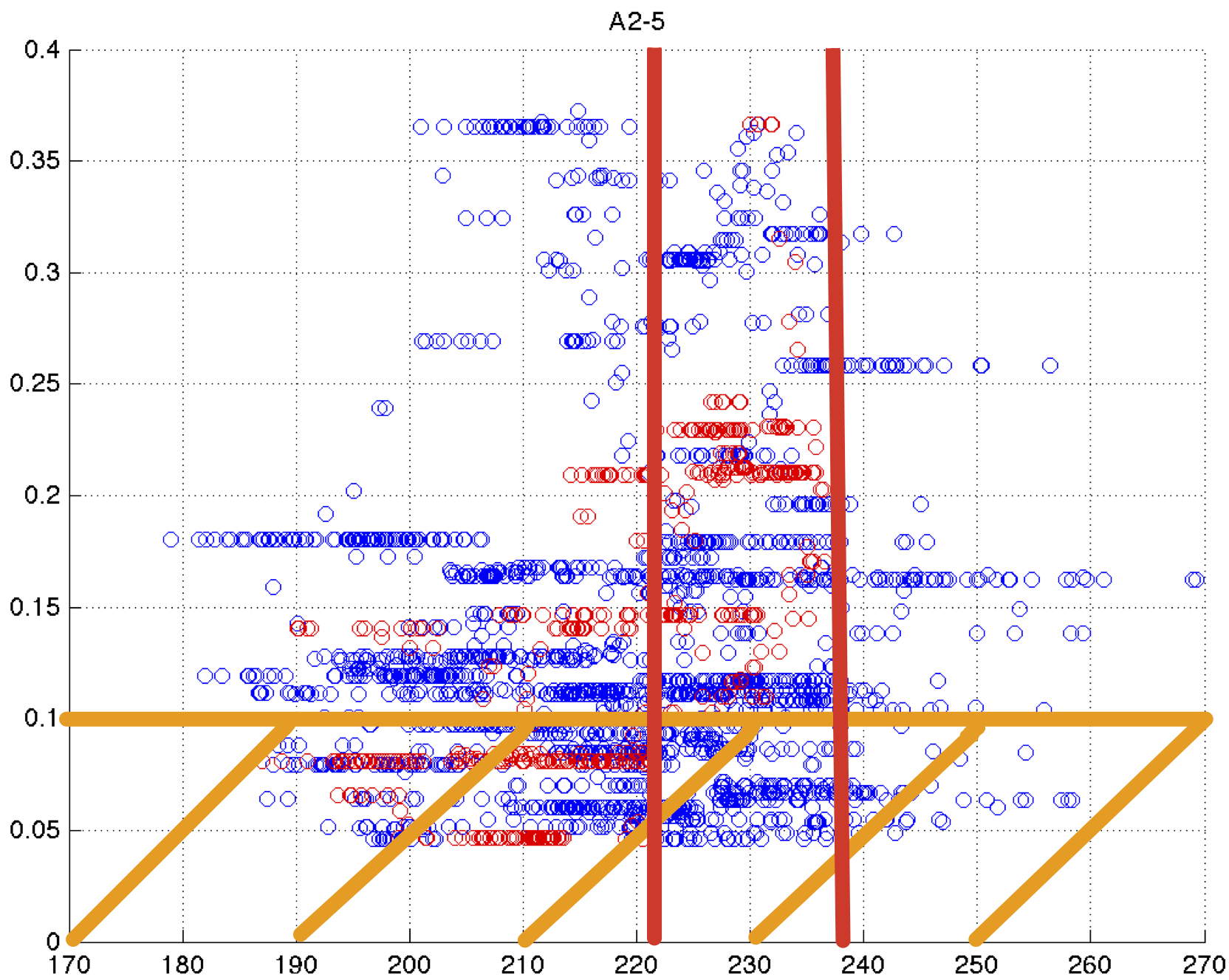}\\
Figure 2b: $\mathscr{T}$ is represented an horizontal orange line (10\% in this example), any point below $\mathscr{T}$ is not taken into consideration. The danger zone is represented as the area between the vertical red lines.
\end{center}

\section{Systematic Trading Strategies}
The 29 financial crisis indicators, both of the $\alpha$-series and of the $\beta$-series are now properly calibrated, which means that they each have discovered where their danger zone is located during their in-sample training period, according to the choice that was made for the value of the MDD threshold $\mathscr{T}$. Like we have already stated, we view those 29 indicators as expert opinions and we try to reach consensus among them at a given date $t_{0}$ in order to make a trading decision. Following the work of Clemen (1989), our intention is to combine multiple individual forecasts, provided by our 29 indicators, in order to improve the quality and focus of our previsions and therefore to increase to chances of success of the systematic trading strategies, which are based on those previsions. Since it was stated in the work of Douady and Kornprobst (2017) that one of the main limitations to the predictive power of each individual financial crisis indicator of either the $\alpha$-series or the $\beta$-series, was the presence of false positives inside the results, our expectation is that combining the signals of many different indicators will reduce the influence of those false positives. The strategies that are going to be considered in this study are applied to a portfolio containing both cash, which earns the riskless Libor rate, and ETF stocks, which replicates an equity index. At each date $t_{0}$, according to the aggregated signal produced by our financial crisis indicators, we can choose to do nothing if the situation is unclear, convert some cash to shares if the market forecast is good or convert some shares to cash if the risk of a crisis happening is increasing. There are no market frictions or transaction fees in our modeling and there are no considerations of liquidity either, since major equity indices ETF (SP500, BE500, SHSZ300, NASDAQ or CAC40) are extremely liquid securities in all market conditions.\\

We illustrate our approach with the drawing in Figure 3. At each date $t_{0}$ of the active trading period that starts at $K+H$, we look 100 days in the past and count for each of the 29 financial crisis indicators the number of times it takes a value which is inside its danger zone. We have talked about the MDD threshold $\mathscr{T}$ in the previous section and we now introduce the second of the two parameters that define a systematic trading strategy in our framework: the indicator sensitivity $\mathscr{S}$. This indicator sensitivity $\mathscr{S}$ is a number between 0 and 100 that defines a threshold above which an indicator gives a \textit{red flag}. For any one of our 29 indicators, if the indicator takes values inside its danger zone more than $\mathscr{S}$ times inside the interval $\llbracket t_{0}-100, t_{0}-1 \rrbracket$, then it produces a red flag at $t_{0}$, which indicates a higher risk of a financial crisis happening (defined as the crossing of the threshold $\mathscr{T}$) during the time period $\llbracket t_{0}+1, t_{0}+H \rrbracket$ in the future. Like the choice of $\mathscr{T}$ before, the choice of $\mathscr{S}$ is a very important part of the eventual success of a systematic trading strategy based on our financial crisis indicators. Indeed the parameter $\mathscr{S}$ controls the "aggressiveness" of a strategy, which is its tendency to react very quickly to the first signs of deteriorating market conditions and convert shares to cash immediately or, on the contrary, its tendency for showing patience and restraint and only act and sell shares when the signs of an impending crisis in the market are more clear.\\

If $\mathscr{S}$ is chosen to be small (around 50\% or 60\%), then red flags are easier to achieve and the strategy is going to be very aggressive and convert the shares to cash at the first sign of danger. This behavior may be considered prudent in some circumstances,  but it will also severely damage the performance prospects of the strategy because there will often be only cash (which earns very little) inside the portfolio and we will fail to take advantage of periods of market growth. On the other hand if  $\mathscr{S}$ is chosen very high (for example 85\% or 90\%), then the red flags are going to be very hard to obtain and the strategy is going to be very patient and keep the shares inside the portfolio until the signs indicative of an impending crisis become impossible to ignore. Such a behavior might maximize profit in some cases because the shares are kept for as long as possible, in particular during the periods of market growth, but the risk is to keep the shares for too long while their value starts to decline and to have a strategy so apathetic that it follows the ETF down during a crash and does not convert the shares into cash when it is needed. Like with the choice of $\mathscr{T}$ before, the choice of the indicator sensitivity $\mathscr{S}$ is about finding the right balance between responsiveness (low $\mathscr{S}$) and apathy (high $\mathscr{S}$). \\

\begin{center}
\includegraphics[scale=0.20]{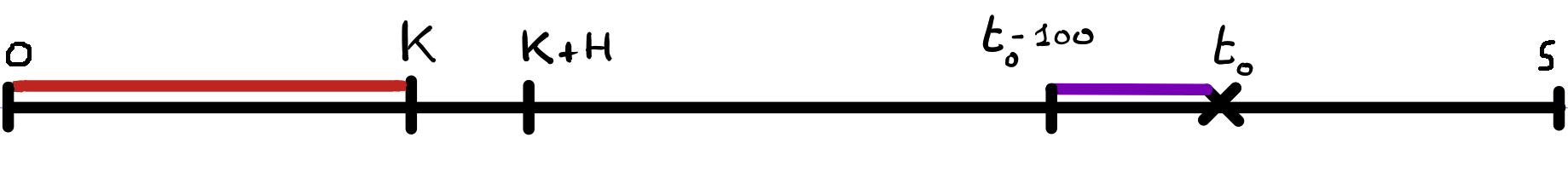}\\
Figure 3: Representation of the calibration period (red) and of the computation period at the date $t_{0}$ (purple).
\end{center}

We then count the total number of red flags produced by the indicators at the date $t_{0}$. It is an integer between 0 and 29 that we call $\Gamma(t_0)$. This number contains the aggregated crisis forecasting signal of all the indicators  of the $\alpha$-series and of the $\beta$-series. Finally we define a set of discrete rules that define what the strategy is going to do regarding the cash and ETF shares mix inside the portfolio.\\

We had initially also pursued an approach where, instead choosing an indicator sensitivity $\mathscr{S}$, then a function $\Gamma$ and finally a set of rules on $\Gamma$, we chose a response function $f: [0,2900] \longrightarrow [0,1]$ that gave the percentage of cash inside the portfolio as a function of the total number of times the 29 indicators were inside their respective danger zone in the time period $\llbracket t_{0}-100, t_{0}-1 \rrbracket$. This approach did not however prove itself viable as the performances of the resulting active strategies were very sensitive to the exact form of the graph of the function $f$ chosen empirically by the operator. This lack of robustness of the approach relying on a response function seemed unacceptable and moreover it would have invited concerns about over-fitting in our framework.\\

The set of discrete rules on $\Gamma$ are very robust and are chosen once and for all for five datasets and for the duration of the study. The rules that we choose for this study may still probably be refined and made more realistic in a real world situation, but they have experimentally proven their worth for the all datasets and the time periods that we have studied. Dealing with a portfolio constituted of a mix of cash and ETF shares, the discrete rules on $\Gamma$ are the following at the date $t_{0}$ of the out-of-sample active trading period:\\

\begin{itemize}
\item $\Gamma(t_{0})=0$ or  $\Gamma(t_{0})=1$: we buy 10\% more of the ETF shares on top of those already present inside the portfolio, unless there is already no more cash inside the portfolio because it has already been entirely previously converted to ETF shares.\\

\item $\Gamma(t_{0}) \in \llbracket 2,4\rrbracket$: we do nothing.\\

\item $\Gamma(t_{0})>4$: we sell 10\% of the ETF shares present inside the portfolio, unless there is already no more shares inside the portfolio because they have already been entirely previously converted to cash.\\

\end{itemize}

Those rules on $\Gamma$ are designed to filter out the false positives in the forecasts provided by our financial crisis indicators, which are the main limitation of our framework, as explained in Douady and Kornprobst (2017). We achieve this goal by combining the opinion of several of them and try to reach a consensus before selling shares and act on the prevision that a market downturn is becoming more likely within the given time horizon $H=100$ days. The occurrence of false negative errors on the other hand is rarer in our framework, but we did also design the rules such that the indicators have to agree among themselves that the risk of a market downturn is low before triggering the purchase of more shares.\\

We may now summarize our framework for the construction of systematic trading strategies based on our financial crisis indicators. Once  the rules for $\Gamma$ have been fixed beforehand for the whole study, the operator decides which parameters are the best given the dataset being studied and the market conditions. The skills and experience of the person making the decisions for the parameters $\mathscr{T}$ and $\mathscr{S}$ are a very important part of the potential for success of a strategy. We recall that:\\

\begin{itemize}
\item The operator first chooses a MDD threshold $\mathscr{T}$ to calibrate the indicators (i.e. find their danger zone) over the in-sample calibration period). A high $\mathscr{T}$ means that we bet on accurately forecasting a small number of large market downturns and a low $\mathscr{T}$ means that we bet on forecasting a large number of small market downturns.

\item The operator then chooses an indicator sensitivity $\mathscr{S}$. It determines the level of aggressiveness of the strategy. A high $\mathscr{S}$ means that the red flags are difficult to obtain and therefore that the strategy is very apathetic. A low $\mathscr{S}$ means that the red flags are easier to obtain and therefore that the strategy is very aggressive ans sells the shares at the first sign of danger.\\
\end{itemize}

Once these choices have been made for a given equity index dataset, the systematic trading strategy that we have designed is able to decide at every date of the out-of sample period what to do regarding the composition of the portfolio constituted of a mix of cash and ETF shares replicating the index. We are now going to compare the performances of our active trading strategies to a passive buy-and-hold strategy as well as to random strategies in order to demonstrate their worth, which is built upon the predictive power of our 29 financial crisis indicators of both the $\alpha$-series and the $\beta$-series.
\section{Numerical Results}

We now switch our attention to the application of our systematic trading strategies inside the framework defined in the previous sections to our five datasets:  SP500, BE500, SHSZ300, NASDAQ and CAC40. As usual, we compute the financial crisis indicators on the components of the index present inside each dataset and the financial instrument that we trade is an ETF share replicating the equity index.\\

For each dataset, we will consider three investors: 
\begin{itemize}
\item A passive investor who hold the portfolio $PP$. That investor starts with 10.000 shares of the ETF replicating the index and 10 millions in cash, which earns the riskless monthly Libor rate. That investor keeps those assets for the duration of the study.

\item An active investor who holds the portfolio $PA$. That investor starts with 10.000 shares of the index and 10 millions in cash and then chooses a set of parameters ($\mathscr{T}$, $\mathscr{S}$) in order to create a systematic trading strategy governing the composition of $PA$ and deciding automatically what to do according the set of discrete rules that was detailed in the previous section. We do not consider fractional ETF shares and round the number of shares to the closest lower integer.

\item An active investor who holds a portfolio $PR$ and who starts with 10.000 shares of the index and 10 millions in cash and adopts a random strategy. At each date, the random strategy governing a path of $PR$ chooses to buy more shares (unless there is already no more cash), do nothing or sell some shares (unless there are already no more shares) in a random fashion, but in such a manner that the proportion of "buy", "do nothing" and "sell" orders is the same as in $PA$. Those random strategies are similar in nature to the \textit{random same proportion} (RSP) strategies described in Douady and Kornprobst (2017). The role of the $PR$ paths is to demonstrate that our active strategy $PA$ does bring added value and (hopefully, as we will discuss below) beats an average of random strategies.

\end{itemize}

For every decision date in the out-of-sample period, the riskless asset $TB$ is chosen as the one month U.S Treasury Bond (US0001M) and the cash, if present inside a portfolio, earns this rate as well. Considering an asset $A$, which can represent either $PP$, $PA$ or one path of $PR$, we define the following benchmarks, Sharpe ratio and Calmar ratio. We also define the investment ratio $IR$ and we recall the Maximum Draw Down (MDD), that we have already defined in Equation (9). Here, we compute the MDD(A) over the entire period of study and not at a 100 days horizon like when we were calibrating the indicators.\\

\begin{itemize}
\item The investment ratio $IR$, that is computed for $PA$ only, is simply a measure of the proportion of riskless cash inside the active portfolio. It measures whether the strategy is choosing to convert the cash into ETF shares or the ETF shares into cash. $IR=1$ means we are fully invested and  $IR=0$ means the strategy has decided to convert all the ETF shares into cash because, according to its parameters, it sees a high risk of a crisis happening within the 100 days forecast horizon of the financial crisis indicators. $$IR = \dfrac{PA-cash}{PA} = \dfrac{shares}{PA}\quad (11)$$

\item To compute the yearly Sharpe ratio, we proceed in the following manner. The Sharpe ratio measures the quotient of the excess performance with respect to a riskless asset over the volatility. To achieve the desired highest Sharpe possible, a strategy has to maximize return while at the same time minimize volatility, which represents risk.\\

\begin{itemize}
\item At every date $t$ of the out-of-sample period, $ExcessReturn_{A}(t) =\frac{A(t)}{A(t-1)} - \frac{TB(t-1)}{12} -1$
\item With  the notations of Figure 3, we compute the annualized performance, assuming 30 days per month $Perf(A) = (\frac{A(S)}{A(K+H)})^{\frac{360}{S-(K+H)}} - 1$\\

\item $Vol(A)=stdev(ExcessReturn_{A}).\sqrt{12}$ (Annualized volatility)\\

\end{itemize}
$$Sharpe(A) = \frac{Perf(A) - mean(TB)}{Vol} \quad (12)$$

\item We define the yearly Calmar ratio as the quotient of the performance by the maximum draw down (MDD).
$$Calmar(A) = \frac{Perf(A)}{MDD(A)}\quad (13)$$ 
\end{itemize}

Regarding the set of parameters  ($\mathscr{T}$, $\mathscr{S}$)  that we choose, we divide the five equity indices studied into three groups, detailed below, with similar characteristics. Our intention is to standardize the choice of $\mathscr{T}$ and $\mathscr{T}$ and therefore demonstrate that our framework does not suffer from over-fitting and the selection of ad-hoc parameters that would work only for a specific index in a particular situation. Indeed the results that will be provided by our successful systematic trading strategies are robust and able to be generalized to other datasets and many market regimes.

\begin{itemize}
\item The first group contains the BE500, the CAC40 and the SP500. Those are are equity indices containing stocks belonging to companies from all the sectors of activity in large mature economies.  We choose $\mathscr{T}=20$ and $\mathscr{S}=75$. According to the discussion in the previous sections, it means that we intend to bet on the detection of a small number of large crises and that we want the strategy to be relatively calm and patient. Indeed, market events representing a MDD of $20\%$ or more are already very significant and asking the indicators to return a red flag when they are at least  $75\%$ of the time inside their danger zone in the 100 days preceding a decision makes the red flags relatively hard to achieve and therefore it makes the strategy less likely to sell the shares too early at the first sign of danger.

\item The second group contains the NASDAQ index only. This equity index contains many high tech companies, which  are often younger and sometimes more prone to volatility in the prices of their respective shares. We choose $\mathscr{T}=15$ and $\mathscr{S}=80$. We therefore decide to bet on relatively large market event that will however still be more numerous and smaller than those we had decided to bet on for the large generalist equity indices like the SP500. The choice of  $\mathscr{S}$ reflects our desire to have very patient strategies, that will not panic at the first sign of danger and wait until the signs of an impending crisis are undeniable to start converting the ETF shares into cash.

\item Finally, the third group is constituted of the SHSZ300 index alone. Like we are going to see when we will comment the results obtained, building a winning systematic trading strategy for the Chinese index has been a challenging task. Indeed, the Chinese market has undergone a lot of profound structural transformations, from being a large emerging market to being the world's second largest, during the period of study covered by Dataset-SHSZ300 (June 2006 to March 2016). This inherent instability is reflected in the very composition of Dataset-SHSZ300, since it contains only 147 of the current 300 components, meaning that more than half of the companies that are currently part of the index were not present ten years ago. Because of all those uncertainties creating instability and volatility inside the SHSZ300 index, we chose $\mathscr{T}=10$ and $\mathscr{S}=70$. Our choice is to bet on the detection of a large number of small crises (MDD threshold = 10\%) and to have relatively aggressive strategies, with an indicator sensitivity of only 70\% that will react quickly to signs indicative of a higher risk of a crisis happening withing the forecast horizon and not wait to too long to take action in a structurally chaotic financial market.
\end{itemize}

The results that we obtain for Dataset-BE500 with our systematic trading strategy, defined by all the parameters chosen as discussed above ($\mathscr{T}=20$ , $\mathscr{S}=75$), are presented in Figure 4a, Figure 4b and Figure 4c. We recall that the rolling window is of 461 days in length and that the calibration period is 511 days. Over the course of the out-of-sample prediction period, the active strategy governing $PA$ issued 983 buy orders, 357 stay orders and 556 sell orders. By examining all the calibration graphs provided in appendix for the 29 indicators of both the $\alpha$-series and of the $\beta$-series, we notice that most of them have the expected bell-shaped structure, both the in-sample calibration period (red dots) and the out-of-sample test period (blue dots). This guarantees a proper calibration of the indicators and therefore the quality of the active strategy based on their aggregated signals. In Figure 4a, the active portfolio $PA$  does much better than the passive portfolio $PP$ in terms of Sharpe ratio (increase to 0.43 from 0.27), performance (increase to 8.5\% from 5.2\%). $PA$ does slightly better than $PP$ in terms of Calmar ratio as well, but it is less significant (increase to 0.28 from 0.26). This is due to the fact that $PA$ has an higher MDD over the course of the out-of-sample period than $PP$. This drawback of the active strategy results from the fact than whenever $IR=1$ ($PA$ is fully invested), if there is drop in the ETF share prices, $PA$ suffers more locally than $PP$, which still contains its cash. Excluding the considerations on the MDD, $PA$  still beats $PP$ on all the other market benchmarks. The graph of $\Gamma$ is represented in Figure 4c and has to be studied in concert with the graph of $IR$ in Figure 4a. We notice that the active strategy, with its rules on $\Gamma$ that we have detailed earlier, correctly anticipates the the big crises, $IR$ falls to zero at mostly the rights times, but it is a bit too bad that the strategy decides to sell all the shares too early in 2014, which results in a performance that does not reach its full potential. We talked a lot about the limitations in our framework induced by the false positive signals sent by the indicators and combining the signals of 29 of them to compute the function $\Gamma$ did help a lot, but there is still room for improvement. In figure 4b, we have represented $PA$ against 50,000 paths of $PR$. The average Sharpe of the random strategies is only 0.29, while $PA$ has a Sharpe of 0.43. $PA$ beats 100\% of the random paths on that important benchmark. While the average MDD of the paths of $PR$ is slightly smaller than the MDD of $PA$ for the reasons we have given just before (the presence of cash in $PA$ does boost the MDD at some point, unfortunately), the active portfolio $PA$ still beats nearly 69\% of the random paths in terms of Calmar because $PA$ features a better overall performance than most random paths $PR$. The examination of the plane area (in green on Figure 4b) created by the random paths underlines the fact that the active strategy governing $PA$ takes the right decisions at the right time, especially in late 2011, but decides to sell the ETF shares too soon in early 2014, which somewhat reduced the overall performance of $PA$.\\

The results that we obtain for Dataset-CAC40 ($\mathscr{T}=20$ , $\mathscr{S}=75$) are presented in Figure 5a, Figure 5b and Figure 5c. We recall that the rolling window was 41 days in length and that the calibration period was 500 days. Over the course of the out-of-sample forecast period, the strategy governing $PA$ has issued 1152 buy orders, 711 stay orders and only 36 sell orders, but as we are going to discover, those few sell orders were given just at the right time, before a major drop in price and have made all the difference, making $PA$ on Dataset-CAC40 very successful. The success of $PA$ is rooted in the excellent calibration of the indicators, as highlighted by the 29 scatter plots provided in appendix. The danger zones were very accurately defined and the structure of the (MDD vs. Indicator Value) scatter plots is for most indicators very regular, both in the in-sample and out-of-sample periods. Moreover the small number of assets inside Dataset-CAC40, resulting in a small rolling window, allowed the indicators the have many usable calibration points inside the calibration period (the red dots in the scatter plots), which boosted the accuracy of the determination of the danger zones even more for all the 29 financial crisis indicators. In Figure 5a, we see that the Sharpe of $PA$ is 0.40, while the Sharpe of $PP$ is only 0.19. While $PA$ has an annualized volatility similar to the one of $PP$, the overall performance of $PA$ jumps to 11\%, while it was only 6\% for $PP$. Even though $PA$ and $PA$ have a similar MDD (the MDD of $PA$ is slightly better: 0.31 against 0.32), the Calmar ratio of $PA$ (0.37) is much better than the Calmar ratio of $PP$ (0.20). We also notice by examining $IR$ that the active strategy governing $PA$ takes the right decisions at the right time. It does not sell the ETF shares often, but when it does it is in anticipation of crises and drop in ETF prices that did happen, especially in 2011 and late 2015. It is also fully invested ($TR=1$) at the right times of market growth, which boosted the performance of $PA$. This is confirmed by Figure 5c, which shows the spikes in the value of $\Gamma$ happening at the right times. When we switch our attention to Figure 5b which shows $PA$ and 50,000 paths of $PR$, the success of our active systematic trading strategy is very impressive. Because the active strategy sold the shares in $PA$ just at the right moment in anticipation of a significant drop in the CAC40 index in late 2011, probably a consequence of the  the European Debt Crisis, the value of the active portfolio soars above the random paths. $PA$ beats 100\% of the $PR$ paths in terms of Sharpe ratio, and the average of the Sharpe for the random paths is only 0.19. $PA$ is also less volatile than all the random paths, even though the volatility of $PA$ at 0.26 is close to the average volatility of $PR$, which is 0.28. $PA$ performs better than all the random paths in terms of overall performance and in terms of Calmar ratio. Even when we examine the MDD, which is usually the weak point of $PA$ because it may contain no cash at all, while $PR$ does often keep some (we recall that $PP$ always keep its cash), the success of $PA$ is striking. Indeed, $PA$ has an MDD of 0.31 while the average of the random paths' MDD is 0.33 and it beats the $PR$ paths in terms of MDD 99.9\% of the time.\\

The results obtained for Dataset-SP500 ($\mathscr{T}=20$ , $\mathscr{S}=75$) are presented in Figure 6a, Figure 6b and Figure 6c. For this dataset, the rolling window was made of 462 days and  the calibration period was therefore 512 days, according to Formula (10). The active trading strategy governing $PA$ issued 934 buy orders, 748 stay orders and 162 sell orders. The calibration of the financial crisis indicators was generally good, as we can see with the graphs in appendix for the SP500 index and the indicators of both the $\alpha$-series and the $\beta$-series. Therefore, the forecasts of the indicators were often accurate, which resulted in the right decisions taken at the right times by the active strategy. Since there were few large drops in the value of the SP500 during our out-of-sample prediction period, there were few sell orders given by the strategy and the $IR$ graph on Figure 6a shows that we are most of time fully invested (i.e no cash left in $PA$). In the rare instances when the strategy did decide the sell the ETF shares, those decisions were taken at the right time, as seen in the graph of $\Gamma$ in Figure 6c as well, in anticipation of significant drops in the index, which is good and underlines the predictive power of the aggregated signal produced by our 29 financial crisis indicators and the added value of the systematic trading strategy based on them. The performance of $PA$ is 19.3\% and beats the performance of $PP$ which is only 16.6\%.  $PA$ is slightly less volatile than $PP$, which again is a good result and the Sharpe ratio of $PA$ reaches 0.90 while $PP$ had only 0.72 for a Sharpe ratio. As we have seen with the previous indices, in the case of the SP500 too $PA$ has a slightly higher MDD than $PP$. This is due to the fact that $PA$ is most of the time fully invested, while $PP$ still holds its cash. As a result, the Calmar of $PA$ is similar and only marginally higher than the Calmar of $PP$. Switching our attention to Figure 6b, we can compare $PA$ to 50,000 paths of $PR$. $PA$ beats the random paths in terms of Sharpe ratio 100\% of the time, is less volatile than all the random paths as well and achieves a better performance 99.97\% of the time. Those are very reassuring results. In terms of MDD, as we expected, $PA$ tends to be a little disappointing and beats the random paths of $PR$ only 22\% of the time. However, since $PA$ performs overall so much better than the paths of $PR$, it still manages to beat the random paths 97.18\% of the time in terms of Calmar, which is a remarkable result.\\

We present the results for Dataset-NASDAQ ($\mathscr{T}=15$ , $\mathscr{S}=80$)  in Figure 7a, Figure 7b and Figure 7c. For this dataset, the rolling window was made of 76 days and the calibration period was therefore 500 days, according to Formula (10). The active trading strategy governing $PA$ issued 1533 buy orders, 287 stay orders and 36 sell orders. The calibration of the 29 financial crisis indicators was mostly good, as we can see in appendix for the NASDAQ, with most scatter plots (MDD vs. Indicator Value) featuring the usual bell-shape. Besides, the structure of the red dots (in-sample) and blue-dots (out-of-sample) is similar, which validates the possibility of forecasting the future by studying the past in the case of the Dataset-NASDAQ. The NASDAQ index was mostly in a pattern of growth from 2009 to 2016, without any large scale market downturn, therefore our active investment strategy has issued mostly buy and stay orders and very few sell orders, mostly in the first half of 2009, as we can see on Figure 7a and Figure 7c. The consequence of this is that $PA$ and $PP$ perform in a similar fashion, with $PA$ still keeping a clear advantage, because the few sell orders were issued at an appropriate time. Those few sell orders issued at the right time did not however have a massive positive impact on the quality of the results provided by $PA$, like it was the case when considering the CAC40 index. It is a little disappointing that the active strategy governing $PA$ did not issue sell orders in late 2015 and early 2016, when a possible larger drop in the value of the NASDAQ becomes a possibility, but it did not change much the fact that $PA$ and $PP$ perform mostly in a similar fashion because the out-of-sample forecasting period that we chose did not feature major market events that we could have accurately predicted and acted upon. The overall performance of $PA$ and $PP$ is very close, with a slight advantage for $PA$, essentially explained by the fact that $PA$ was fully invested ($IR=1$) for most of the out-of-sample period, while $PP$ was static and kept its cash, which earned only the Libor. $PA$ and $PP$ have almost identical volatility as well, which explains that they also have similar Sharpe ratios: 0.88 for $PA$ and 0.84 for $PP$. $PP$ has a slightly lower MDD than $PA$ because of the cash it contains which acts as a stabilizer, and that explains the slight advantage in terms of MDD that $PP$ holds over $PA$, while $PA$ holds a slight performance advantage as we have explained. The study of the random paths of $PR$ and their comparison to $PA$ confirms our analysis that $PA$ did not get a chance, for this index over this time period, to predict many crises because there were indeed few to predict. The active strategy still manages to do better than the random paths in a significant and reproducible manner, proving the validity of our approach. Indeed, $PA$ beats the random paths 99.9\% of the time in terms of Sharpe ratio, has an overall performance better than the $PR$ paths in 82\% of the cases and achieves a better Calmar ratio than the $PR$ paths more than $66\%$ of the time.\\

Finally, the results obtained for Dataset-SHSZ300 ($\mathscr{T}=10$ , $\mathscr{S}=70$) are presented in Figure 8a, Figure 8b and Figure 8c. For this dataset, the rolling window was made of 162 days and the calibration period was therefore 500 days. The active trading strategy governing $PA$ issued 1063 buy orders, 484 stay orders and 228 sell orders. Even though $PA$ manages to produce fairly good results over the whole period of the out-of-sample forecasting period, the study of the graphs in Figure 8a and the graph of $\Gamma$ in Figure 8c reveals that the quality of the financial crisis forecasts provided by our 29 financial crisis indicators can probably be made better in the case of the SHSZ300 index. Indeed, the large drop in the value of the index in 2015 is not anticipated and the times when $IR=0$ are not always perfectly well in sync with the times when the SHSZ300 index experiences a drop in value. This relatively poor performance of our financial crisis indicators when applied to the Chinese index, in comparison to their excellent behavior when applied to the other four equity indices of this study, takes its roots in the poor quality of the calibration of the indicators, as we can see in the appendix where the scatter plots (MDD vs. Indicator value) are presented for all our 29 financial crisis indicators of both the $\alpha$-series and the $\beta$-series when applied to the SHSZ300 index. Indeed, many of those scatter plots do not feature the usual bell shape, as we have discussed before, and the structure of the in-sample graph (red dots) is not stable and preserved into the out-of-sample period (blue dots), thus rendering the calibration useless and the determination of the danger zone of many of the indicators meaningless for forecasting purposes. Fortunately, some of our financial crisis indicators are properly calibrated, but many are not and extreme cases of deformities in the scatter plots for our financial crisis indicators applied to the SHSZ300 index can be found, for example, in the  $\alpha$-series, graph (n) (reference : $\mathscr{R}_{3}$; matrix: correlation weighted by market capitalization) or in the $\alpha$-series, graph (h) (trace of the correlation matrix weighted by market capitalization). In those instances the scatter plot is almost bi-modal, with the high values of MDD in the in-sample calibration period corresponding to completely different values of the indicator than in the out-of-sample forecasting period. Of course, when there is little stability for a given index in the behavior of our indicators over time, then accurately forecasting the future by looking at the past becomes an impossibility. Regardless of the difficulties encountered in the calibration of several of our financial crisis indicators, $PA$ in the case of the SHSZ300 index still manages to produce fairly good results. That is the advantage of using the aggregated signal coming from our 29 financial crisis indicators. Even though a large amount of them might in some cases fail to be calibrated properly, enough remain useful to provide the investor holding $PA$ with enough accurate previsions to beat $PP$ and most the random paths $PR$. Indeed the Sharpe of $PA$ still reaches 0.49 while $PP$ only had a Sharpe of 0.29, $PA$ is slightly less volatile than $PP$ and the performance of $PA$ (15\%) beats the performance of $PP$ (10\%). While the MDD of $PA$ is marginally higher than the MDD of $PP$, the Calmar ratio of $PA$ (0.32) still shows a significant improvement with respect to the Calmar ratio of $PP$ (0.23). When considering random paths of $PR$, as shown in Figure 8b, the results that we obtain are very reassuring and attest to the resilience of our framework that is still able to produce useful active trading strategies, even when many of the financial crisis indicators upon which they rely are giving flawed signals. Indeed, $PA$ beats the $PR$ paths 99.9\% of the time in terms of Sharpe ratio and volatility and 99.5\% of the time in terms of performance. The MDD of the random paths are usually better than the MDD of $PA$, but the advantage that $PA$ hold over the 50,000 random paths in terms of performance still permit our active strategy to beat the $PR$ paths 97.2\% of the time in terms of Calmar ratio.

\begin{landscape}
\begin{figure}[p!]
\centering
\includegraphics[scale=0.59]{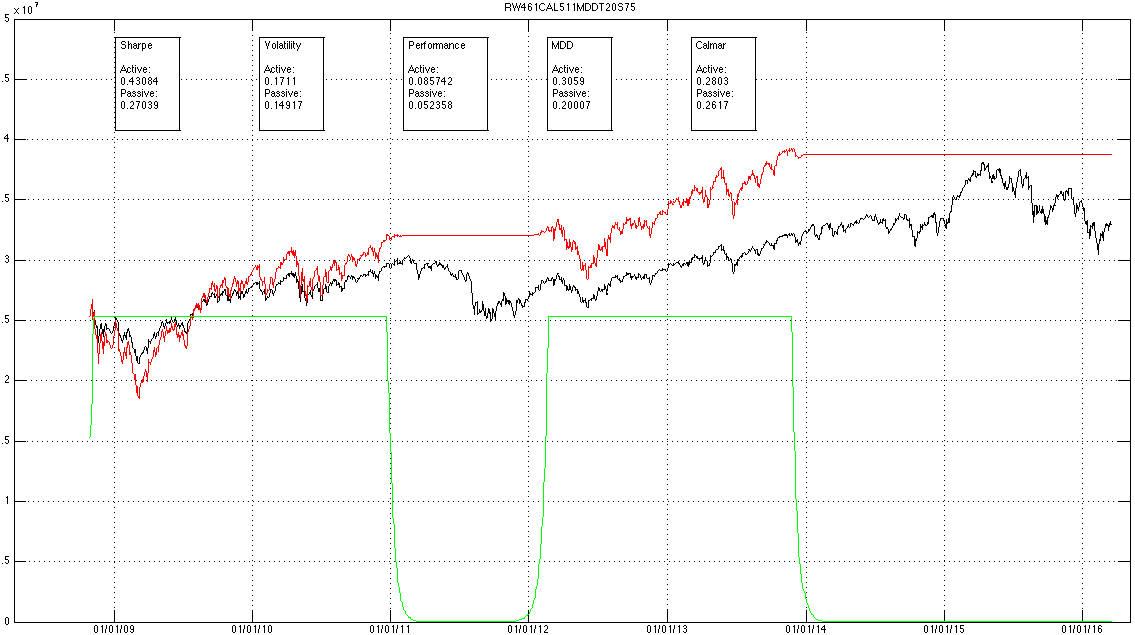}
\captionsetup{labelformat=empty}
\caption{Figure 4a: Dataset-BE500. Red: PA; Black: PP; Green: IR}
\end{figure}
\end{landscape}

\begin{landscape}
\begin{figure}[p!]
\centering
\includegraphics[scale=0.59]{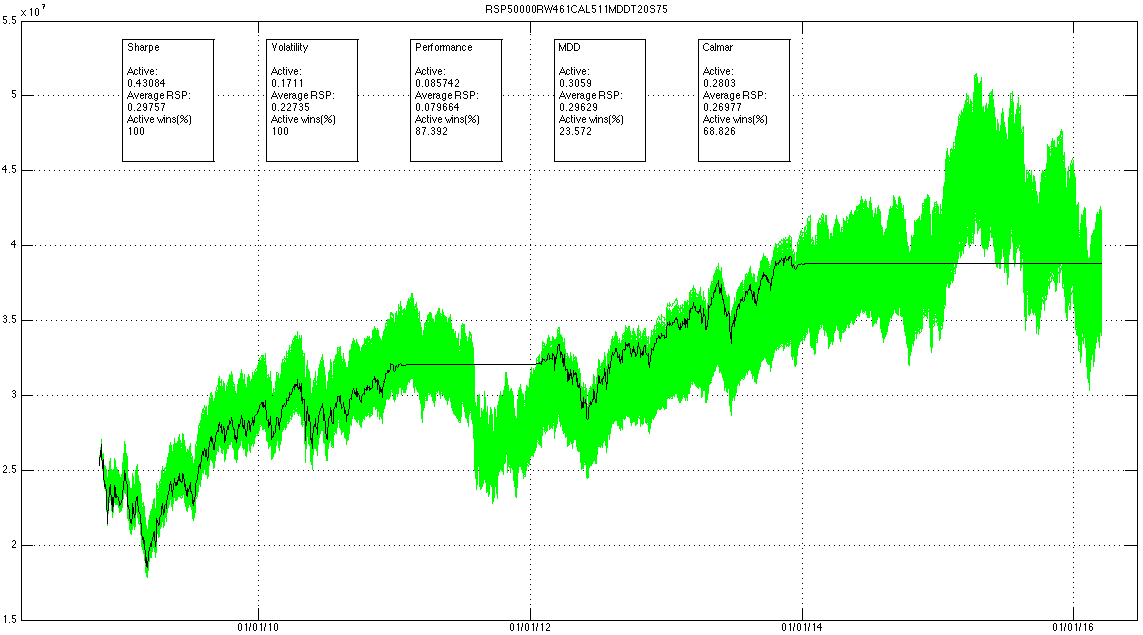}
\captionsetup{labelformat=empty}
\caption{Figure 4b: Dataset-BE500. Black: PA; Green: 50,000 paths of PR}
\end{figure}
\end{landscape}

\begin{landscape}
\begin{figure}[p!]
\centering
\includegraphics[scale=0.59]{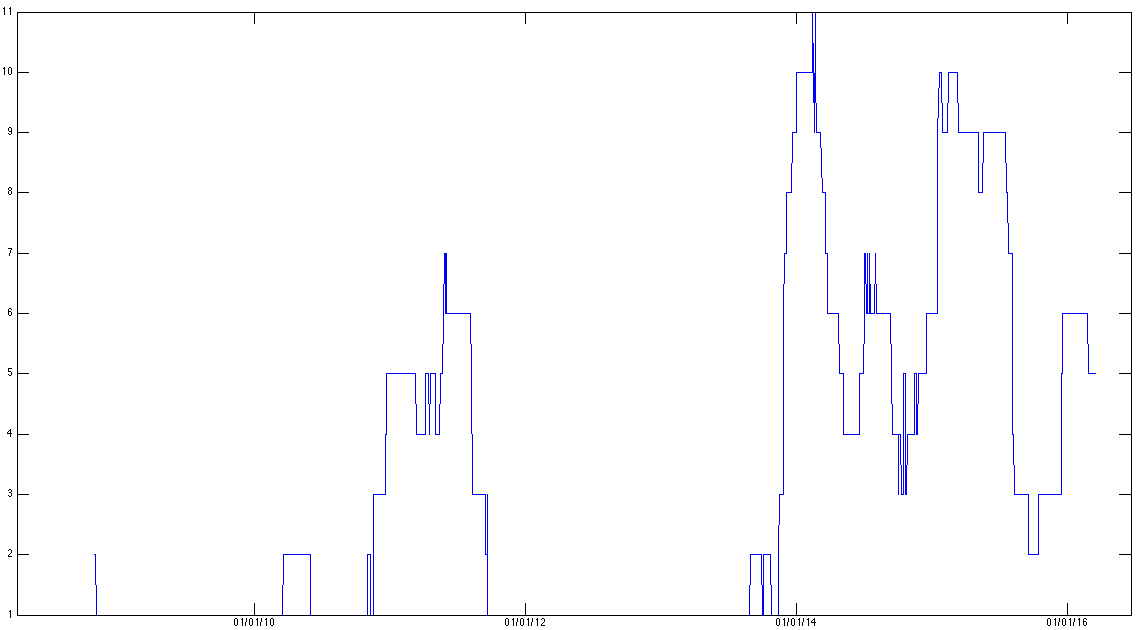}
\captionsetup{labelformat=empty}
\caption{Figure 4c: Graph of $\Gamma$ for PA on Dataset-BE500}
\end{figure}
\end{landscape}

\begin{landscape}
\begin{figure}[p!]
\centering
\includegraphics[scale=0.59]{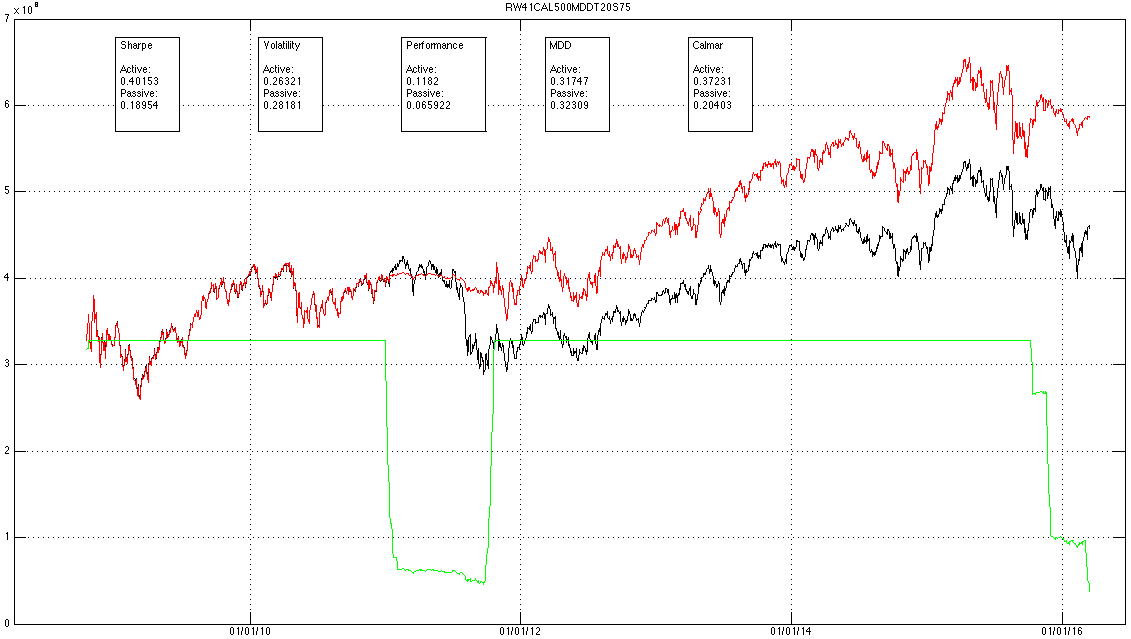}
\captionsetup{labelformat=empty}
\caption{Figure 5a: Dataset-CAC40. Red: PA; Black: PP; Green: IR}
\end{figure}
\end{landscape}

\begin{landscape}
\begin{figure}[p!]
\centering
\includegraphics[scale=0.59]{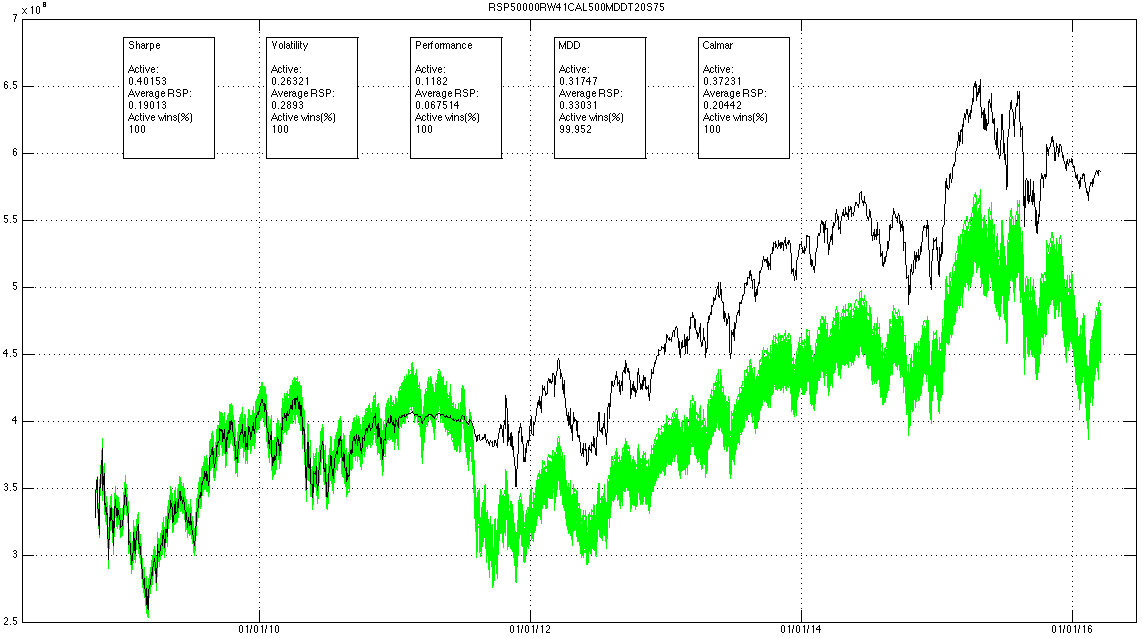}
\captionsetup{labelformat=empty}
\caption{Figure 5b: Dataset-CAC40. Black: PA; Green: 50,000 paths of PR}
\end{figure}
\end{landscape}

\begin{landscape}
\begin{figure}[p!]
\centering
\includegraphics[scale=0.59]{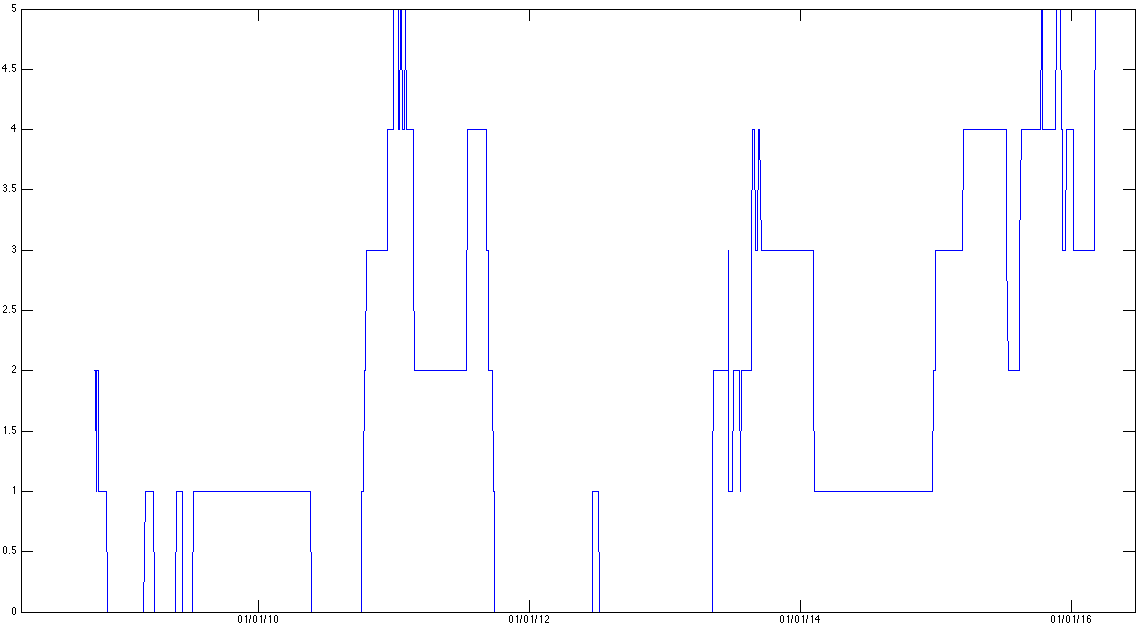}
\captionsetup{labelformat=empty}
\caption{Figure 5c: Graph of $\Gamma$ for PA on Dataset-CAC40}
\end{figure}
\end{landscape}

\begin{landscape}
\begin{figure}[p!]
\centering
\includegraphics[scale=0.59]{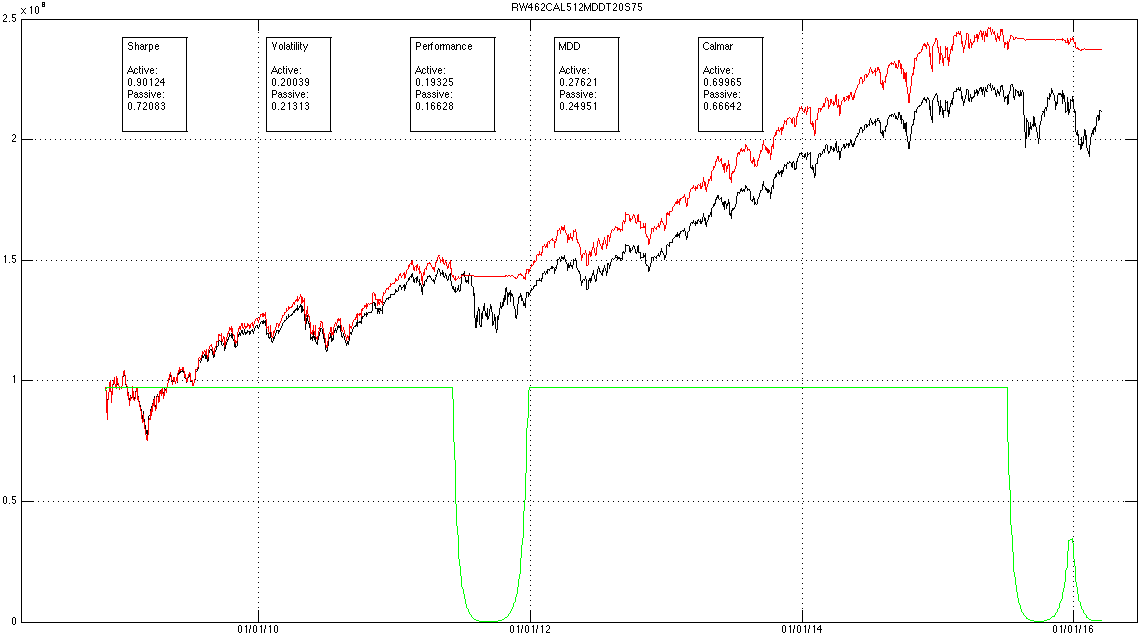}
\captionsetup{labelformat=empty}
\caption{Figure 6a: Dataset-SP500. Red: PA; Black: PP; Green: IR}
\end{figure}
\end{landscape}

\begin{landscape}
\begin{figure}[p!]
\centering
\includegraphics[scale=0.59]{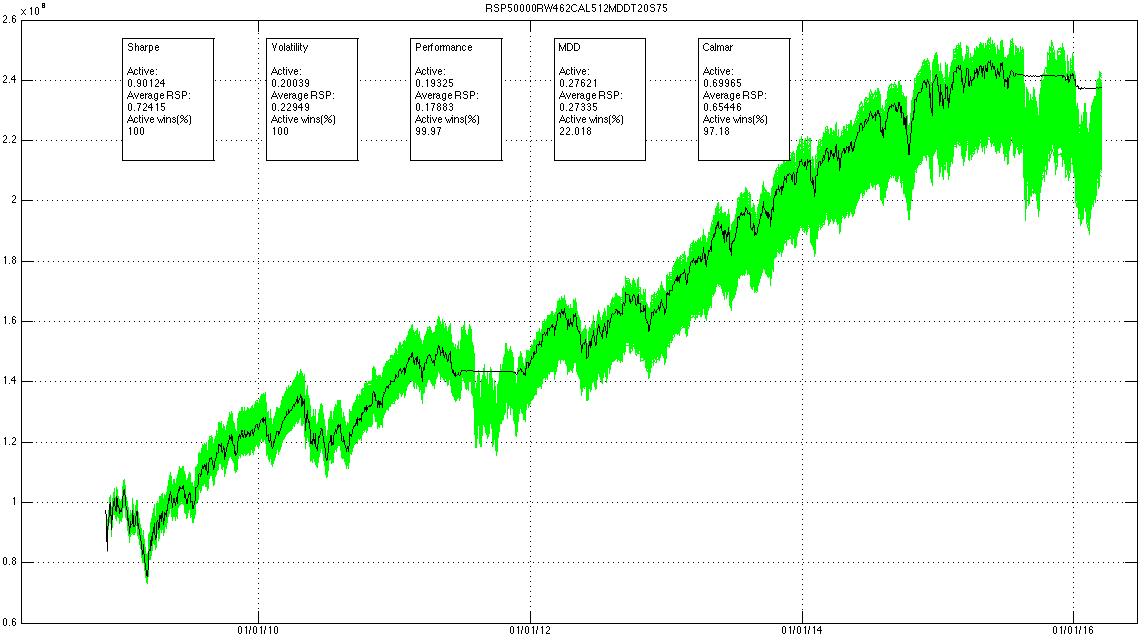}
\captionsetup{labelformat=empty}
\caption{Figure 6b: Dataset-SP500. Black: PA; Green: 50,000 paths of PR}
\end{figure}
\end{landscape}

\begin{landscape}
\begin{figure}[p!]
\centering
\includegraphics[scale=0.59]{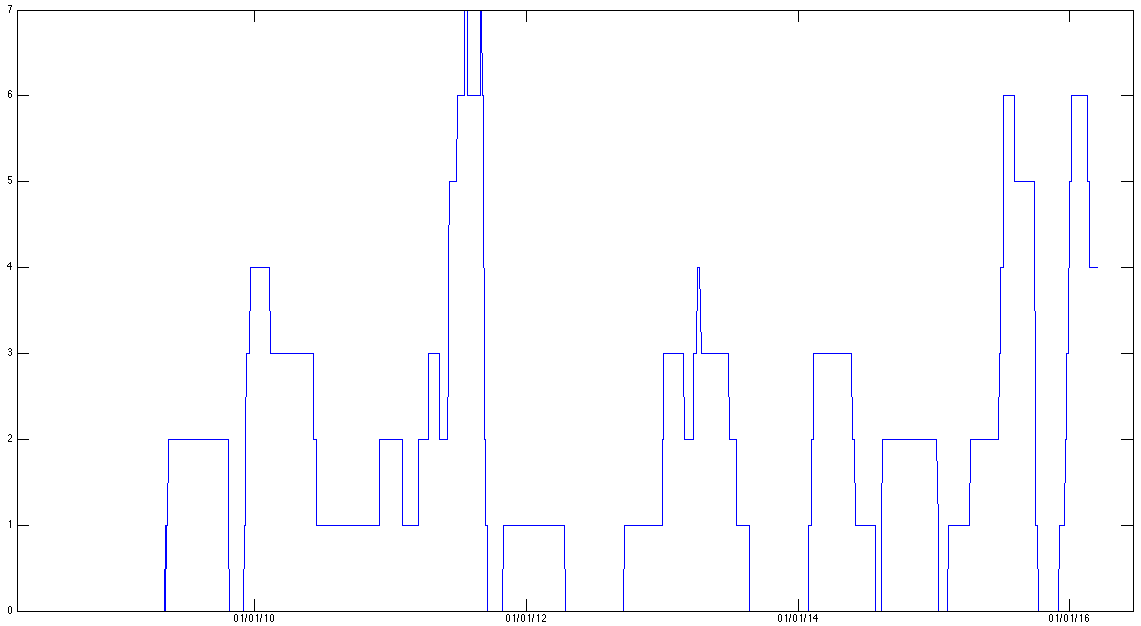}
\captionsetup{labelformat=empty}
\caption{Figure 6c: Graph of $\Gamma$ for PA on Dataset-SP500}
\end{figure}
\end{landscape}

\begin{landscape}
\begin{figure}[p!]
\centering
\includegraphics[scale=0.59]{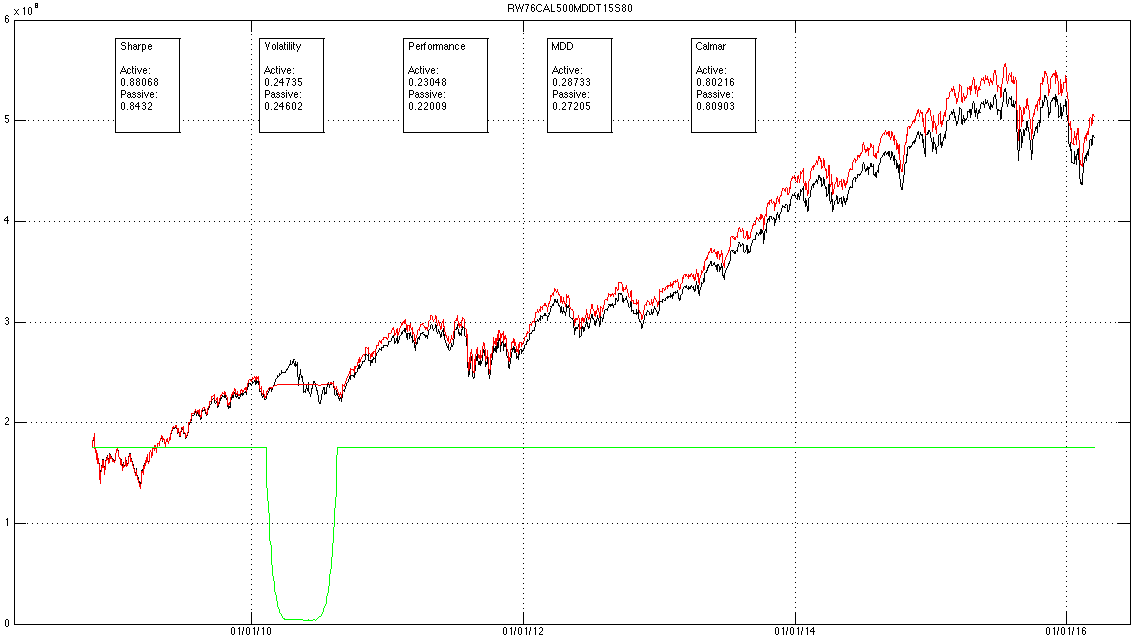}
\captionsetup{labelformat=empty}
\caption{Figure 7a: Dataset-NASDAQ. Red: PA; Black: PP; Green: IR}
\end{figure}
\end{landscape}

\begin{landscape}
\begin{figure}[p!]
\centering
\includegraphics[scale=0.59]{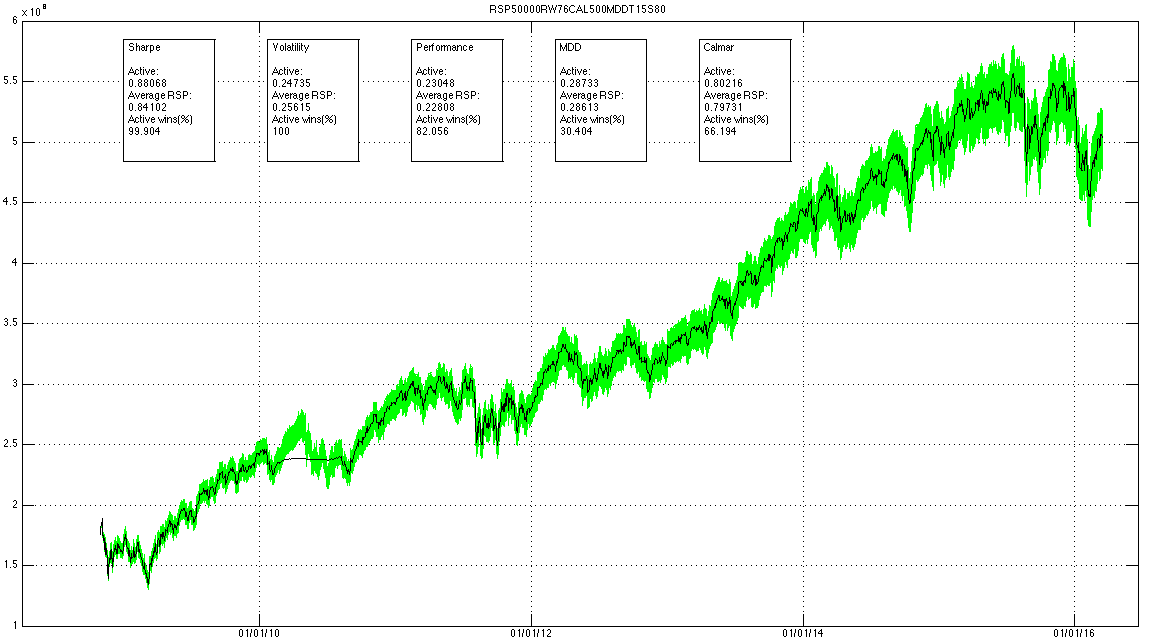}
\captionsetup{labelformat=empty}
\caption{Figure 7b: Dataset-NASDAQ. Black: PA; Green: 50,000 paths of PR}
\end{figure}
\end{landscape}
\begin{landscape}
\begin{figure}[p!]
\centering
\includegraphics[scale=0.59]{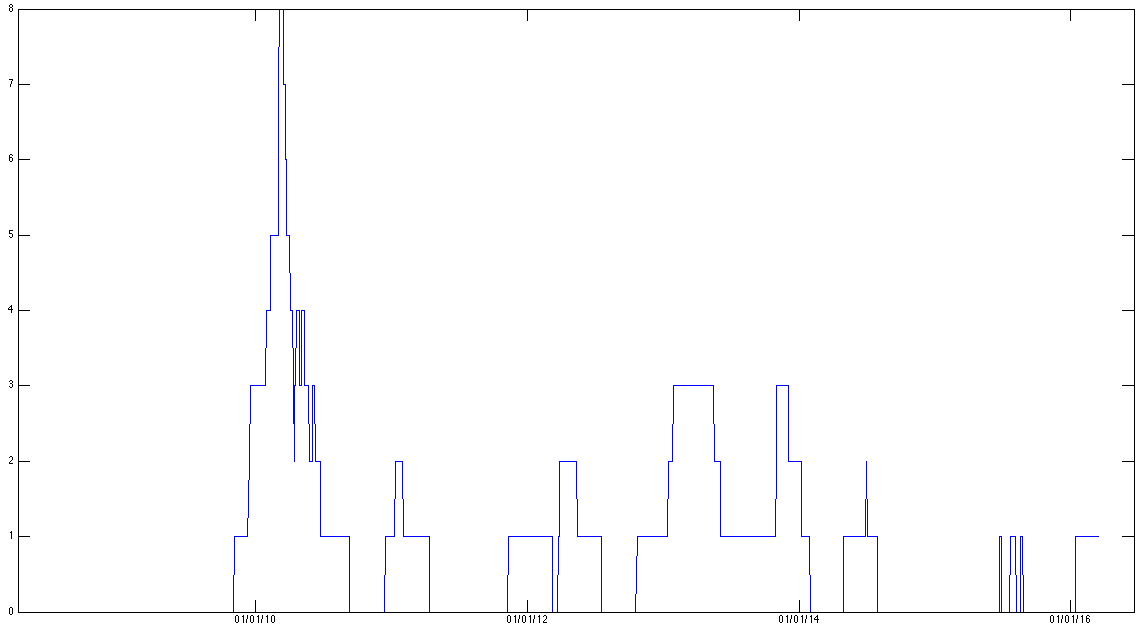}
\captionsetup{labelformat=empty}
\caption{Figure 7c: Graph of $\Gamma$ for PA on Dataset-NASDAQ}
\end{figure}
\end{landscape}

\begin{landscape}
\begin{figure}[p!]
\centering
\includegraphics[scale=0.59]{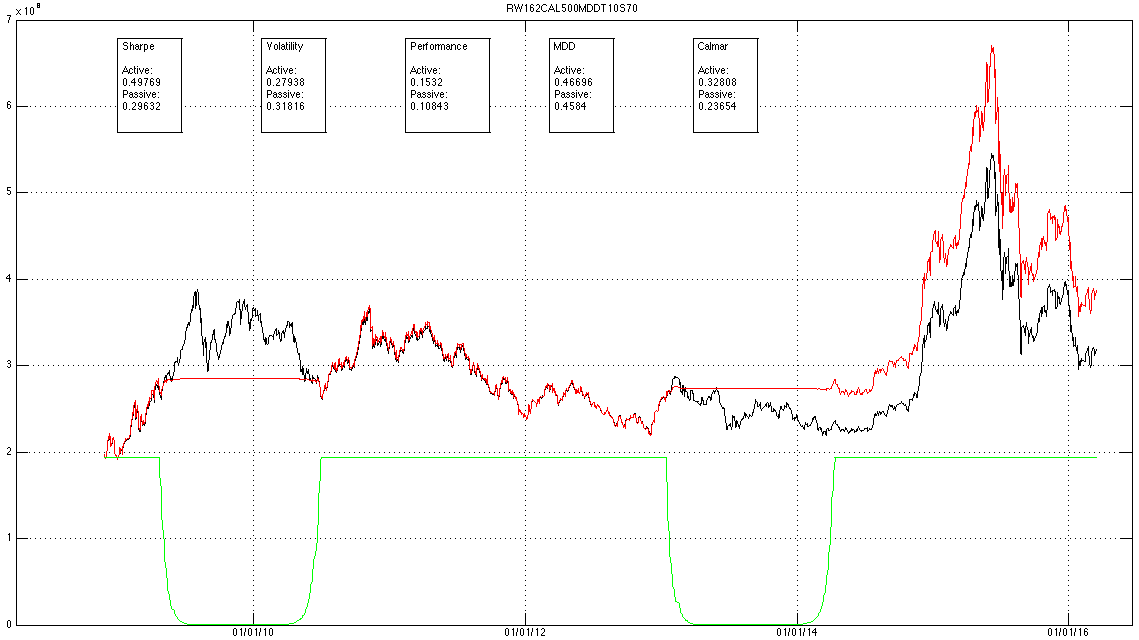}
\captionsetup{labelformat=empty}
\caption{Figure 8a: Dataset-SHSZ300. Red: PA; Black: PP; Green: IR}
\end{figure}
\end{landscape}

\begin{landscape}
\begin{figure}[p!]
\centering
\includegraphics[scale=0.59]{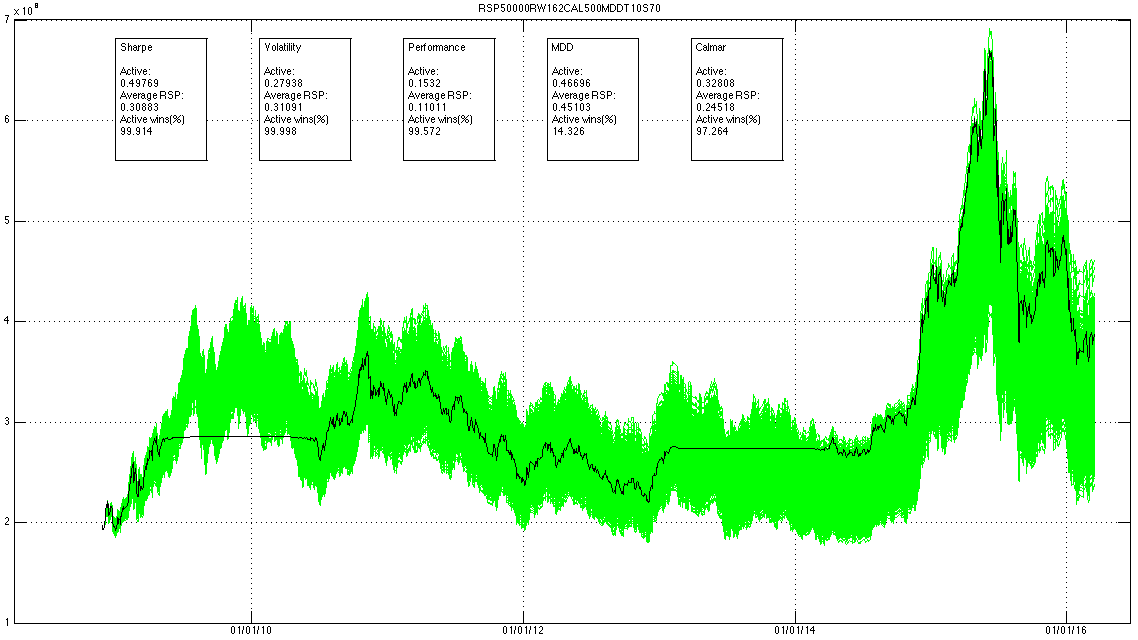}
\captionsetup{labelformat=empty}
\caption{Figure 8b: Dataset-SHSZ300. Black: PA; Green: 50,000 paths of PR}
\end{figure}
\end{landscape}

\begin{landscape}
\begin{figure}[p!]
\centering
\includegraphics[scale=0.59]{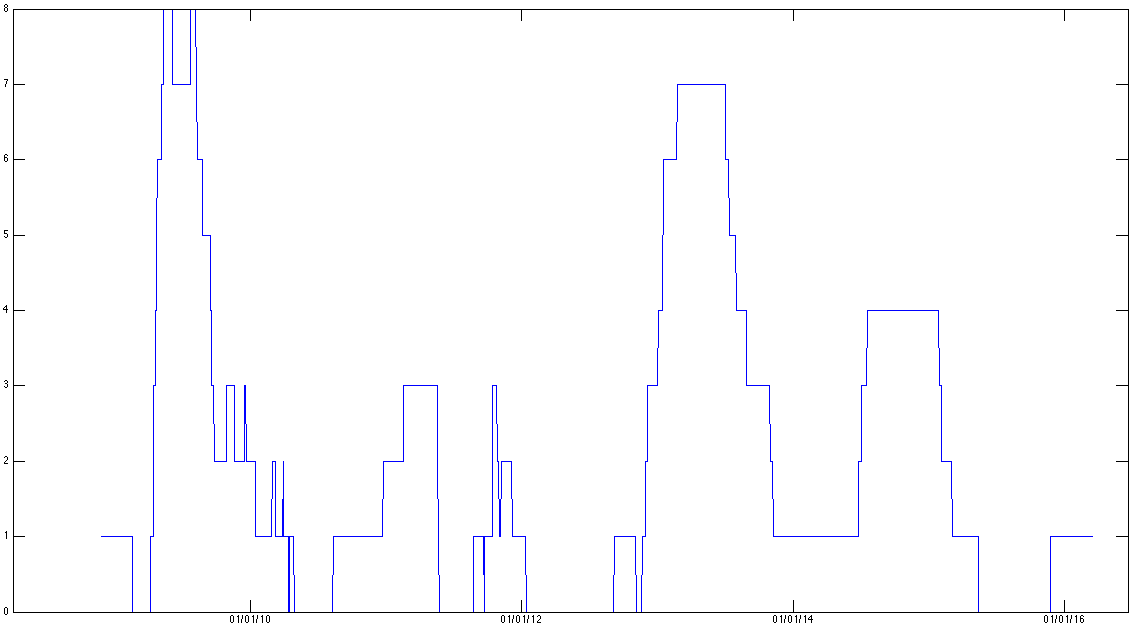}
\captionsetup{labelformat=empty}
\caption{Figure 8c: Graph of $\Gamma$ for PA on Dataset-SHSZ300}
\end{figure}
\end{landscape}

 \section{Conclusion}
 
As a conclusion, we would like first to underline the excellent results provided by our systematic trading strategies based on the forecasting power of our 29 financial crisis indicators, both of the $\alpha$-series and of the $\beta$-series. For every one of the five datasets that we have built by considering a major equity index and its respective stock components (BE500, CAC40, SP500, NASDAQ, SHSZ300), our systematic trading strategies are always able to beat in a clear and reproducible fashion a passive buy-and-hold strategy. Our active systematic investment strategy is also able to beat random strategies, which feature the same proportion of 'buy', 'sell' and 'stay' orders, in an overwhelming majority of the cases.\\

Indeed, for the equity indices that we have considered in this study, the active portfolio $PA$ beats the passive portfolio $PP$ and the random paths $PR$ in terms of Sharpe ratio, performance, volatility and Calmar ratio. Only in terms of maximum-draw-dawn do the passive and random strategies sometimes give better results than the active one but this is only because, by design in this study,  $PP$ always contains some cash and a $PR$ path usually contains cash as well, while $PA$ may be fully invested most of the time. Only in the case of the Chinese SHSZ300 index, did some of our financial crisis indicators provide flawed predictions, regardless of the skills deployed by the operator in choosing the right parameters $\mathscr{T}$ and $\mathscr{S}$. That may have been due to the poor quality of the historical data on the Chinese market or the fact, which definitely introduced survivorship bias in our study,  that we had to consider less than half of the current 300 components because of the profound transformations of the Chinese market over the last 10 years.\\

To summarize our method, we start by establishing a simple rule for the length of the rolling window (Formula (1)) for all of our five datasets and we proceed to establishing another simple rule for the length of the calibration period (Formula (10)). We then choose two parameters that govern the behavior of a systematic trading strategy. The first parameter is the MDD Threshold $\mathscr{T}$, the value of which determines whether we wish to bet on accurately forecasting a large number of small crises or a small number of large crises. The second parameter, the Indicator Sensitivity $\mathscr{S}$ is then chosen and its value determines the level of aggressiveness of a systematic trading strategy. Low value of $\mathscr{S}$ will produce very aggressive strategies that will start converting the ETF shares into cash inside the active portfolio at the first sign of danger, because the red flags provided by the indicators will be easier to obtain. Higher values of $\mathscr{S}$ will make a systematic trading strategy more calm and patient because the red flags provided by the financial crisis indicators will be harder to achieve and the strategy will therefore wait to take action and start converting the shares into cash until the risk of a crisis happening within the 100 days forecasting horizon of our financial crisis indicators becomes impossible to ignore.\\

The choice of the two parameters  $\mathscr{T}$ and  $\mathscr{S}$ is robust and once an operator has chosen a value for $\mathscr{T}$ and $\mathscr{S}$, using his or her experience and knowledge of the equity index being considered, then those parameters may be used for similar equity indices over large periods of time, excluding any possibility of over-fitting our model. In other words, it is the skill of the operator setting-up those strategies, not luck, which is the determining factor that makes the difference between a winning and a failing active trading strategy.\\
 
Future developments of this work could include designing a real-time system of ratings in order to give more weight in the decision and the computation of $\Gamma$ at a given time to the indicators, among the 29 we have built, that have had the most accurate forecasts in a given past period. Indeed, in our work in its present form, all those expert's opinions carry the same weight, regardless of the past accuracy of their prediction and their respective  proportions of false positive (more rarely false negative) errors regarding financial crisis prediction. In a future work, one could envision giving all those indicators a rating on a given scale and then modulate in a strategy's decision process the importance of each indicator with respect to its rating. We also plan to incorporate transaction costs and market frictions in our study. Indeed, those transaction costs are especially important for smaller transactions and providing a proper modeling of their influence is important to give our trading strategies a better chance of being useful regardless of the size of the portfolio that they are being applied to. Concerning the question of scalability again, we also plan to take into account the impact that the investors have on the market when they execute sell or buy orders. Even though they are price takers and not market makers, their actions do have a small influence on the order book and may in particular create slippage, no matter how small the orders are in comparison to the size of the market. That effect will in the future be integrated to our approach in order to make it fully scalable.

\appendices
\begin{small}
\section*{Dataset-SP500}
A UN ;								AA UN ;								AAP UN ;								AAPL UW ;								ABC UN ;								ABT UN ;								ACN UN ;								ADBE UW ;								ADM UN ;								ADS UN ;								ADSK UW ;								AEE UN ;								AEP UN ;								AES UN ;								AET UN ;								AFL UN ;								AGN UN ;								AIG UN ;								AIV UN ;								AIZ UN ;								AKAM UW ;								ALL UN ;								ALXN UW ;								AMAT UW ;								AME UN ;								AMG UN ;	\\
							AMGN UW ;								AMP UN ;								AMT UN ;								AMZN UW ;								AN UN ;								ANTM UN ;								AON UN ;								APA UN ;		\\
													APC UN ;								APD UN ;								APH UN ;								ARG UN ;								ATVI UW ;								AVB UN ;								AVY UN ;								AXP UN ;								AZO UN ;								BA UN ;								BAC UN ;								BAX UN ;								BBBY UW ;								BBT UN ;								BBY UN ;								BCR UN ;								BDX UN ;								BEN UN ;								BF/B UN ;								BHI UN ;								BIIB UW ;								BK UN ;								BLK UN ;								BLL UN ;								BMY UN ;								BRK/B UN ;								BSX UN ;								BWA UN ;								BXP UN ;								C UN ;								CAG UN ;								CAH UN ;								CAM UN ;								CAT UN ;								CB UN ;								CBG UN ;	\\
																				CBS UN ;								CCE UN ;								CCI UN ;								CCL UN ;								CELG UW ;								CERN UW ;								CF UN ;								CHD UN ;								CHK UN ;								CHRW UW ;								CI UN ;								CINF UW ;								CL UN ;								CLX UN ;								CMA UN ;								CMCSA UW ;								CMG UN ;	\\
																											CMI UN ;								CMS UN ;								CNP UN ;								COF UN ;								COG UN ;								COH UN ;								COL UN ;								COP UN ;								COST UW ;								CPB UN ;								CRM UN ;								CSCO UW ;								CTAS UW ;								CTL UN ;								CTSH UW ;								CTXS UW ;								CVC UN ;		\\
																																	CVS UN ;								CVX UN ;								D UN ;								DD UN ;								DE UN ;								DGX UN ;								DHI UN ;								DHR UN ;								DIS UN ;\\
																																									DISCA UW ;								DLTR UW ;								DNB UN ;								DO UN ;								DOV UN ;								DOW UN ;								DRI UN ;								DTE UN ;		\\
																																															DUK UN ;								DVA UN ;								DVN UN ;								EA UW ;								EBAY UW ;								ECL UN ;								ED UN ;								EFX UN ;								EIX UN ;	\\
																																																						EL UN ;								EMC UN ;								EMN UN ;								EMR UN ;								ENDP UW ;								EOG UN ;								EQIX UW ;								EQR UN ;			\\
																																																											EQT UN ;								ES UN ;								ESRX UW ;								ESS UN ;								ESV UN ;								ETN UN ;								ETR UN ;								EW UN ;								EXC UN ;	\\
																																																																		EXPD UW ;								EXPE UW ;								EXR UN ;								F UN ;								FAST UW ;								FCX UN ;								FDX UN ;								FE UN ;		\\
																																																																								FFIV UW ;								FIS UN ;								FISV UW ;								FITB UW ;								FLIR UW ;								FLR UN ;								FLS UN ;								FMC UN ;								FRT UN ;								FTI UN ;								GAS UN ;								GD UN ;								GE UN ;								GILD UW ;								GIS UN ;								GLW UN ;								GME UN ;								GOOGL UW ;								GPC UN ;								GPS UN ;								GRMN UW ;								GS UN ;								GWW UN ;								HAL UN ;								HAR UN ;								HBAN UW ;	\\
																																																																															HCN UN ;
																																																																															HCP UN ;								HD UN ;								HES UN ;								HIG UN ;								HOG UN ;								HON UN ;								HOT UN ;								HP UN ;		\\
																																																																																					HPQ UN ;								HRB UN ;								HRL UN ;								HRS UN ;								HSIC UW ;								HST UN ;								HSY UN ;								HUM UN ;								IBM UN ;								ICE UN ;								IFF UN ;								ILMN UW ;								INTC UW ;								INTU UW ;								IP UN ;								IPG UN ;								IR UN ;								IRM UN ;\\
																																																																																													ISRG UW ;								ITW UN ;								IVZ UN ;								JBHT UW ;								JCI UN ;								JEC UN ;								JNJ UN ;								JPM UN ;								JWN UN ;	\\
																																																																																																				K UN ;								KEY UN ;								KIM UN ;								KLAC UW ;								KMB UN ;								KMX UN ;								KO UN ;								KR UN ;								KSS UN ;				\\
																																																																																																								KSU UN ;
																																																																																																												L UN ;								LB UN ;								LEG UN ;								LEN UN ;								LH UN ;								LLL UN ;								LLTC UW ;								LLY UN ;								LM UN ;								LMT UN ;								LNC UN ;								LOW UN ;								LRCX UW ;								LUK UN ;								LUV UN ;								M UN ;								MA UN ;								MAC UN ;		\\
																																																																																																																		MAS UN ;								MCD UN ;								MCHP UW ;								MCK UN ;								MCO UN ;								MDT UN ;								MET UN ;								MHFI UN ;		\\
																																																																																																																								MHK UN ;								MKC UN ;								MLM UN ;								MMC UN ;								MMM UN ;								MNST UW ;								MO UN ;								MON UN ;		\\
																																																																																																																														MOS UN ;								MRK UN ;								MRO UN ;								MS UN ;								MSFT UW ;								MSI UN ;								MTB UN ;								MUR UN ;								NBL UN ;								NDAQ UW ;								NEE UN ;								NEM UN ;								NFLX UW ;								NFX UN ;								NI UN ;								NKE UN ;								NOC UN ;								NOV UN ;								NRG UN ;								NSC UN ;								NTAP UW ;								NTRS UW ;								NUE UN ;								NVDA UW ;								NWL UN ;								O UN ;								OI UN ;								OKE UN ;								OMC UN ;								ORLY UW ;								OXY UN ;								PAYX UW ;								PBCT UW ;								PBI UN ;								PCAR UW ;	\\
																																																																																																																																					PCG UN ;								PCLN UW ;								PDCO UW ;								PEG UN ;								PEP UN ;								PFE UN ;								PFG UN ;								PG UN ;	\\
																																																																																																																																												PGR UN ;								PH UN ;								PHM UN ;								PKI UN ;								PLD UN ;								PNC UN ;								PNR UN ;								PNW UN ;								POM UN ;	\\
																																																																																																																																																			PPG UN ;								PPL UN ;								PRU UN ;								PSA UN ;								PVH UN ;								PWR UN ;								PX UN ;								PXD UN ;	\\
																																																																																																																																																										QCOM UW ;								R UN ;								RAI UN ;								RCL UN ;								REGN UW ;								RF UN ;								RHI UN ;								RIG UN ;								RL UN ;	\\
																																																																																																																																																																	ROK UN ;								ROP UN ;								ROST UW ;								RRC UN ;								RSG UN ;								RTN UN ;								SBUX UW ;								SCG UN ;\\
																																																																																																																																																																									SEE UN ;								SHW UN ;								SIG UN ;								SJM UN ;								SLB UN ;								SLG UN ;								SNA UN ;								SNDK UW ;								SO UN ;\\
																																																																																																																																																																																	SPG UN ;								SPLS UW ;								SRCL UW ;								SRE UN ;								STI UN ;								STJ UN ;								STT UN ;								STZ UN ;								SWK UN ;								SWKS UW ;								SWN UN ;								SYK UN ;								SYMC UW ;								SYY UN ;								T UN ;								TAP UN ;								TE UN ;								TGNA UN ;								TGT UN ;								THC UN ;								TIF UN ;								TJX UN ;								TMK UN ;								TMO UN ;								TROW UW ;								TRV UN ;	\\
																																																																																																																																																																																								TSCO UW ;								TSN UN ;								TSO UN ;								TSS UN ;								TWX UN ;								TXT UN ;								TYC UN ;								UDR UN ;								UHS UN ;								UNH UN ;								UNM UN ;								UNP UN ;								UPS UN ;								URBN UW ;								URI UN ;								USB UN ;								UTX UN ;								VAR UN ;								VFC UN ;								VLO UN ;								VMC UN ;								VNO UN ;								VRSN UW ;								VRTX UW ;								VTR UN ;								VZ UN ;								WAT UN ;								WEC UN ;								WFC UN ;								WFM UW ;								WHR UN ;								WM UN ;								WMB UN ;								WMT UN ;								WY UN ;\\
																																																																																																																																																																																																WYNN UW ;								XEC UN ;								XEL UN ;								XL UN ;								XLNX UW ;								XOM UN ;								XRAY UW ;								XRX UN ;	\\
																																																																																																																																																																																																							YHOO UW ;								YUM UN ;								ZBH UN ;								ZION UW

\section*{Dataset-BE500}
						A2A IM ;								AAL LN ;								AALB NA ;								ABBN VX ;								ABE SM ;								ABF LN ;								ABI BB ;								AC FP ;								ACA FP ;\\
														ACS SM ;								ADEN VX ;								ADM LN ;								ADN LN ;								ADS GR ;								AGK LN ;								AGN NA ;								AGS BB ;								AH NA ;	\\
																					AHT LN ;								AI FP ;								AIR FP ;								AKE FP ;								AKZA NA ;								ALFA SS ;								ALO FP ;								ALU FP ;								ALV GR ;	\\
																												AMEAS FH ;								ANA SM ;								ANDR AV ;								ARM LN ;								ASC LN ;								ASML NA ;								ASSAB SS ;								ATCOA SS ;\\
																																				ATL IM ;								ATLN VX ;								ATO FP ;								AV/ LN ;								AXFO SS ;								AZN LN ;								BA/ LN ;								BAB LN ;								BALDB SS ;	\\
																																			BALN VX ;								BARC LN ;								BAS GR ;								BATS LN ;								BAYN GR ;								BBVA SM ;								BDEV LN ;								BEI GR ;\\																											
																																				BG LN ;								BILL SS ;								BKG LN ;								BKIR ID ;								BKT SM ;								BLND LN ;								BMED IM ;								BMW GR ;								BN FP ;								BNP FP ;								BNZL LN ;								BOK LN ;								BOKA NA ;								BOL FP ;								BOL SS ;								BOSS GR ;								BP LN ;								BRBY LN ;								BT/A LN ;								BTG LN ;								BWY LN ;								CA FP ;								CAP FP ;								CARLB DC ;								CBK GR ;								CFR VX ;								CLN VX ;								CLS1 GR ;								CNA LN ;								CNP FP ;								CO FP ;								COB LN ;								COLOB DC ;								COLR BB ;								CON GR ;								CPG LN ;								CPI LN ;								CPR IM ;								CRDA LN ;								CRH ID ;								CS FP ;								CSGN VX ;								CWC LN ;								DAI GR ;	\\
																																											DANSKE DC ;								DB1 GR ;								DBK GR ;								DCC LN ;								DEC FP ;								DELB BB ;								DG FP ;								DGE LN ;								DLG IM ;								DLN LN ;								DMGT LN ;								DNB NO ;								DPW GR ;								DSM NA ;								DSV DC ;								DSY FP ;								DTE GR ;	\\
																																																		DUFN SW ;								DWNI GR ;								EBS AV ;								EDF FP ;								EDP PL ;								EI FP ;								ELE SM ;								ELI1V FH ;		\\
																																																								ELUXB SS ;								EMG LN ;								EN FP ;								ENEL IM ;								ENG SM ;								ENGI FP ;								ENI IM ;								EO FP ;								EOAN GR ;								ERICB SS ;								ETL FP ;								EZJ LN ;								FER SM ;								FGR FP ;								FME GR ;								FNC IM ;								FP FP ;								FR FP ;	\\
																																																															FRA GR ;								FRE GR ;								FUM1V FH ;								G IM ;								G1A GR ;								GAM SM ;								GAS SM ;								GBLB BB ;	\\
																																																																						GEBN VX ;								GEN DC ;								GETIB SS ;								GFS LN ;								GKN LN ;								GLB ID ;								GLE FP ;								GNK LN ;								GPOR LN ;								GSK LN ;								GTO NA ;								HAV FP ;								HEI GR ;								HEIA NA ;								HEIO NA ;								HER IM ;								HEXAB SS ;								HGG LN ;								HLMA LN ;								HMB SS ;								HMSO LN ;								HNR1 GR ;								HO FP ;								HOT GR ;								HSBA LN ;								HTO GA ;\\
																																																																														HUH1V FH ;								HUSQB SS ;								HWDN LN ;								IAG LN ;								IAP LN ;								IBE SM ;								ICA SS ;								IFX GR ;								IGG LN ;								IHG LN ;								III LN ;								IMB LN ;								IMI LN ;								INCH LN ;								INDUA SS ;								INF LN ;								ING FP ;								INTU LN ;	\\
																																																																																					INVEB SS ;								ISAT LN ;								ISP IM ;								IT IM ;								ITRK LN ;								ITV LN ;								ITX SM ;								JD/ LN ;								JMAT LN ;	\\
																																																																																												JMT PL ;								JYSK DC ;								KBC BB ;								KER FP ;								KESBV FH ;								KGF LN ;								KINVB SS ;								KN FP ;	\\
																																																																																																			KNEBV FH ;								KNIN VX ;								KPN NA ;								KSP ID ;								KU2 GR ;								KYG ID ;								LAND LN ;								LGEN LN ;		\\
																																																																																																									LHA GR ;								LHN VX ;								LI FP ;								LIN GR ;								LLOY LN ;								LONN VX ;								LR FP ;								LSE LN ;								LUN DC ;	\\
																																																																																																																LUPE SS ;								LUX IM ;								LXS GR ;								MAP SM ;								MB IM ;								MC FP ;								MCRO LN ;								MEDAA SS ;								MEO GR ;								MEO1V FH ;								MGGT LN ;								MHG NO ;								MKS LN ;								ML FP ;								MMB FP ;								MRK GR ;								MRW LN ;\\
																																																																																																																								MS IM ;								MT NA ;								MTX GR ;								MUV2 GR ;								NCCB SS ;								NDA SS ;								NESN VX ;								NESTE FH ;	\\
																																																																																																																															NG LN ;								NHY NO ;								NIBEB SS ;								NOKIA FH ;								NOS PL ;								NOVN VX ;								NOVOB DC ;								NRE1V FH ;	\\
																																																																																																																																						NXT LN ;								NZYMB DC ;								OERL SW ;								OML LN ;								OMV AV ;								OR FP ;								ORA FP ;								ORK NO ;								ORP FP ;								PFC LN ;								PFG LN ;								PHIA NA ;								PLT IM ;								PMI IM ;								PNN LN ;								POM FP ;								POP SM ;								PPB ID ;\\
																																																																																																																																														PROX BB ;								PRU LN ;								PSM GR ;								PSN LN ;								PSON LN ;								PUB FP ;								RAND NA ;								RB/ LN ;								RBI AV ;								RBS LN ;								RCO FP ;								RDSA LN ;								REC IM ;								REE SM ;								REL LN ;								REN NA ;								REP SM ;								REX LN ;\\
																																																																																																																																																						RGU LN ;								RI FP ;								RIO LN ;								RMV LN ;								RNO FP ;								ROG VX ;								RPC LN ;								RR/ LN ;								RRS LN ;\\
																																																																																																																																																														RSA LN ;								RTO LN ;								RWE GR ;								RYA ID ;								SAABB SS ;								SAB LN ;								SAB SM ;								SAF FP ;\\
																																																																																																																																																																						SAMAS FH ;								SAN FP ;								SAN SM ;								SAND SS ;								SAP GR ;								SBMO NA ;								SBRY LN ;								SCAB SS ;	\\
																																																																																																																																																																													SCHA NO ;								SCHP VX ;								SCMN VX ;								SCR FP ;								SDF GR ;								SDR LN ;								SEBA SS ;								SECUB SS ;	\\
																																																																																																																																																																																				SGE LN ;								SGO FP ;								SGRO LN ;								SHB LN ;								SHBA SS ;								SHP LN ;								SIE GR ;								SKAB SS ;								SKFB SS ;\\
																																																																																																																																																																																												SKY LN ;								SLHN VX ;								SMDS LN ;								SMIN LN ;								SN LN ;								SNH GR ;								SOLB BB ;								SOON VX ;								SPM IM ;								SPR GR ;								SPSN SW ;								SREN VX ;								SRG IM ;								SSE LN ;								STAN LN ;								STERV FH ;								STJ LN ;	\\
																																																																																																																																																																																																			STL NO ;								STM IM ;								SU FP ;								SVT LN ;								SW FP ;								SWEDA SS ;								SWMA SS ;								SYNN VX ;								SZU GR ;								TATE LN ;								TDC DC ;								TEC FP ;								TEF SM ;								TEL NO ;								TEL2B SS ;								TEMN SW ;								TEN IM ;	\\
																																																																																																																																																																																																										TIT IM ;								TKA AV ;								TKA GR ;								TL5 SM ;								TLSN SS ;								TNET BB ;								TPK LN ;								TRELB SS ;								TRN IM ;								TRYG DC ;								TSCO LN ;								TW/ LN ;								UBI IM ;								UBM LN ;								UBSG VX ;								UCB BB ;								UCG IM ;								UG FP ;								UHR VX ;								ULVR LN ;								UMI BB ;								UPM1V FH ;								US IM ;								UTDI GR ;								UU/ LN ;								VER AV ;								VIE FP ;								VIG AV ;								VIV FP ;								VOD LN ;								VOE AV ;								VOLVB SS ;								VOW GR ;								VPK NA ;								VWS DC ;								WCH GR ;								WDH DC ;								WDI GR ;								WG/ LN ;								WKL NA ;								WMH LN ;								WOS LN ;								WPP LN ;								WRT1V FH ;	\\
																																																																																																																																																																																																																	WTB LN ;								YAR NO ;								ZC FP ;								ZOT SM ;								ZURN VX					

\section*{Dataset-SHSZ-CSI300}
000001 CH ;								000002 CH ;								000009 CH ;								000027 CH ;								000039 CH ;								000046 CH ;								000060 CH ;								000061 CH ;								000063 CH ;								000069 CH ;								000100 CH ;								000157 CH ;								000400 CH ;								000402 CH ;								000413 CH ;								000415 CH ;								000423 CH ;								000425 CH ;								000503 CH ;								000538 CH ;								000539 CH ;								000540 CH ;								000559 CH ;								000568 CH ;								000581 CH ;								000598 CH ;								000625 CH ;								000629 CH ;								000651 CH ;								000686 CH ;								000709 CH ;								000712 CH ;								000725 CH ;								000728 CH ;								000729 CH ;								000738 CH ;								000768 CH ;								000778 CH ;								000792 CH ;								000793 CH ;								000800 CH ;								000825 CH ;								000826 CH ;								000858 CH ;								000876 CH ;								000883 CH ;								000898 CH ;								000917 CH ;								000937 CH ;								000983 CH ;								000999 CH ;								002007 CH ;								002008 CH ;								002024 CH ;								002038 CH ;								600000 CH ;								600005 CH ;								600008 CH ;								600009 CH ;								600010 CH ;								600011 CH ;								600015 CH ;								600016 CH ;								600019 CH ;								600021 CH ;								600028 CH ;								600029 CH ;								600030 CH ;								600031 CH ;								600036 CH ;								600038 CH ;								600050 CH ;								600060 CH ;								600066 CH ;								600068 CH ;								600085 CH ;								600089 CH ;								600100 CH ;								600104 CH ;								600109 CH ;\\
								600111 CH ;						600115 CH ;								600118 CH ;								600150 CH ;								600157 CH ;								600166 CH ;								600170 CH ;\\
								600177 CH ;								600188 CH ;								600196 CH ;								600208 CH ;								600221 CH ;								600252 CH ;								600256 CH ;								600271 CH ;\\
								600276 CH ;								600309 CH ;								600317 CH ;								600332 CH ;								600340 CH ;								600350 CH ;								600352 CH ;								600362 CH ;\\
								600373 CH ;								600398 CH ;								600406 CH ;								600415 CH ;								600485 CH ;								600489 CH ;								600518 CH ;								600519 CH ;\\
								600535 CH ;								600547 CH ;								600549 CH ;								600570 CH ;								600578 CH ;								600583 CH ;								600585 CH ;								600588 CH ;\\
								600600 CH ;								600642 CH ;								600648 CH ;								600660 CH ;								600663 CH ;								600674 CH ;								600688 CH ;								600690 CH ;	\\
							600717 CH ;								600718 CH ;								600739 CH ;								600741 CH ;								600783 CH ;								600795 CH ;								600804 CH ;								600820 CH ;\\
								600827 CH ;								600837 CH ;								600839 CH ;								600863 CH ;								600867 CH ;								600873 CH ;								600875 CH ;								600886 CH ;\\
								600887 CH ;								600895 CH ;								601607 CH ;								601988 CH					
\section*{Dataset-NASDAQ}
AAPL UW ;								ADBE UW ;								ADSK UW ;								AKAM UW ;								ALXN UW ;								AMAT UW ;								AMGN UW ;	\\
							AMZN UW ;								ATVI UW ;								BBBY UW ;								BIDU UW ;								BIIB UW ;								BMRN UW ;								CELG UW ;\\
															CERN UW ;								CHKP UW ;								CMCSA UW ;								COST UW ;								CSCO UW ;								CTRP UW ;								CTSH UW ;\\
																							CTXS UW ;								DISCA UW ;								DISH UW ;								DLTR UW ;								EA UW ;								EBAY UW ;								ENDP UW ;								ESRX UW ;								EXPE UW ;								FAST UW ;								FISV UW ;								GILD UW ;								GOOGL UW ;								HSIC UW ;								ILMN UW ;\\
																															INCY UW ;								INTC UW ;								INTU UW ;								ISRG UW ;								LBTYA UW ;								LBTYK UW ;								LLTC UW ;	\\
																																						LRCX UW ;								MNST UW ;								MSFT UW ;								NFLX UW ;								NTAP UW ;								NTES UW ;								NVDA UW ;\\
																																														ORLY UW ;								PAYX UW ;								PCAR UW ;								PCLN UW ;								QCOM UW ;								QVCA UW ;								REGN UW ;	\\
																																																					ROST UW ;								SBAC UW ;								SBUX UW ;								SIRI UW ;								SRCL UW ;								SWKS UW ;								SYMC UW ;								TSCO UW ;								VRTX UW ;								WFM UW ;								XLNX UW ;								YHOO UW			
\section*{Dataset-CAC40}
	AC FP ;								ACA FP ;								AI FP ;								AIR FP ;								ALO FP ;								BN FP ;								BNP FP ;								CA FP ;								CAP FP ;								CS FP ;	\\
								DG FP ;								EI FP ;								EN FP ;								ENGI FP ;								FP FP ;								FR FP ;								GLE FP ;								KER FP ;								LI FP ;								LR FP ;\\
																MC FP ;								ML FP ;								MT NA ;								OR FP ;								ORA FP ;								PUB FP ;								RI FP ;								RNO FP ;								SAF FP ;								SAN FP ;								SGO FP ;								SOLB BB ;								SU FP ;								TEC FP ;								UG FP ;								VIE FP ;								VIV FP		
 
 \end{small}
 
\begin{figure}[H]
\begin{tabular}{ccc}
\subfloat[$\mathscr{R}_{1}$covar]{\includegraphics[width = 1.7in]{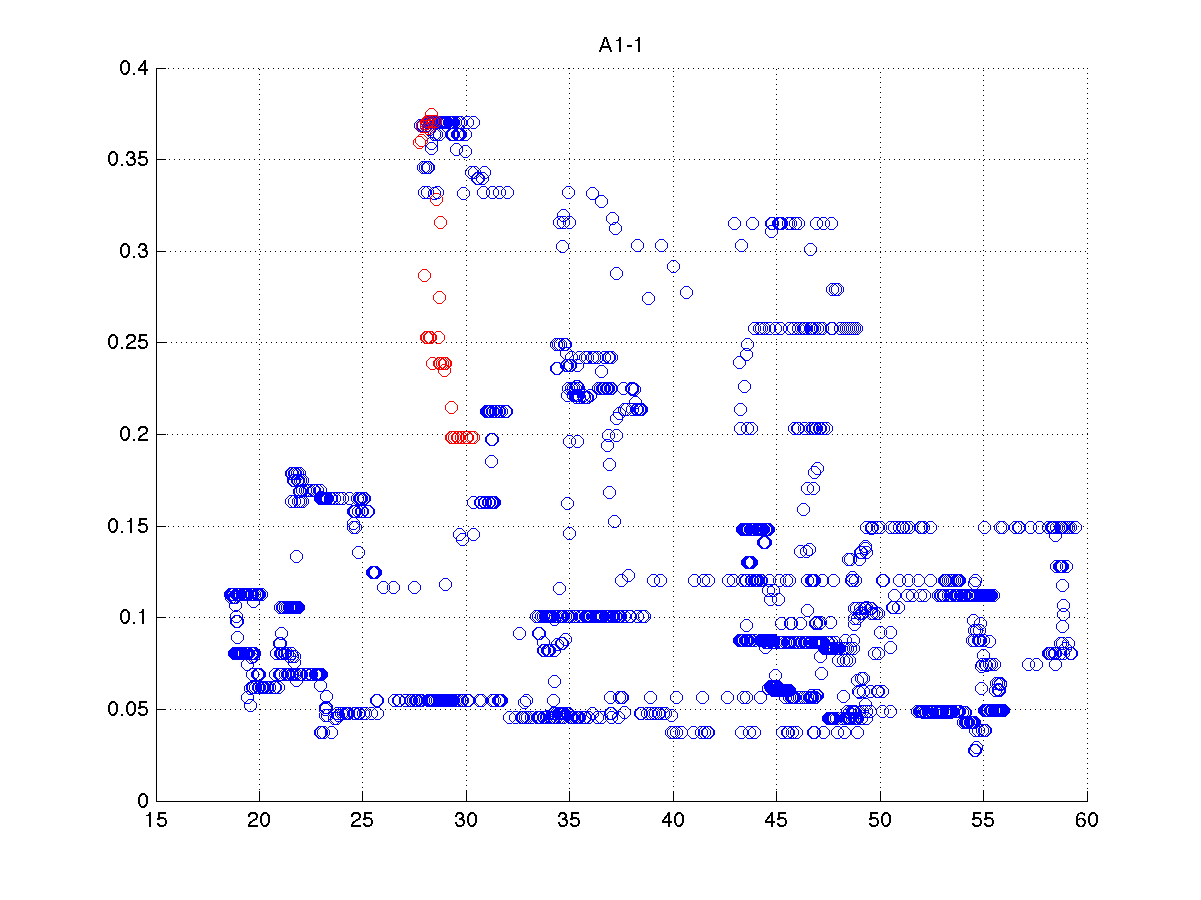}} &
\subfloat[$\mathscr{R}_{1}$correl]{\includegraphics[width = 1.7in]{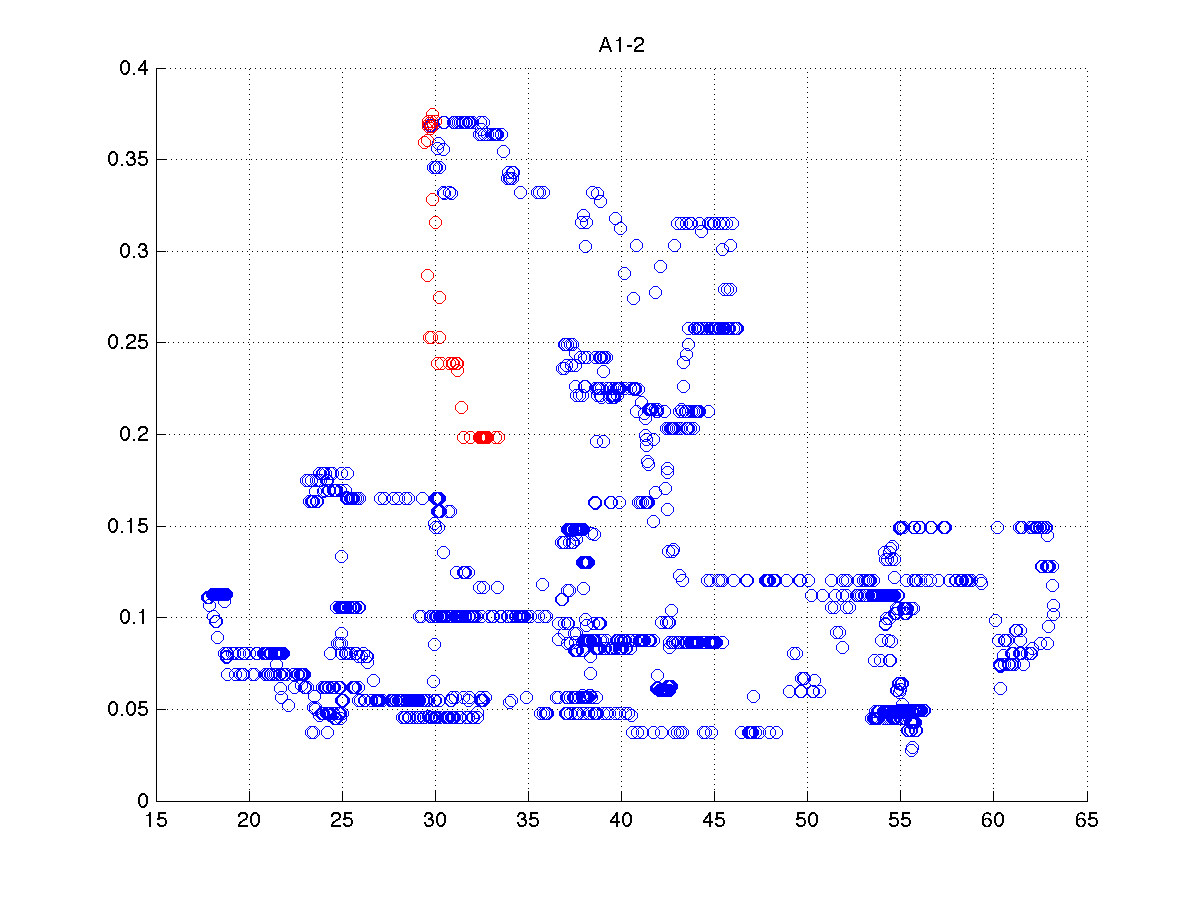}} &
\subfloat[$\mathscr{R}_{1}$correl-volume]{\includegraphics[width = 1.7in]{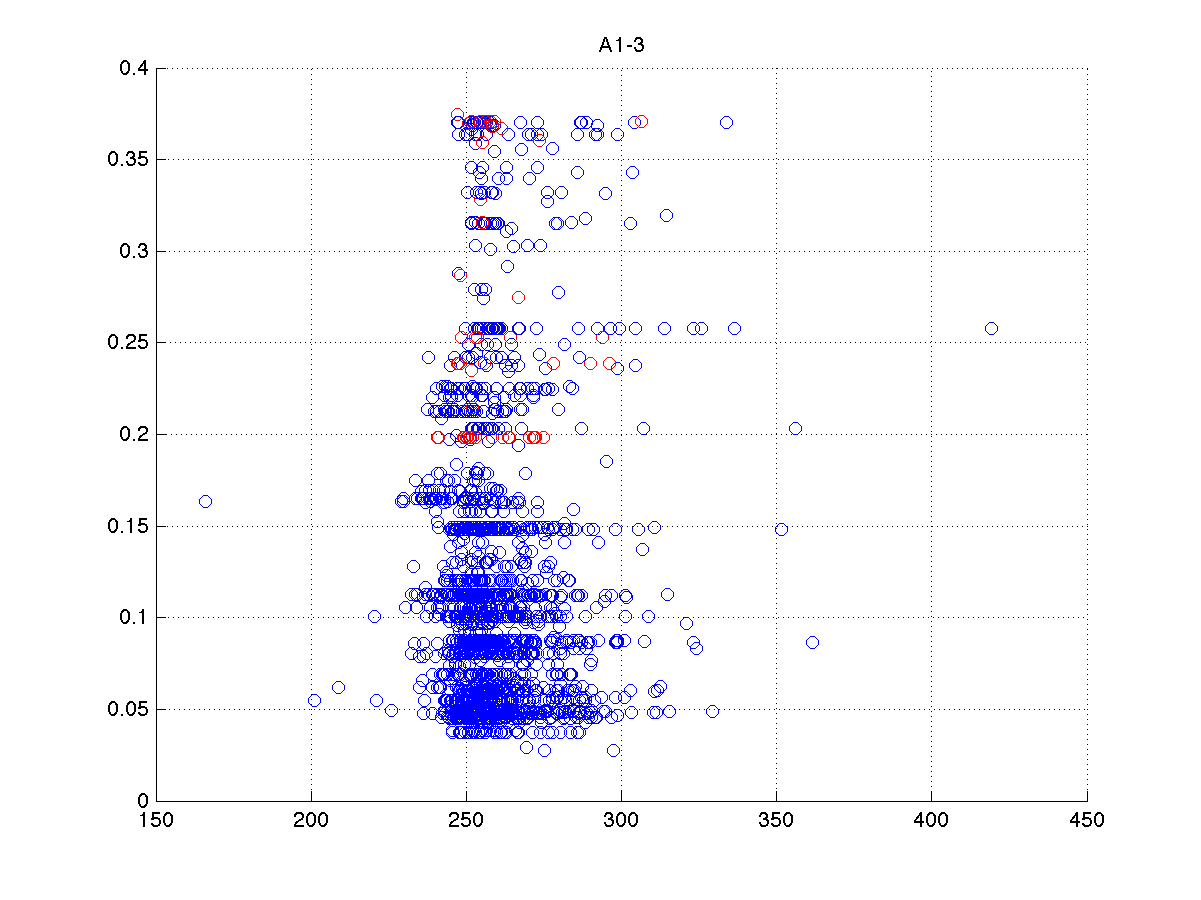}}\\
\subfloat[$\mathscr{R}_{1}$correl-mcap]{\includegraphics[width = 1.7in]{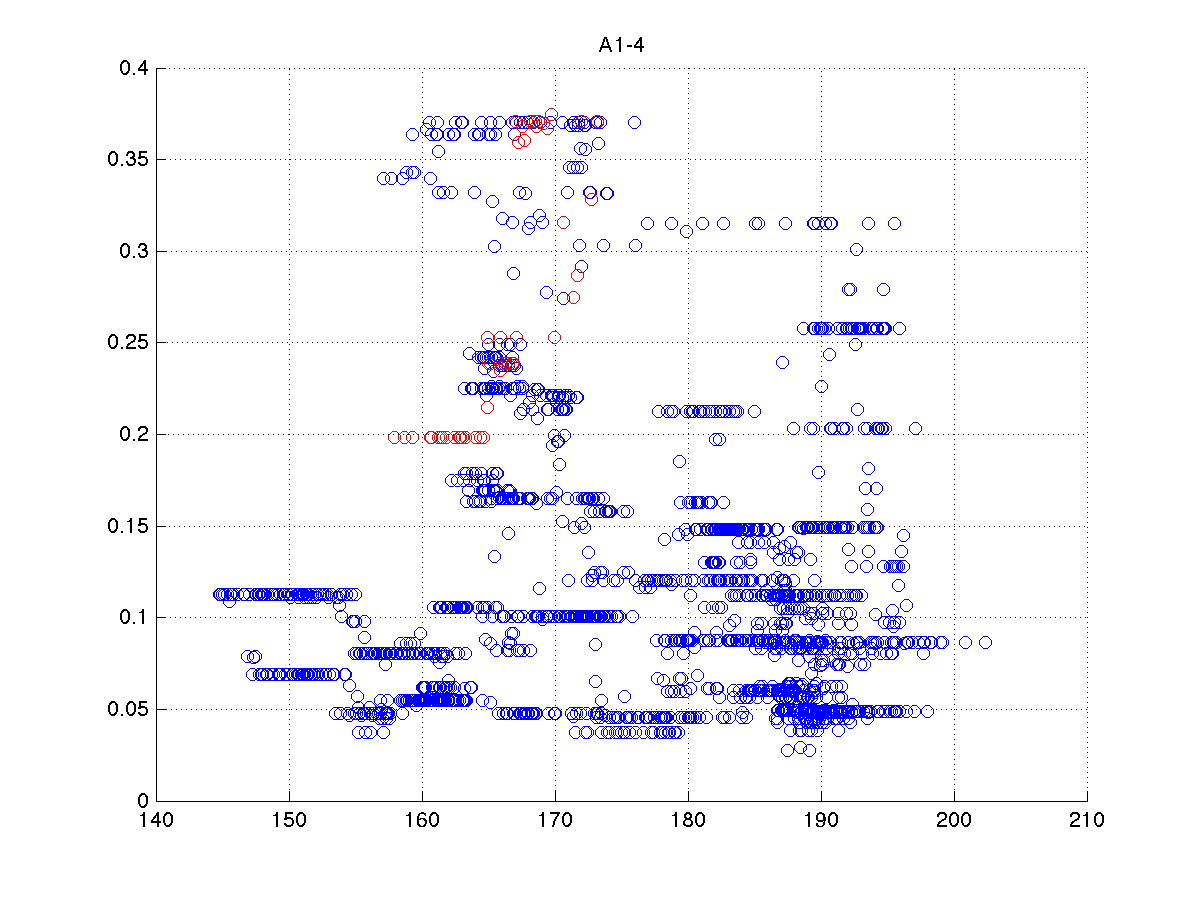}}&
\subfloat[$\mathscr{R}_{1}$correl-leverage]{\includegraphics[width = 1.7in]{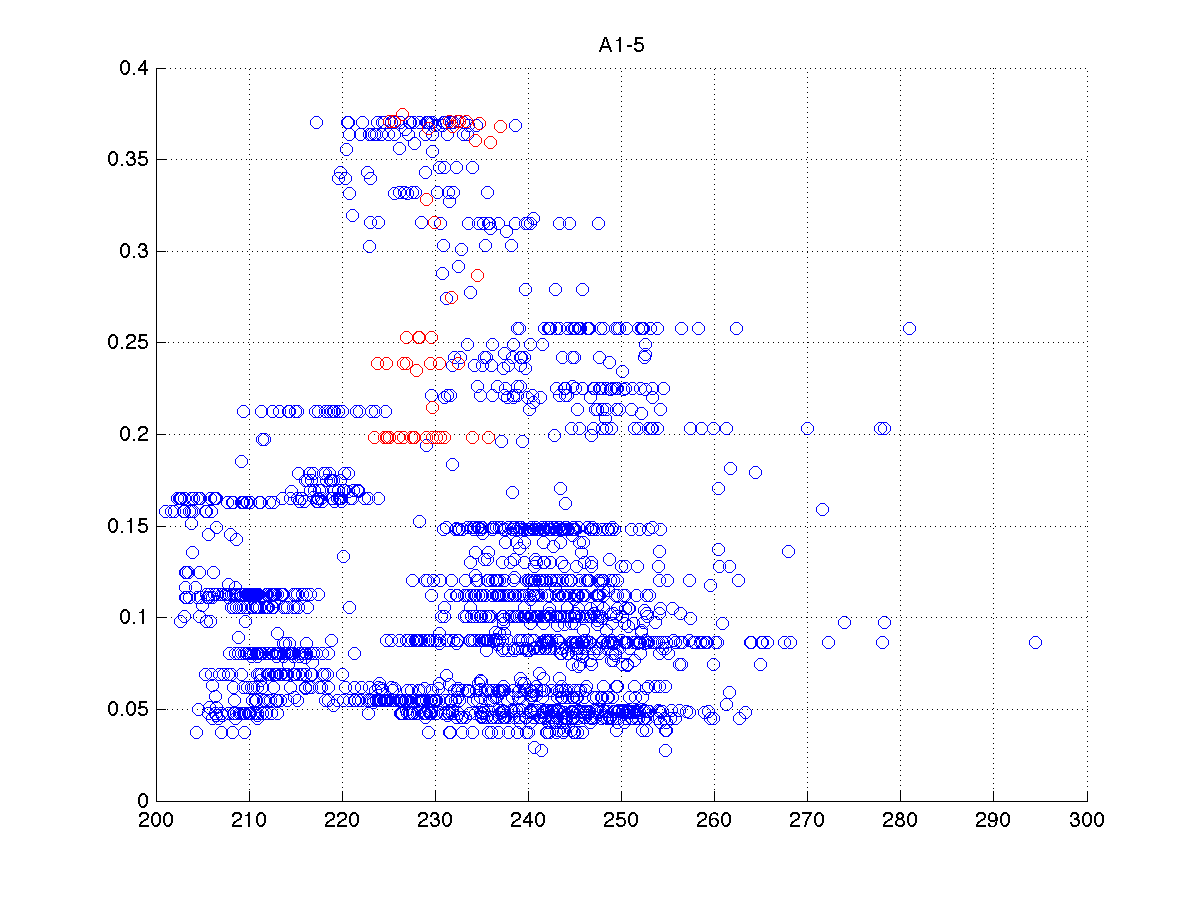}} &
\subfloat[$\mathscr{R}_{2}$covar]{\includegraphics[width = 1.7in]{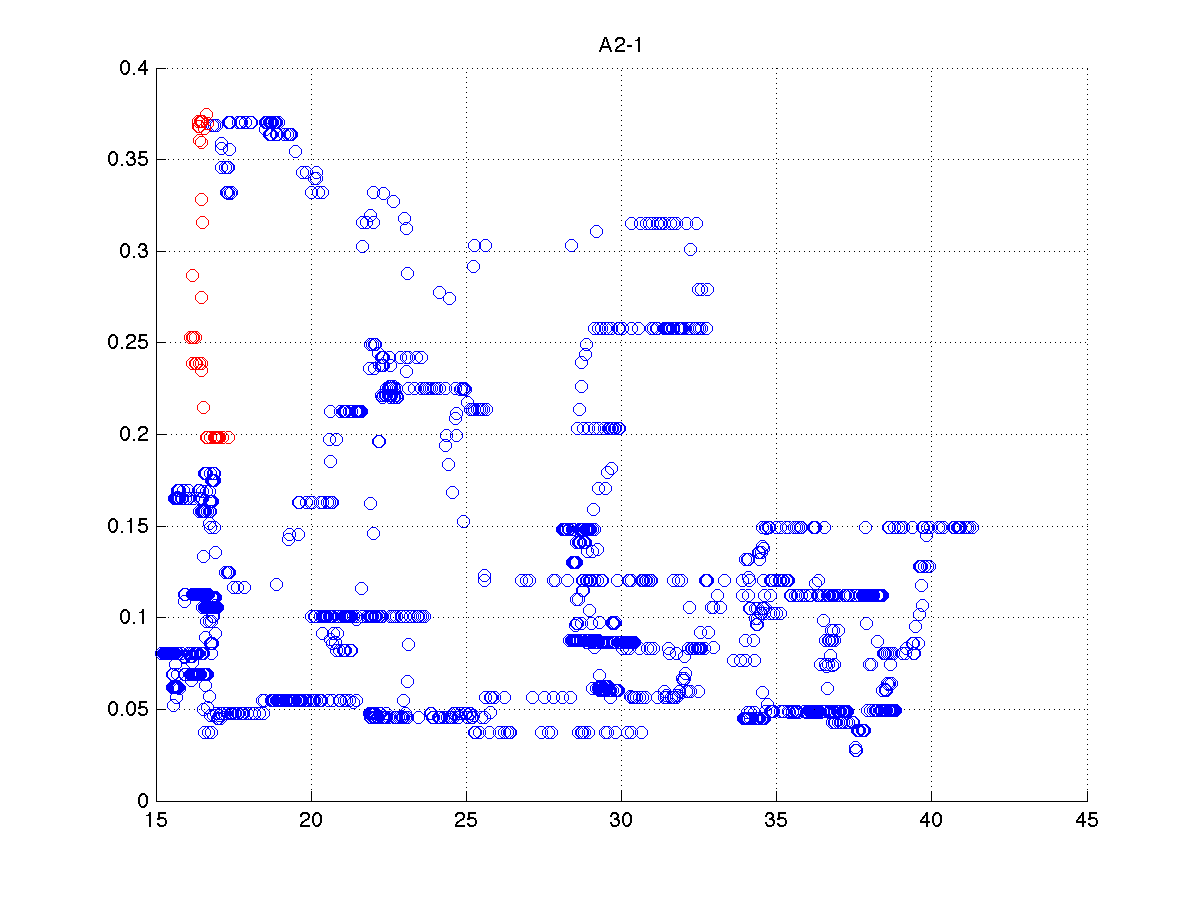}}\\
\subfloat[$\mathscr{R}_{2}$correl]{\includegraphics[width = 1.7in]{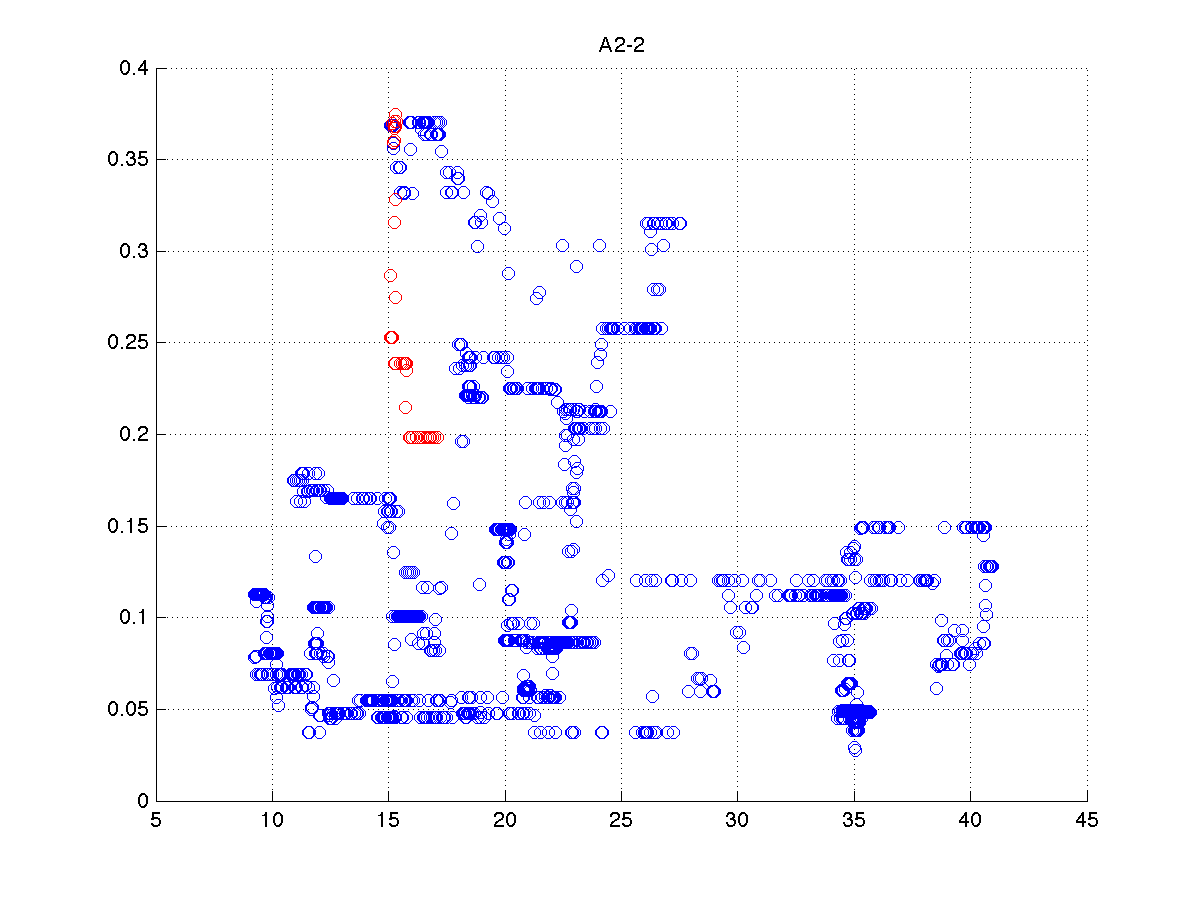}} &
\subfloat[$\mathscr{R}_{2}$correl-volume]{\includegraphics[width = 1.7in]{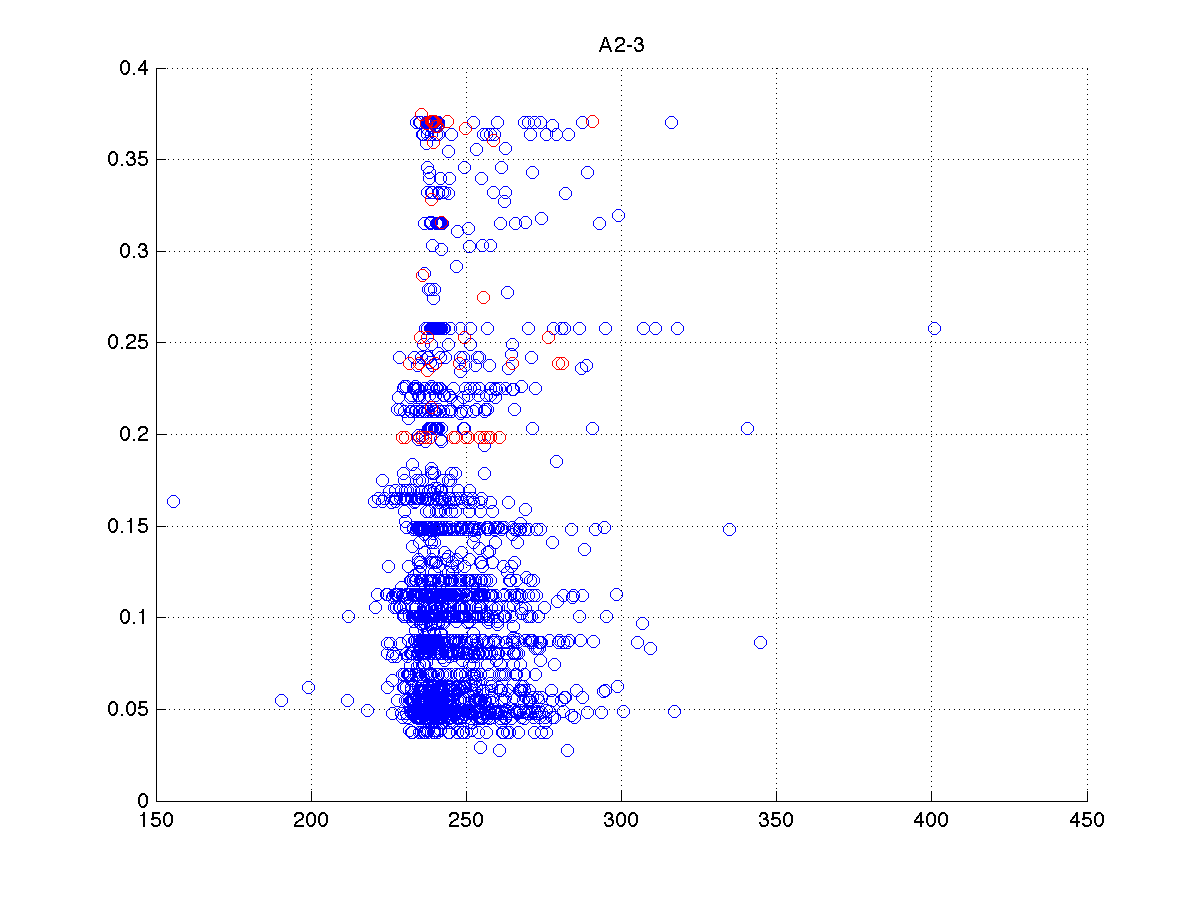}}&
\subfloat[$\mathscr{R}_{2}$correl-mcap]{\includegraphics[width = 1.7in]{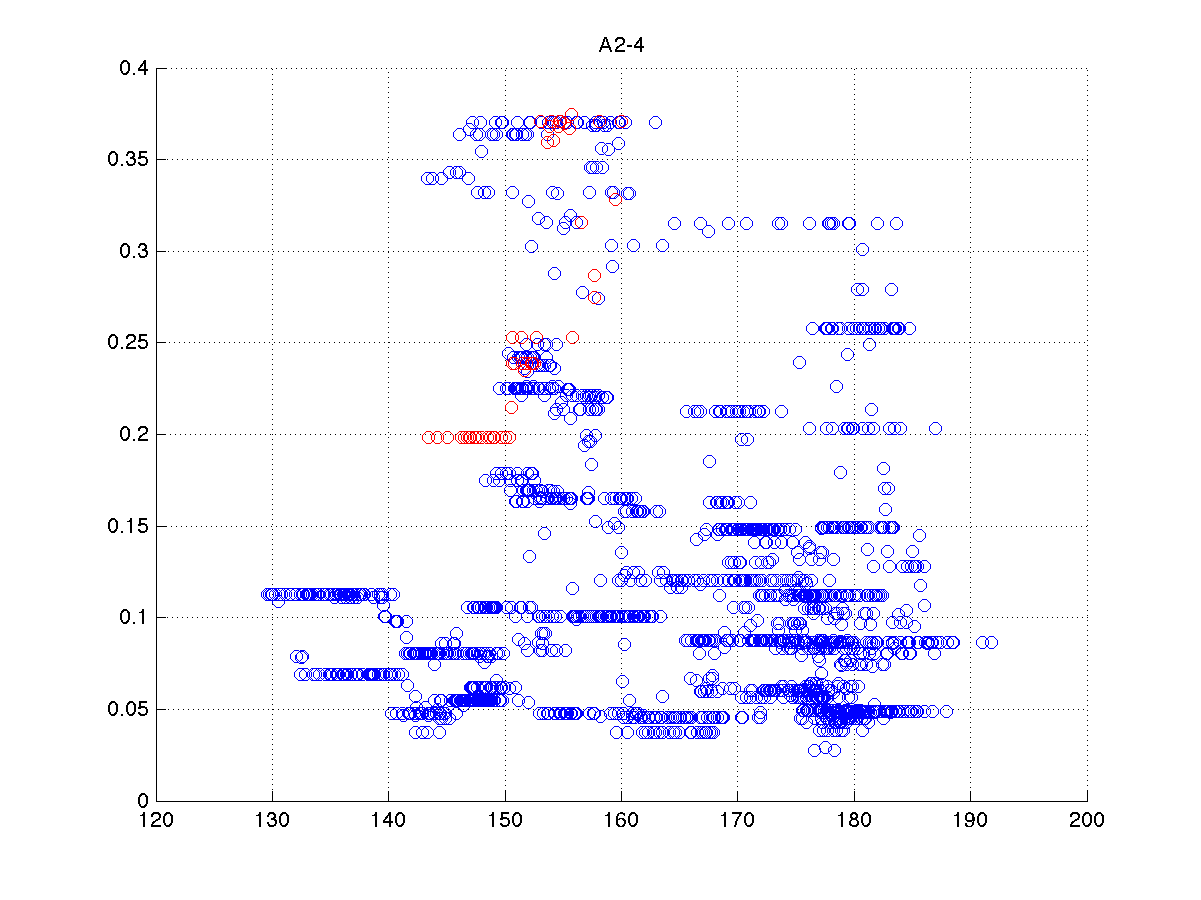}}\\
\subfloat[$\mathscr{R}_{2}$correl-leverage]{\includegraphics[width = 1.7in]{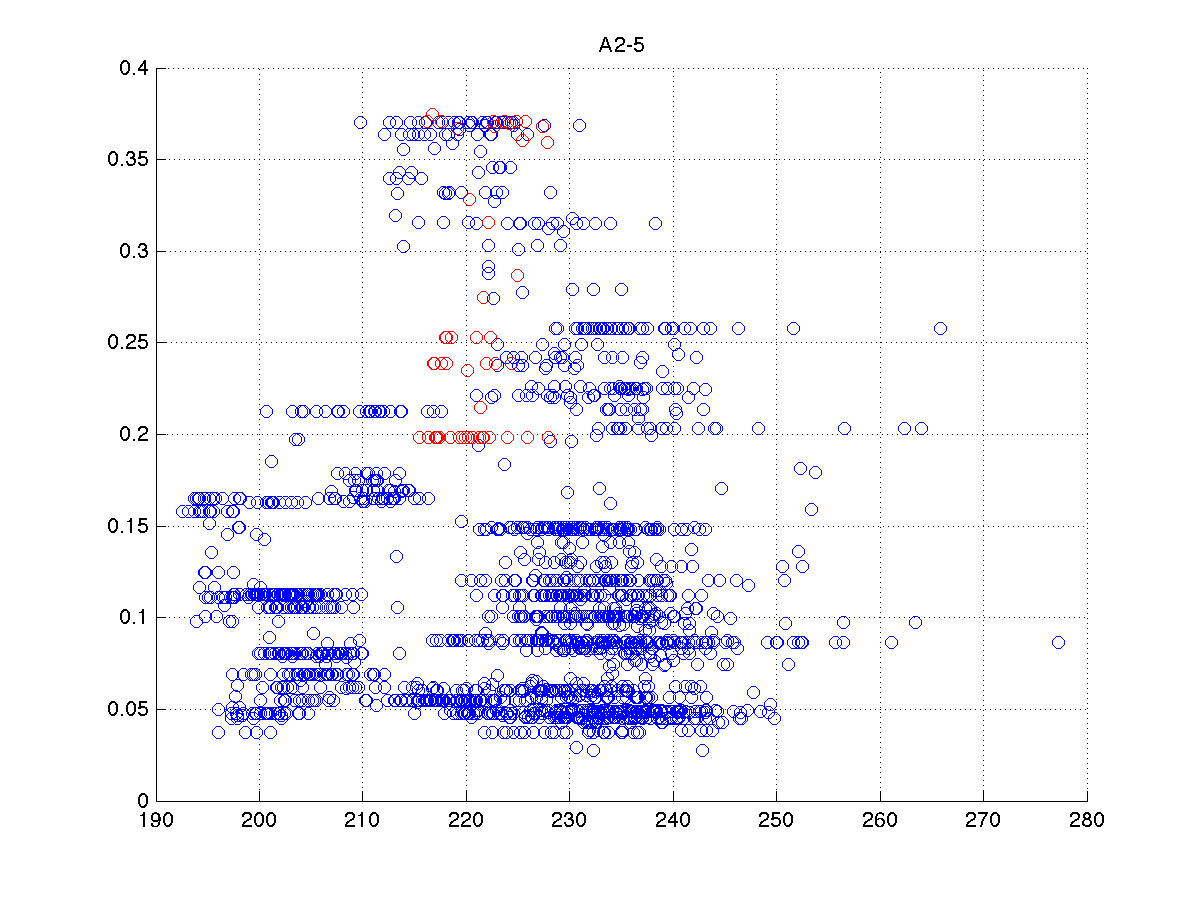}} &
\subfloat[$\mathscr{R}_{3}$covar]{\includegraphics[width = 1.7in]{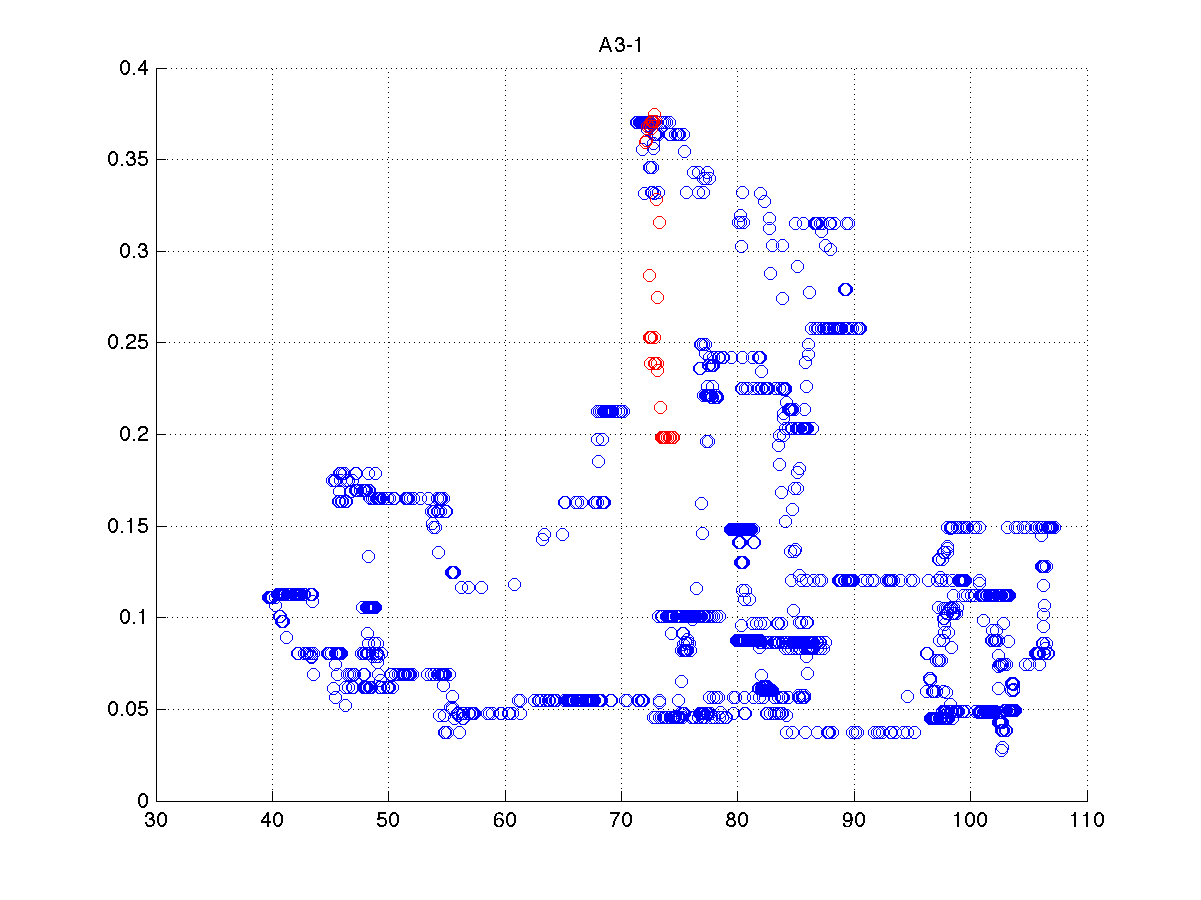}} &
\subfloat[$\mathscr{R}_{3}$correl]{\includegraphics[width = 1.7in]{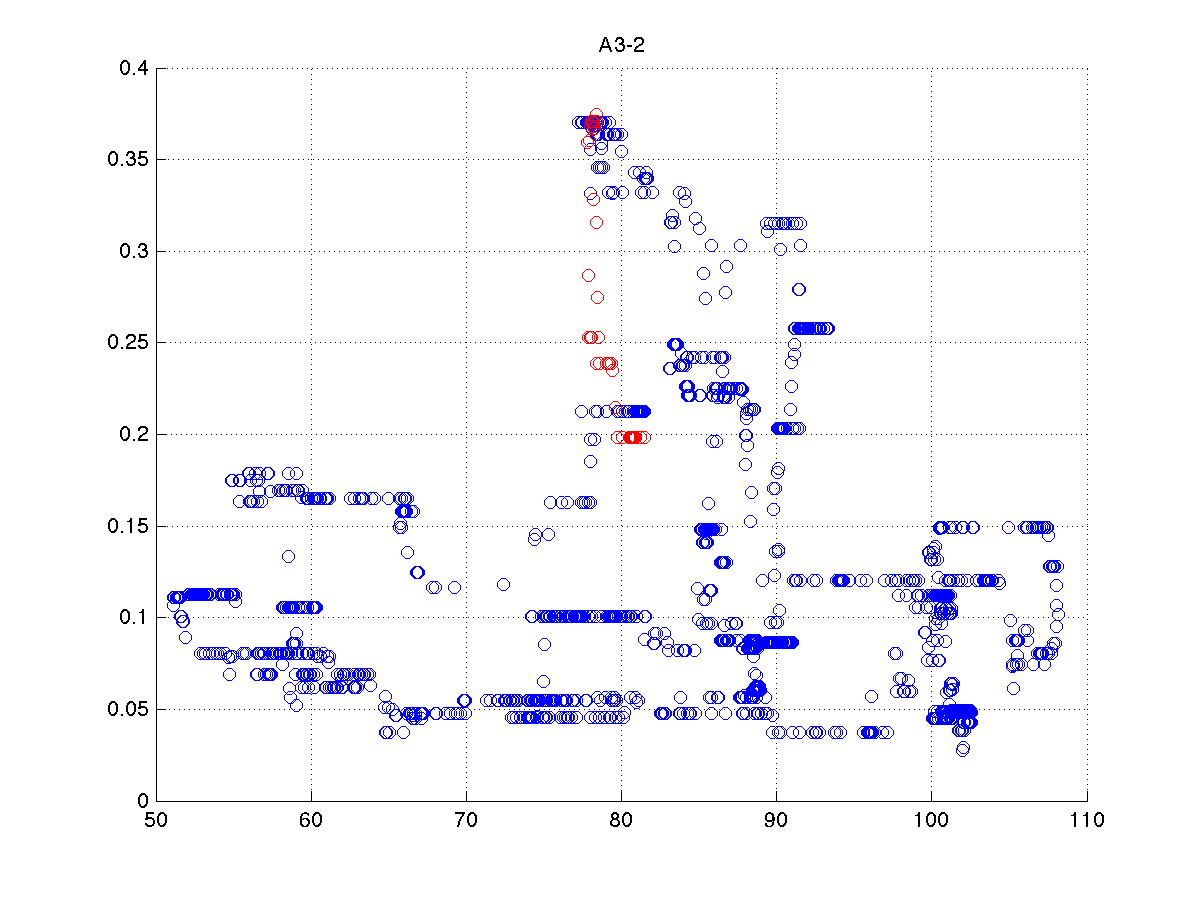}}\\
\subfloat[$\mathscr{R}_{3}$correl-volume]{\includegraphics[width = 1.7in]{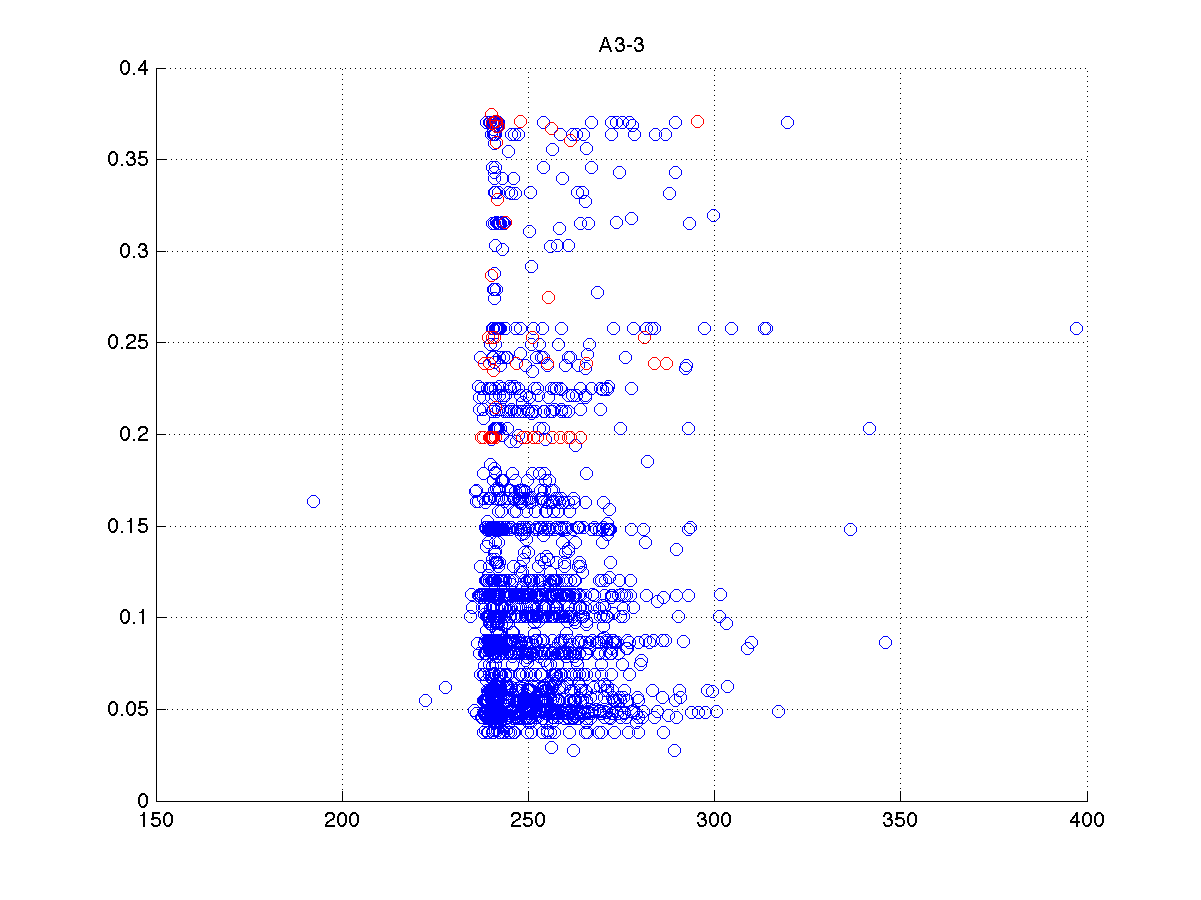}} &
\subfloat[$\mathscr{R}_{3}$correl-mcap]{\includegraphics[width = 1.7in]{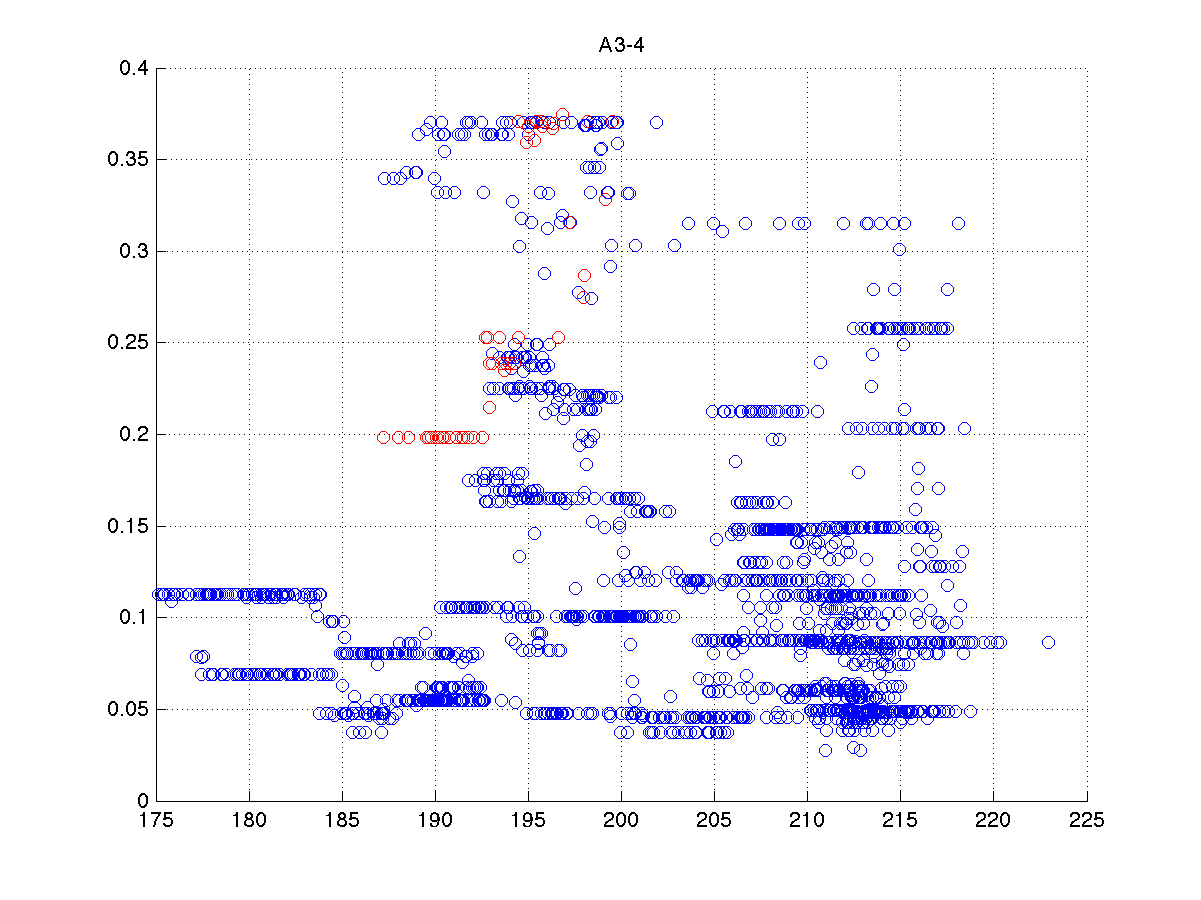}} &
\subfloat[$\mathscr{R}_{3}$correl-leverage]{\includegraphics[width = 1.7in]{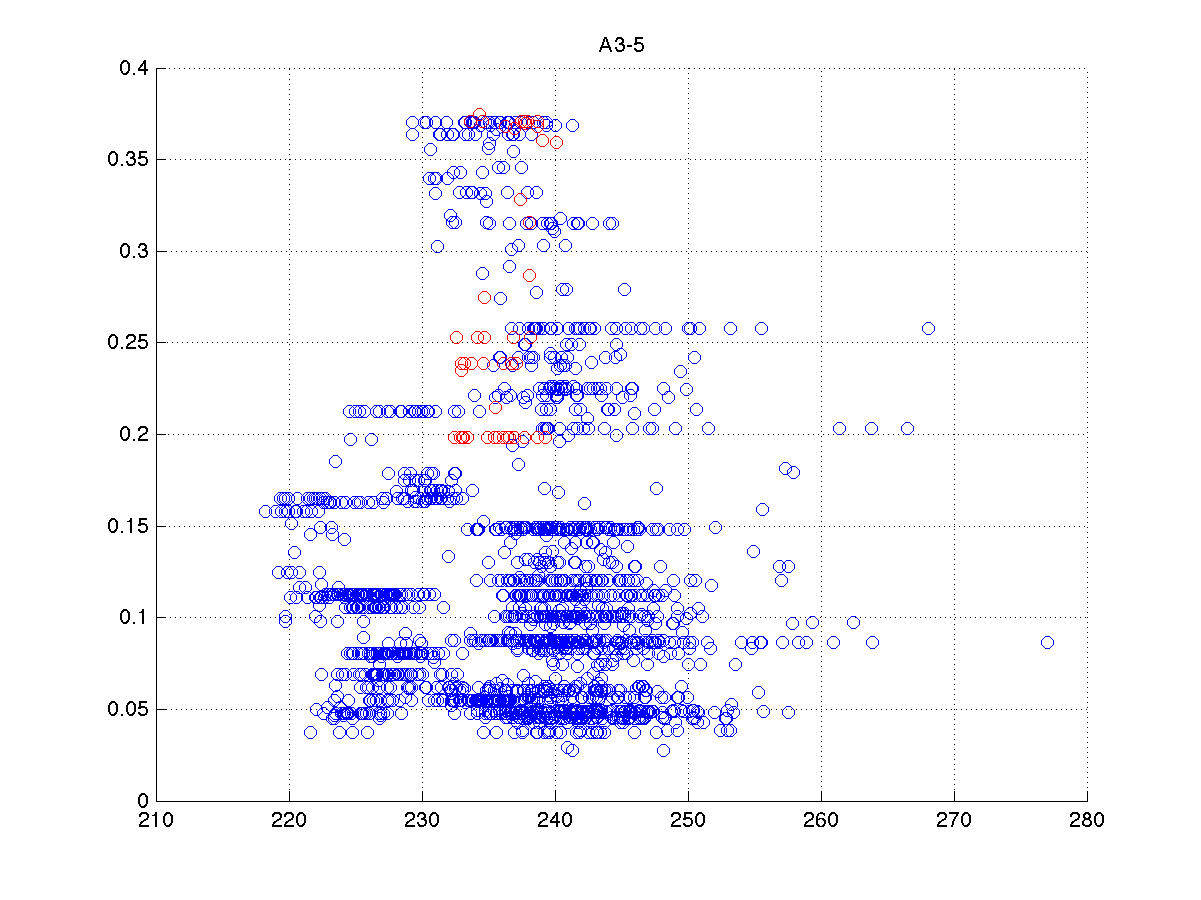}}\\
\end{tabular}
\captionsetup{labelformat=empty}
\caption{BE500: Indicators of the $\alpha$-series. Red: in-sample ; Blue: out-of-sample}
\end{figure}

\begin{figure}[H]
\begin{tabular}{ccc}
\subfloat[rspec-covar]{\includegraphics[width = 1.7in]{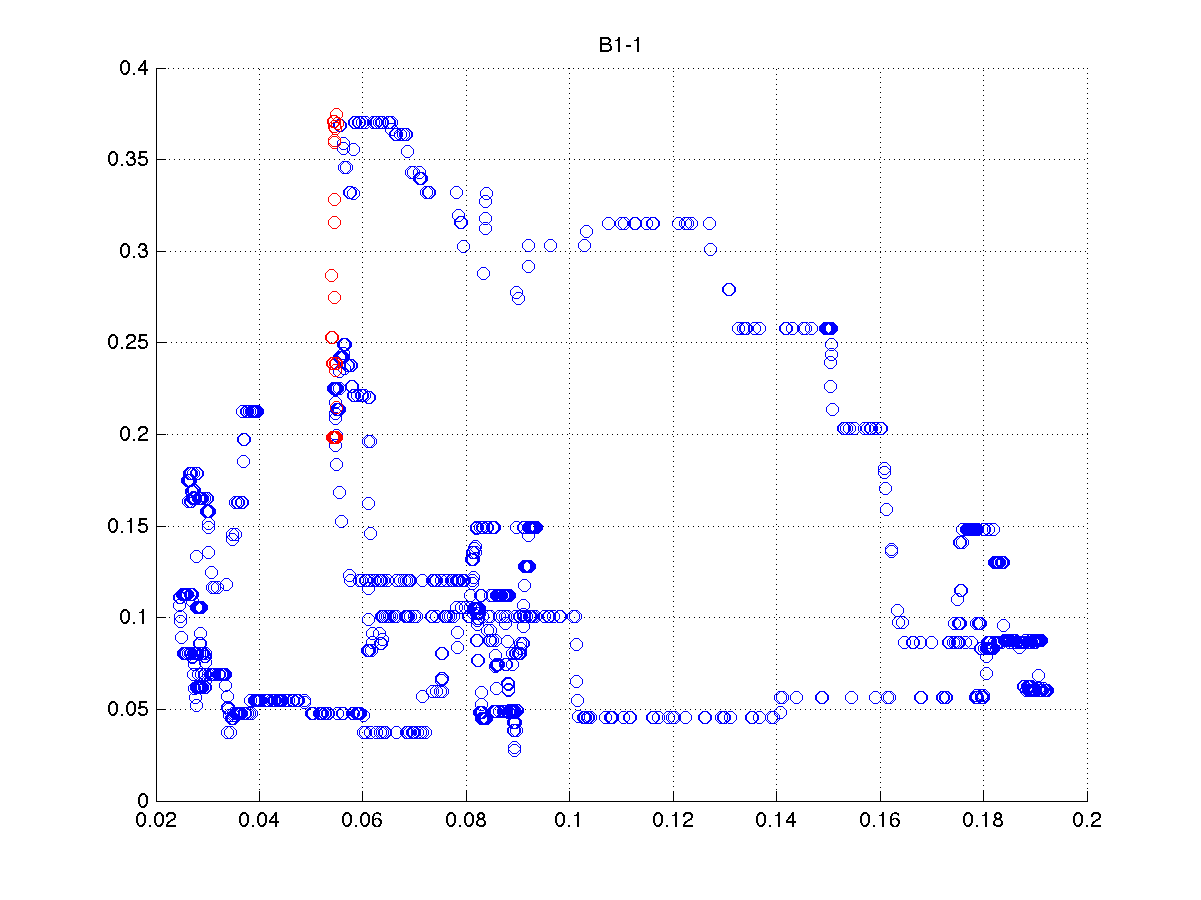}} &
\subfloat[rspec-correl]{\includegraphics[width = 1.7in]{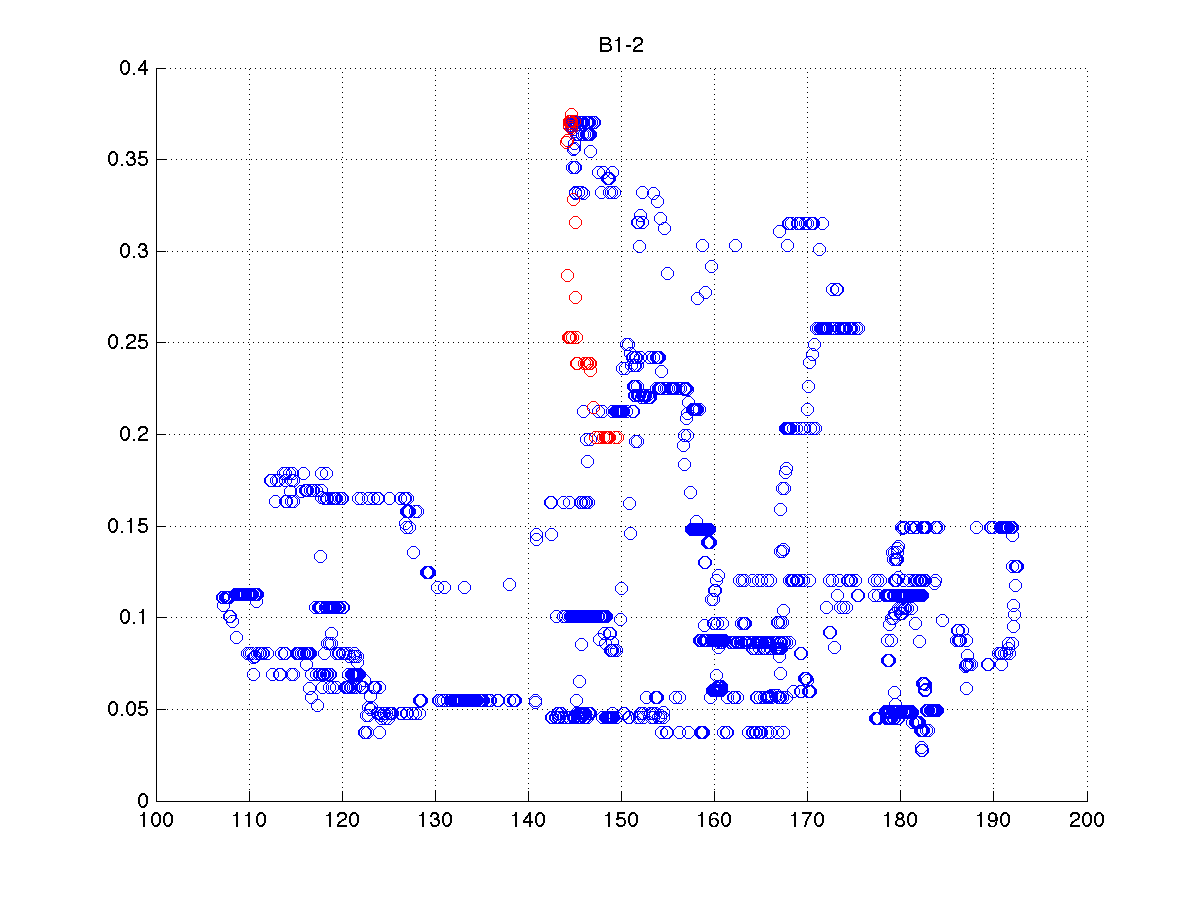}} &
\subfloat[rspec-correl-volume]{\includegraphics[width = 1.7in]{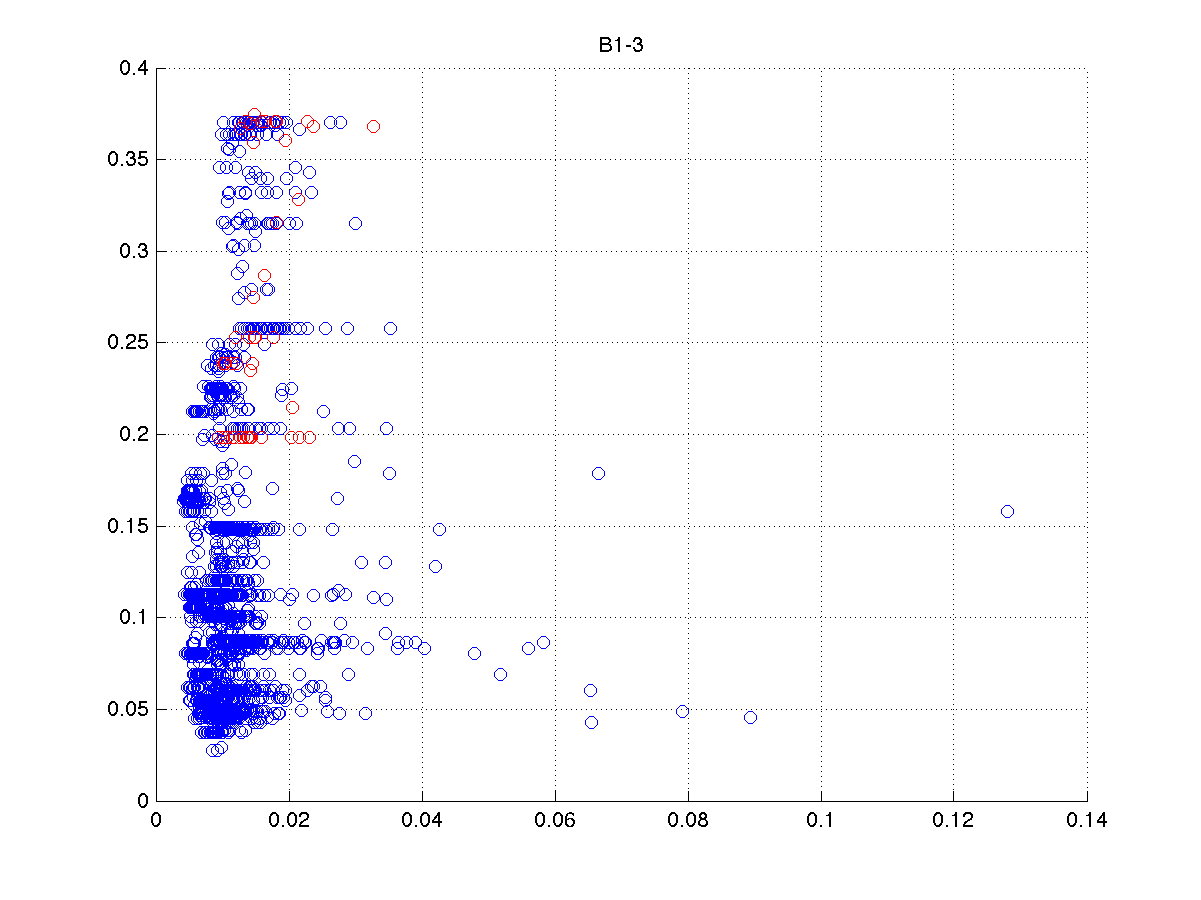}} \\
\subfloat[rspec-correl-mcap]{\includegraphics[width = 1.7in]{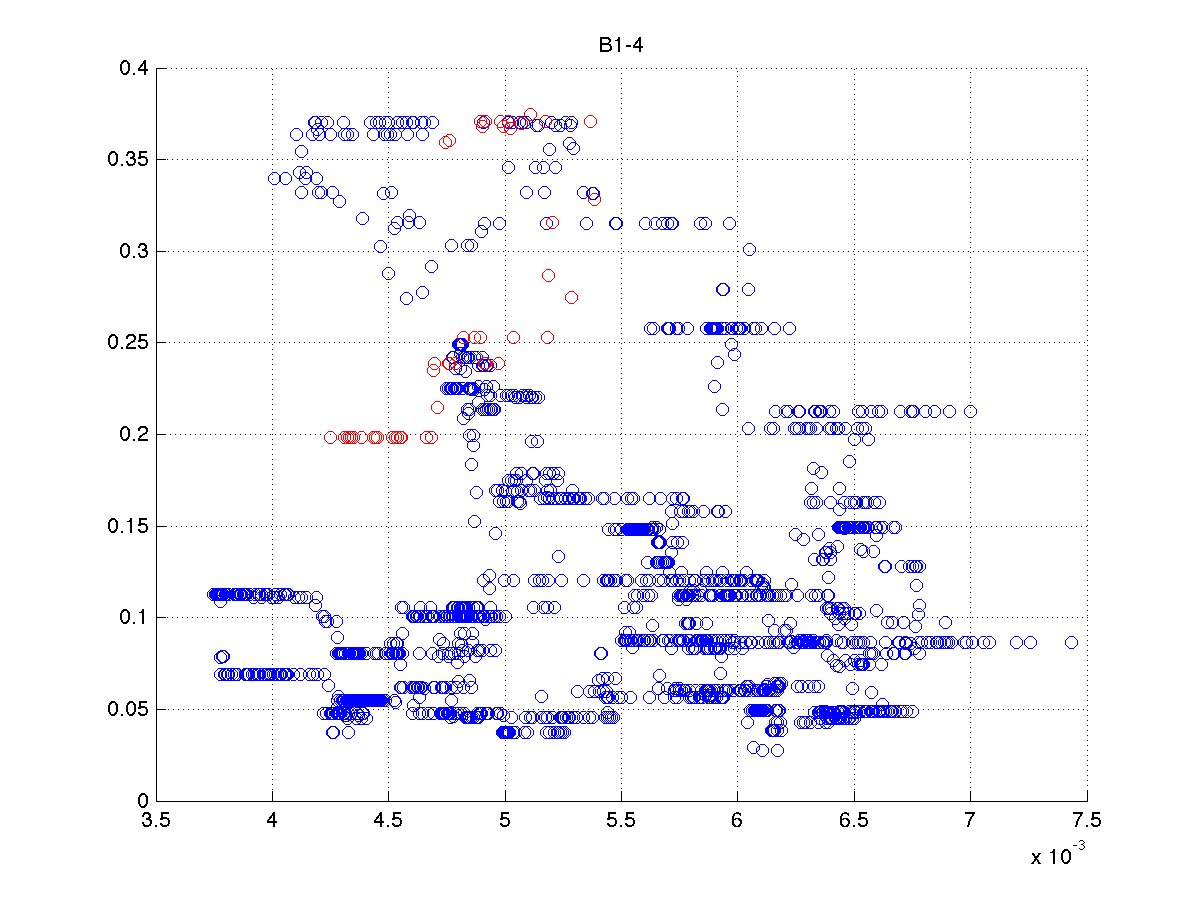}}&
\subfloat[rspec-correl-leverage]{\includegraphics[width = 1.7in]{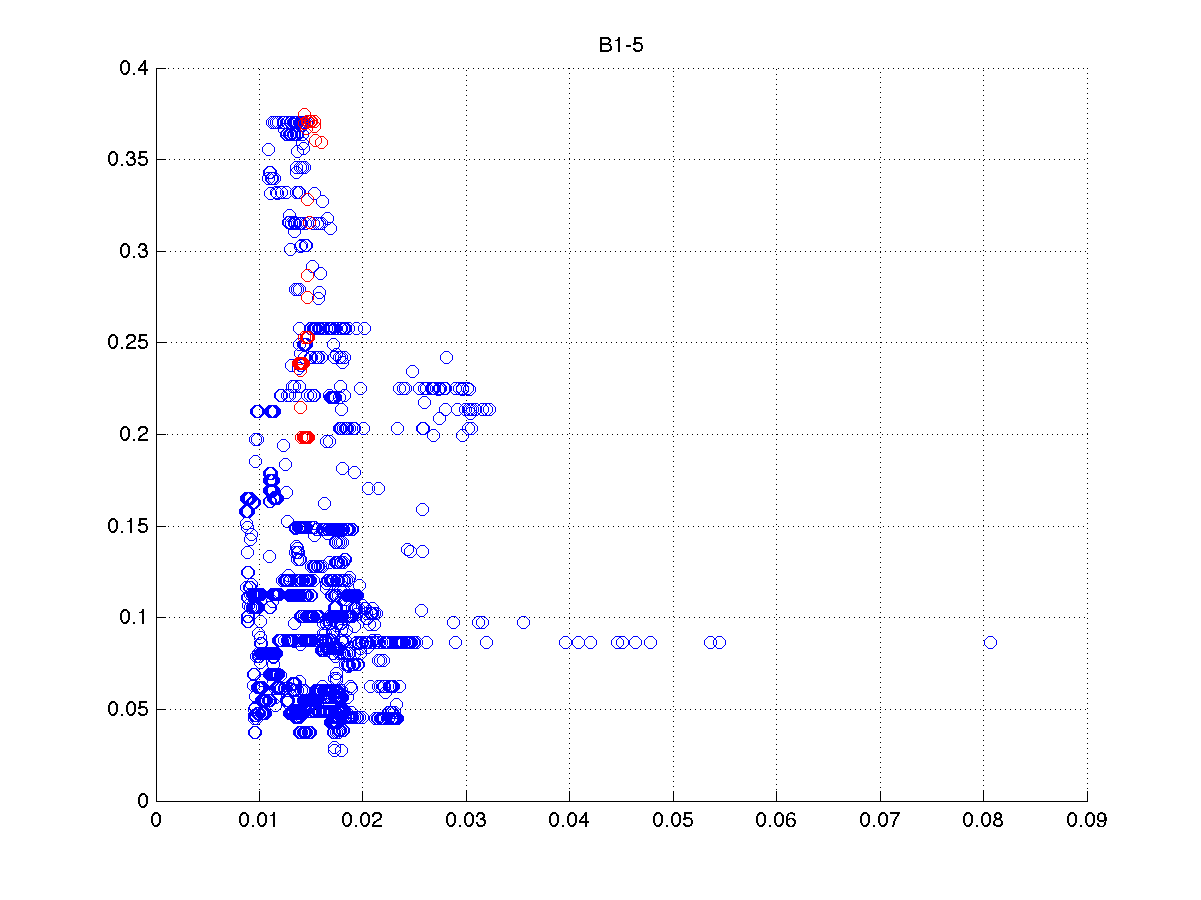}} &
\subfloat[trace-covar]{\includegraphics[width = 1.7in]{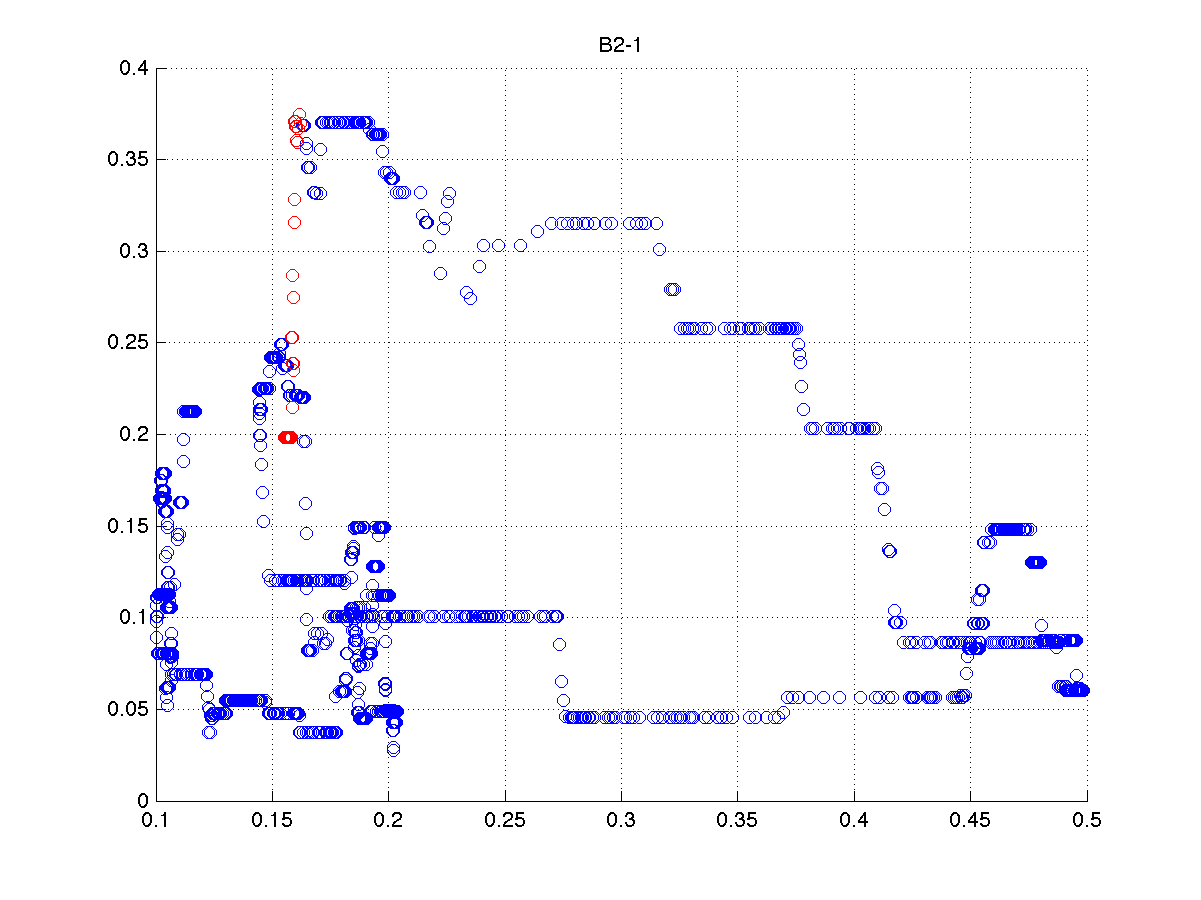}}\\
\subfloat[trace-correl-volume]{\includegraphics[width = 1.7in]{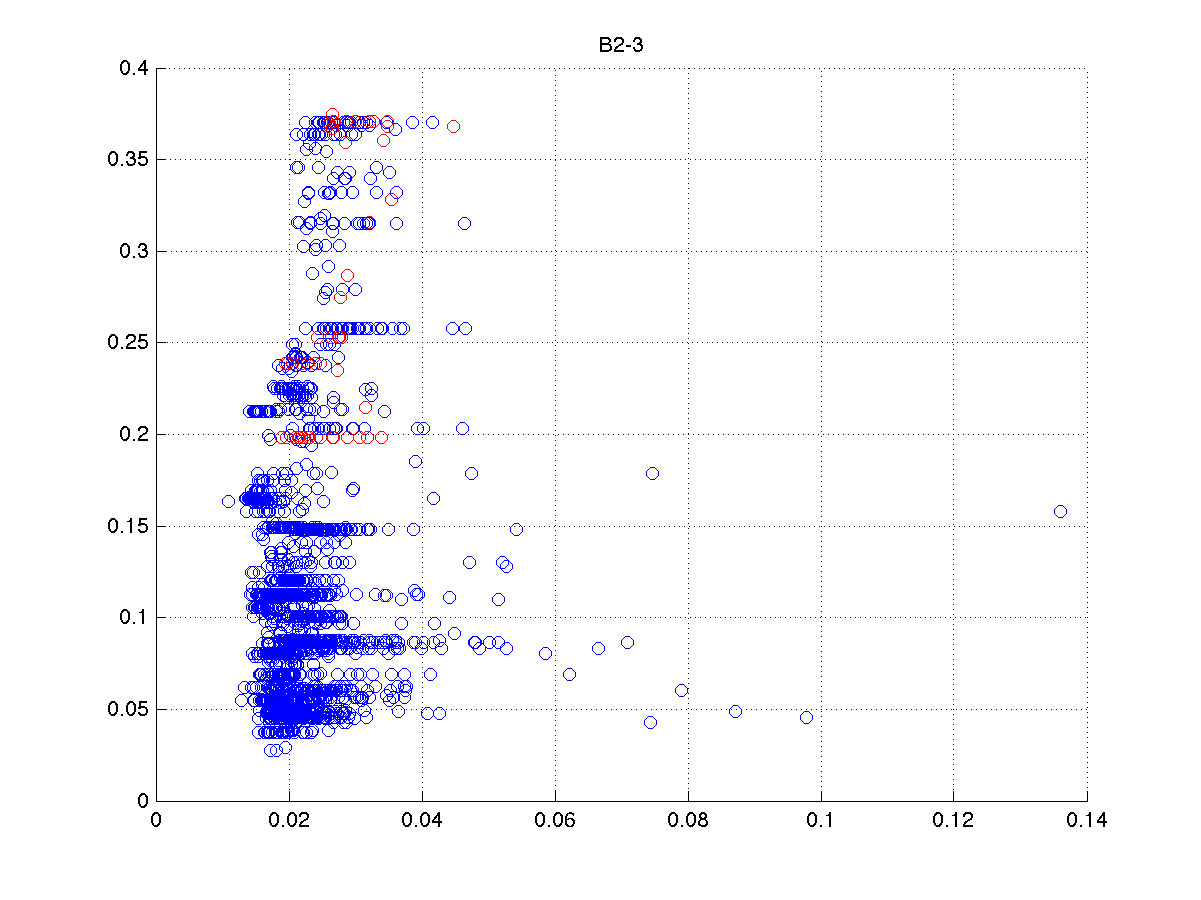}} &
\subfloat[trace-correl-mcap]{\includegraphics[width = 1.7in]{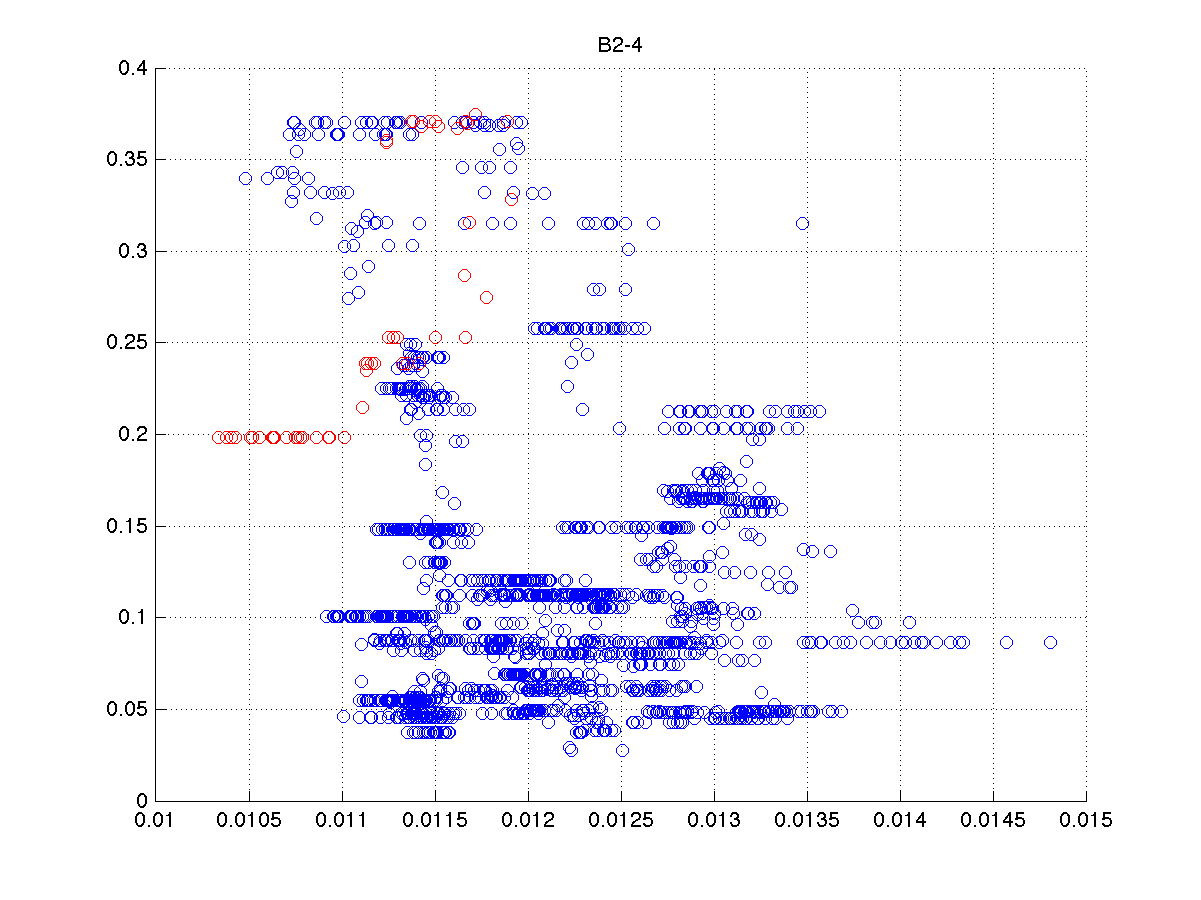}}&
\subfloat[trace-correl-leverage]{\includegraphics[width = 1.5in]{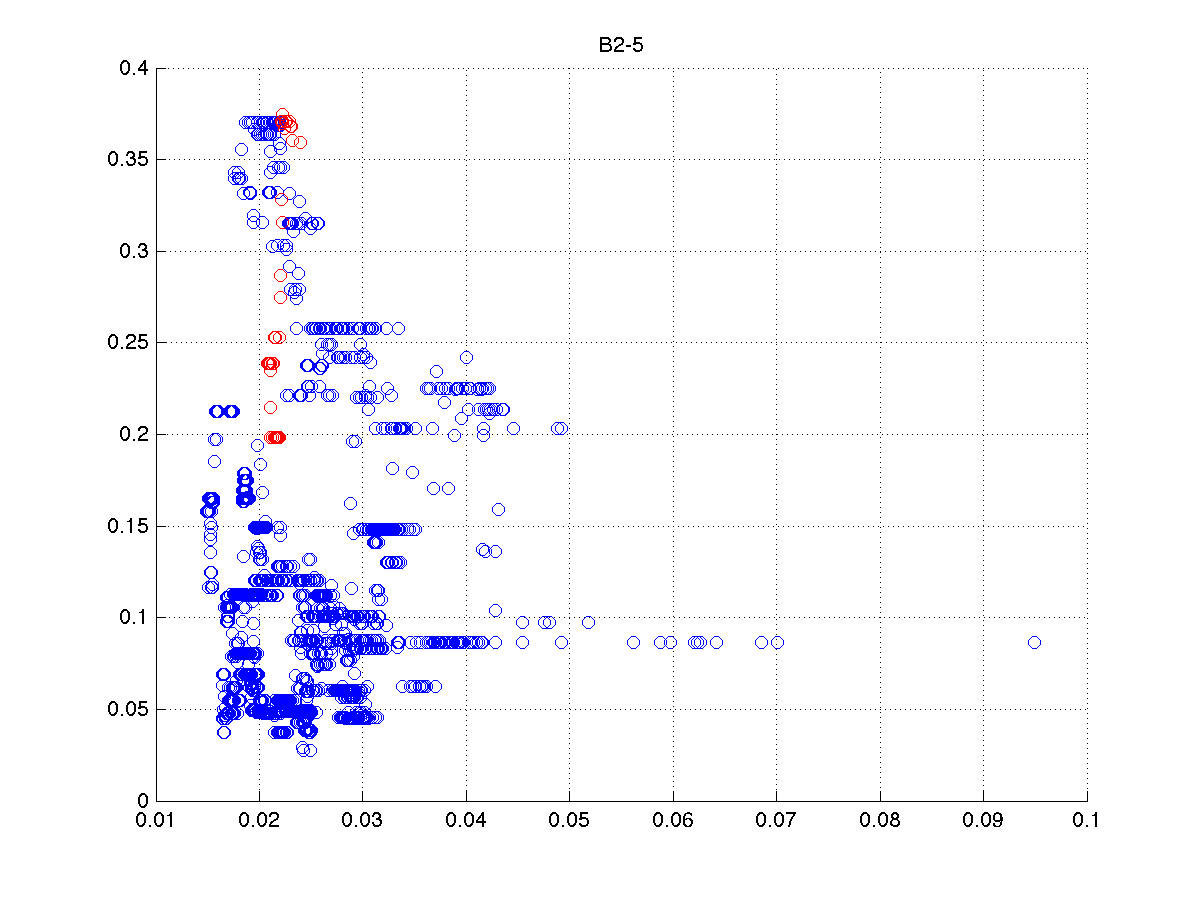}}\\
\subfloat[froben-covar]{\includegraphics[width = 1.7in]{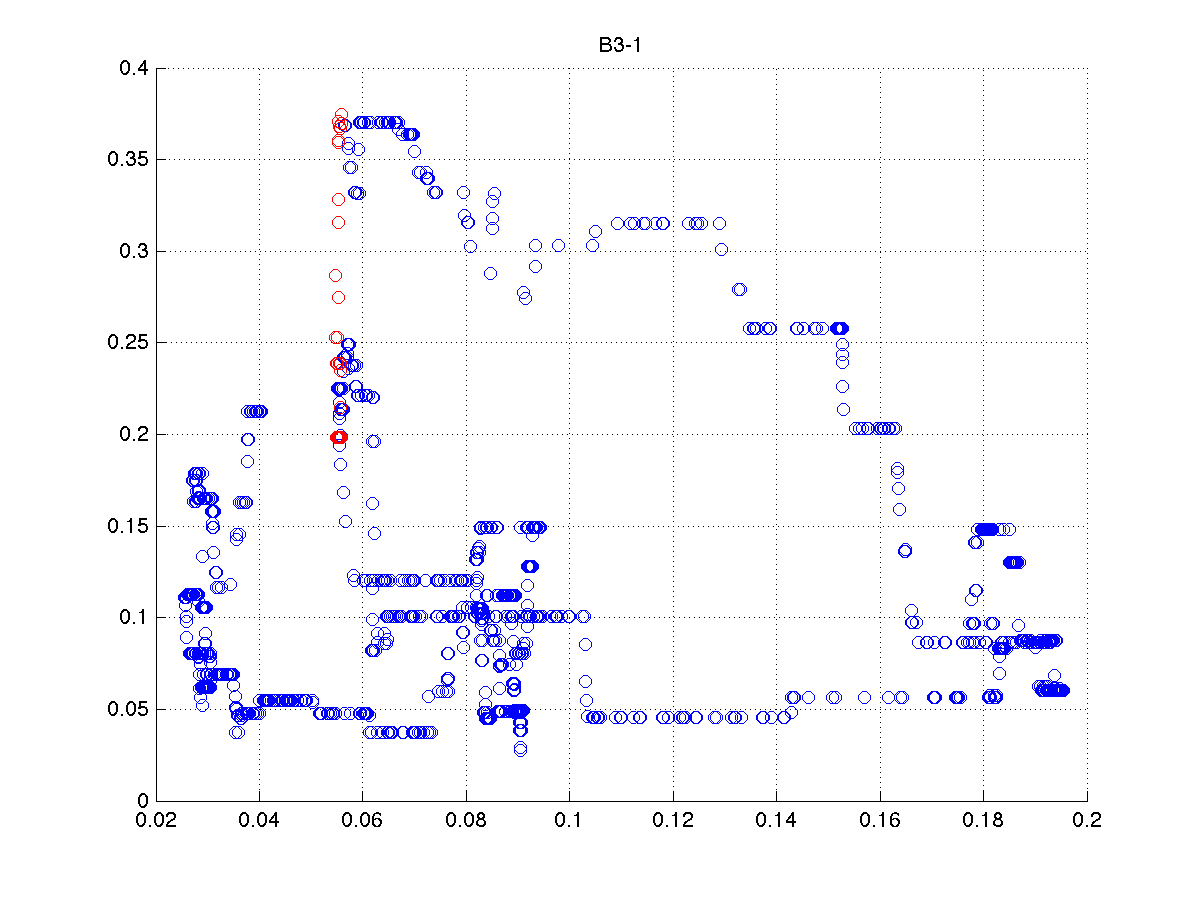}} &
\subfloat[froben-correl]{\includegraphics[width = 1.7in]{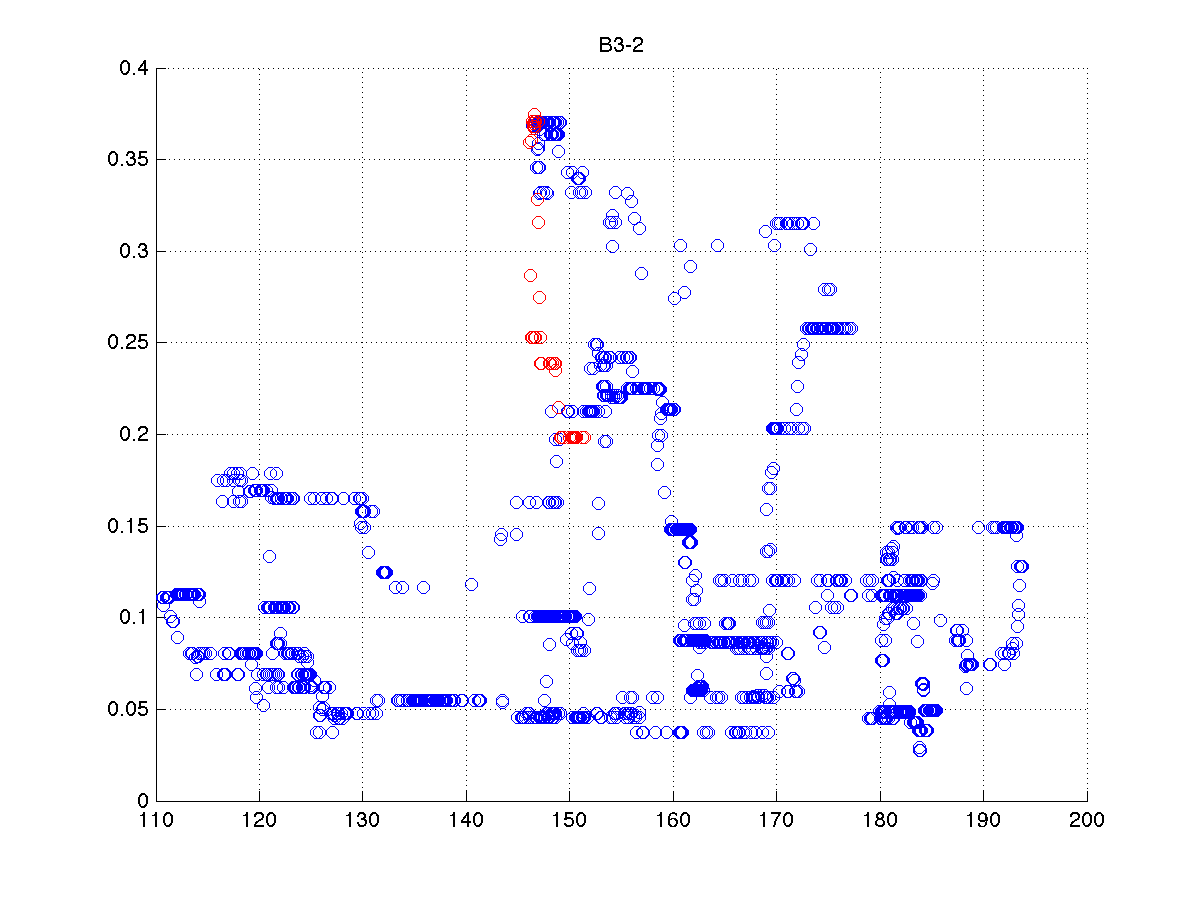}} &
\subfloat[froben-correl-volume]{\includegraphics[width = 1.7in]{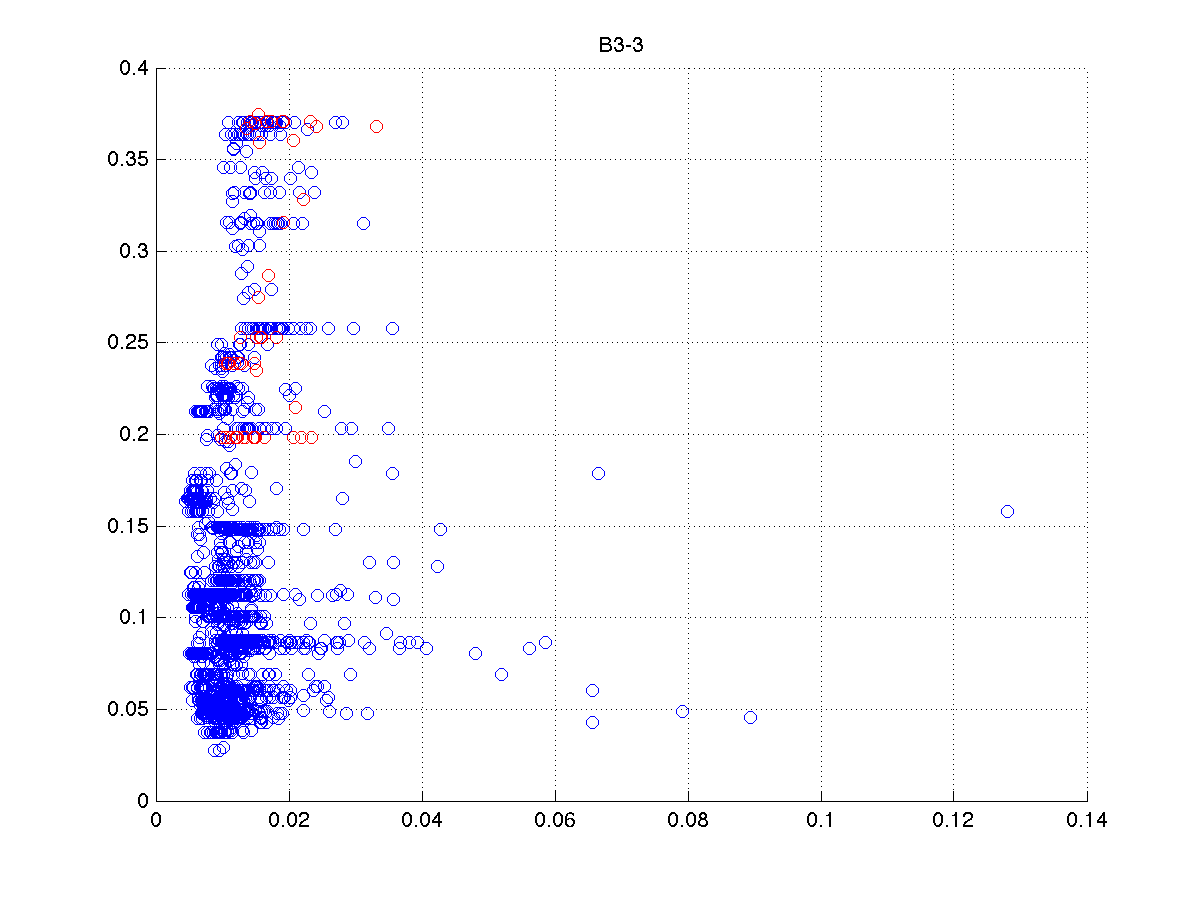}}\\
\subfloat[froben-correl-mcap]{\includegraphics[width = 1.7in]{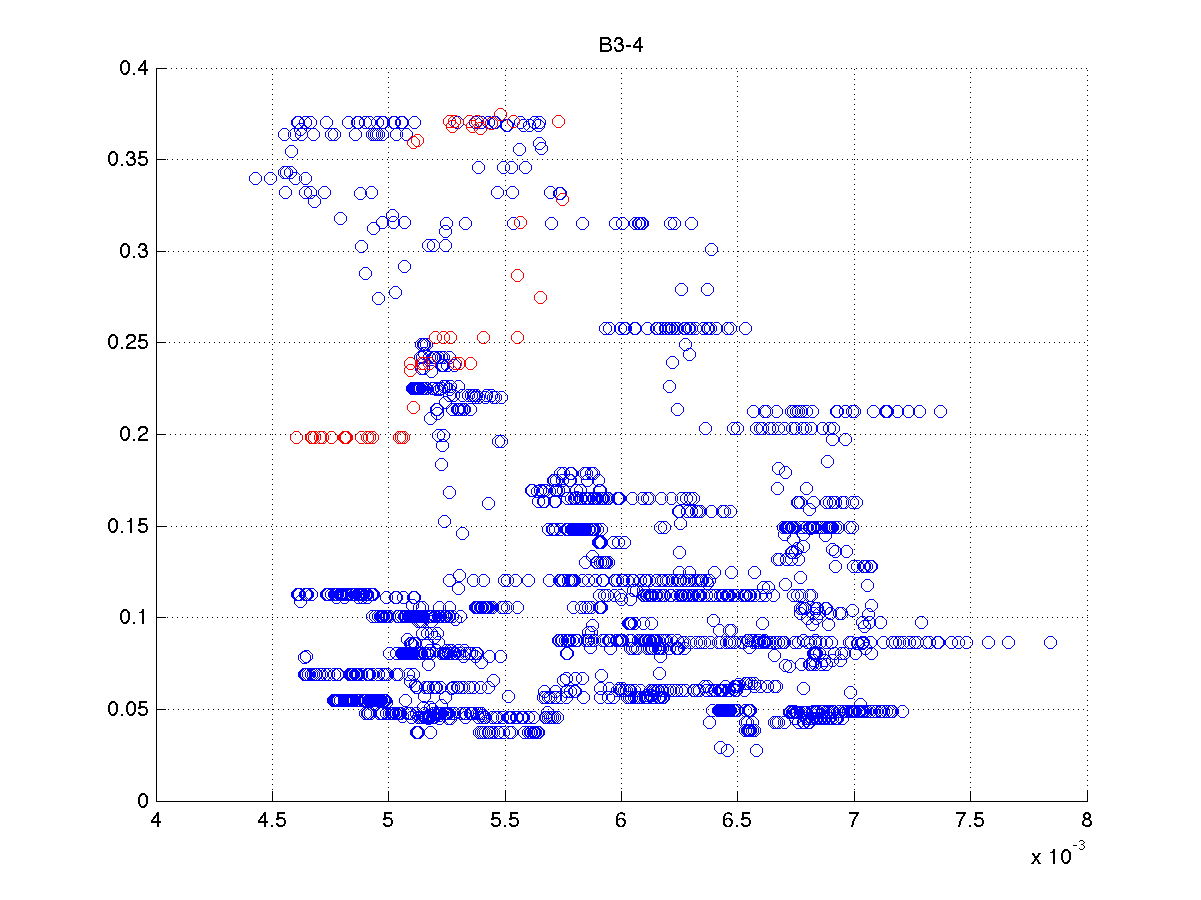}} &
\subfloat[froben-correl-leverage]{\includegraphics[width = 1.7in]{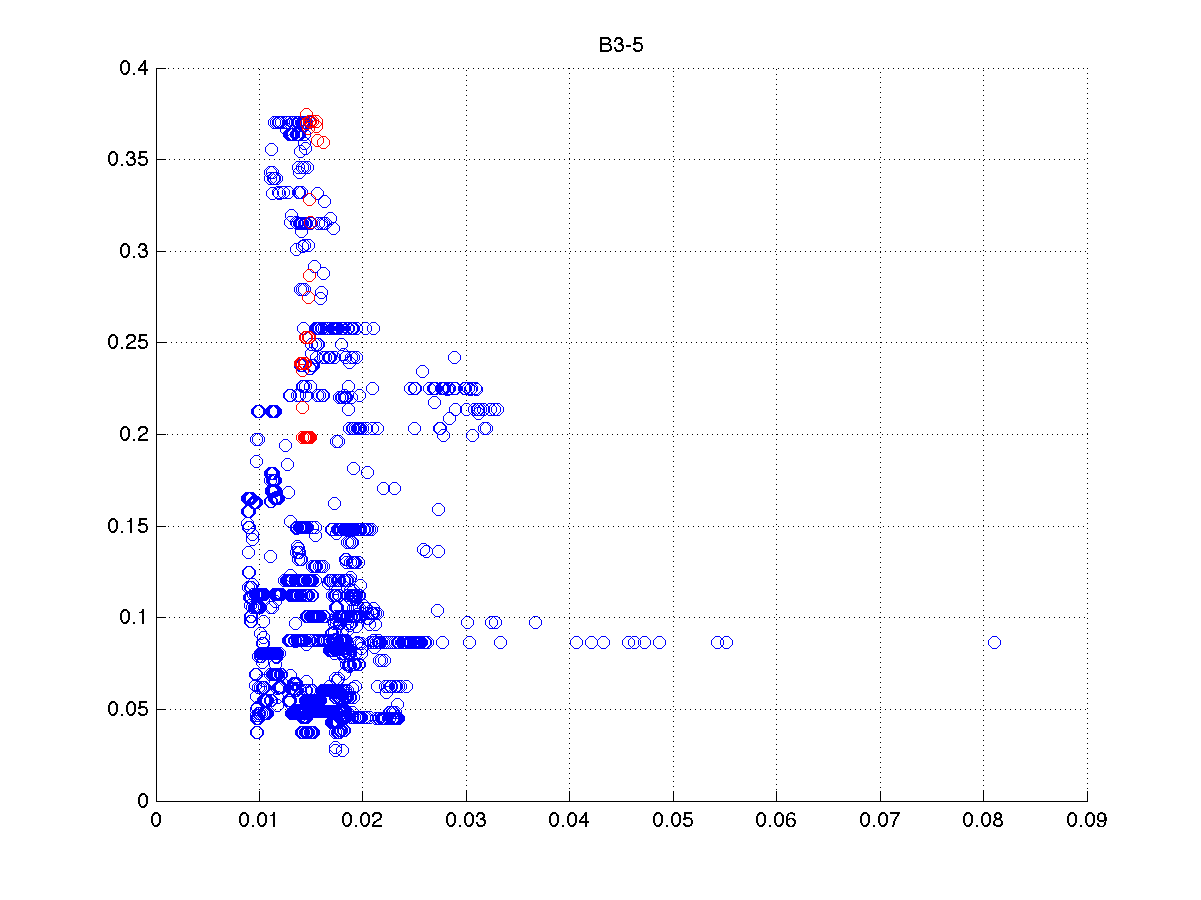}} &
\end{tabular}
\captionsetup{labelformat=empty}
\caption{BE500: Indicators of the $\beta$-series. Red: in-sample ; Blue: out-of-sample}
\end{figure}

\begin{figure}[H]
\begin{tabular}{ccc}
\subfloat[$\mathscr{R}_{1}$covar]{\includegraphics[width = 1.7in]{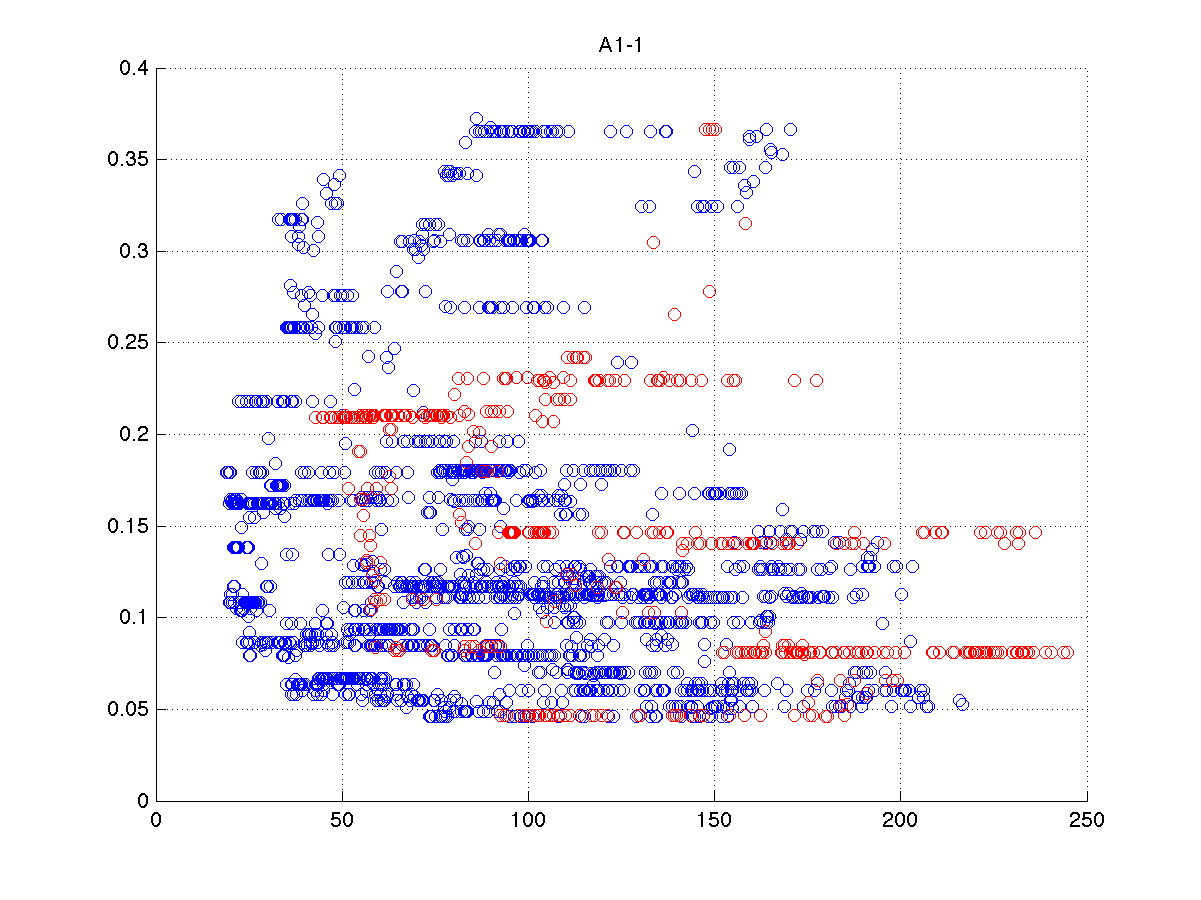}} &
\subfloat[$\mathscr{R}_{1}$correl]{\includegraphics[width = 1.7in]{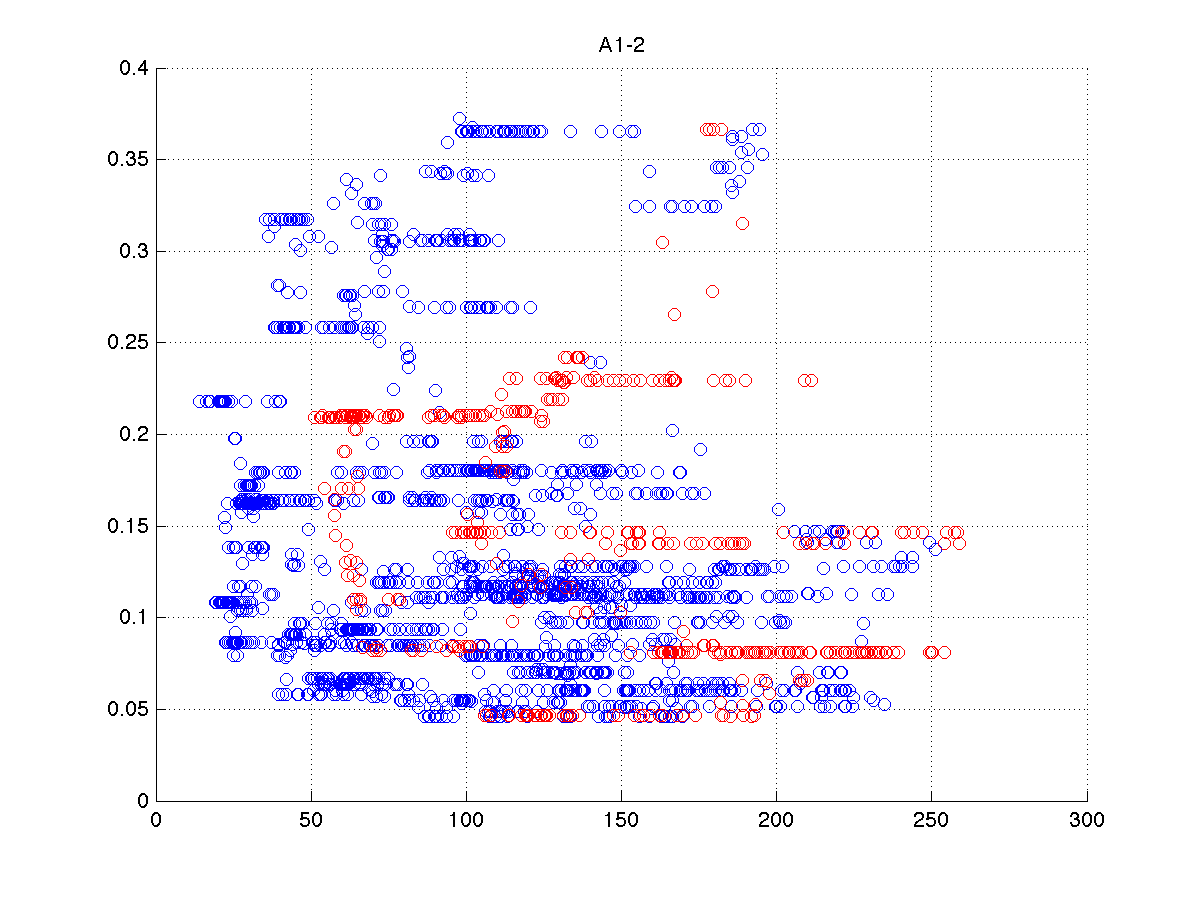}} &
\subfloat[$\mathscr{R}_{1}$correl-volume]{\includegraphics[width = 1.7in]{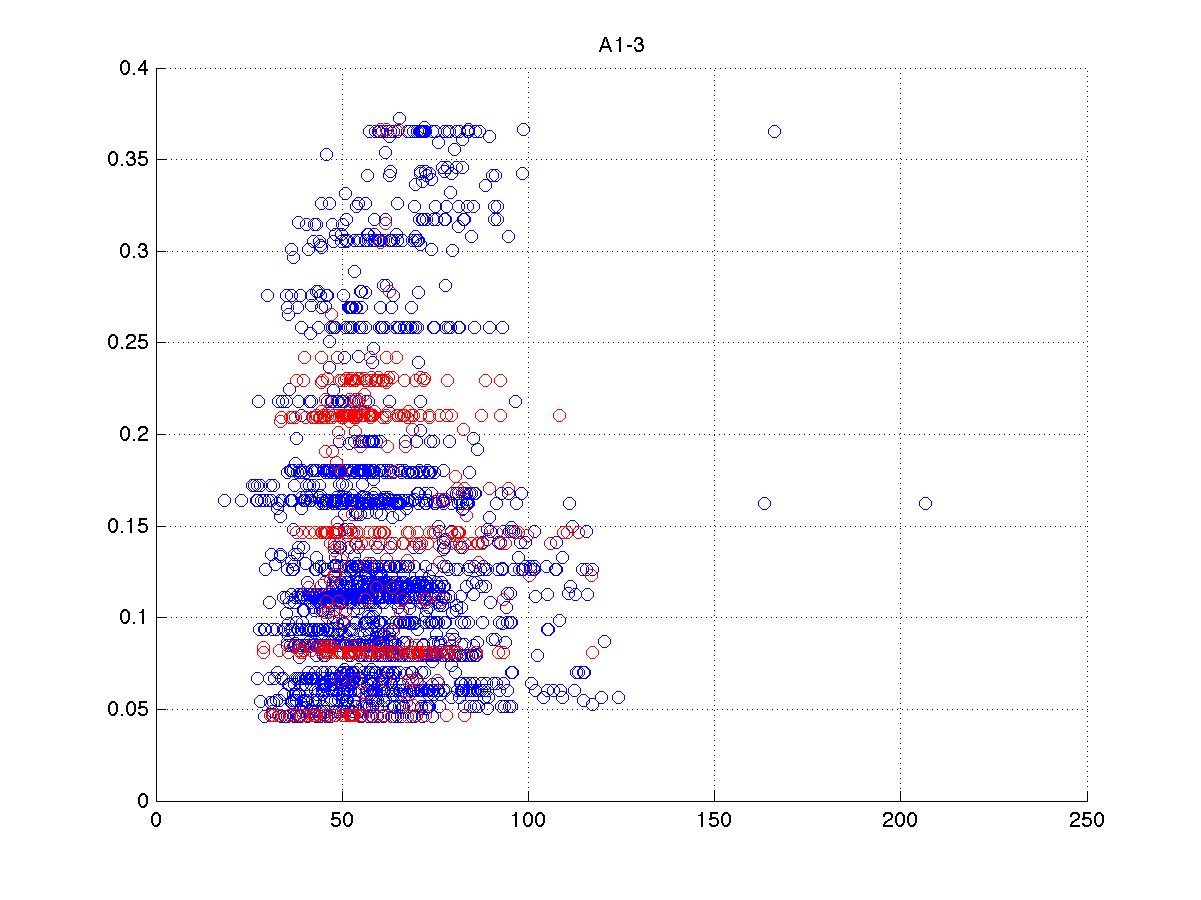}}\\
\subfloat[$\mathscr{R}_{1}$correl-mcap]{\includegraphics[width = 1.7in]{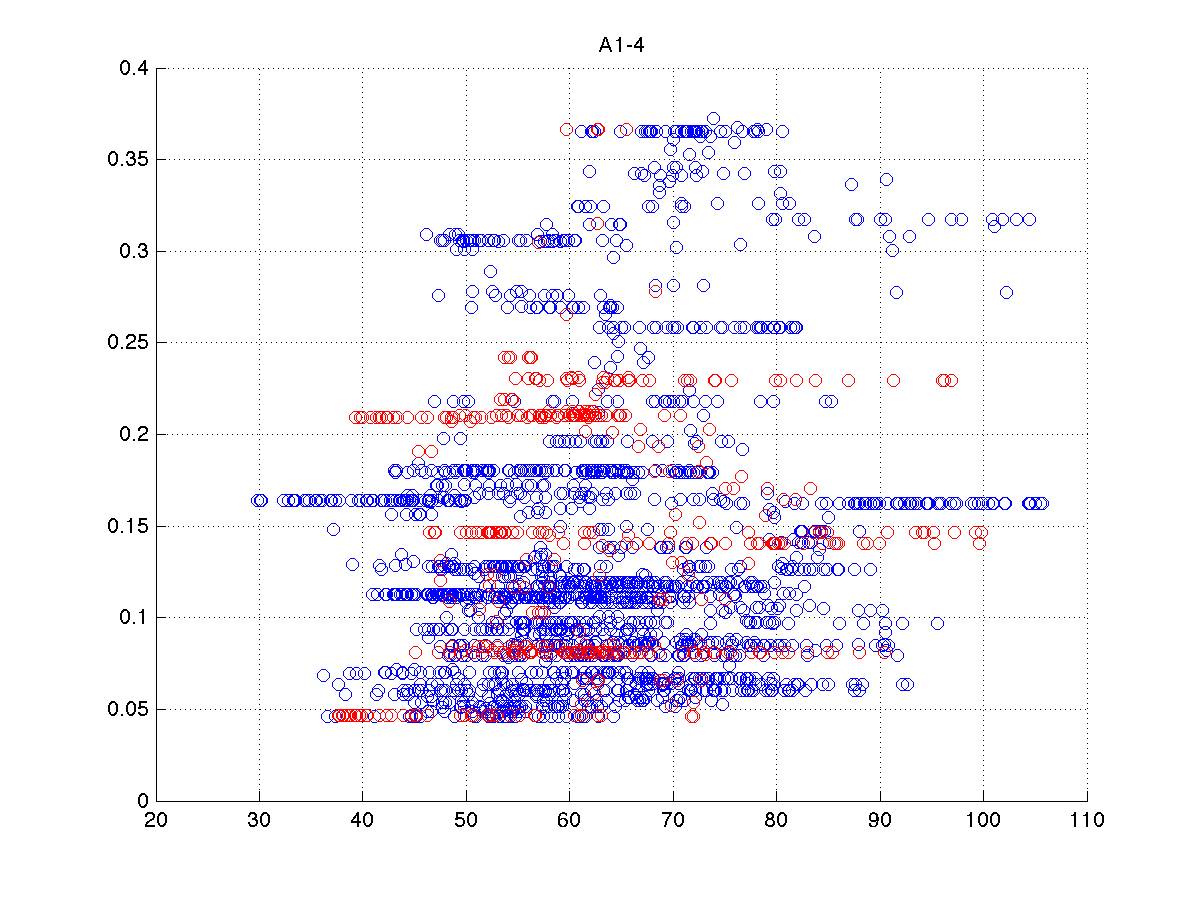}}&
\subfloat[$\mathscr{R}_{1}$correl-leverage]{\includegraphics[width = 1.7in]{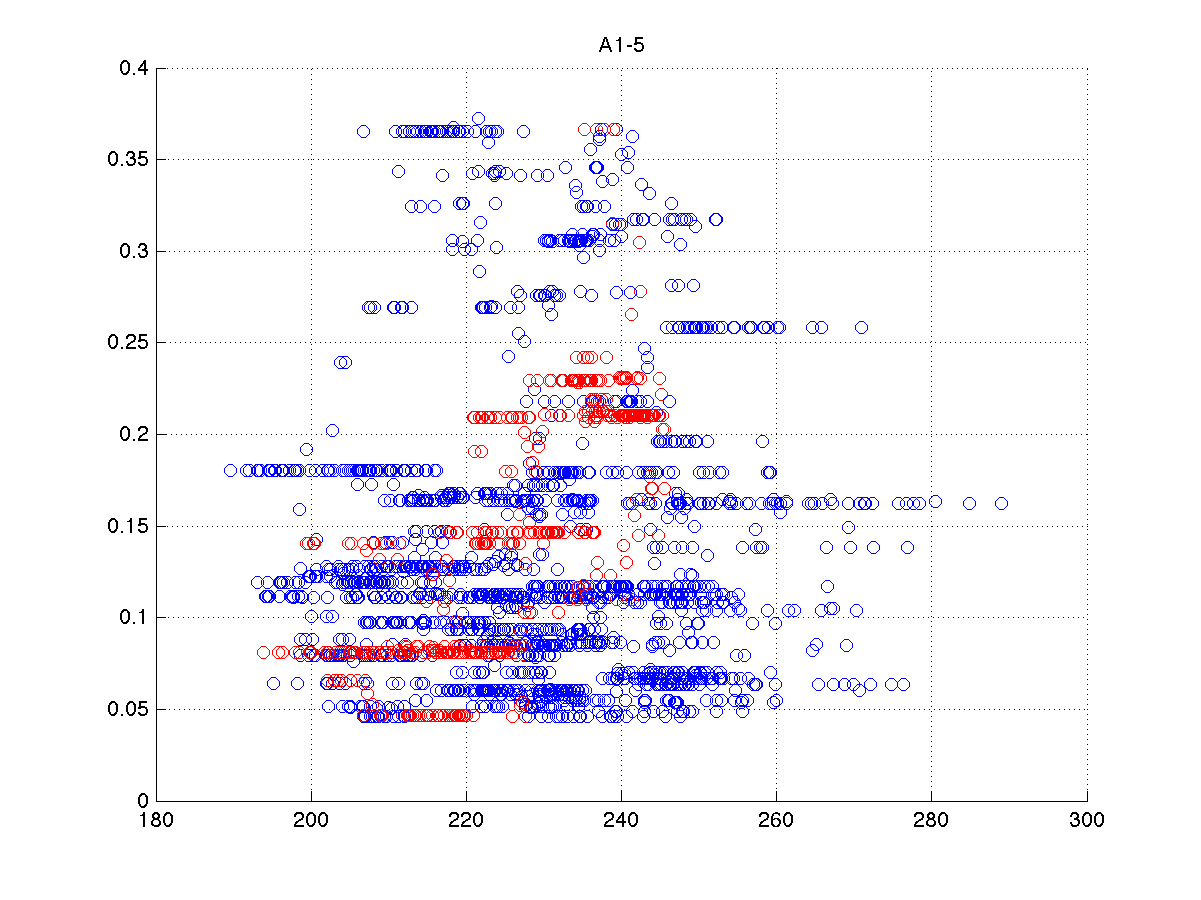}} &
\subfloat[$\mathscr{R}_{2}$covar]{\includegraphics[width = 1.7in]{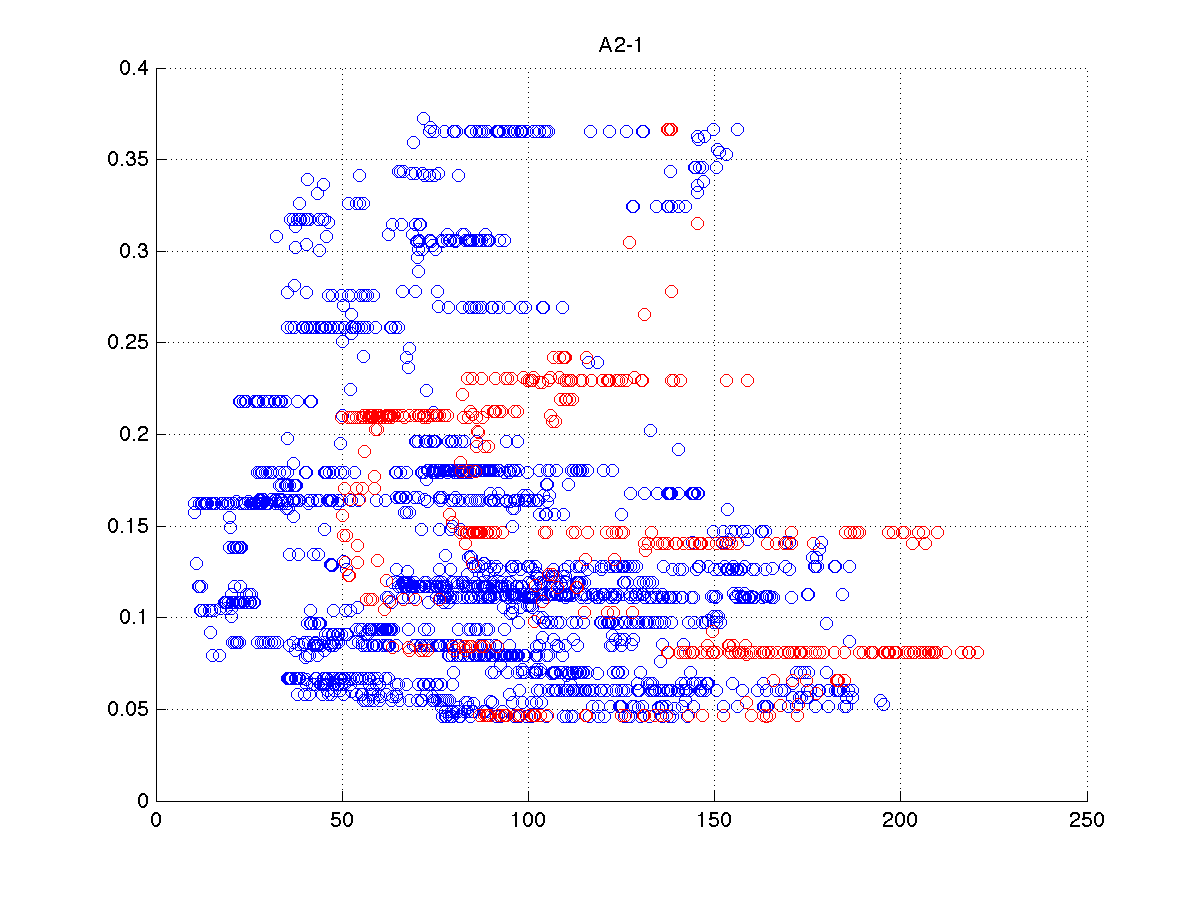}}\\
\subfloat[$\mathscr{R}_{2}$correl]{\includegraphics[width = 1.7in]{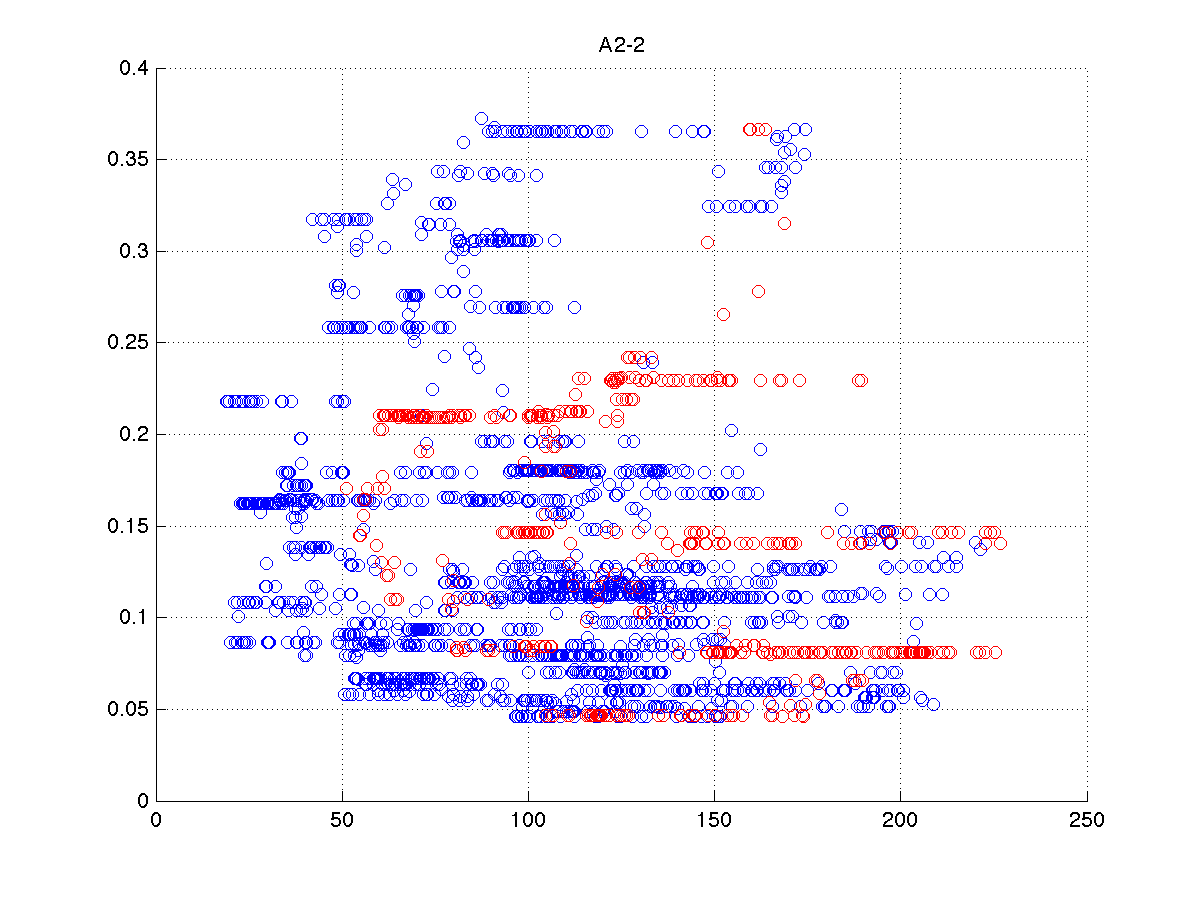}} &
\subfloat[$\mathscr{R}_{2}$correl-volume]{\includegraphics[width = 1.7in]{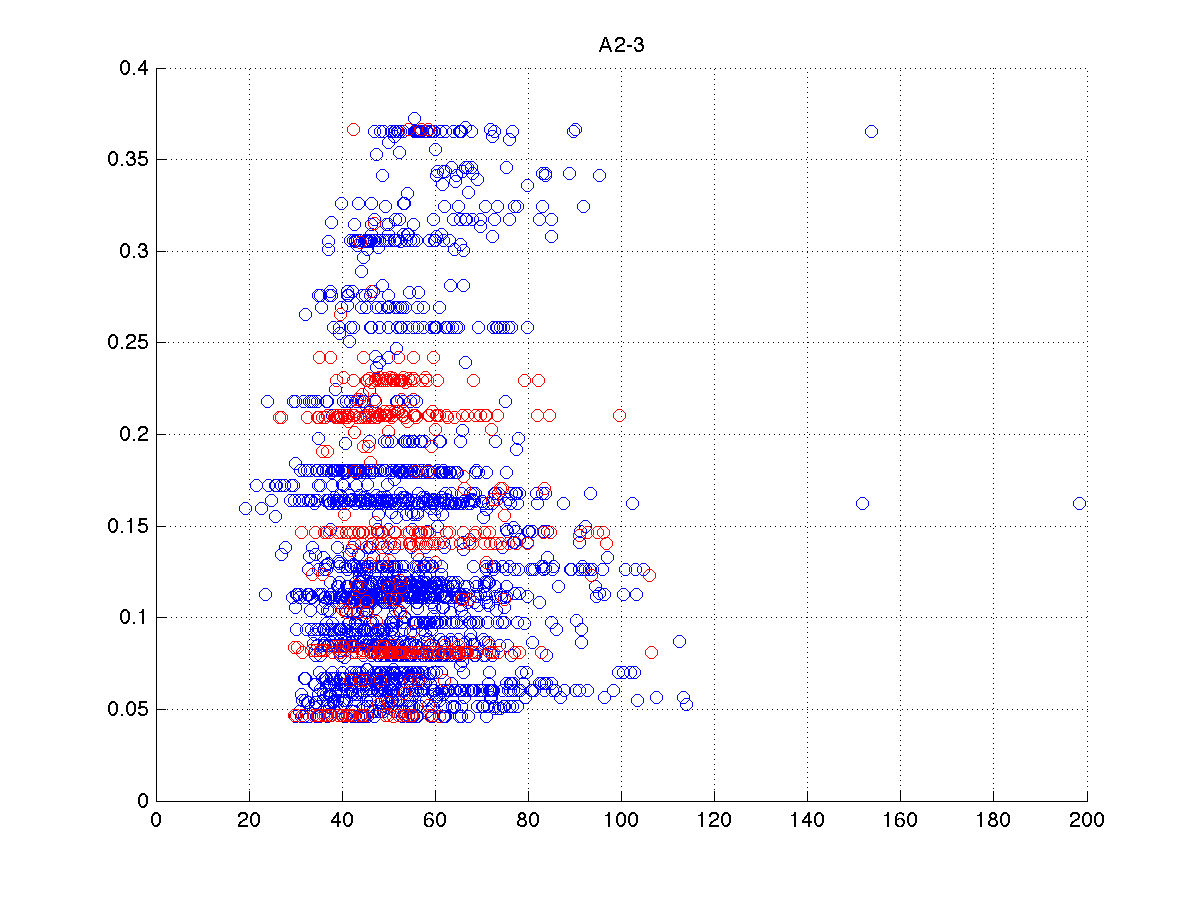}}&
\subfloat[$\mathscr{R}_{2}$correl-mcap]{\includegraphics[width = 1.7in]{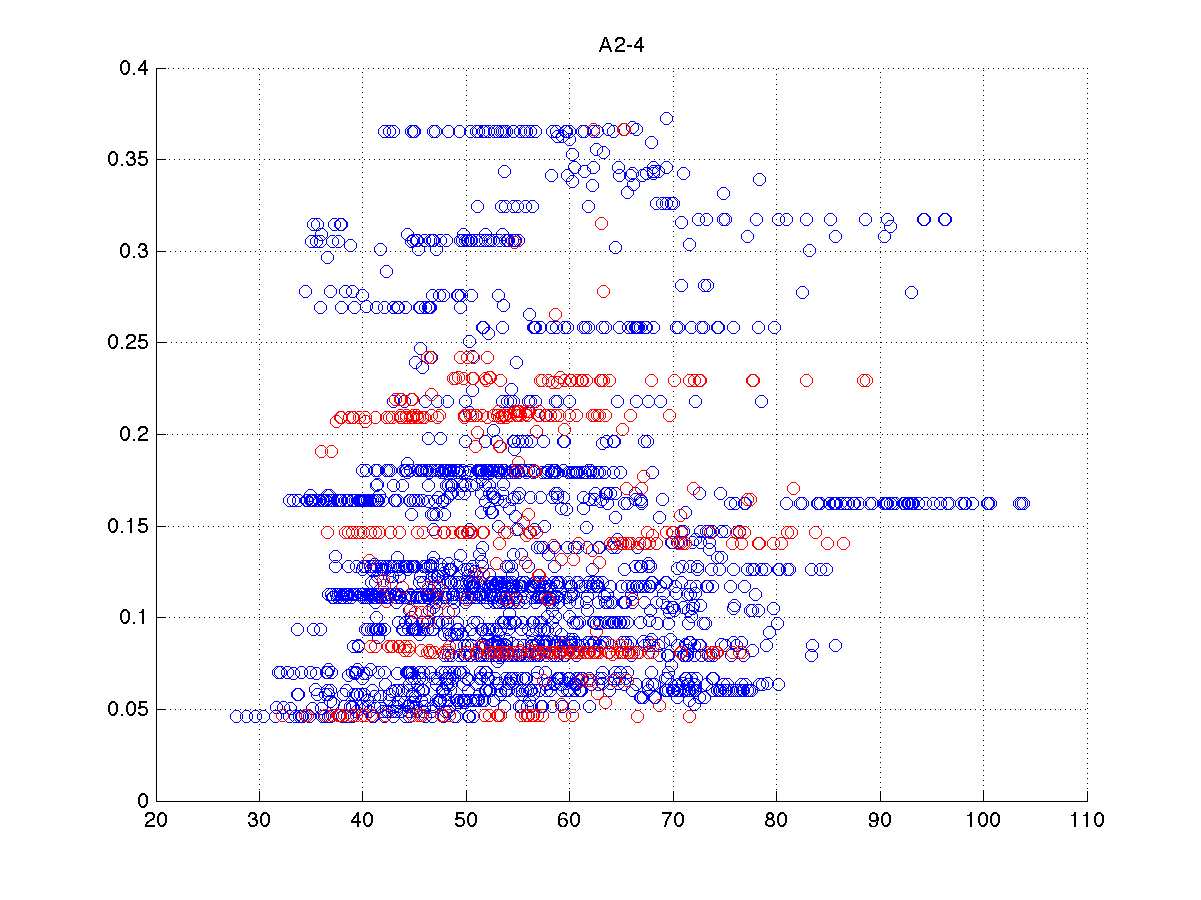}}\\
\subfloat[$\mathscr{R}_{2}$correl-leverage]{\includegraphics[width = 1.7in]{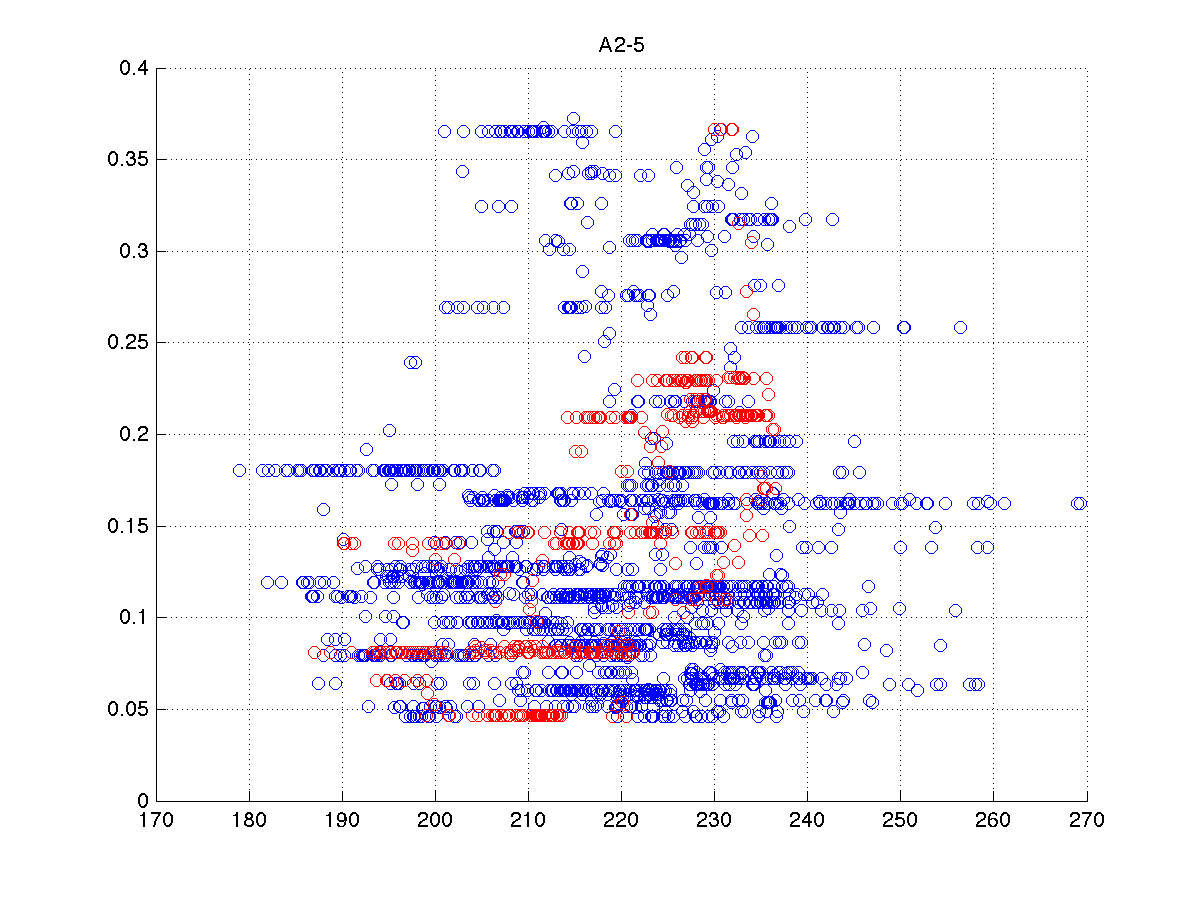}} &
\subfloat[$\mathscr{R}_{3}$covar]{\includegraphics[width = 1.7in]{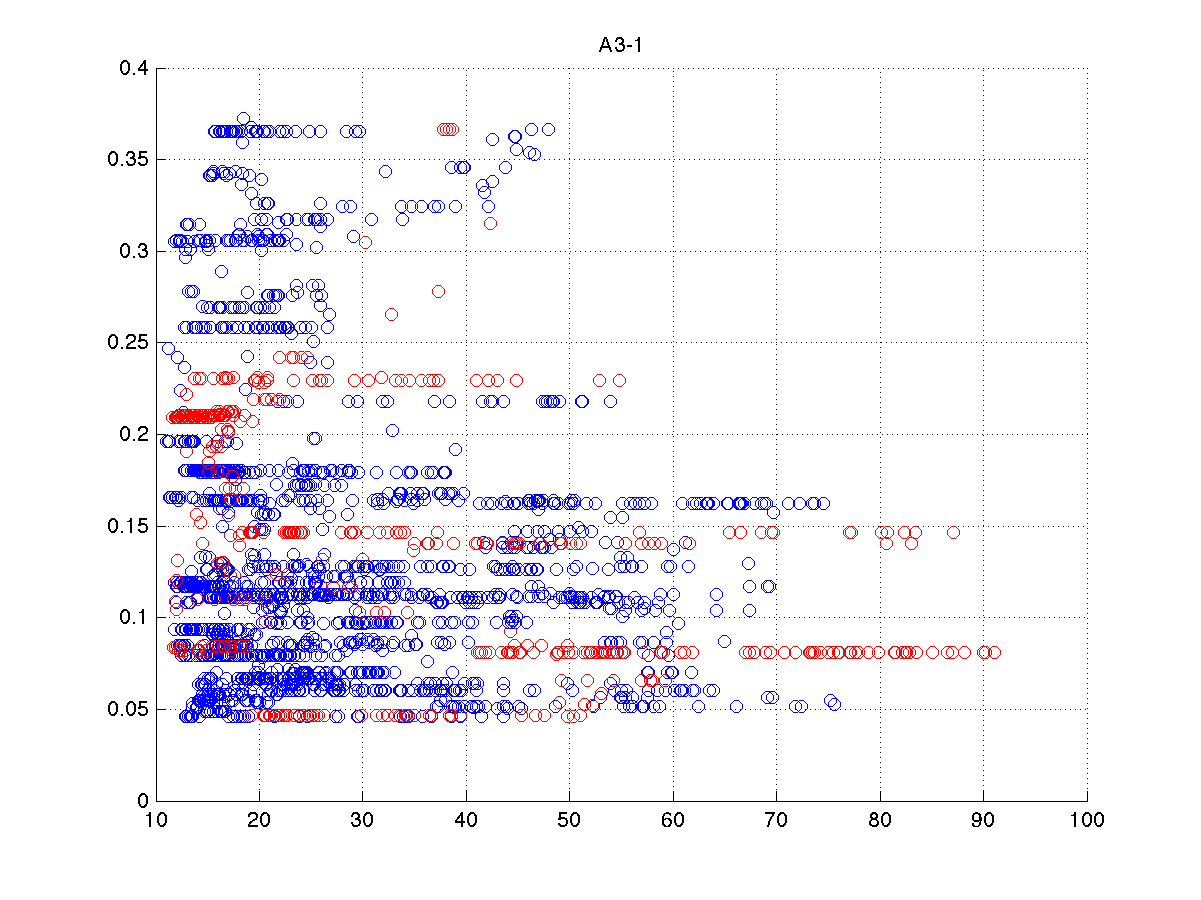}} &
\subfloat[$\mathscr{R}_{3}$correl]{\includegraphics[width = 1.7in]{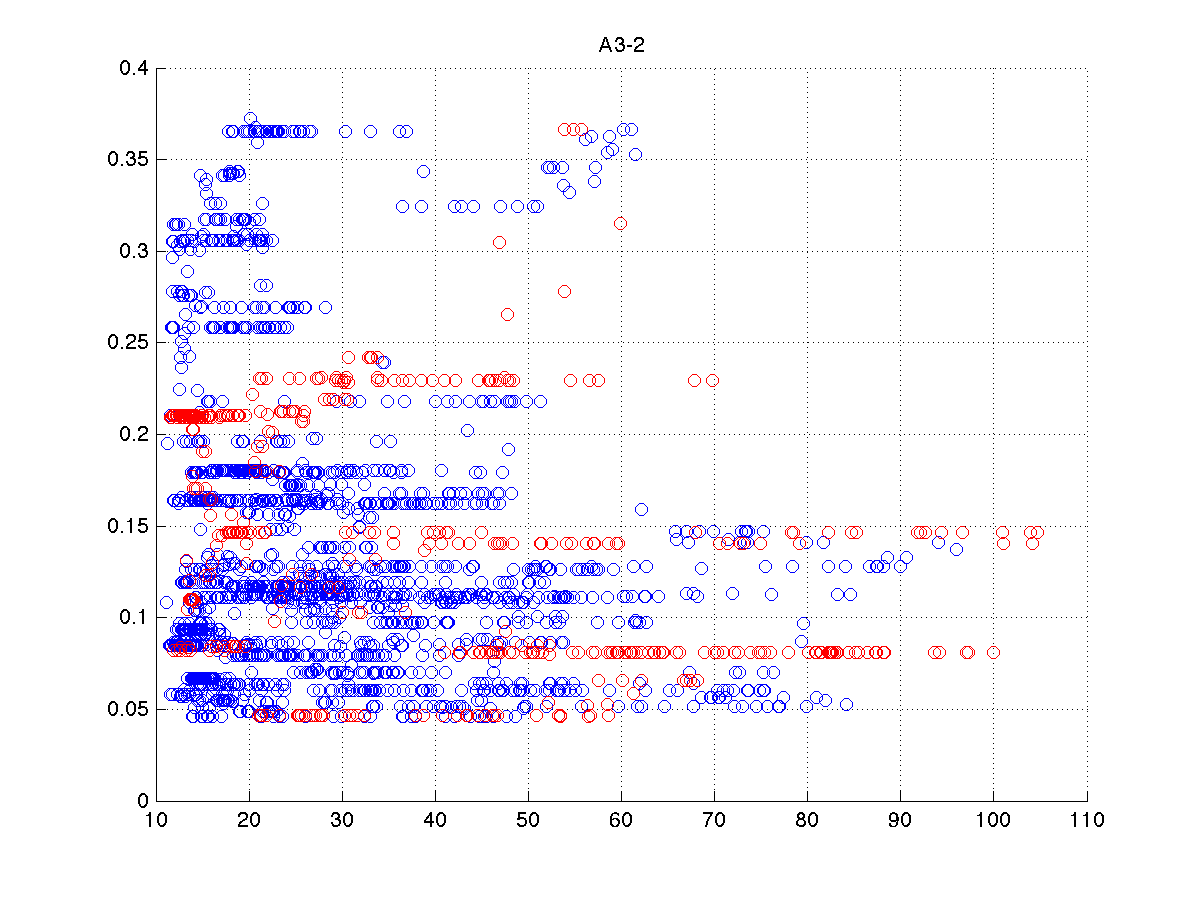}}\\
\subfloat[$\mathscr{R}_{3}$correl-volume]{\includegraphics[width = 1.7in]{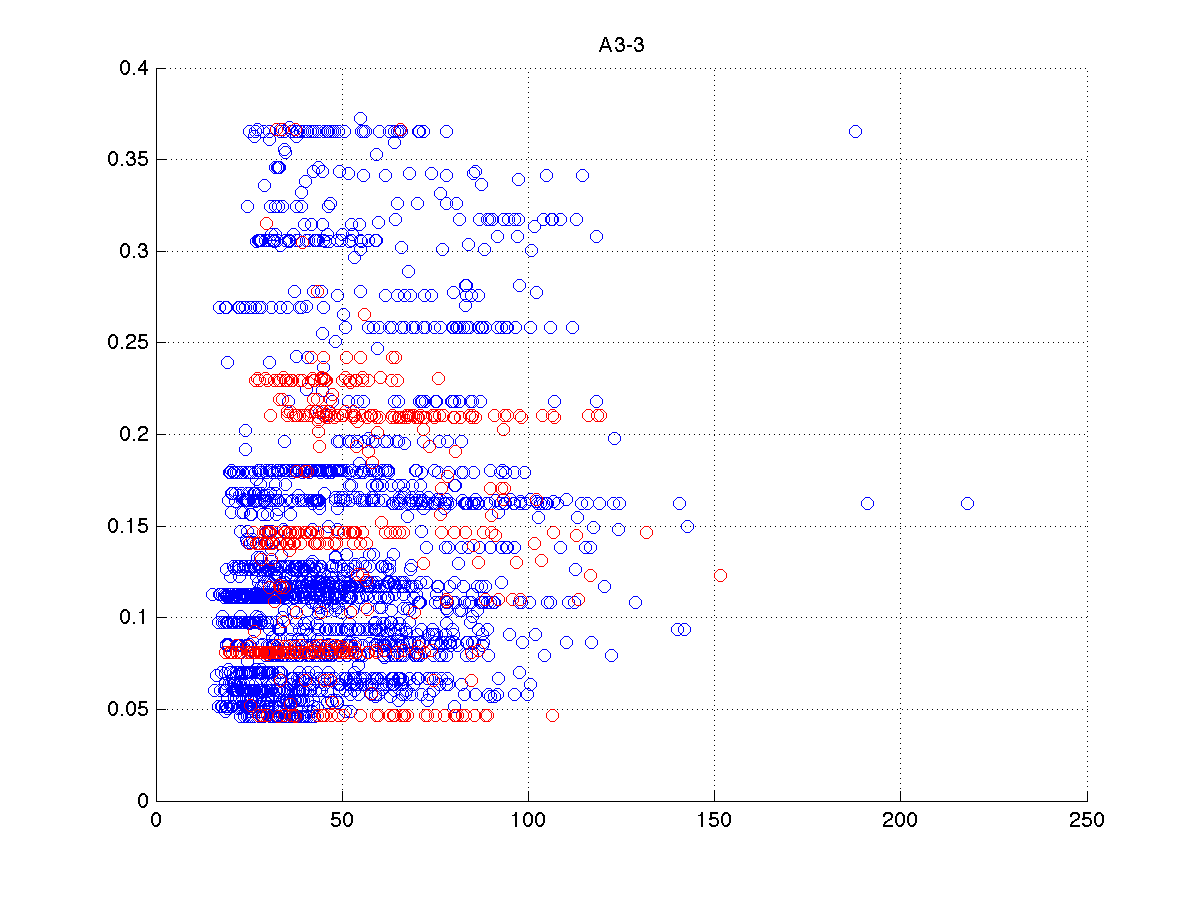}} &
\subfloat[$\mathscr{R}_{3}$correl-mcap]{\includegraphics[width = 1.7in]{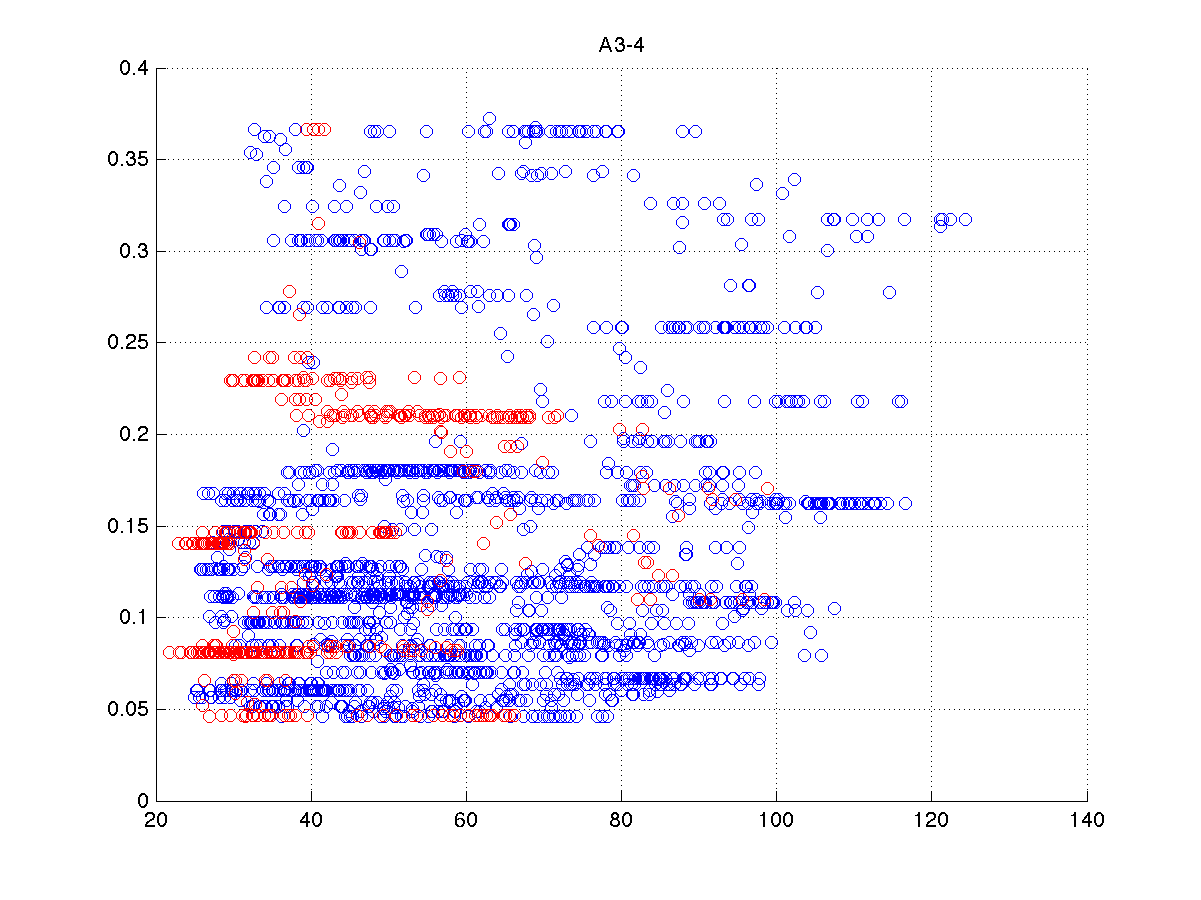}} &
\subfloat[$\mathscr{R}_{3}$correl-leverage]{\includegraphics[width = 1.7in]{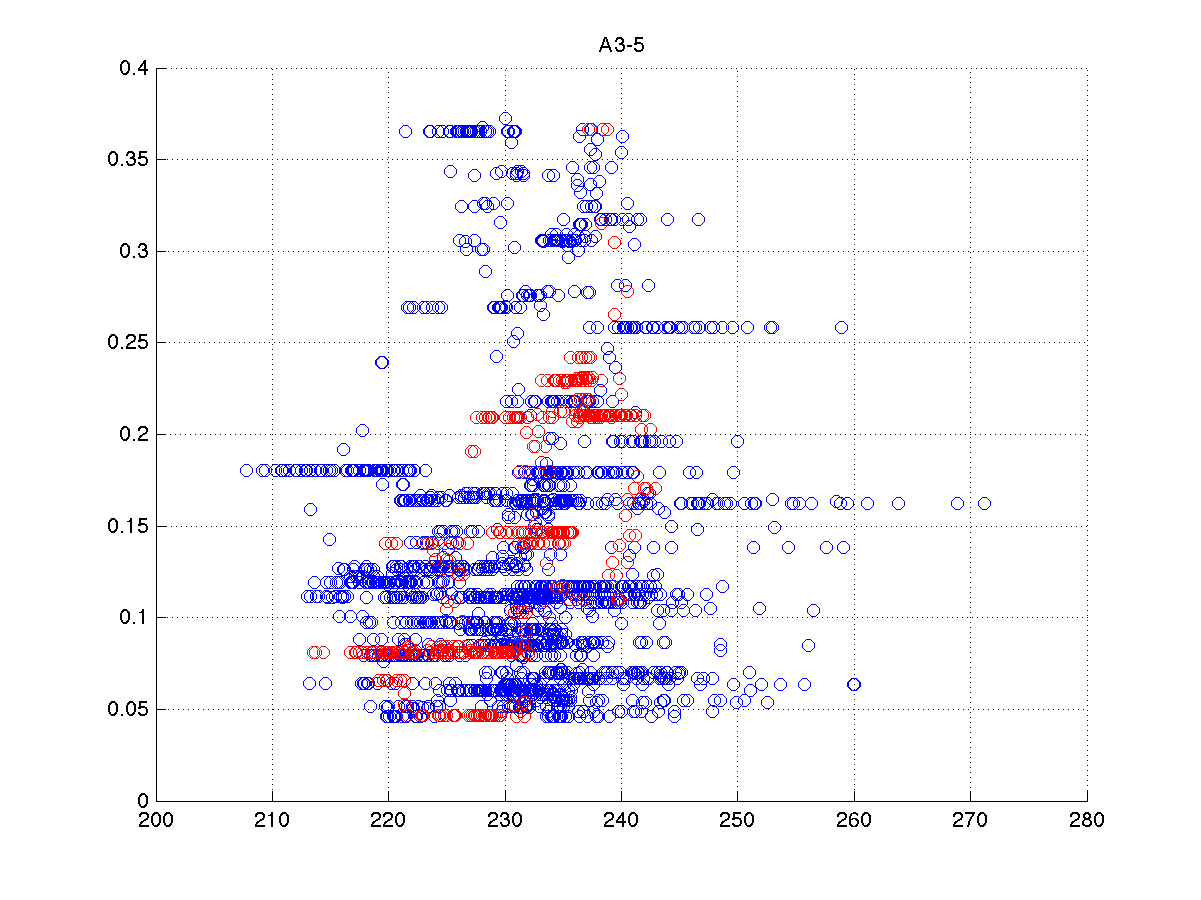}}\\
\end{tabular}
\captionsetup{labelformat=empty}
\caption{CAC40: Indicators of the $\alpha$-series. Red: in-sample ; Blue: out-of-sample}
\end{figure}

\begin{figure}[H]
\begin{tabular}{ccc}
\subfloat[rspec-covar]{\includegraphics[width = 1.7in]{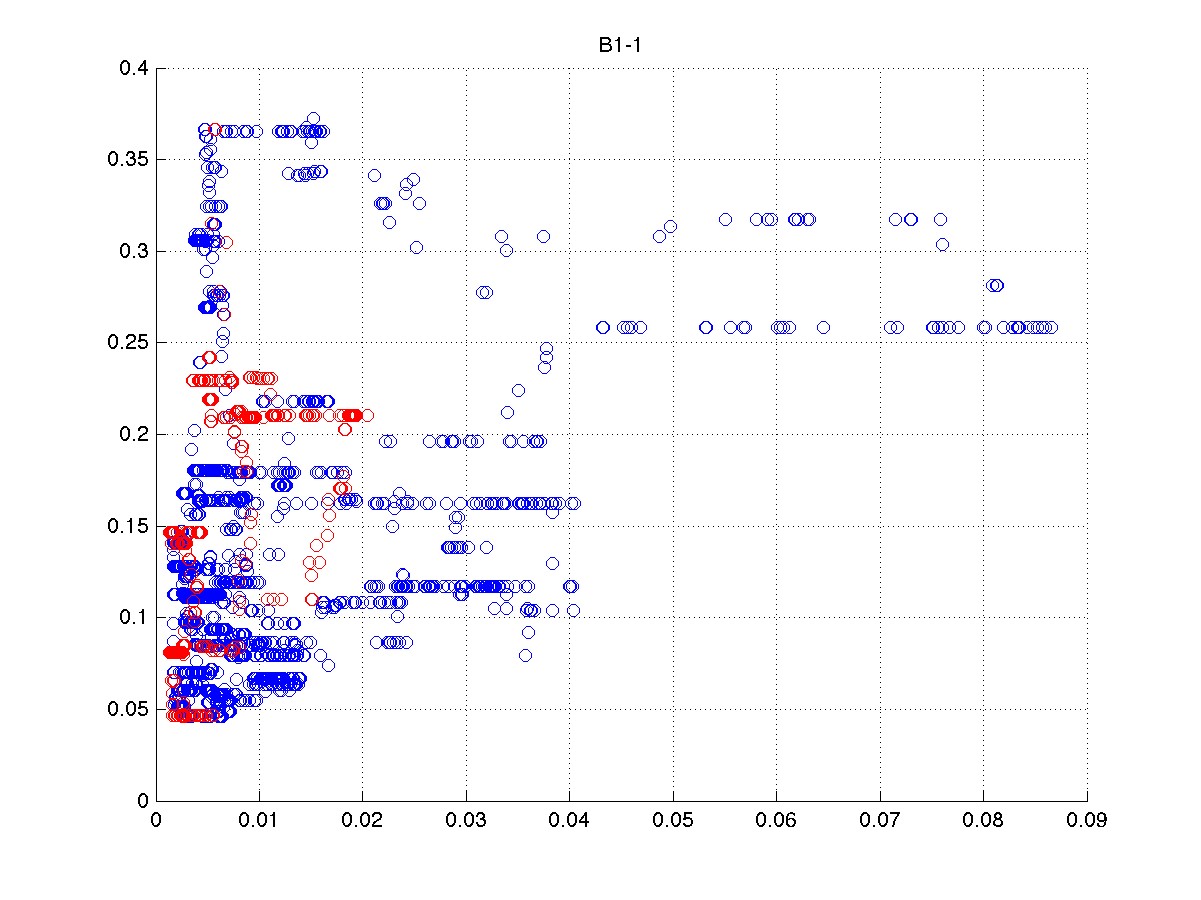}} &
\subfloat[rspec-correl]{\includegraphics[width = 1.7in]{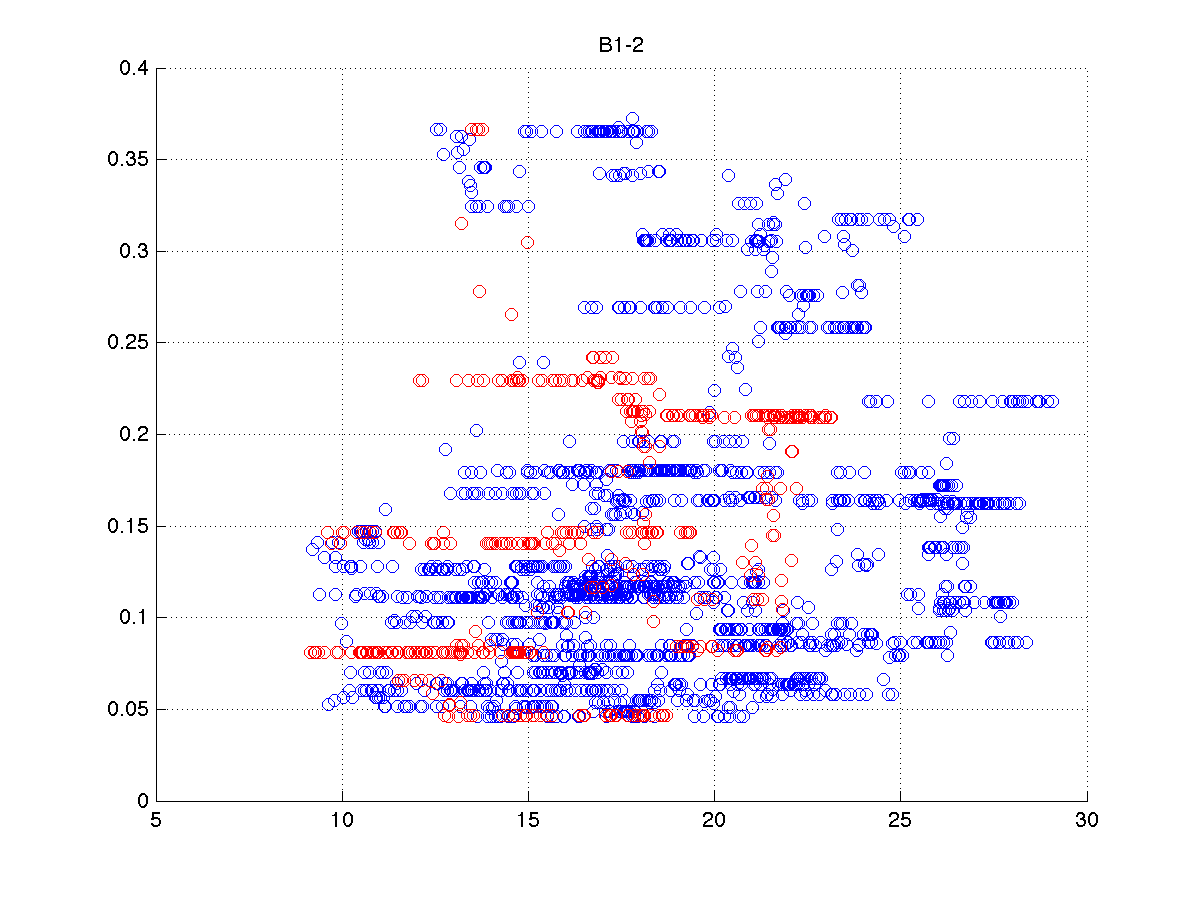}} &
\subfloat[rspec-correl-volume]{\includegraphics[width = 1.7in]{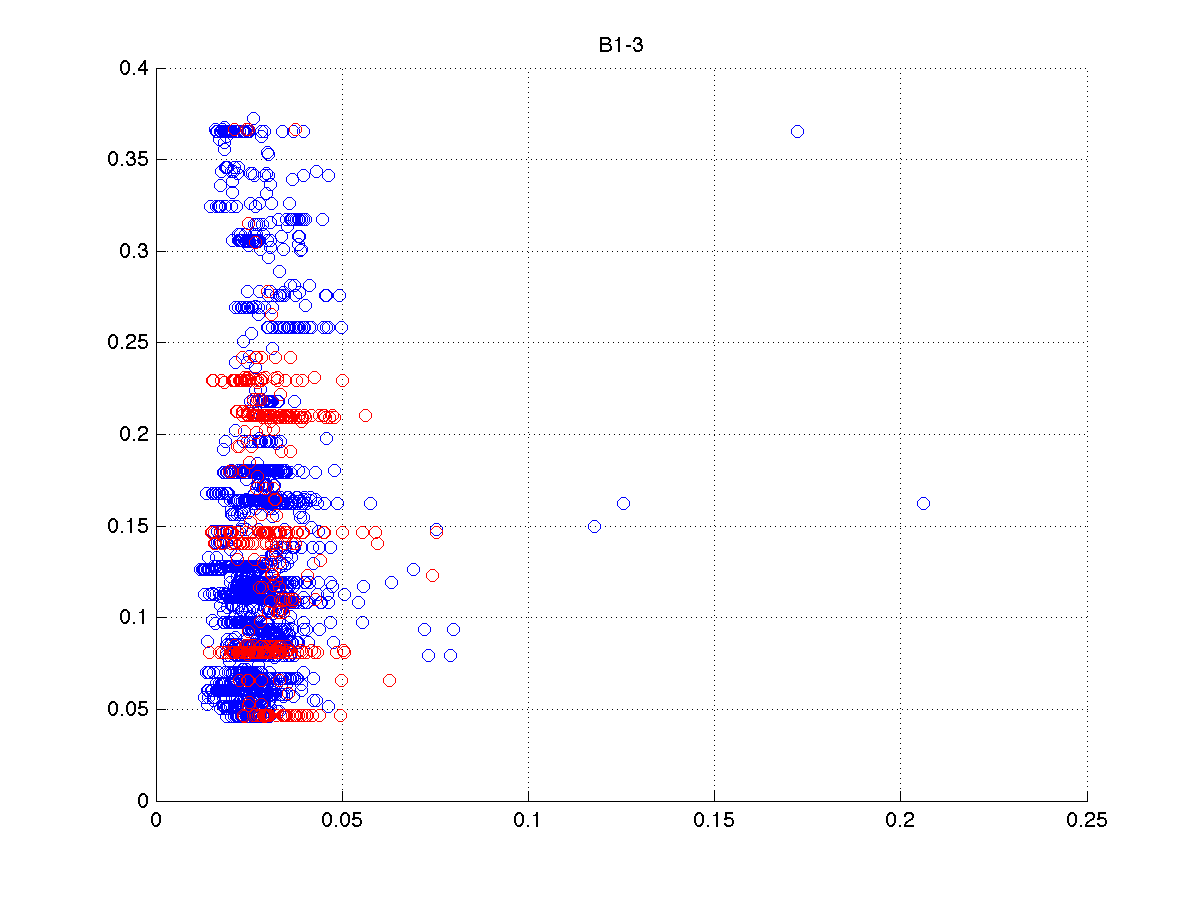}} \\
\subfloat[rspec-correl-mcap]{\includegraphics[width = 1.7in]{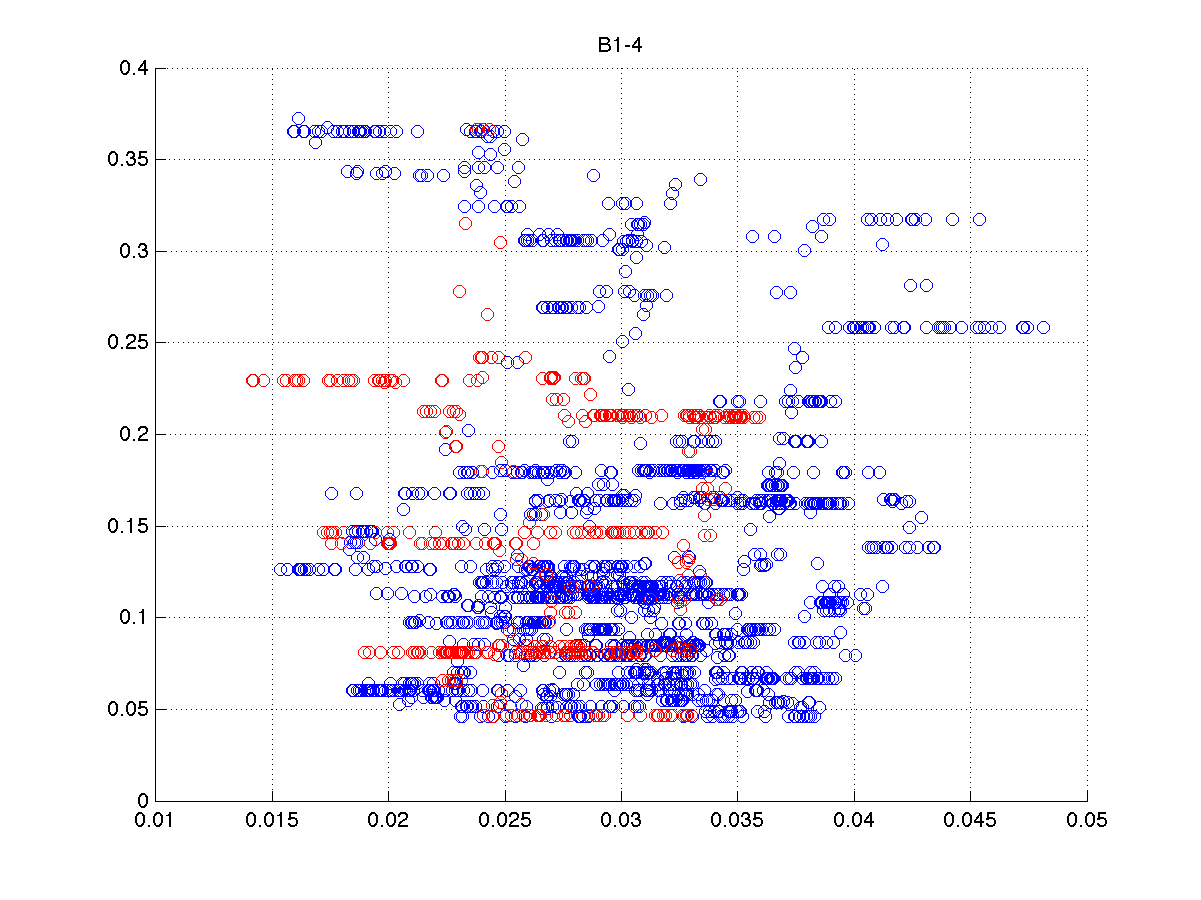}}&
\subfloat[rspec-correl-leverage]{\includegraphics[width = 1.7in]{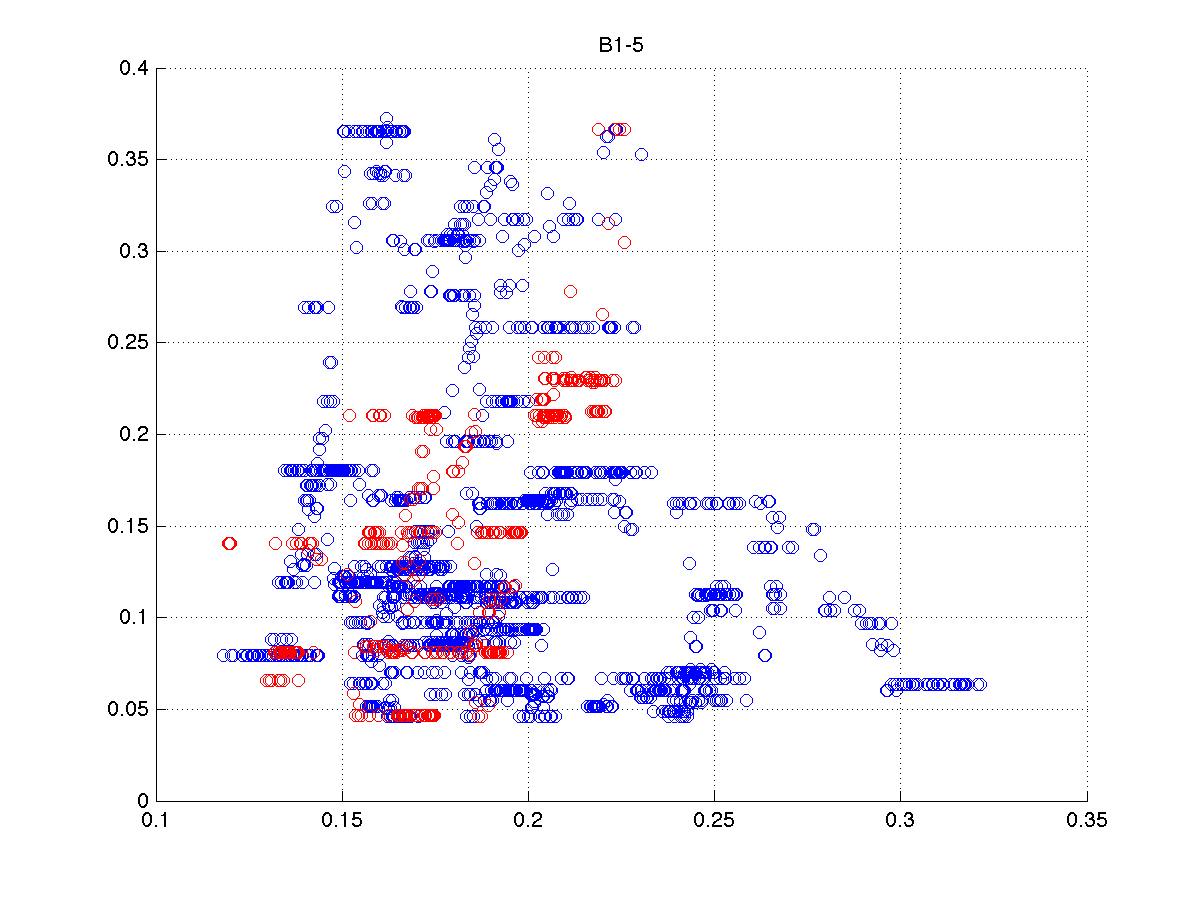}} &
\subfloat[trace-covar]{\includegraphics[width = 1.7in]{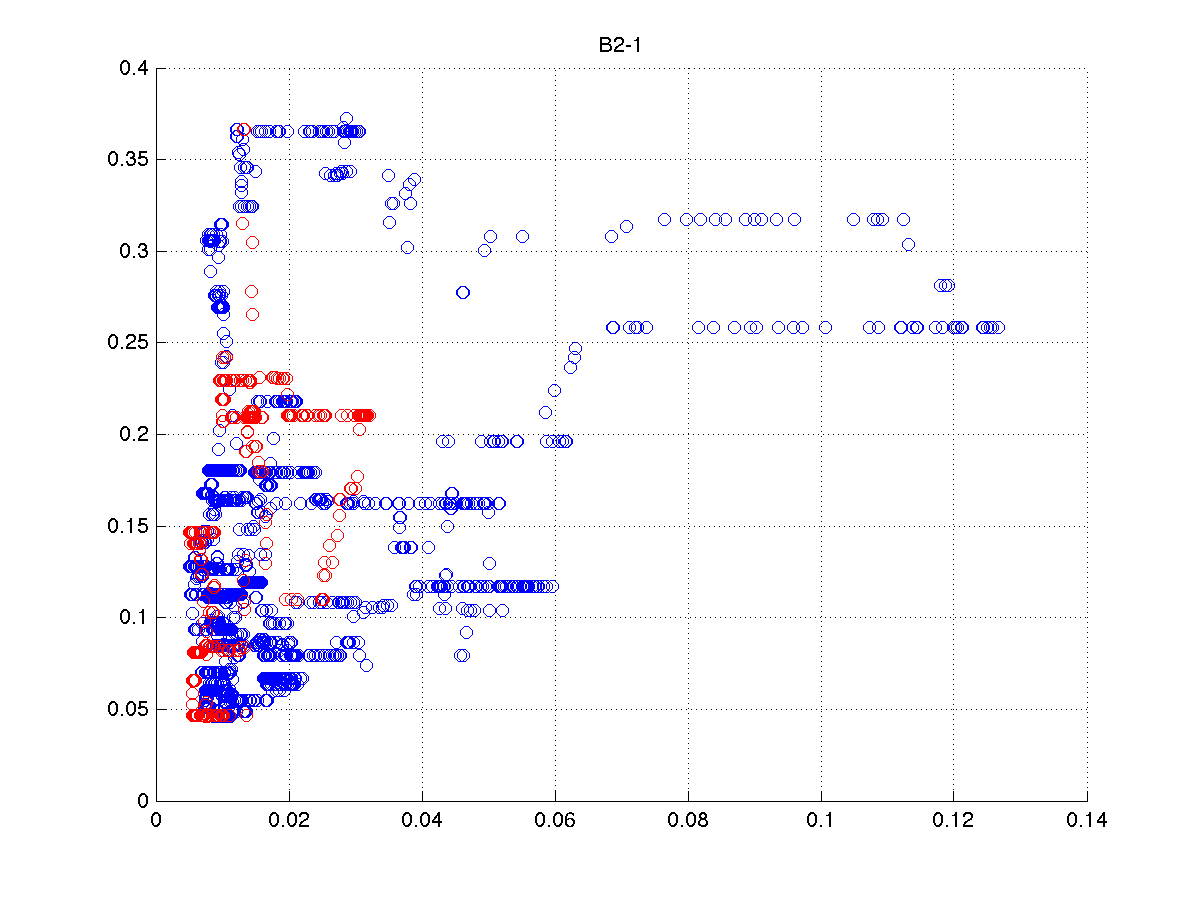}}\\
\subfloat[trace-correl-volume]{\includegraphics[width = 1.7in]{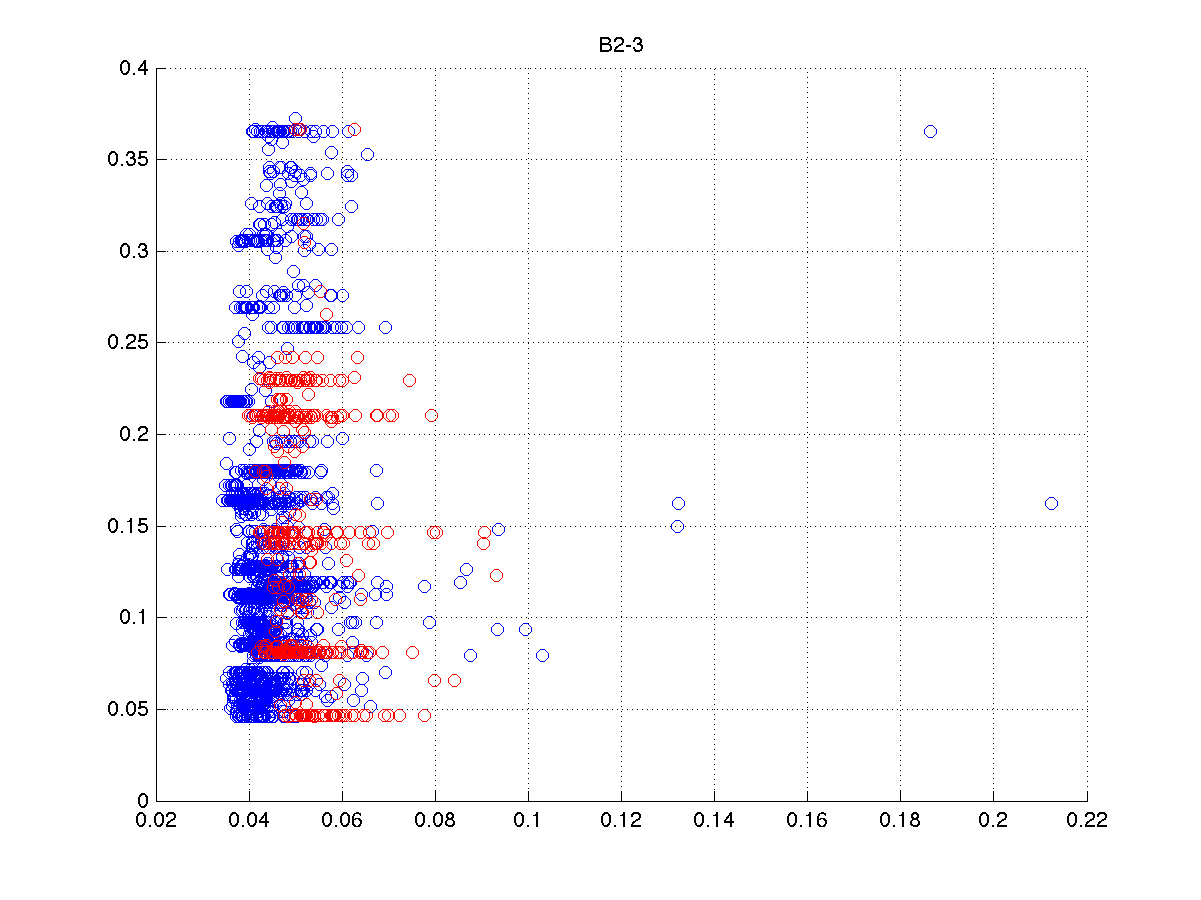}} &
\subfloat[trace-correl-mcap]{\includegraphics[width = 1.7in]{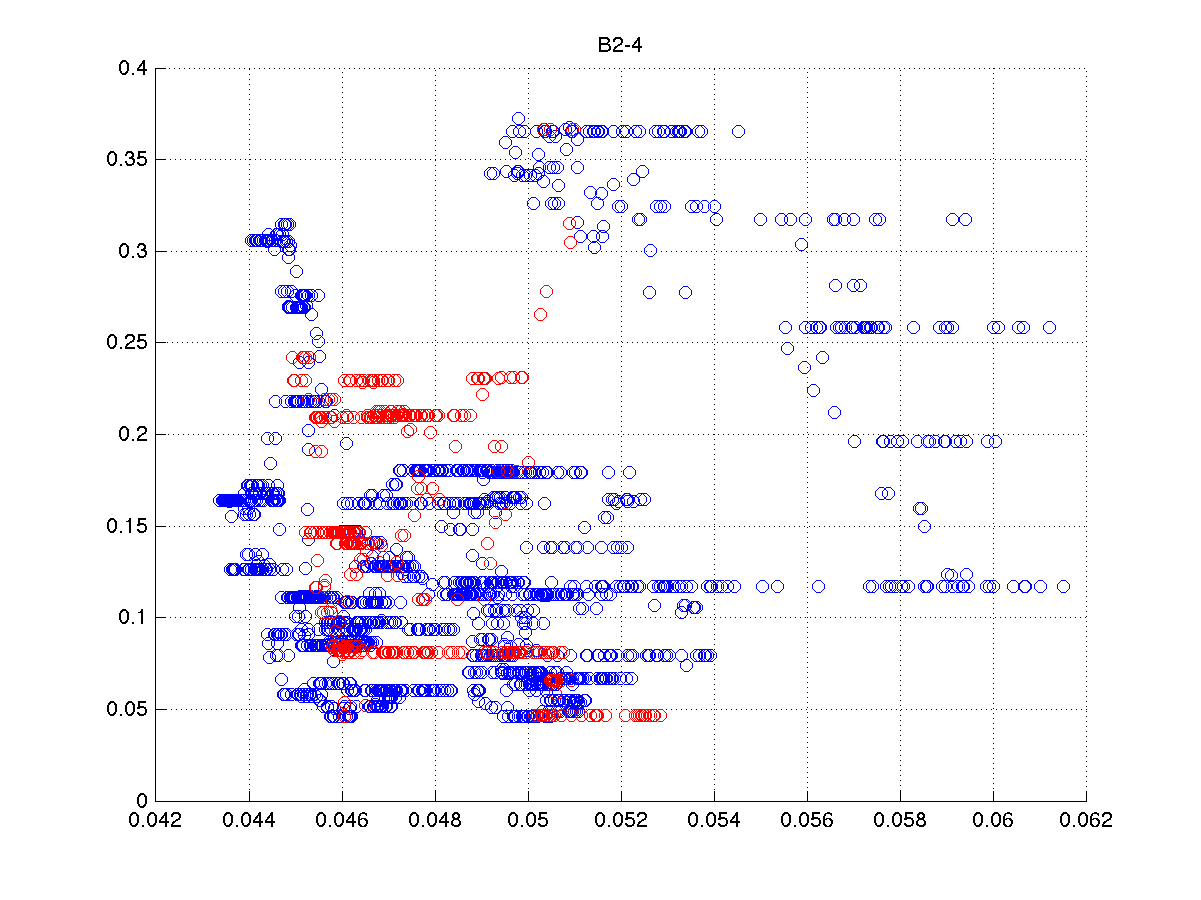}}&
\subfloat[trace-correl-leverage]{\includegraphics[width = 1.5in]{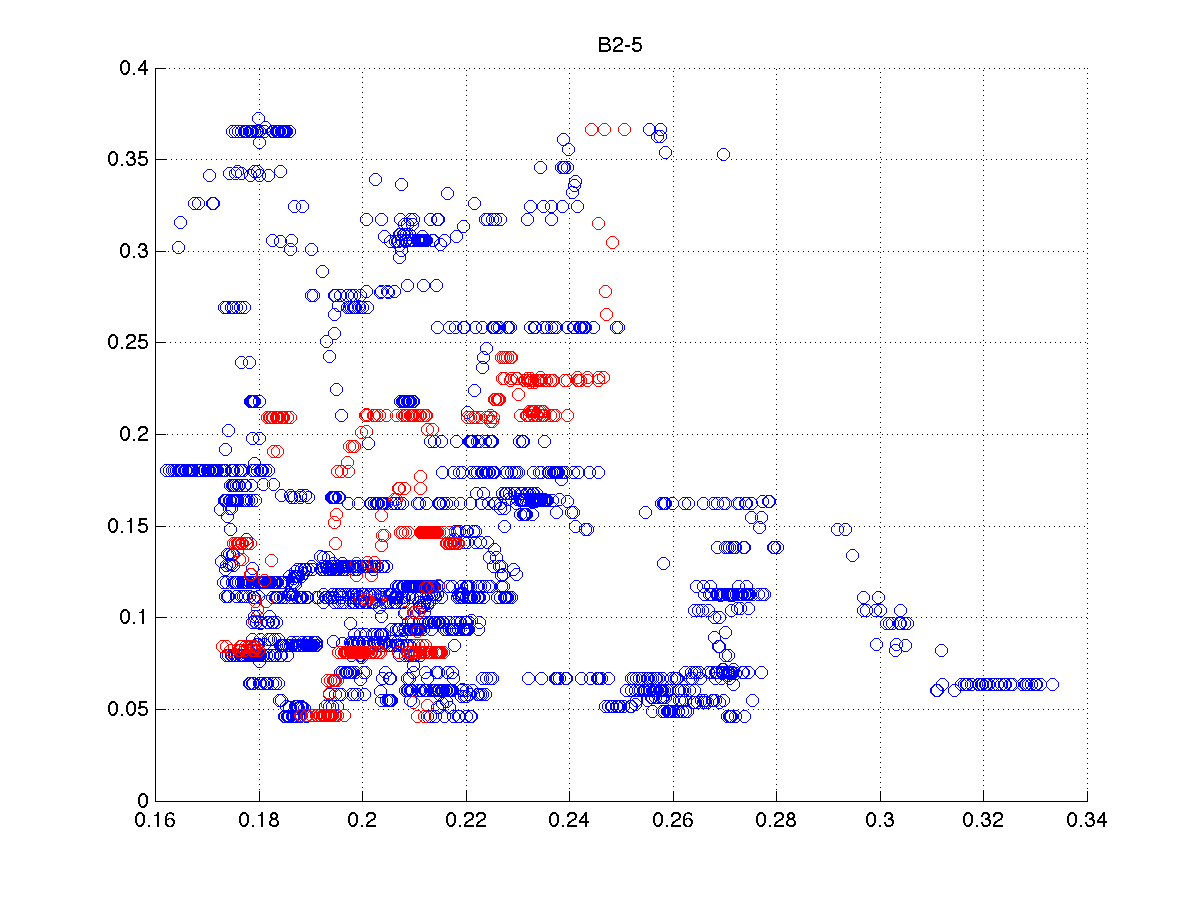}}\\
\subfloat[froben-covar]{\includegraphics[width = 1.7in]{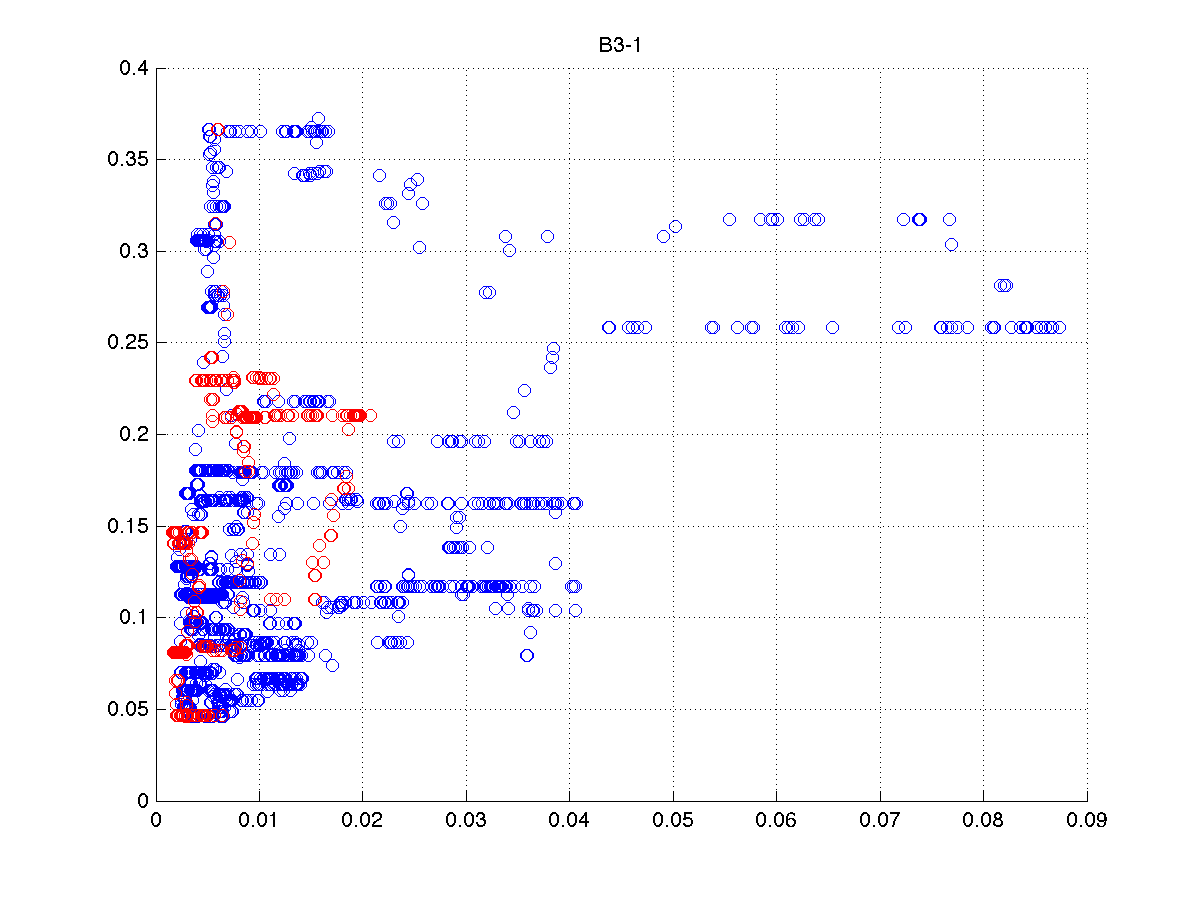}} &
\subfloat[froben-correl]{\includegraphics[width = 1.7in]{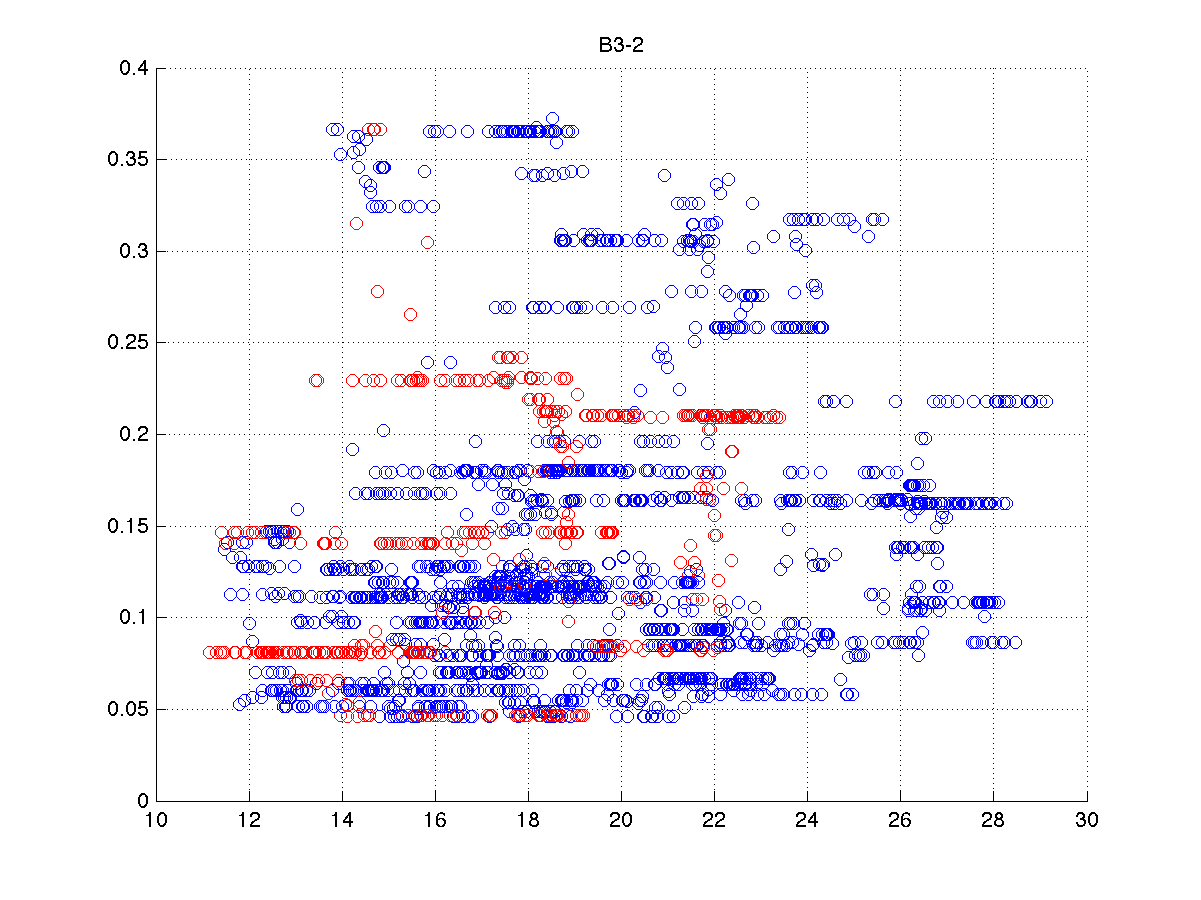}} &
\subfloat[froben-correl-volume]{\includegraphics[width = 1.7in]{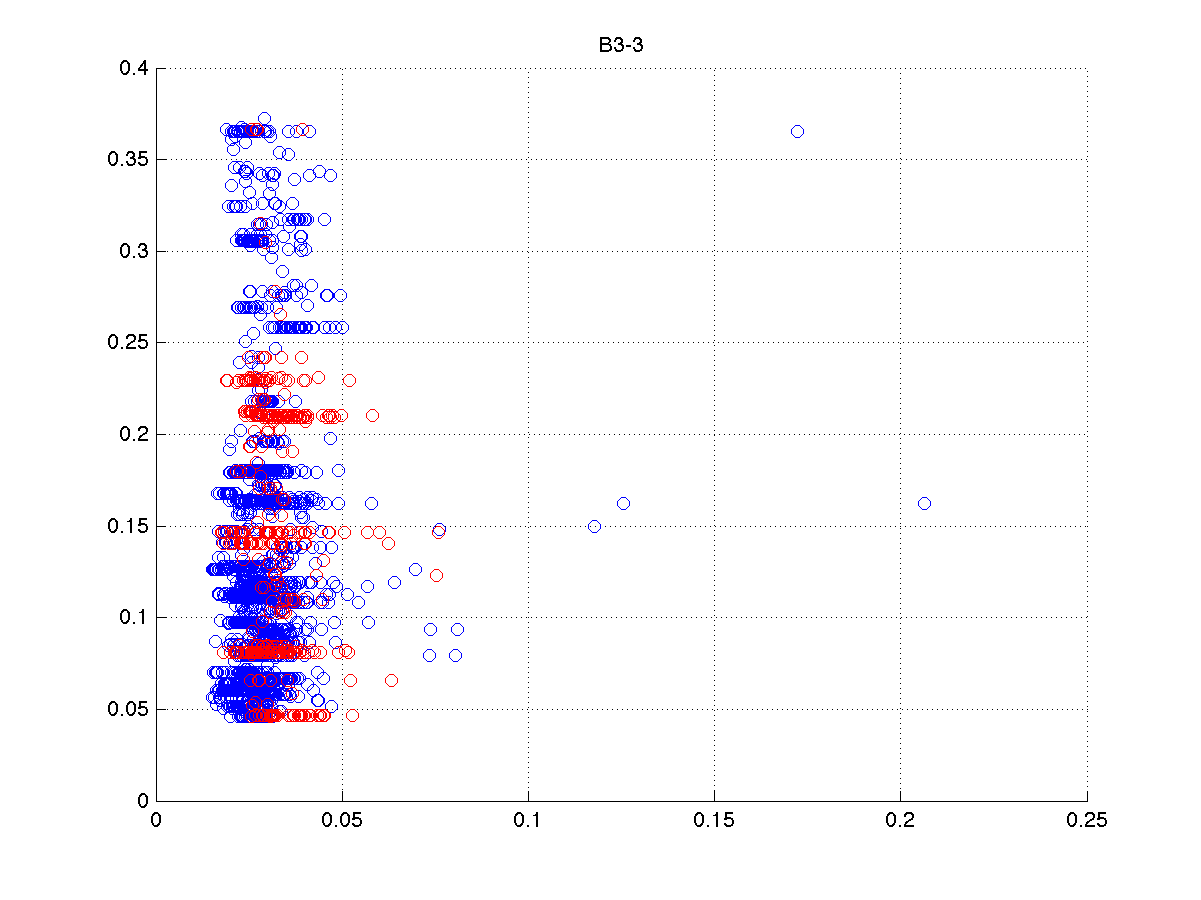}}\\
\subfloat[froben-correl-mcap]{\includegraphics[width = 1.7in]{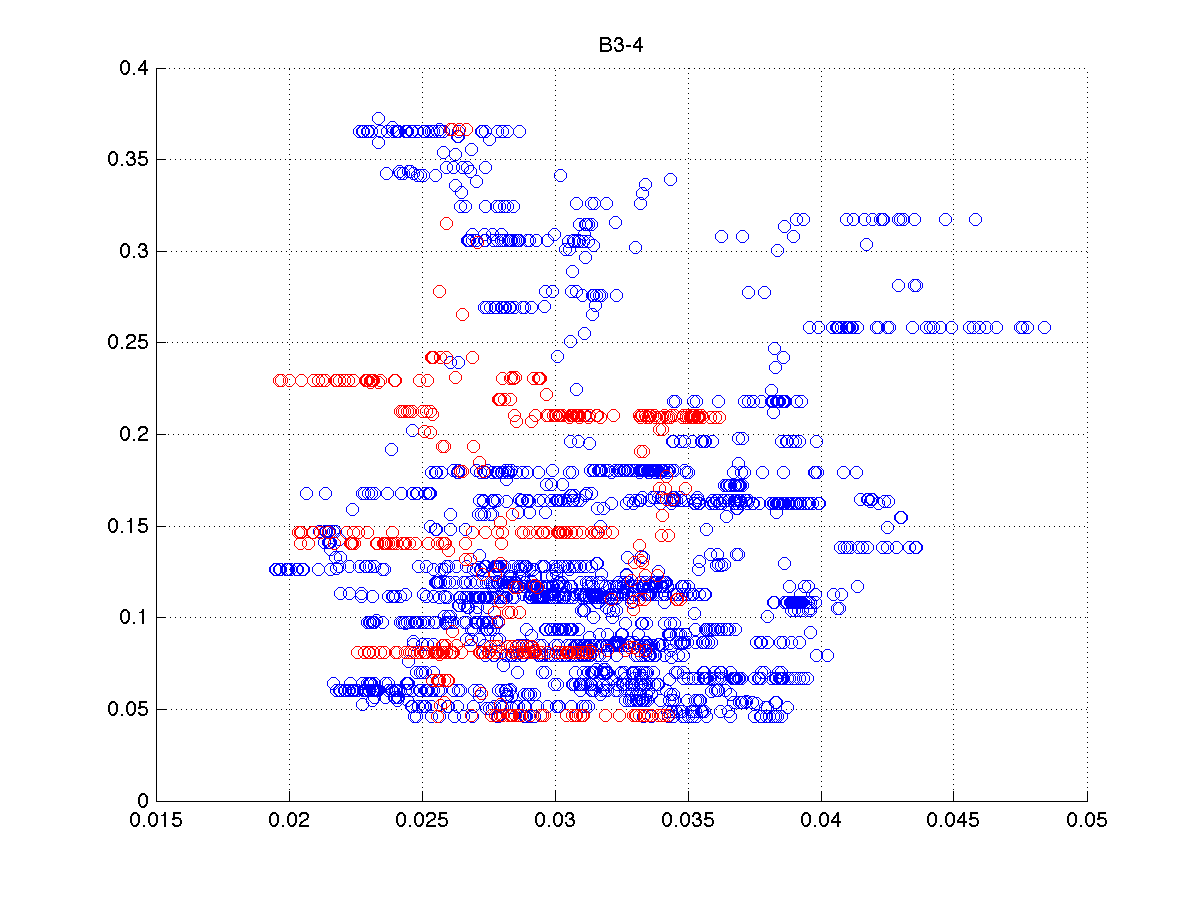}} &
\subfloat[froben-correl-leverage]{\includegraphics[width = 1.7in]{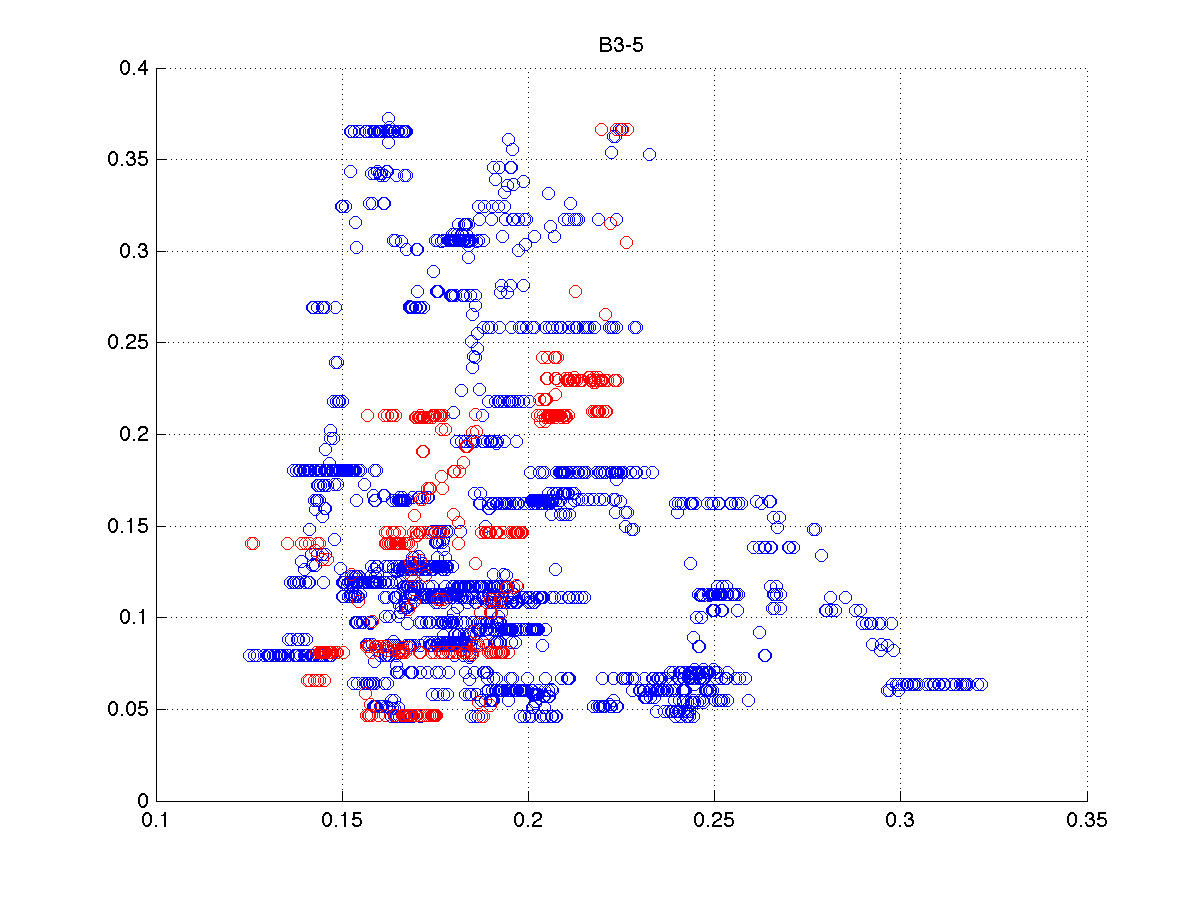}} &
\end{tabular}
\captionsetup{labelformat=empty}
\caption{CAC40: Indicators of the $\beta$-series. Red: in-sample ; Blue: out-of-sample}
\end{figure}

\begin{figure}[H]
\begin{tabular}{ccc}
\subfloat[$\mathscr{R}_{1}$covar]{\includegraphics[width = 1.7in]{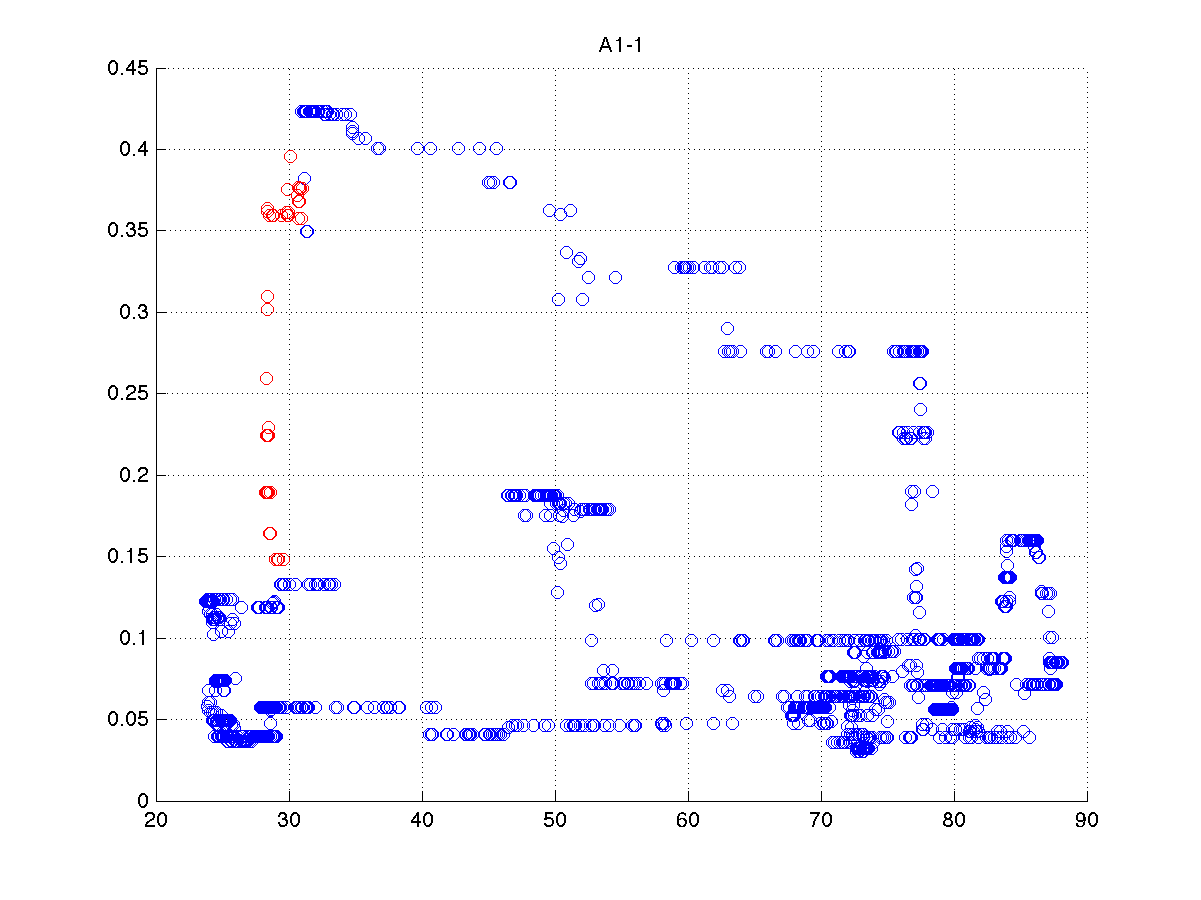}} &
\subfloat[$\mathscr{R}_{1}$correl]{\includegraphics[width = 1.7in]{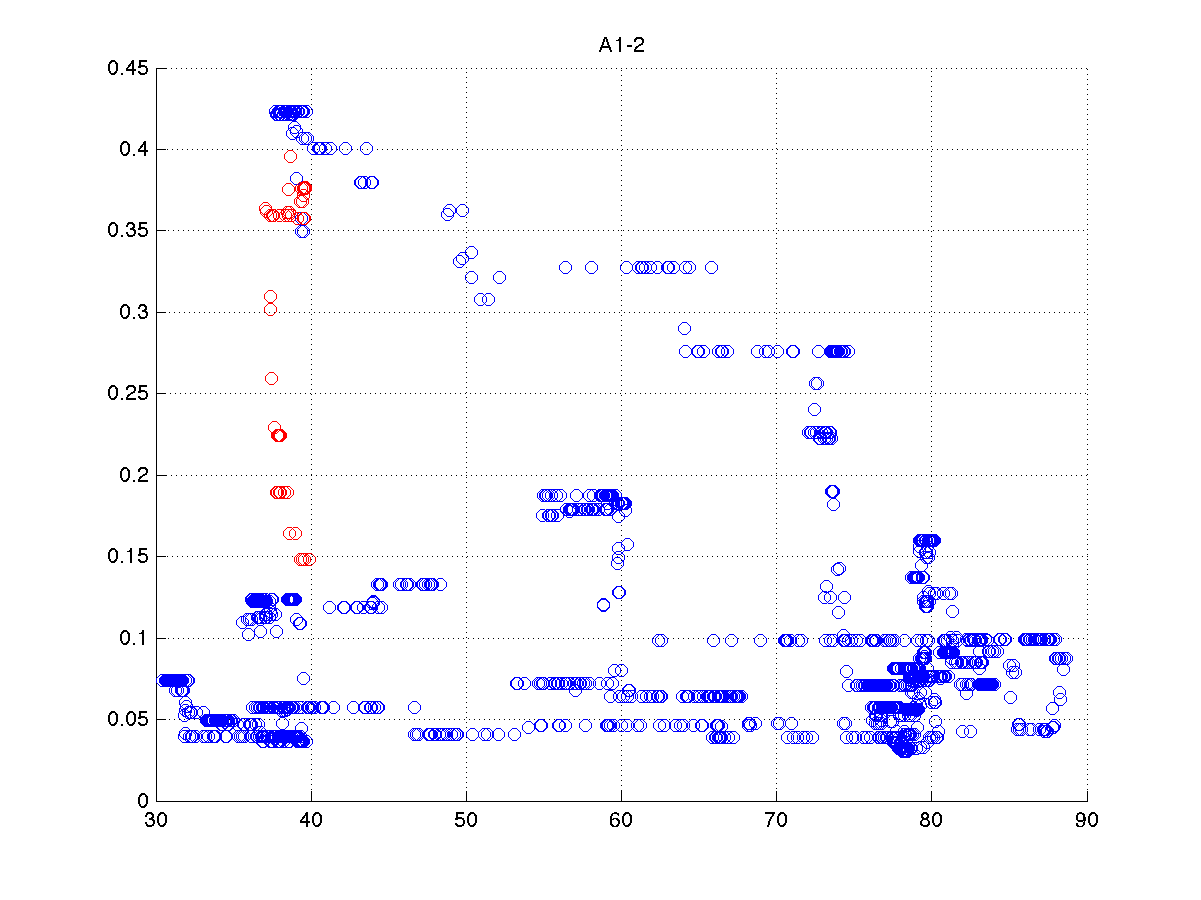}} &
\subfloat[$\mathscr{R}_{1}$correl-volume]{\includegraphics[width = 1.7in]{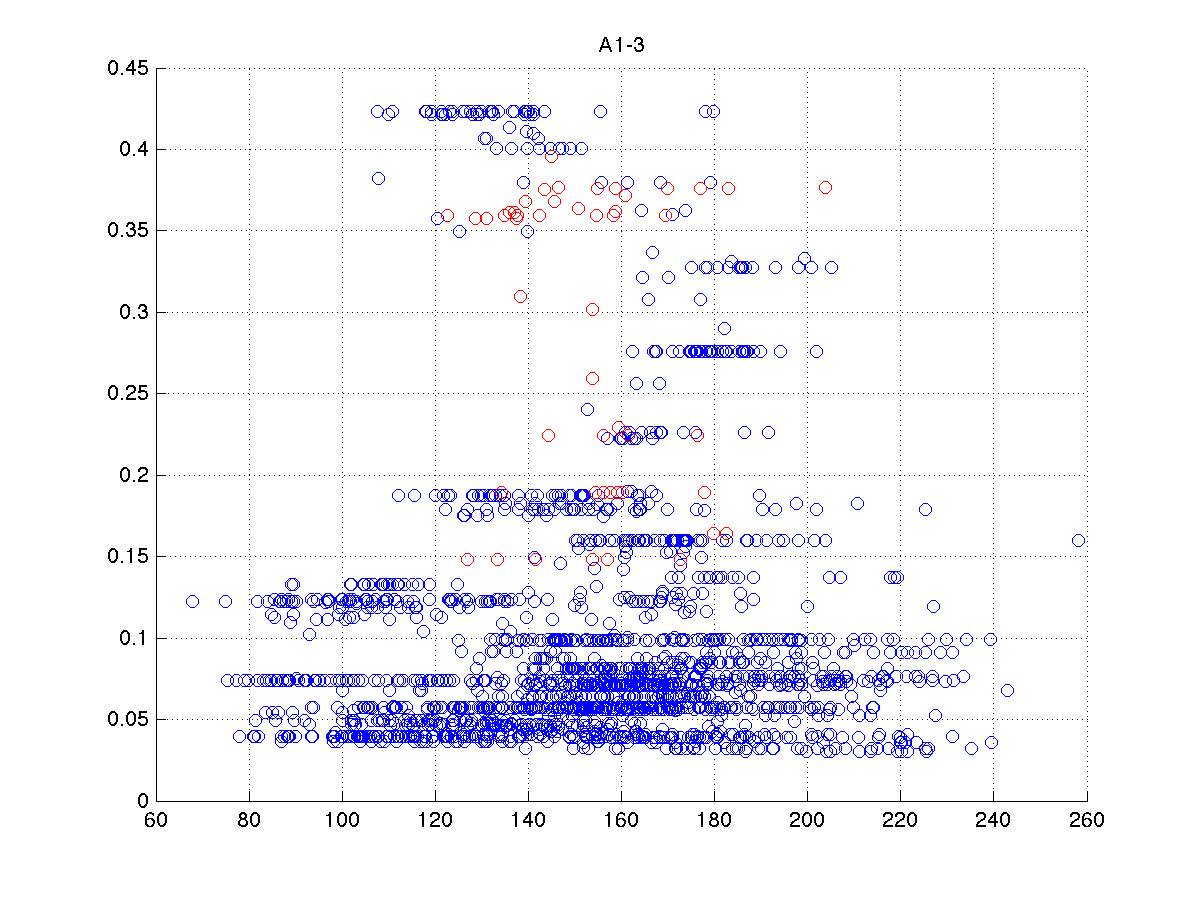}}\\
\subfloat[$\mathscr{R}_{1}$correl-mcap]{\includegraphics[width = 1.7in]{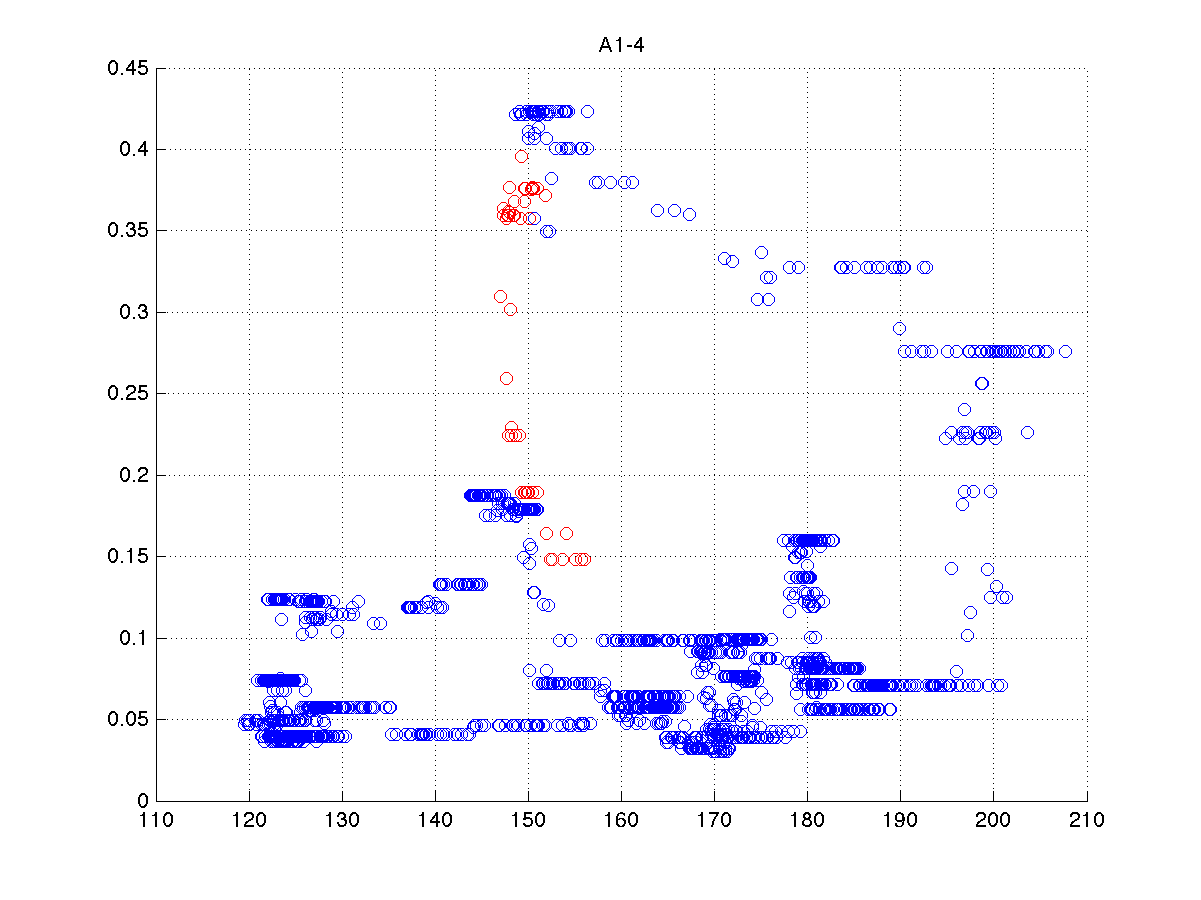}}&
\subfloat[$\mathscr{R}_{1}$correl-leverage]{\includegraphics[width = 1.7in]{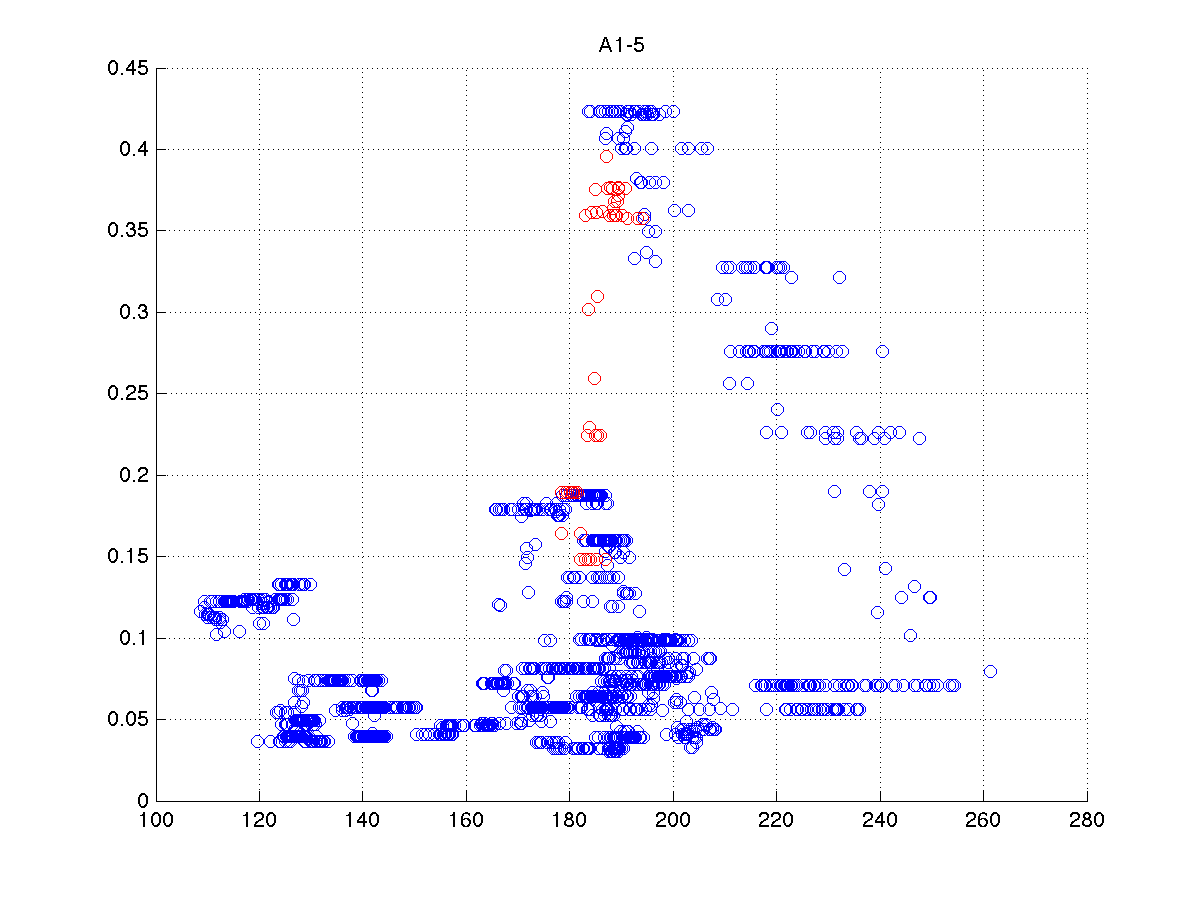}} &
\subfloat[$\mathscr{R}_{2}$covar]{\includegraphics[width = 1.7in]{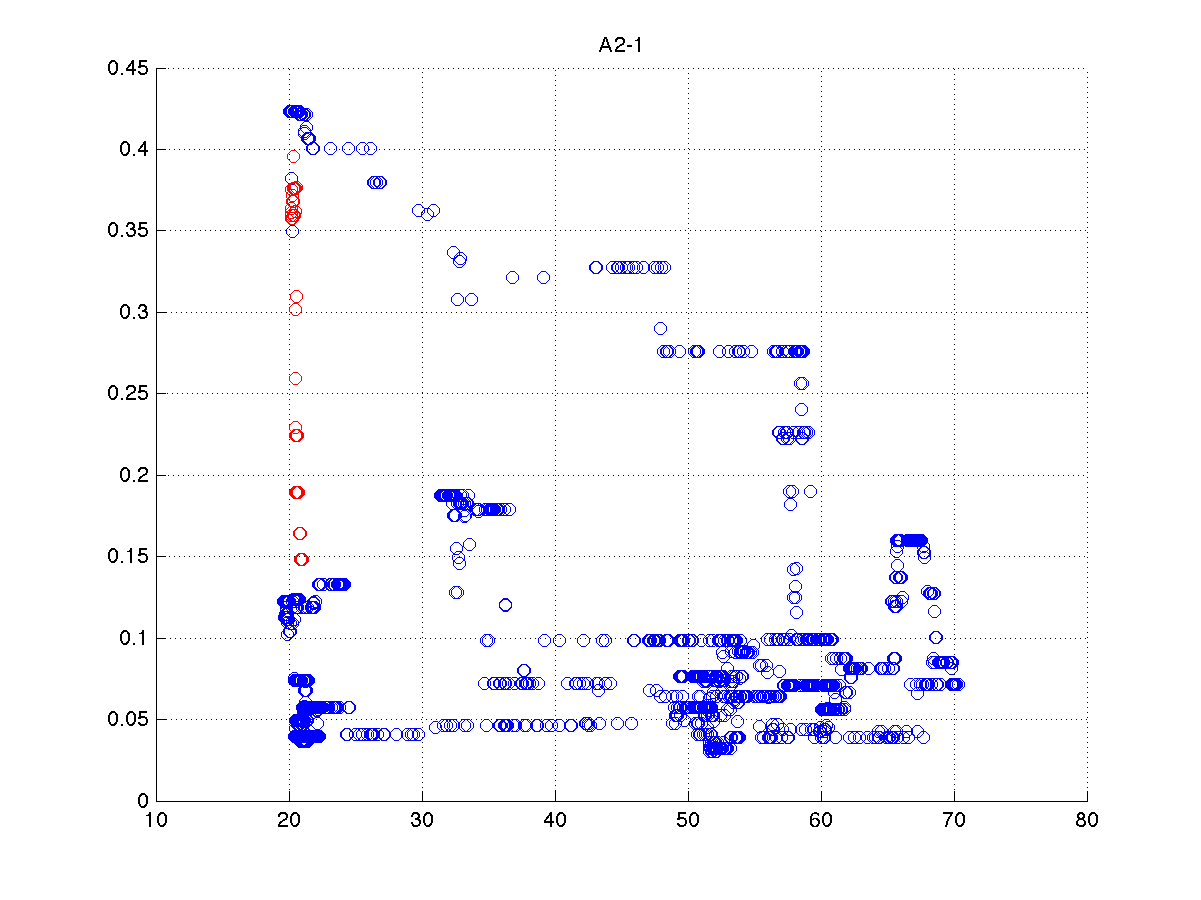}}\\
\subfloat[$\mathscr{R}_{2}$correl]{\includegraphics[width = 1.7in]{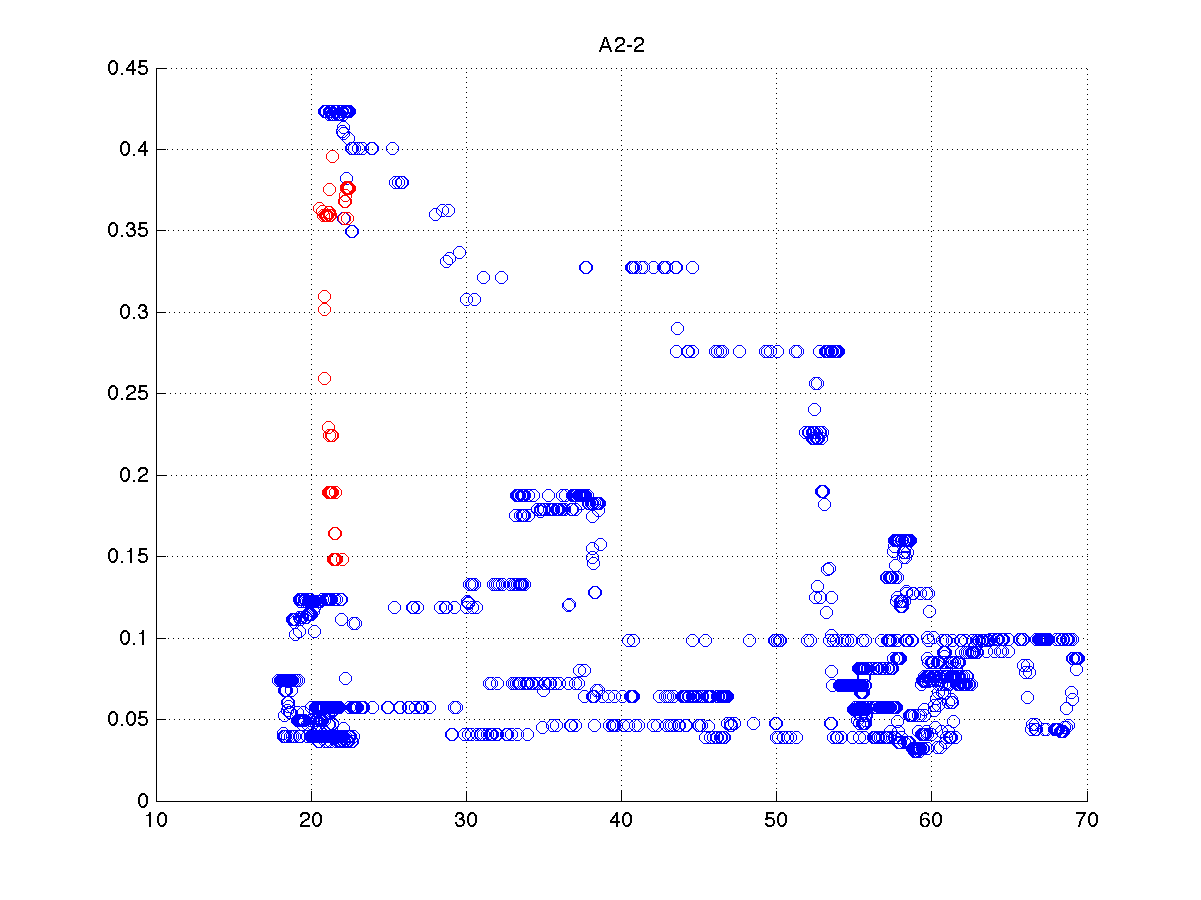}} &
\subfloat[$\mathscr{R}_{2}$correl-volume]{\includegraphics[width = 1.7in]{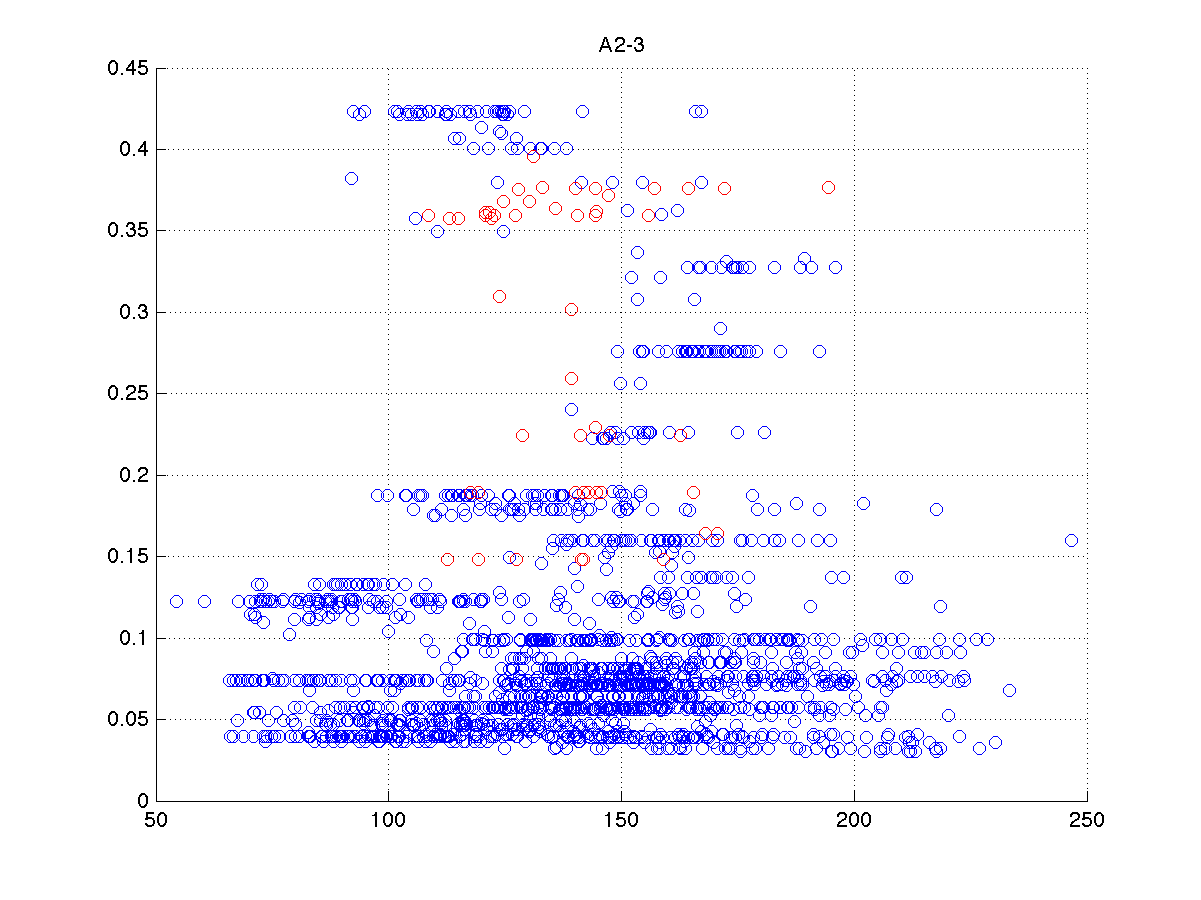}}&
\subfloat[$\mathscr{R}_{2}$correl-mcap]{\includegraphics[width = 1.7in]{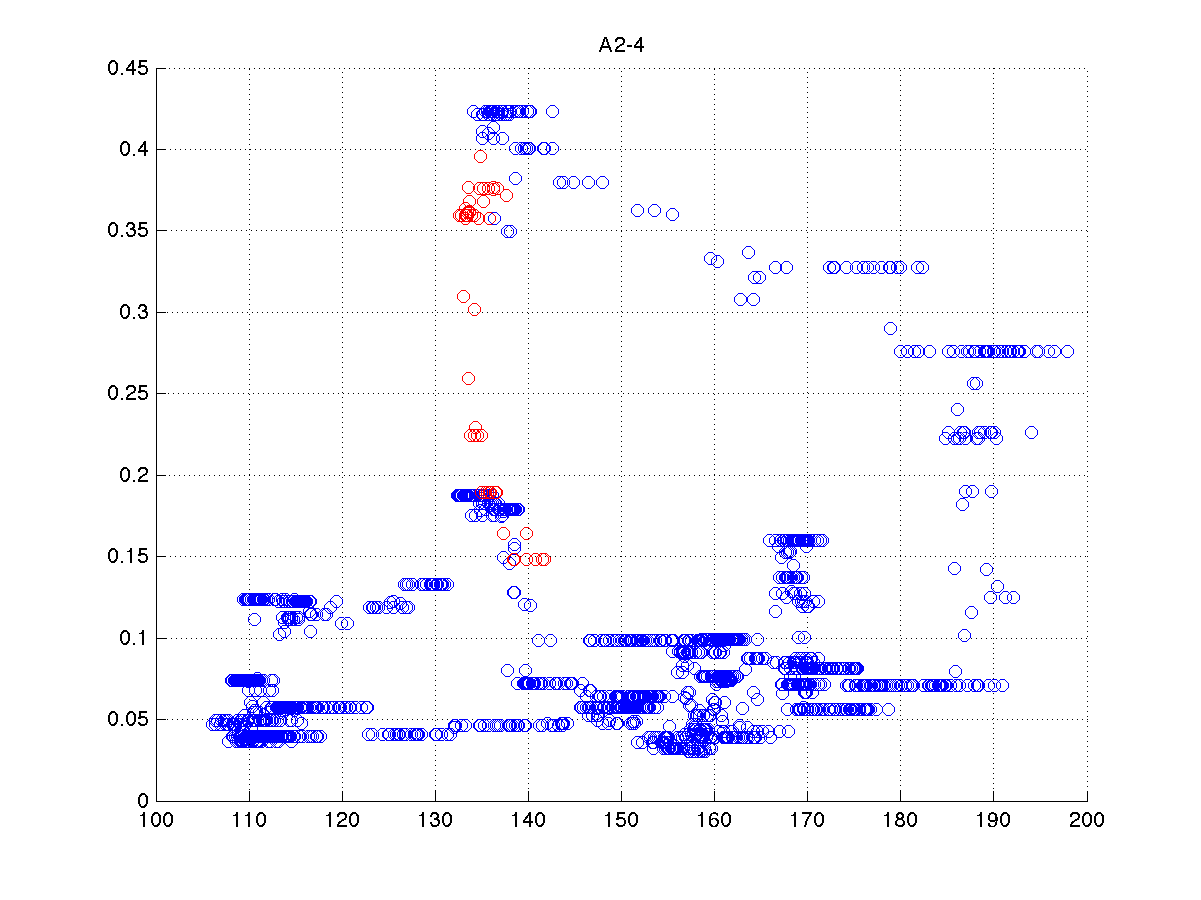}}\\
\subfloat[$\mathscr{R}_{2}$correl-leverage]{\includegraphics[width = 1.7in]{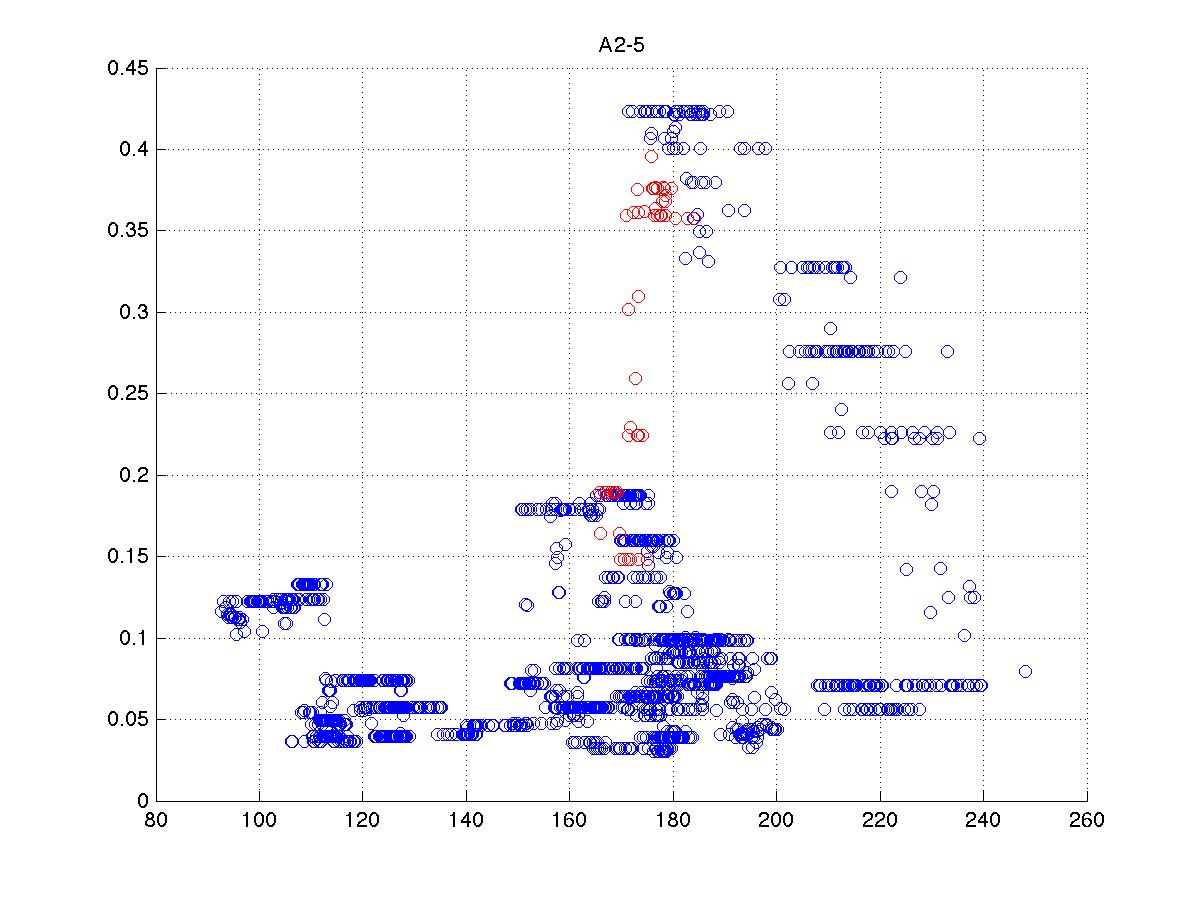}} &
\subfloat[$\mathscr{R}_{3}$covar]{\includegraphics[width = 1.7in]{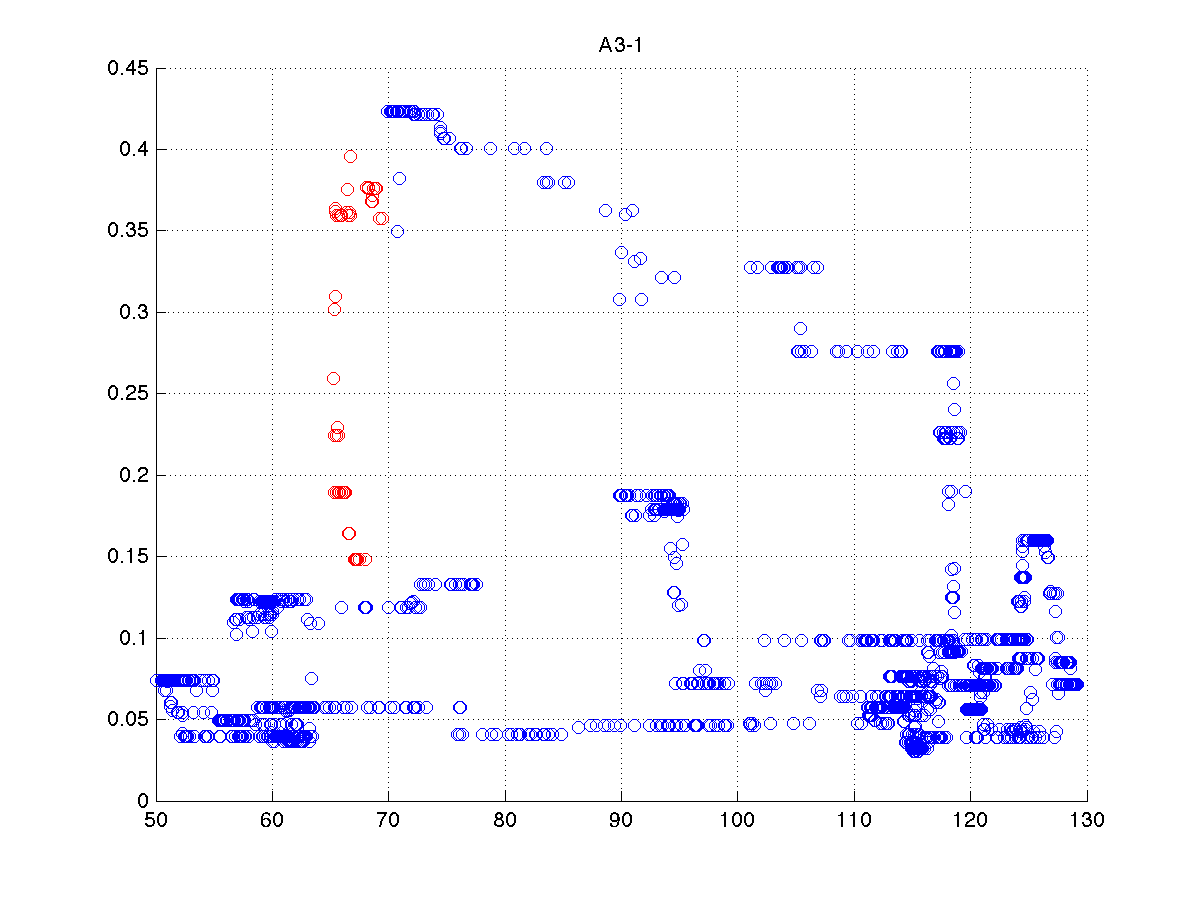}} &
\subfloat[$\mathscr{R}_{3}$correl]{\includegraphics[width = 1.7in]{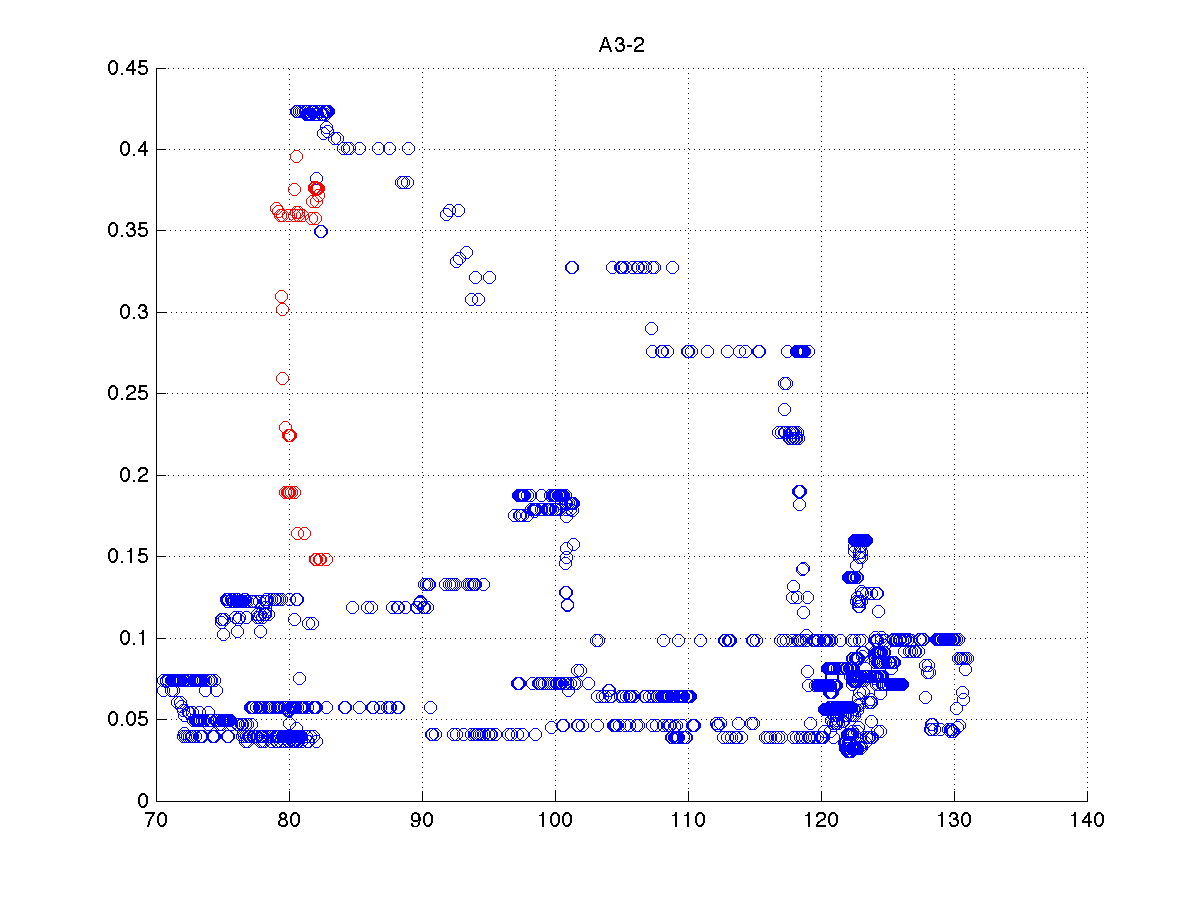}}\\
\subfloat[$\mathscr{R}_{3}$correl-volume]{\includegraphics[width = 1.7in]{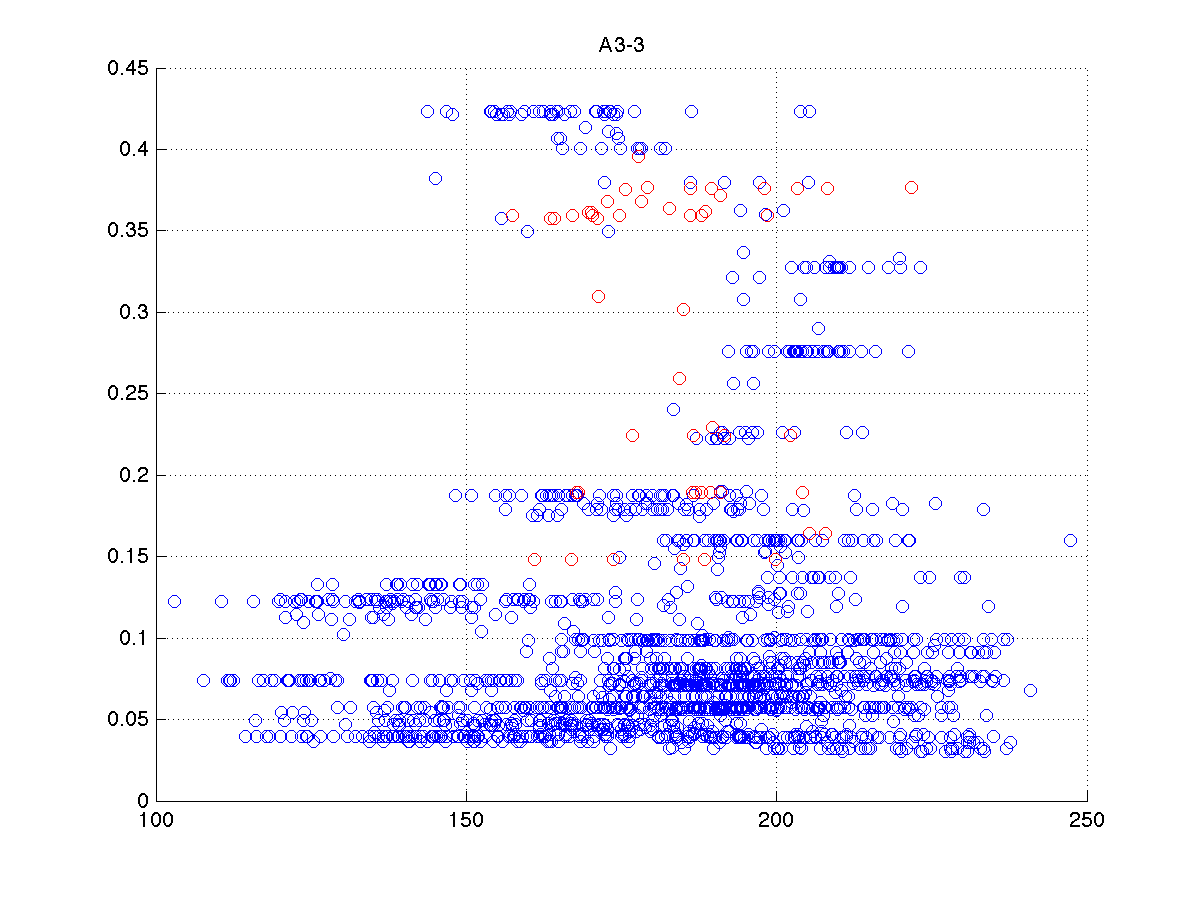}} &
\subfloat[$\mathscr{R}_{3}$correl-mcap]{\includegraphics[width = 1.7in]{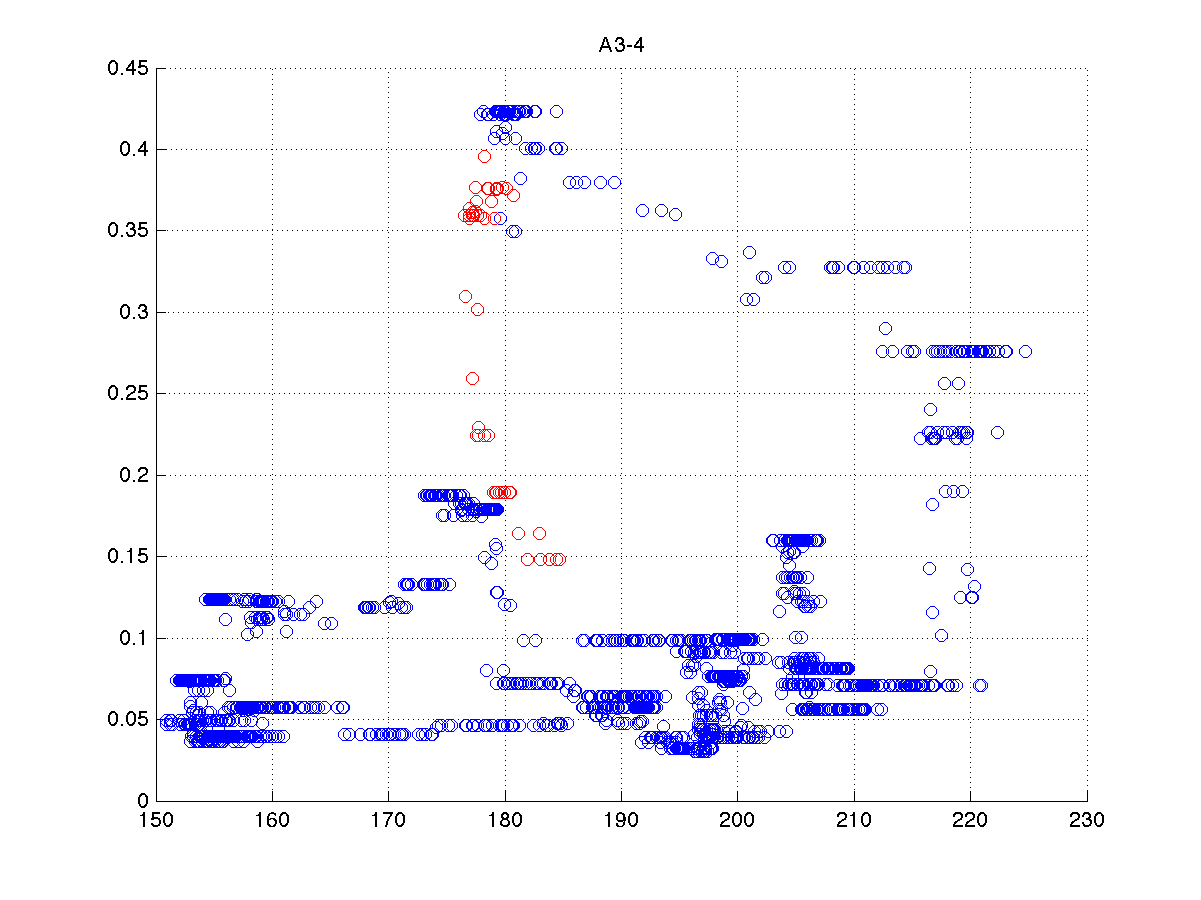}} &
\subfloat[$\mathscr{R}_{3}$correl-leverage]{\includegraphics[width = 1.7in]{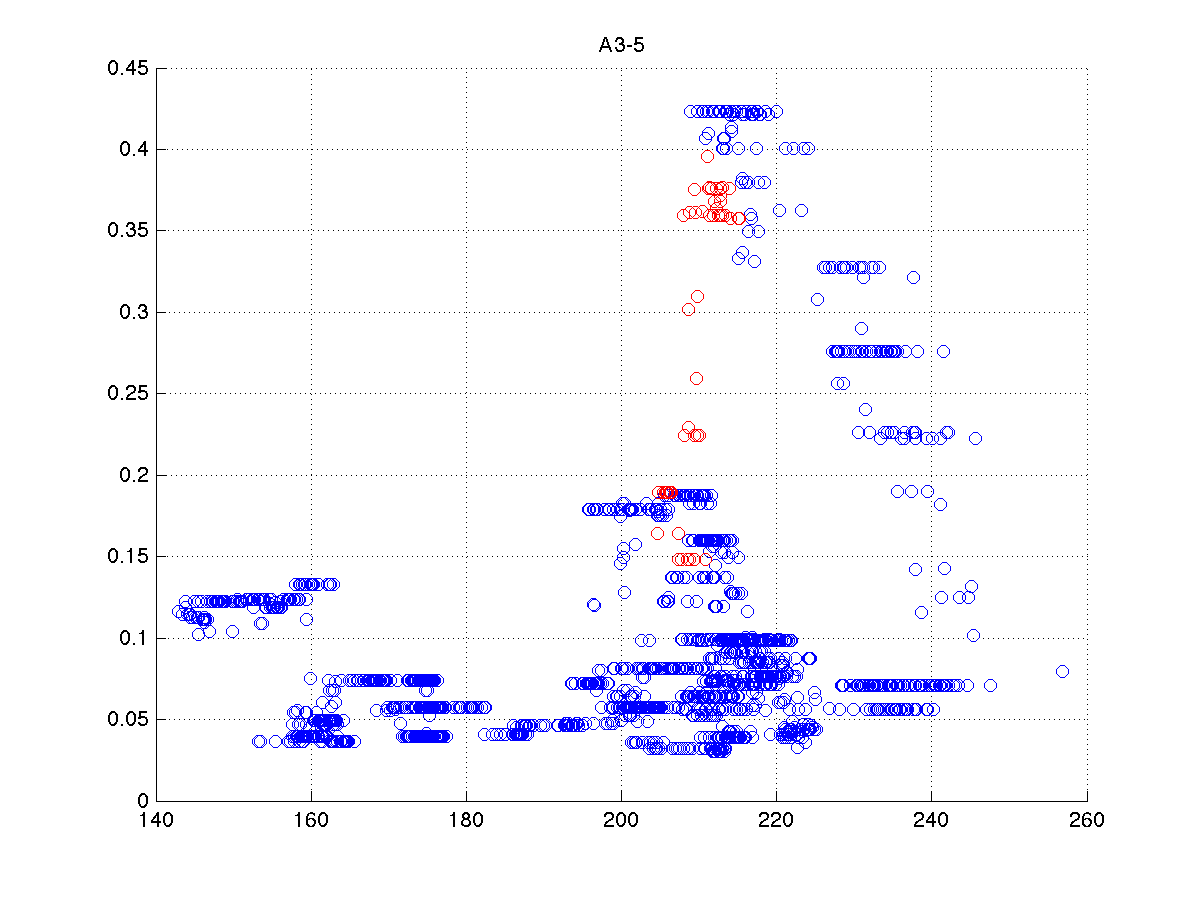}}\\
\end{tabular}
\captionsetup{labelformat=empty}
\caption{SP500: Indicators of the $\alpha$-series. Red: in-sample ; Blue: out-of-sample}
\end{figure}

\begin{figure}[H]
\begin{tabular}{ccc}
\subfloat[rspec-covar]{\includegraphics[width = 1.7in]{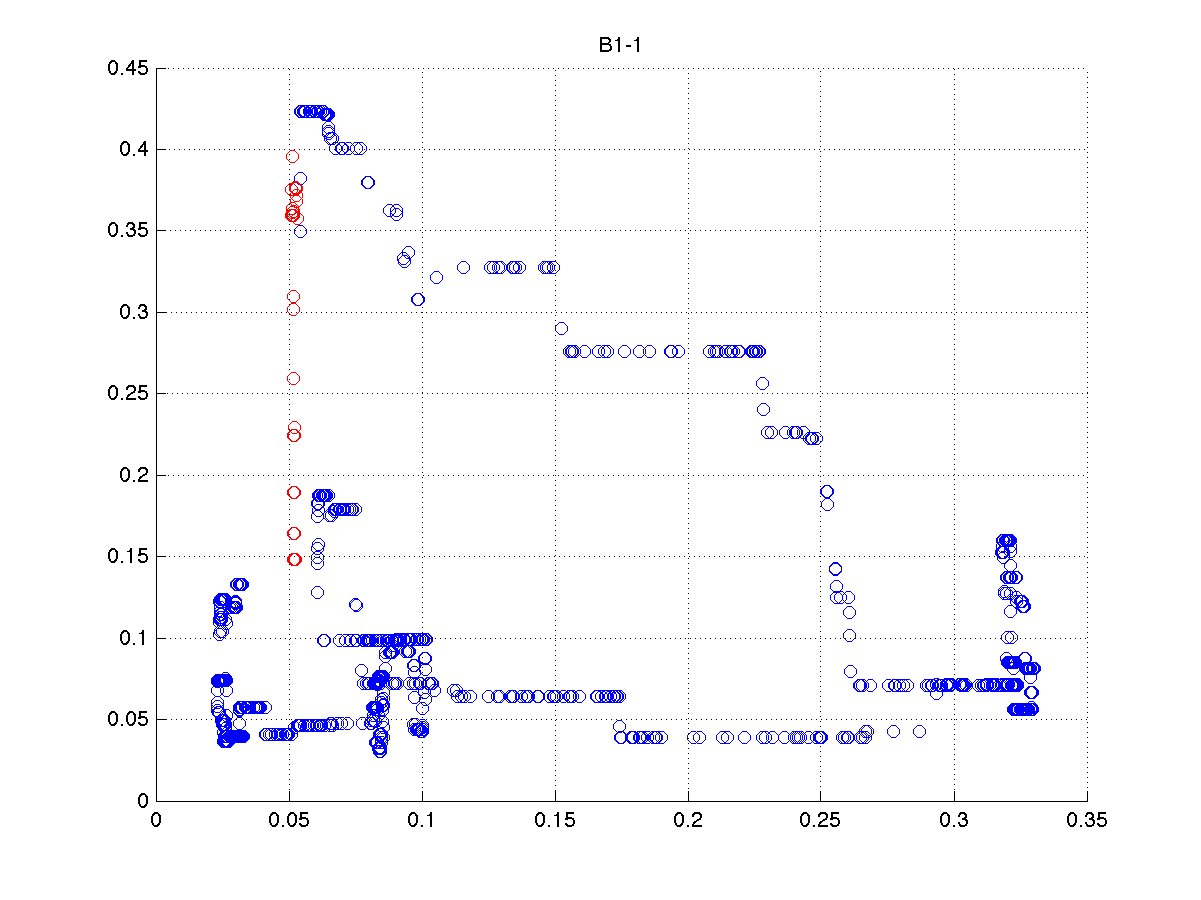}} &
\subfloat[rspec-correl]{\includegraphics[width = 1.7in]{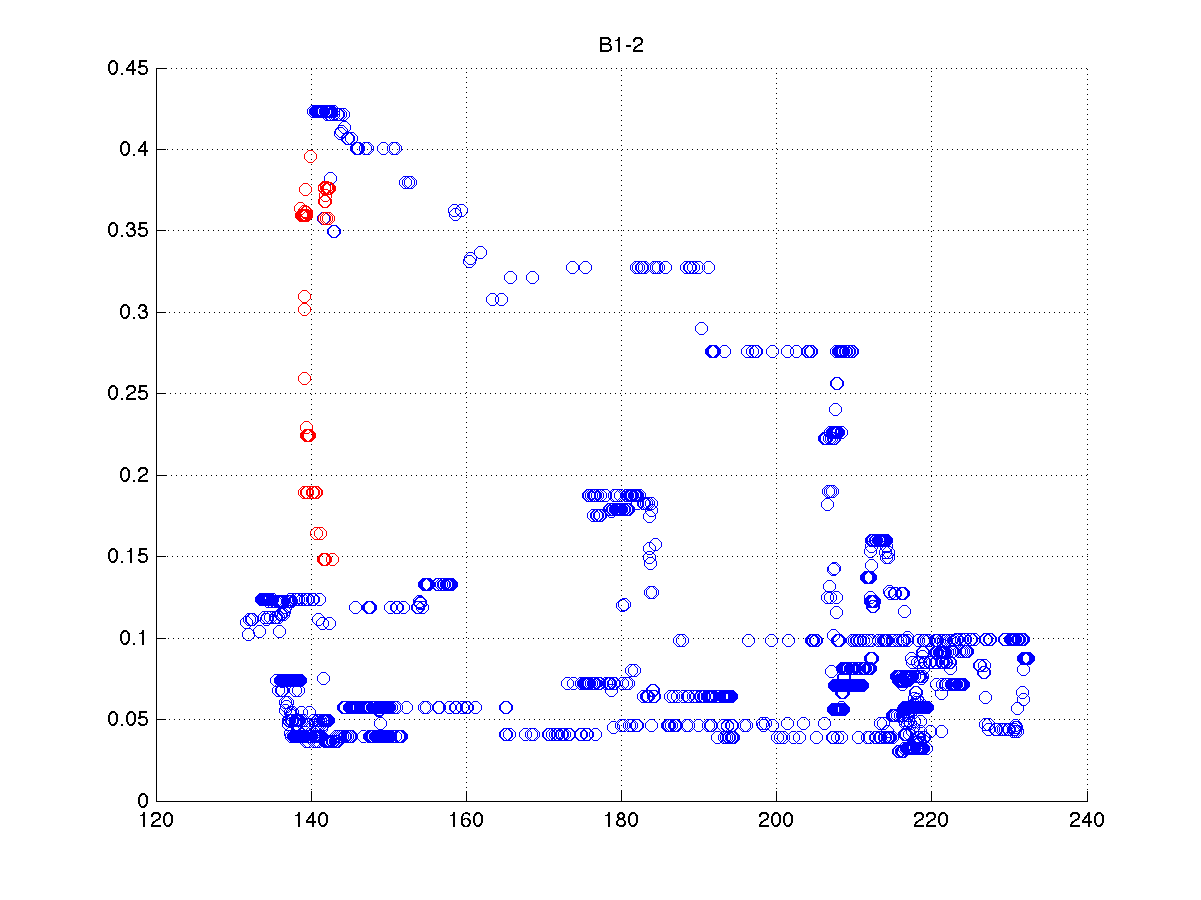}} &
\subfloat[rspec-correl-volume]{\includegraphics[width = 1.7in]{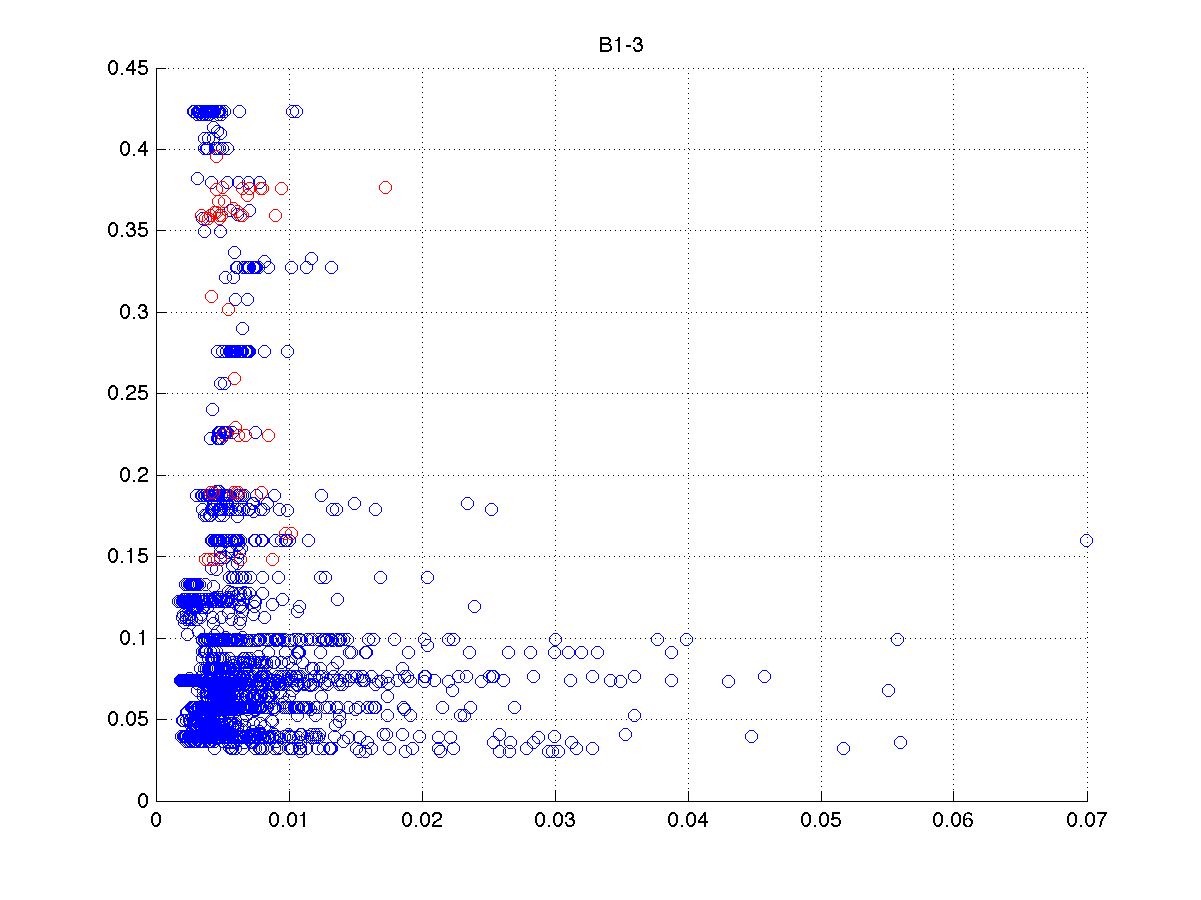}} \\
\subfloat[rspec-correl-mcap]{\includegraphics[width = 1.7in]{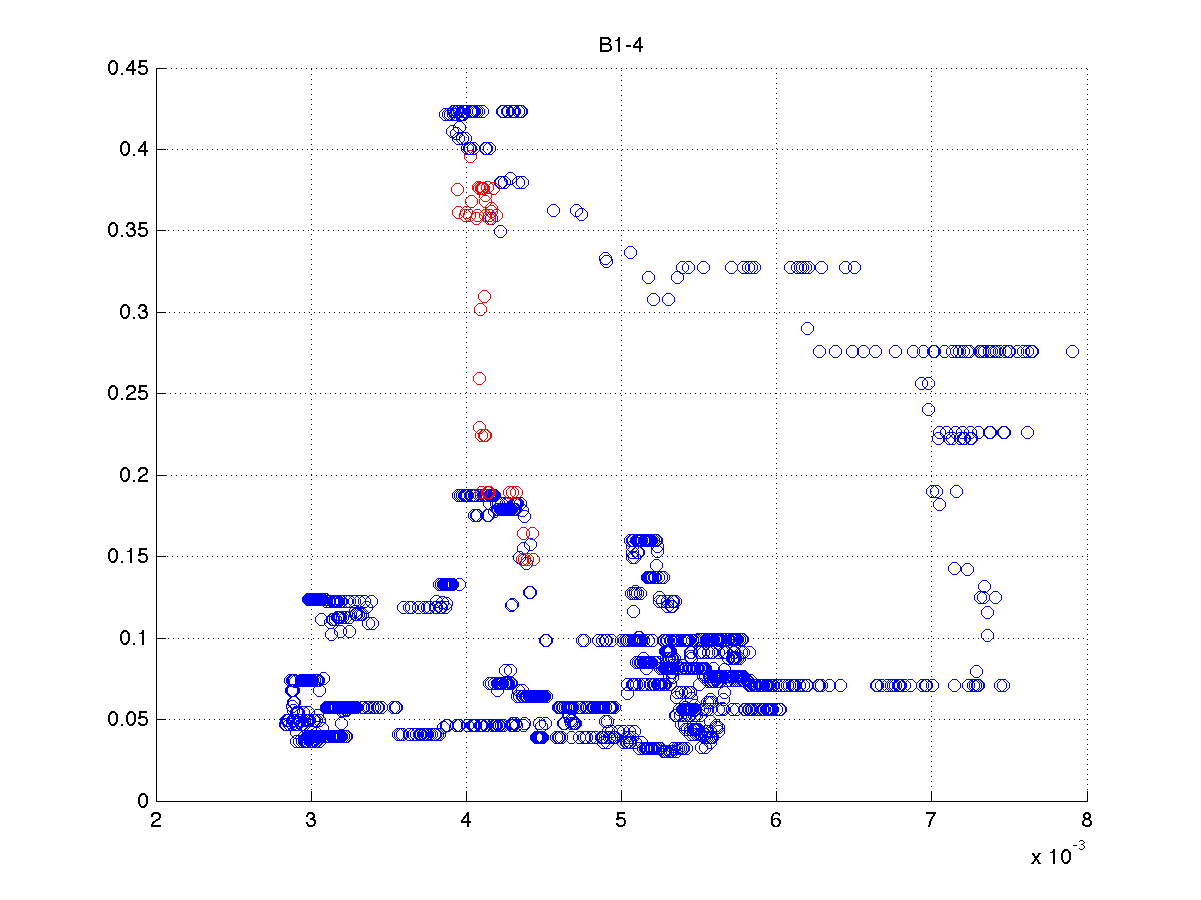}}&
\subfloat[rspec-correl-leverage]{\includegraphics[width = 1.7in]{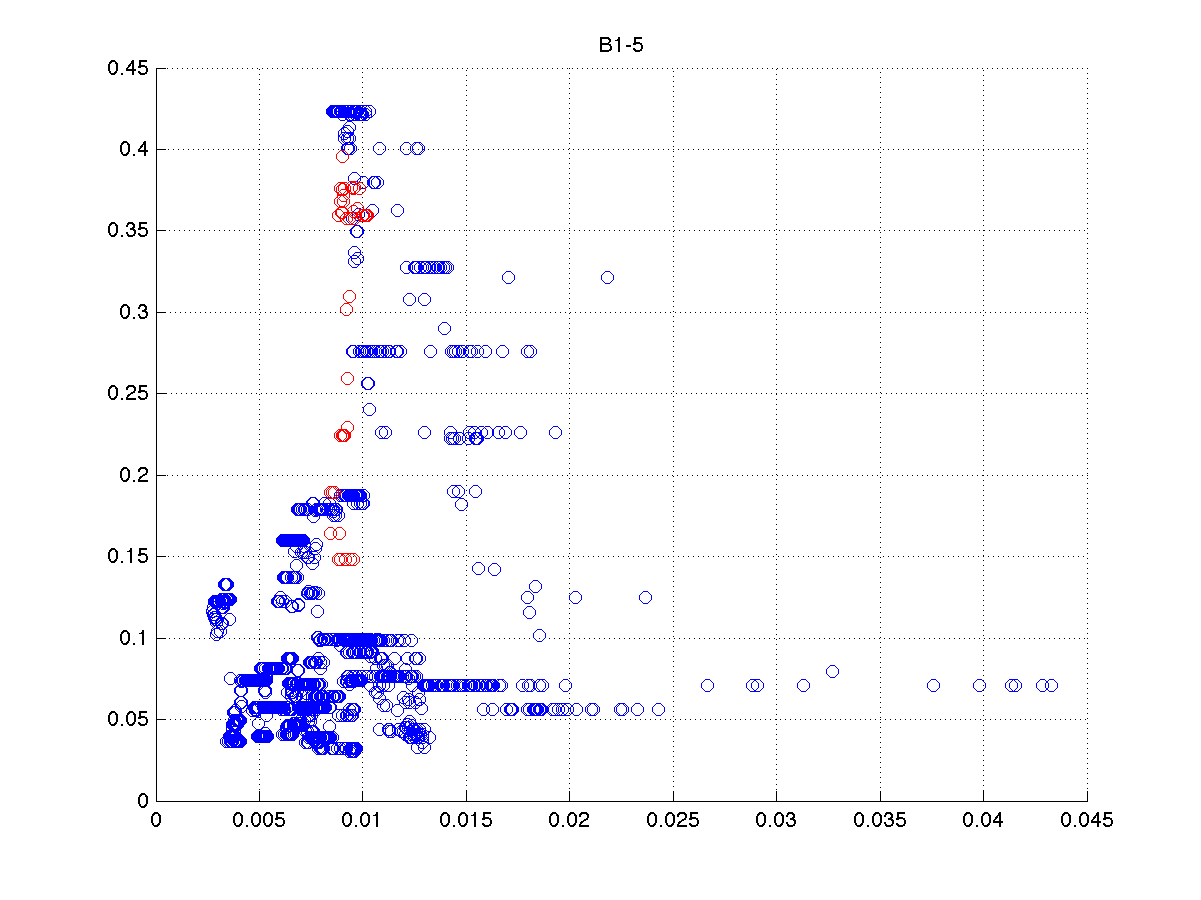}} &
\subfloat[trace-covar]{\includegraphics[width = 1.7in]{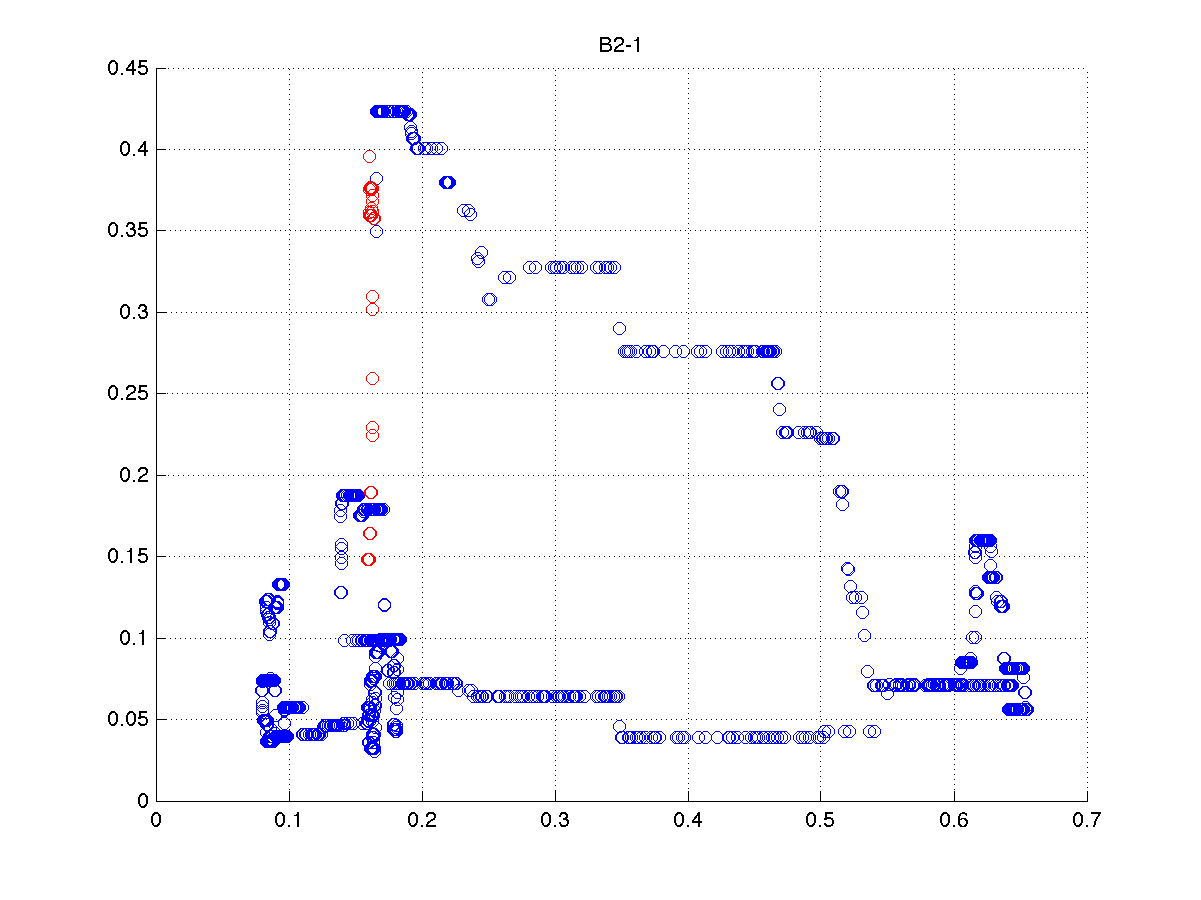}}\\
\subfloat[trace-correl-volume]{\includegraphics[width = 1.7in]{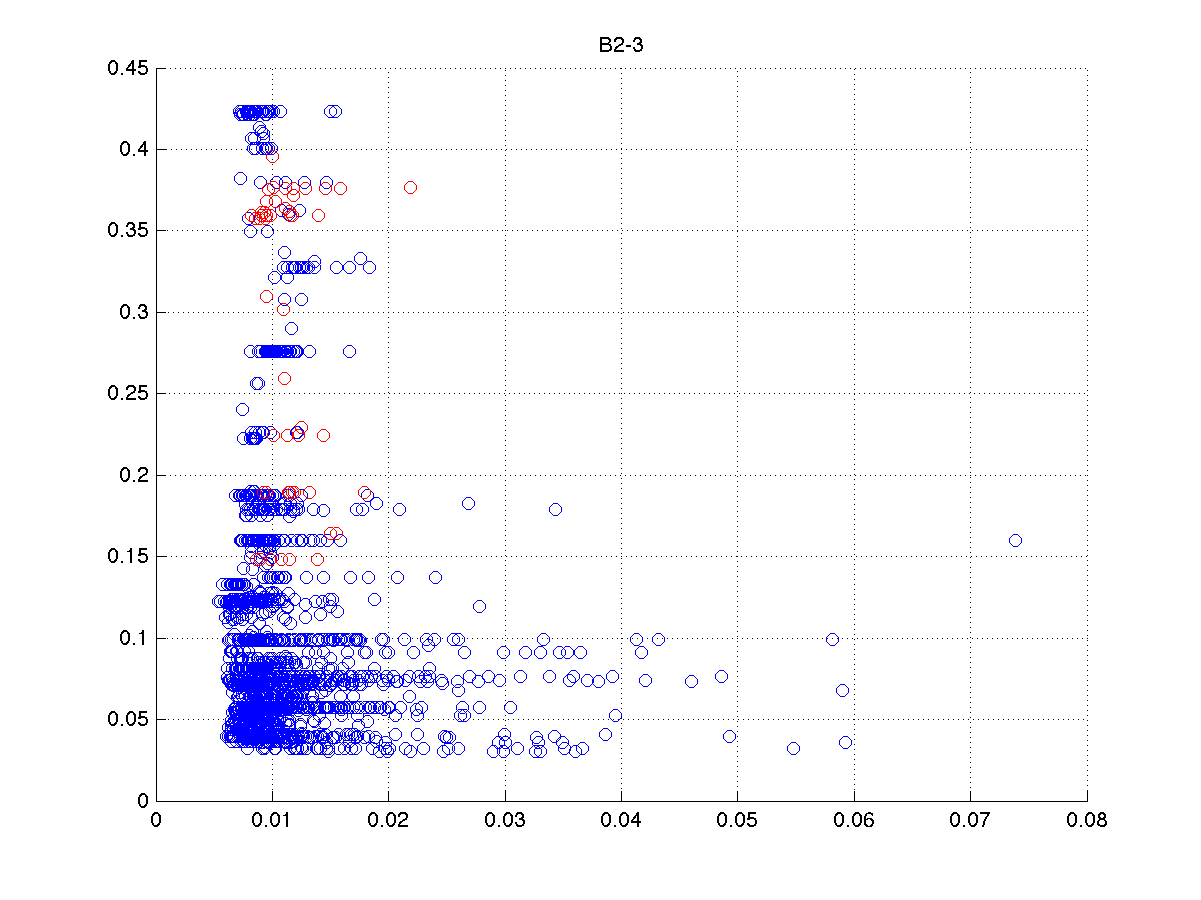}} &
\subfloat[trace-correl-mcap]{\includegraphics[width = 1.7in]{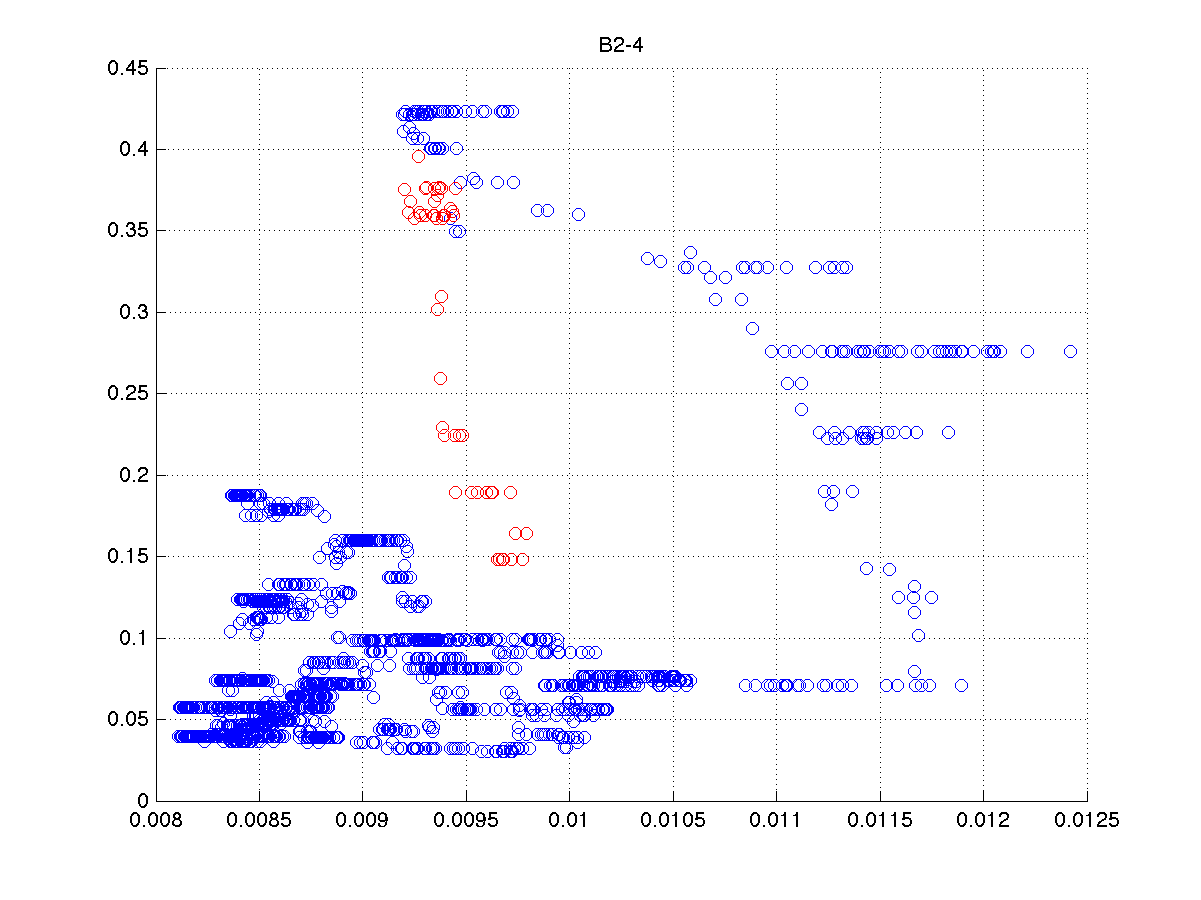}}&
\subfloat[trace-correl-leverage]{\includegraphics[width = 1.5in]{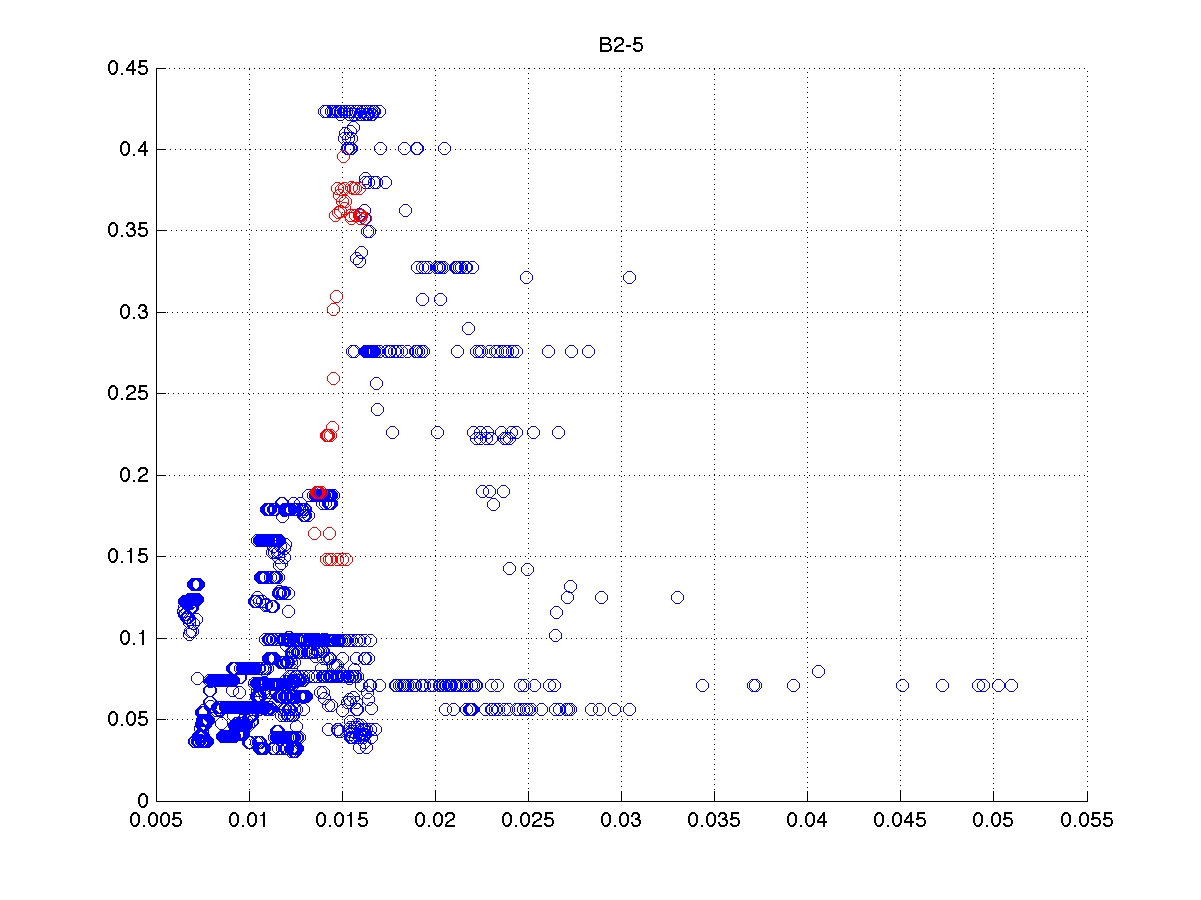}}\\
\subfloat[froben-covar]{\includegraphics[width = 1.7in]{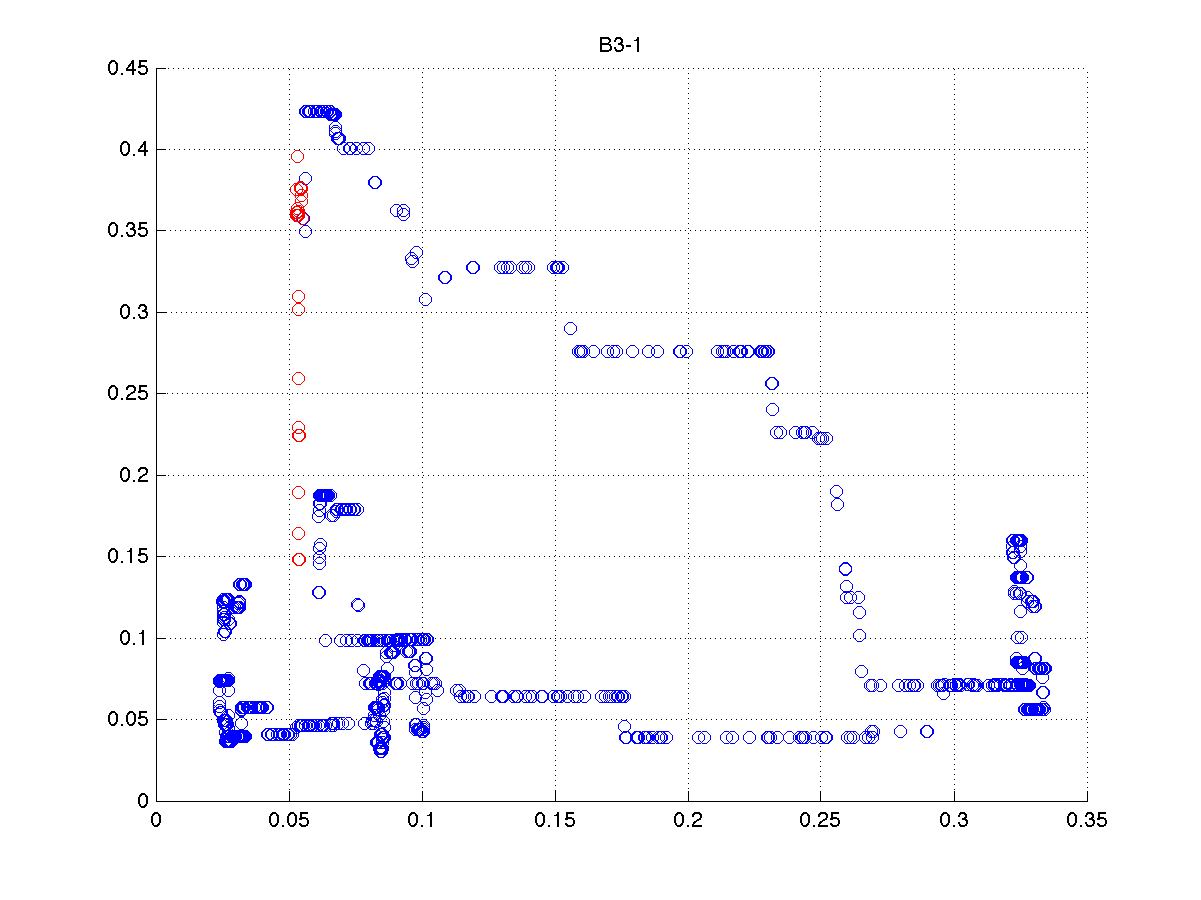}} &
\subfloat[froben-correl]{\includegraphics[width = 1.7in]{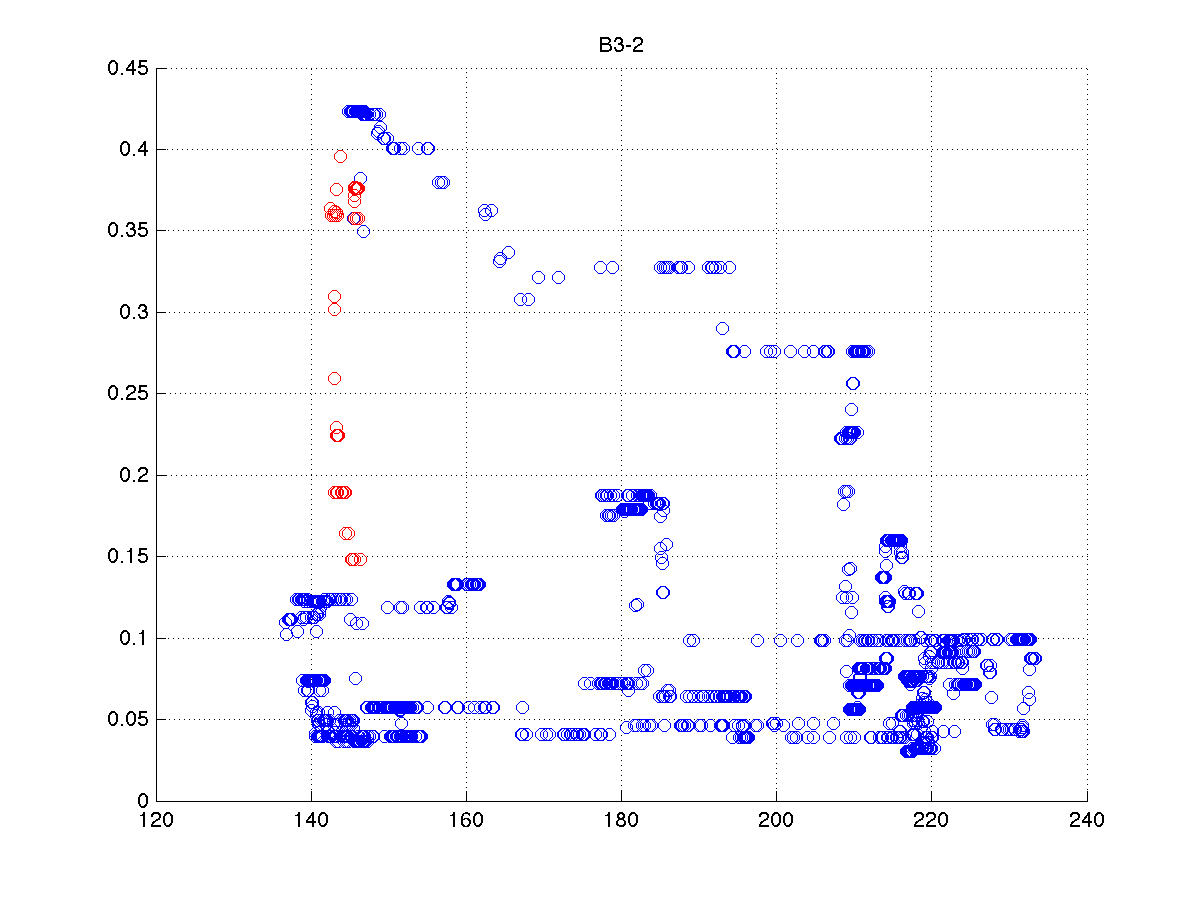}} &
\subfloat[froben-correl-volume]{\includegraphics[width = 1.7in]{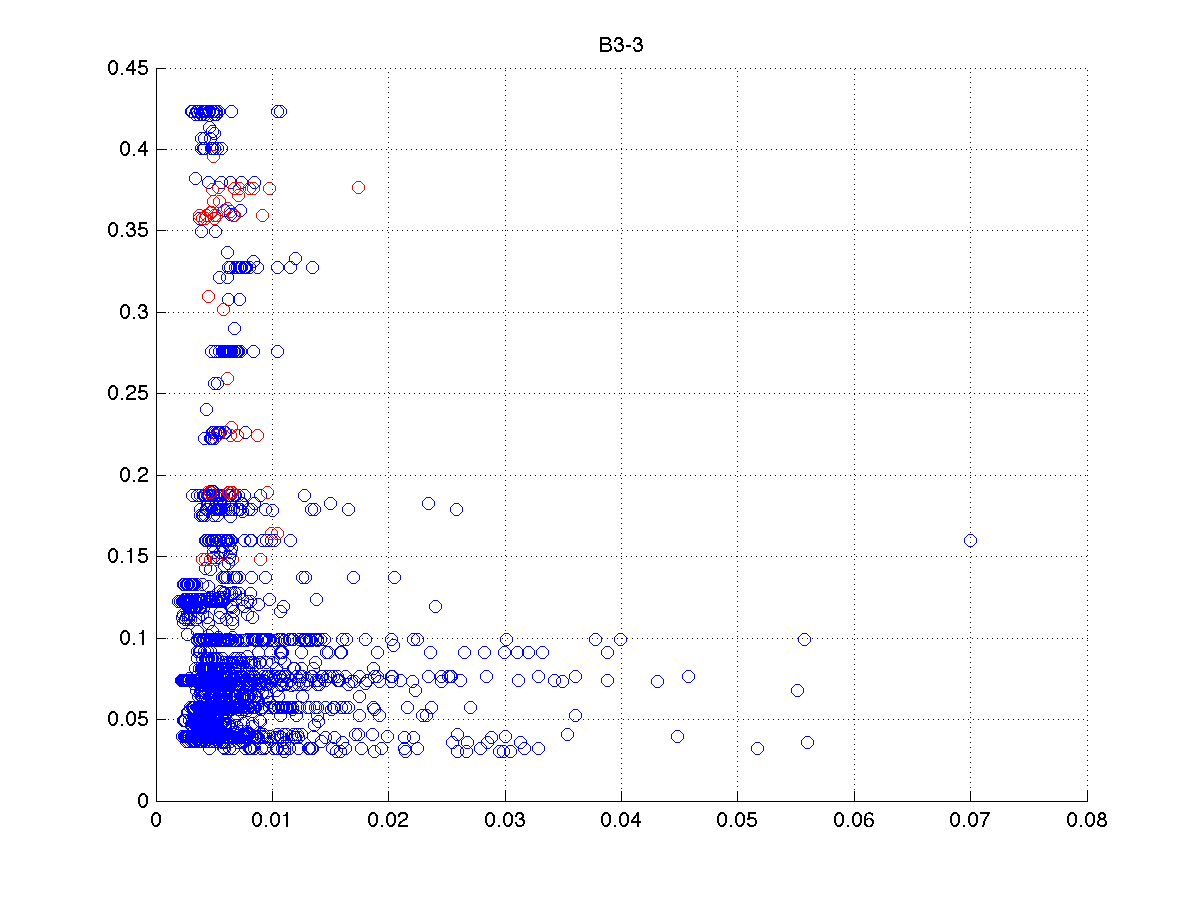}}\\
\subfloat[froben-correl-mcap]{\includegraphics[width = 1.7in]{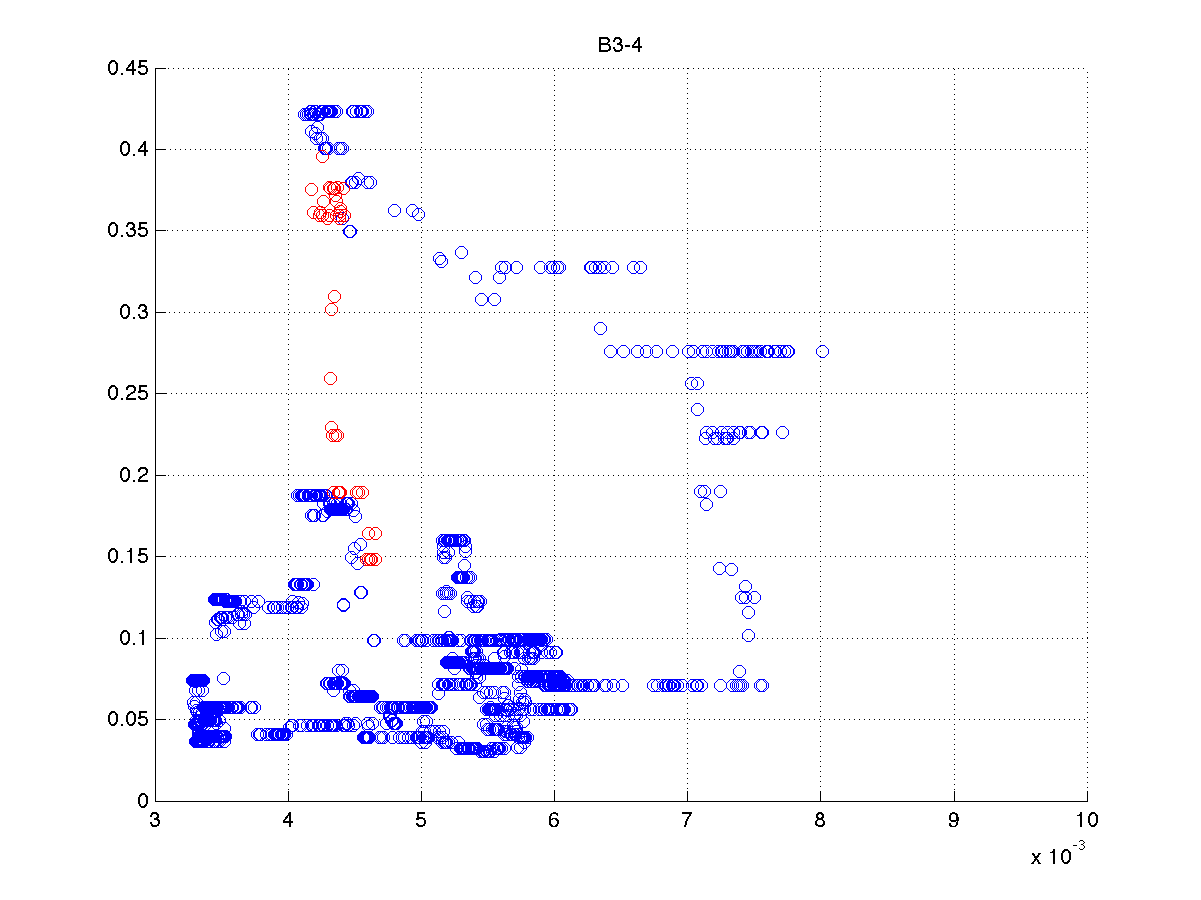}} &
\subfloat[froben-correl-leverage]{\includegraphics[width = 1.7in]{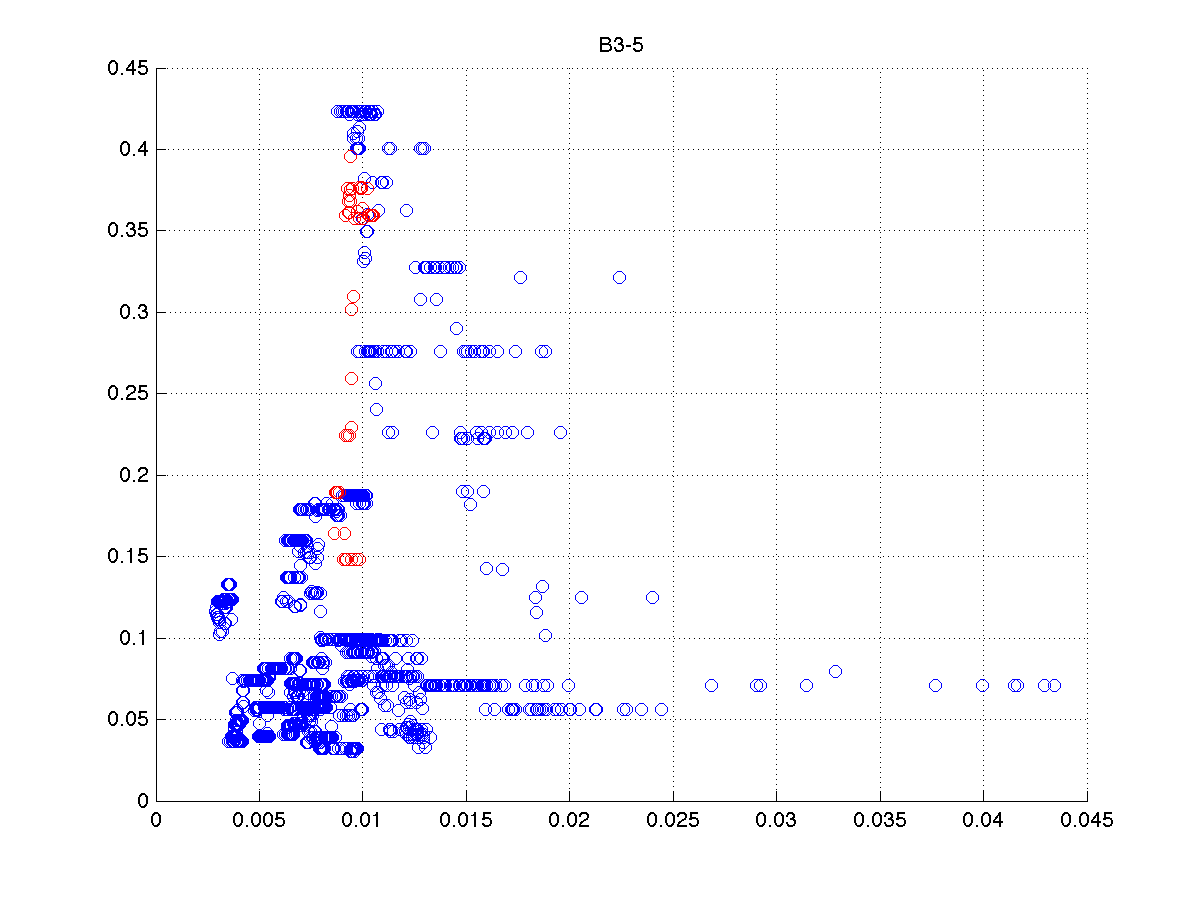}} &
\end{tabular}
\captionsetup{labelformat=empty}
\caption{SP500: Indicators of the $\beta$-series. Red: in-sample ; Blue: out-of-sample}
\end{figure}

\begin{figure}[H]
\begin{tabular}{ccc}
\subfloat[$\mathscr{R}_{1}$covar]{\includegraphics[width = 1.7in]{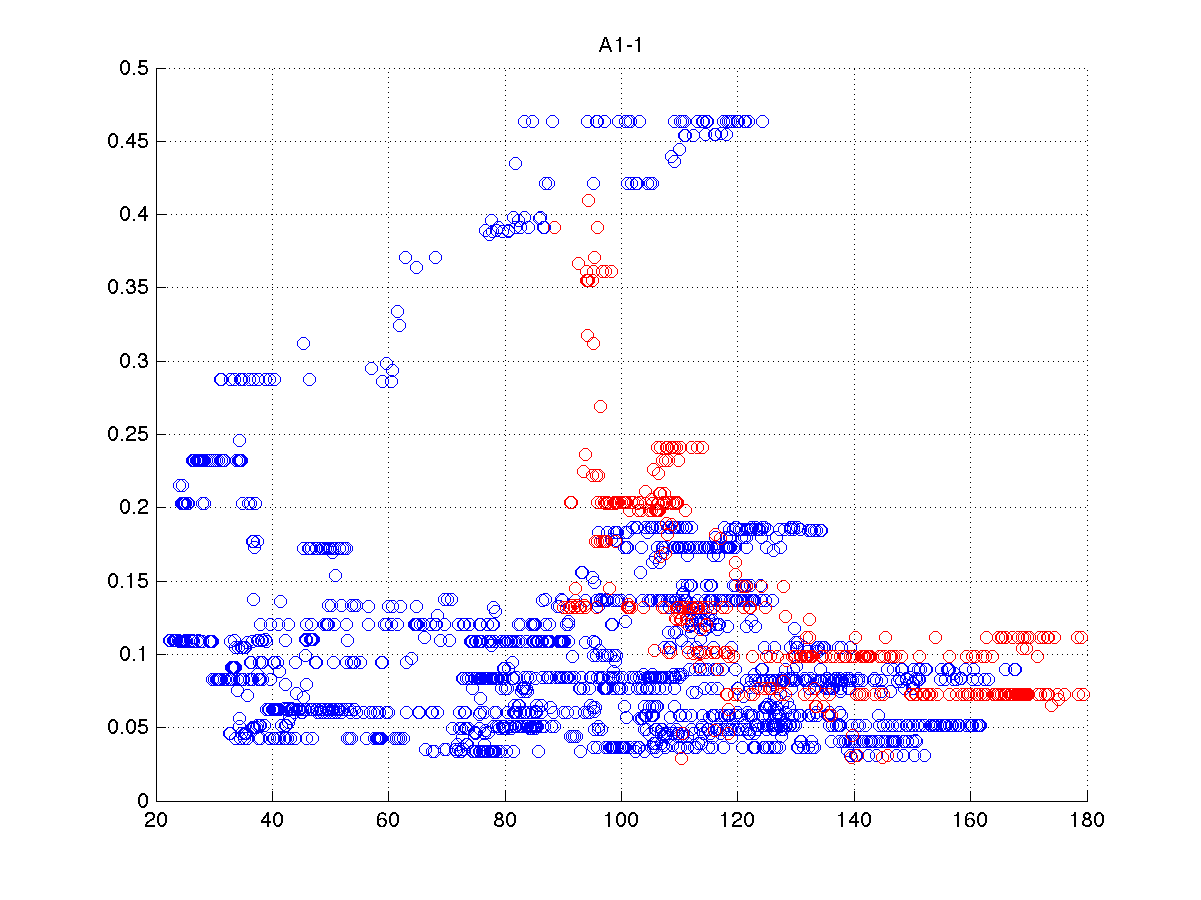}} &
\subfloat[$\mathscr{R}_{1}$correl]{\includegraphics[width = 1.7in]{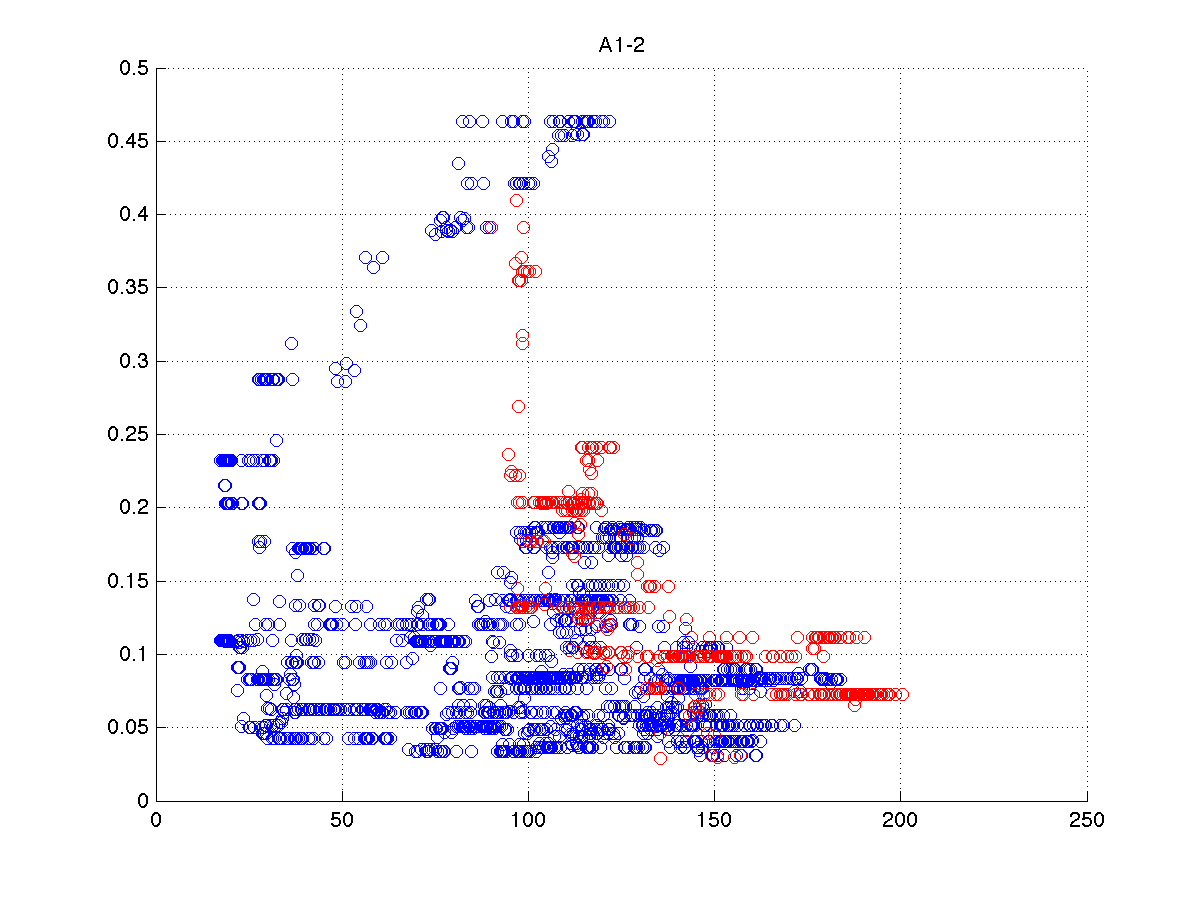}} &
\subfloat[$\mathscr{R}_{1}$correl-volume]{\includegraphics[width = 1.7in]{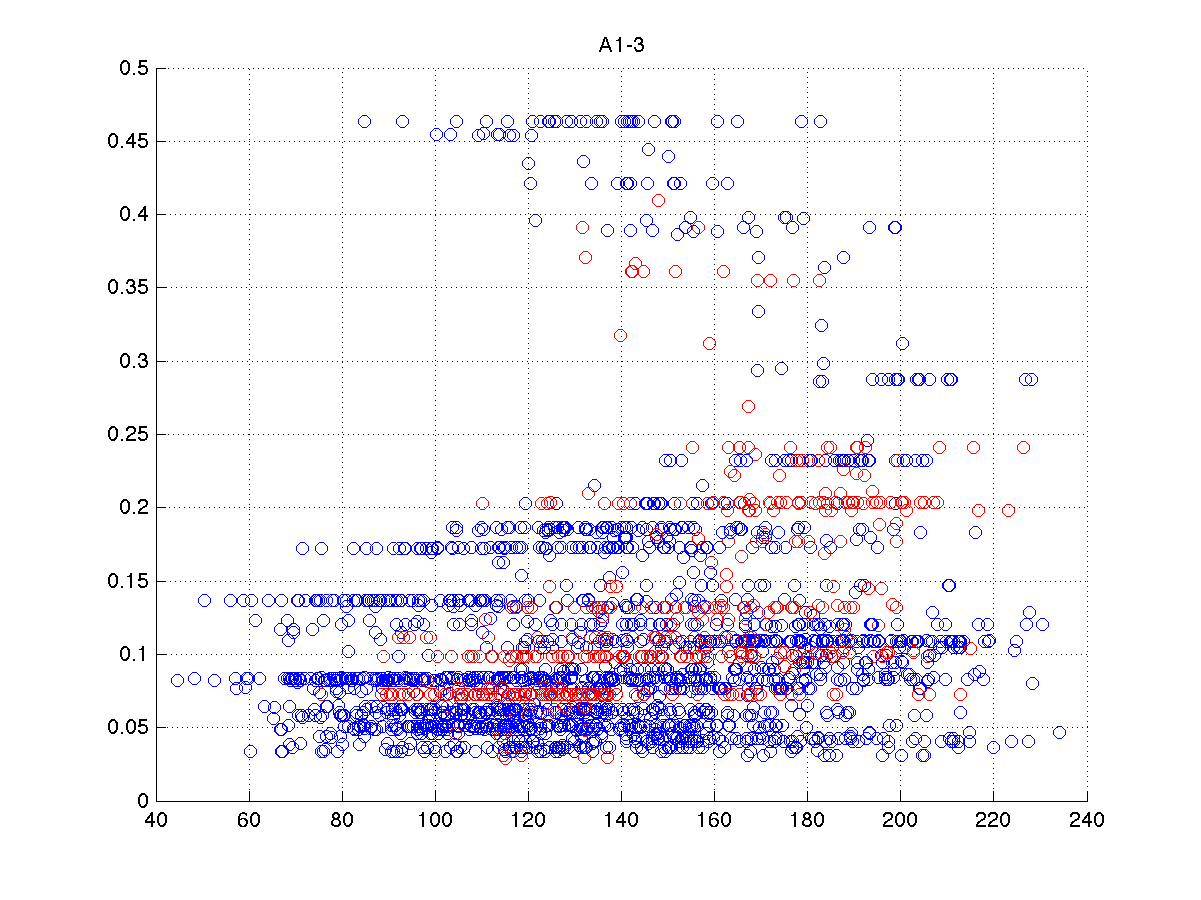}}\\
\subfloat[$\mathscr{R}_{1}$correl-mcap]{\includegraphics[width = 1.7in]{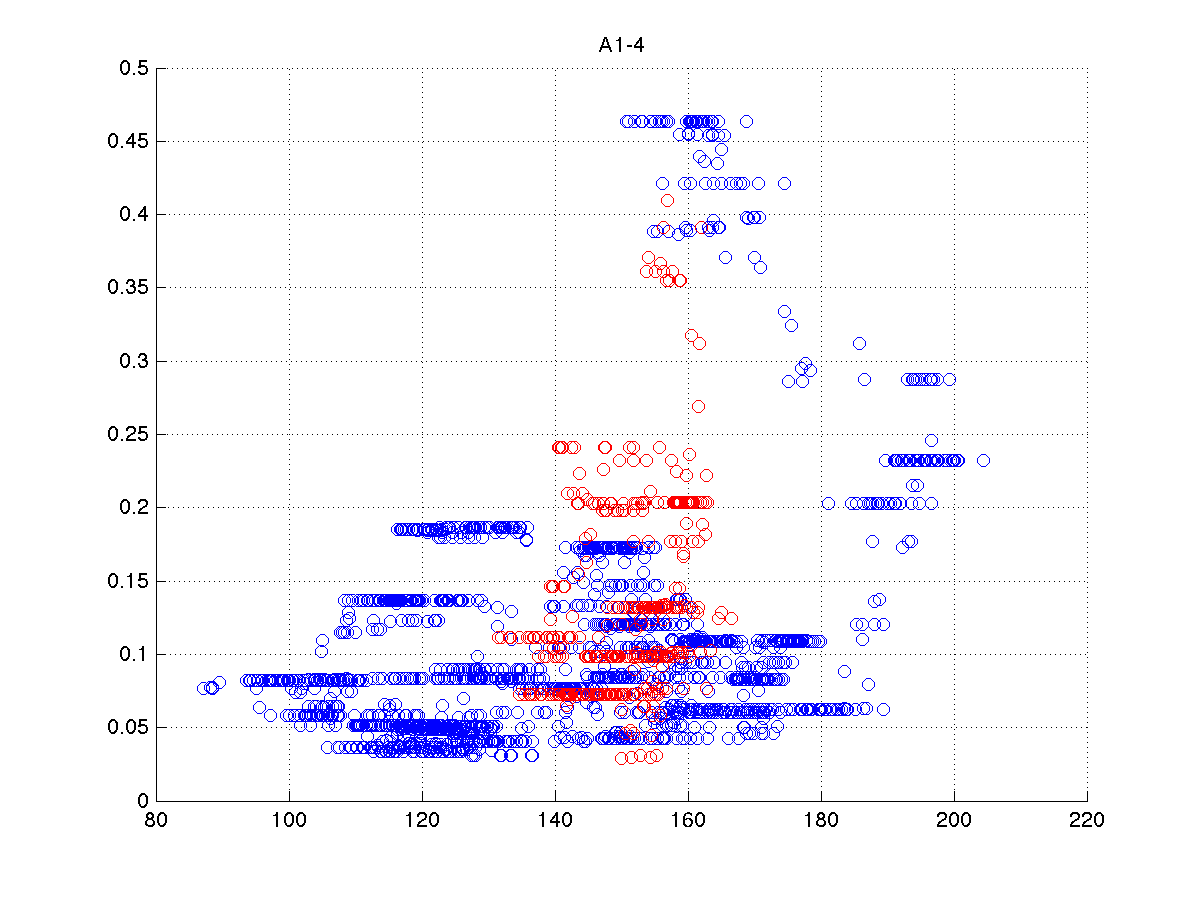}}&
\subfloat[$\mathscr{R}_{1}$correl-leverage]{\includegraphics[width = 1.7in]{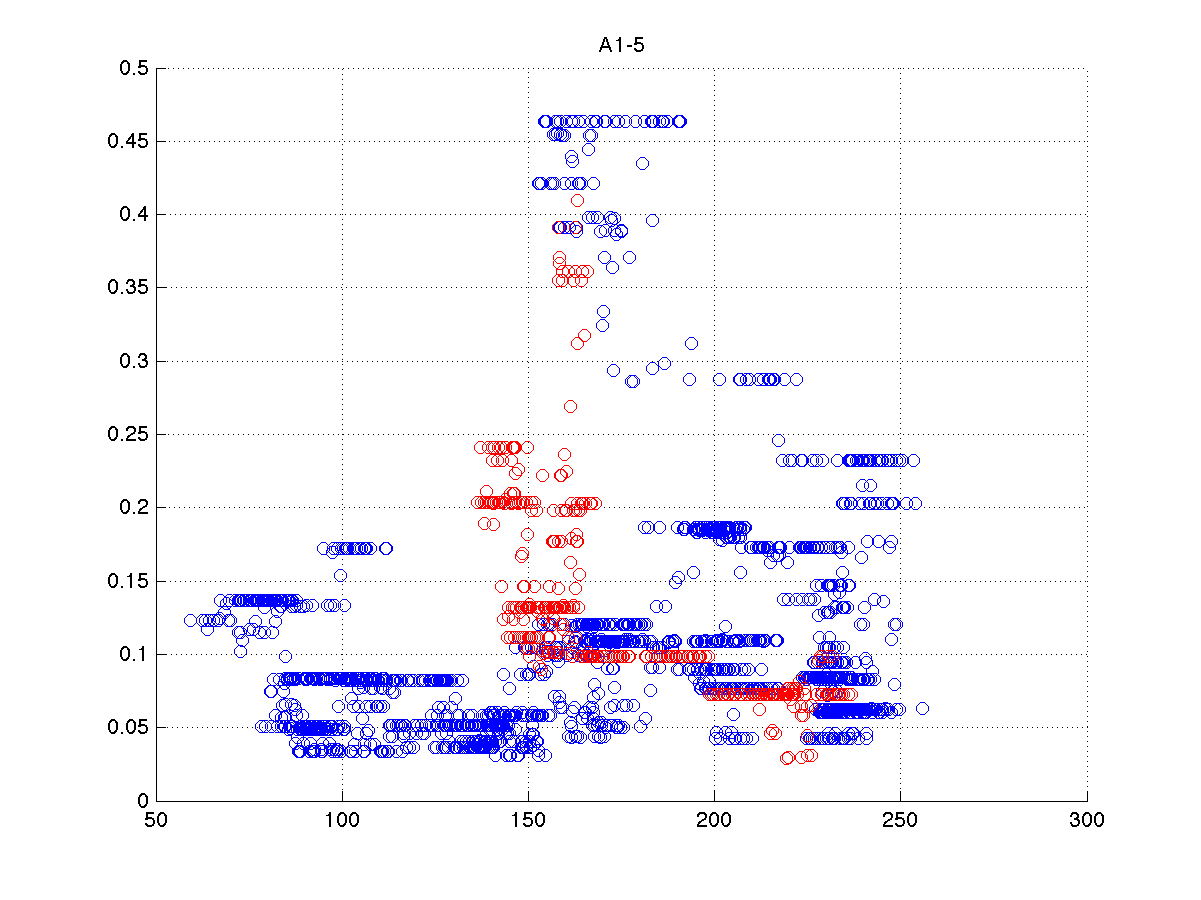}} &
\subfloat[$\mathscr{R}_{2}$covar]{\includegraphics[width = 1.7in]{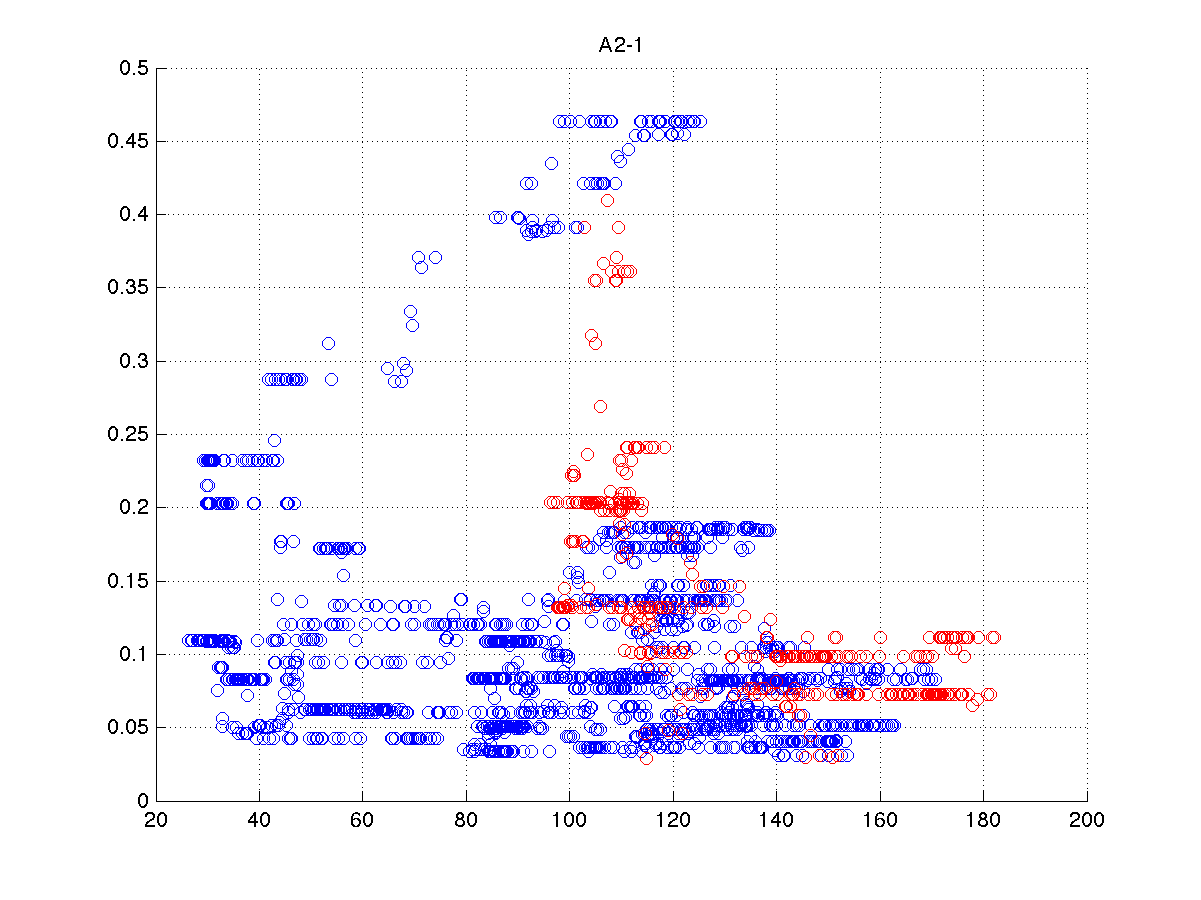}}\\
\subfloat[$\mathscr{R}_{2}$correl]{\includegraphics[width = 1.7in]{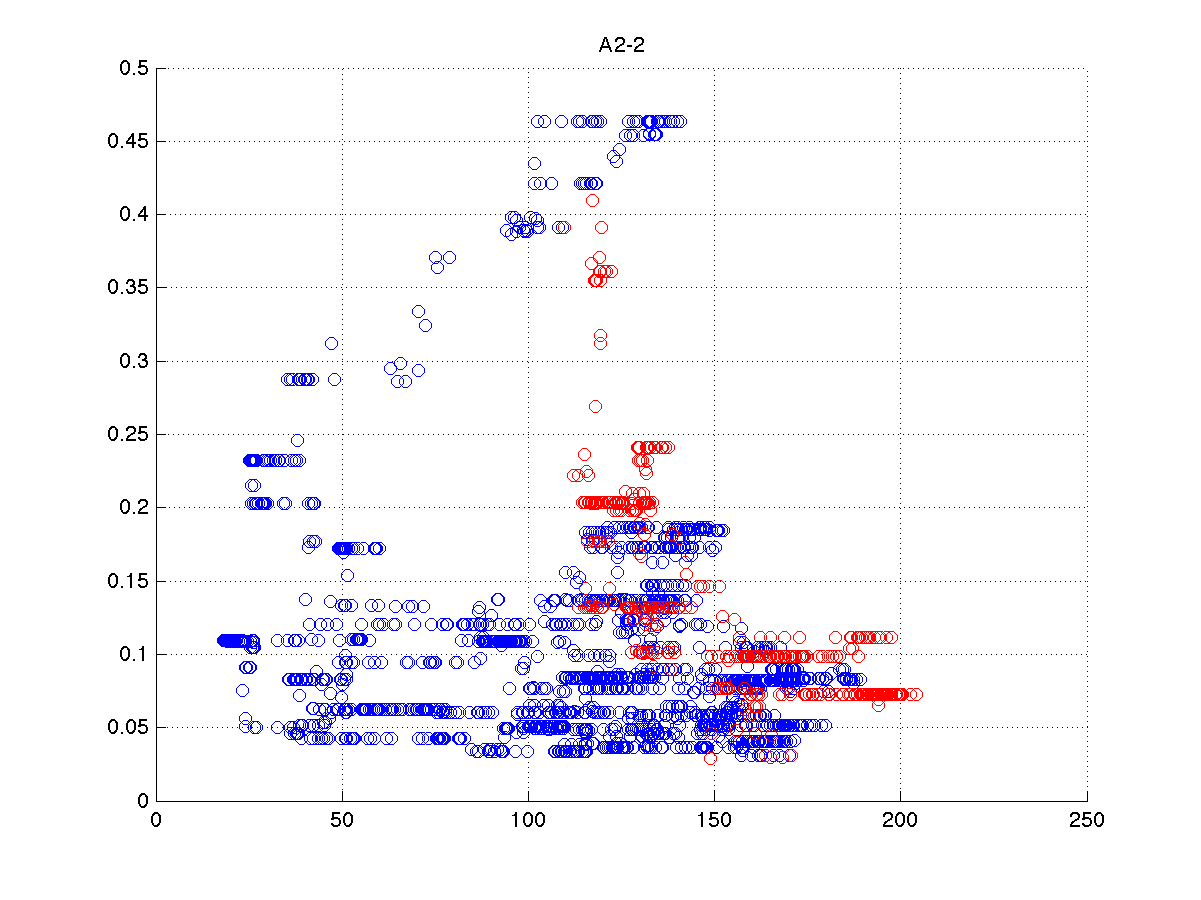}} &
\subfloat[$\mathscr{R}_{2}$correl-volume]{\includegraphics[width = 1.7in]{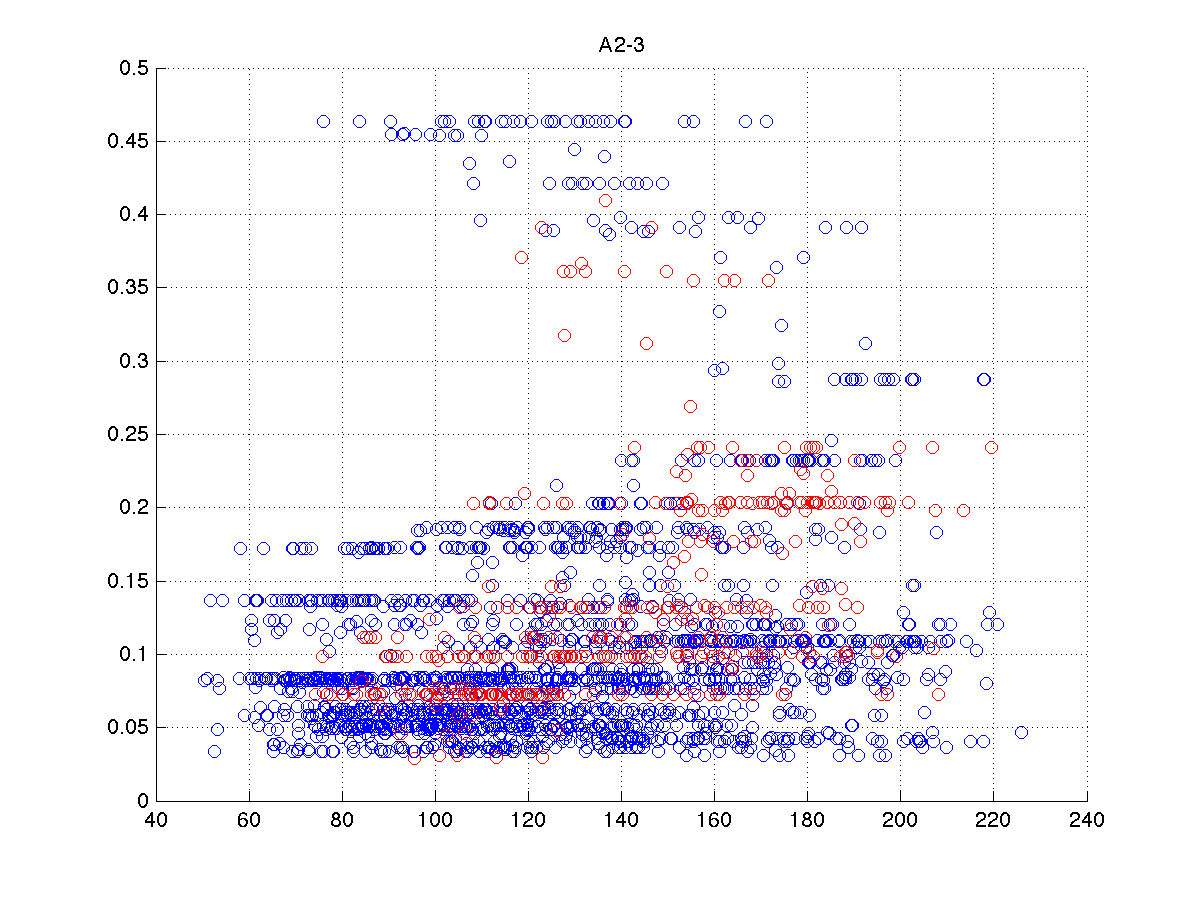}}&
\subfloat[$\mathscr{R}_{2}$correl-mcap]{\includegraphics[width = 1.7in]{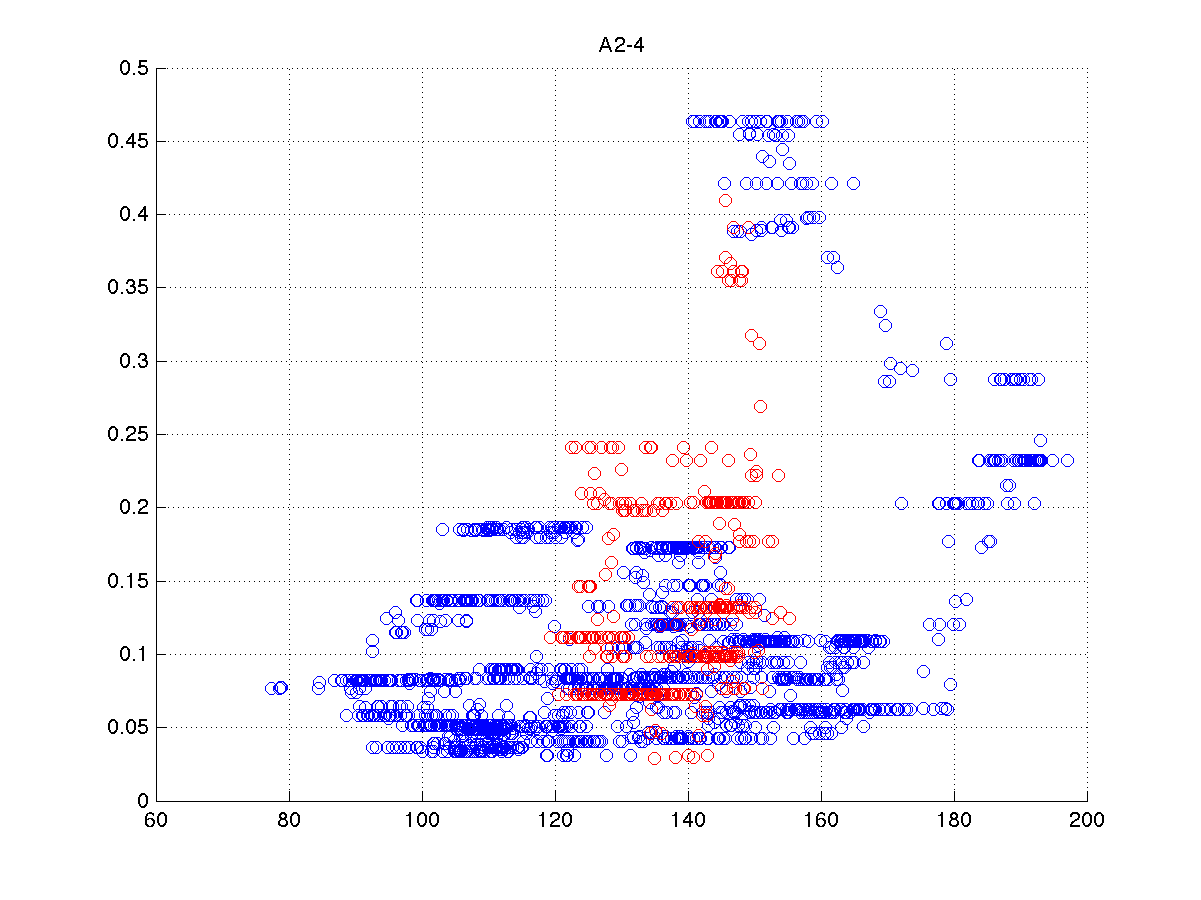}}\\
\subfloat[$\mathscr{R}_{2}$correl-leverage]{\includegraphics[width = 1.7in]{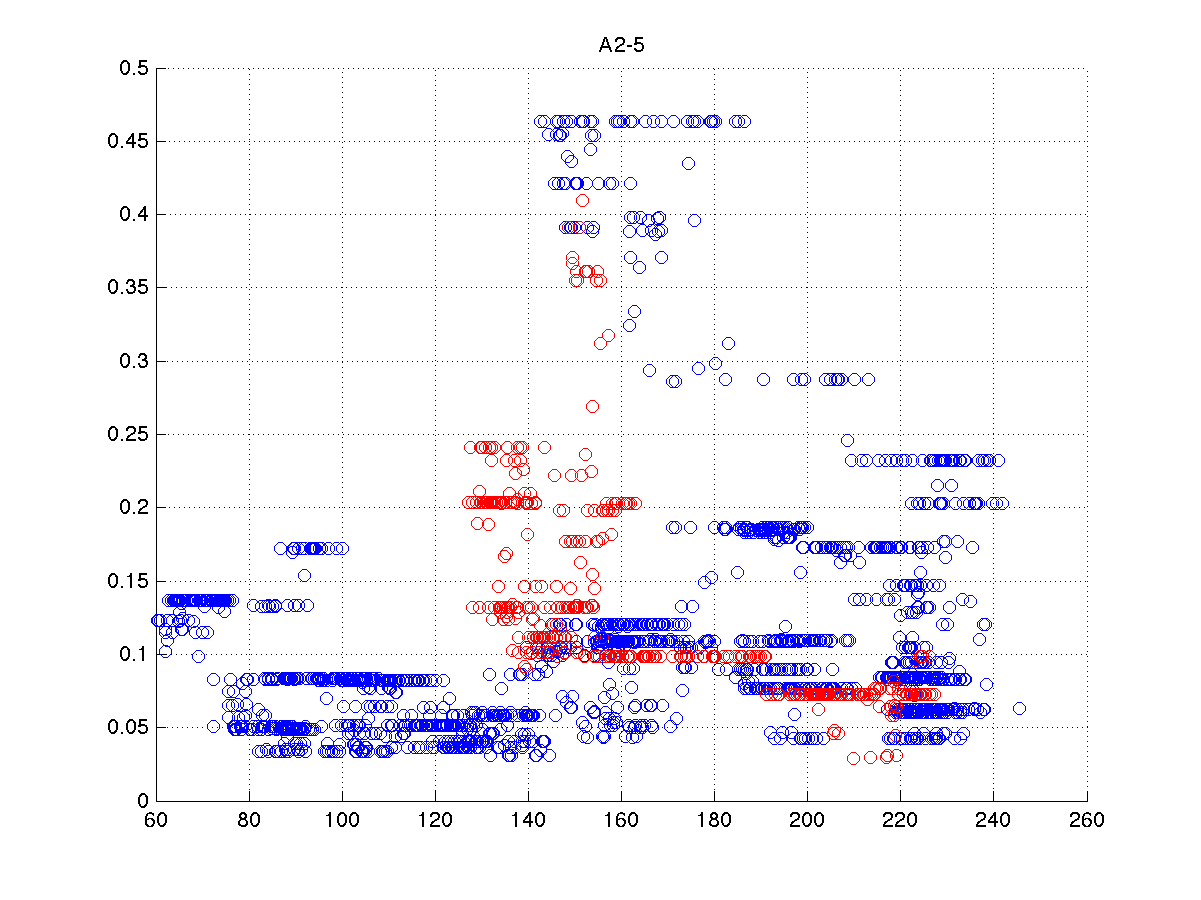}} &
\subfloat[$\mathscr{R}_{3}$covar]{\includegraphics[width = 1.7in]{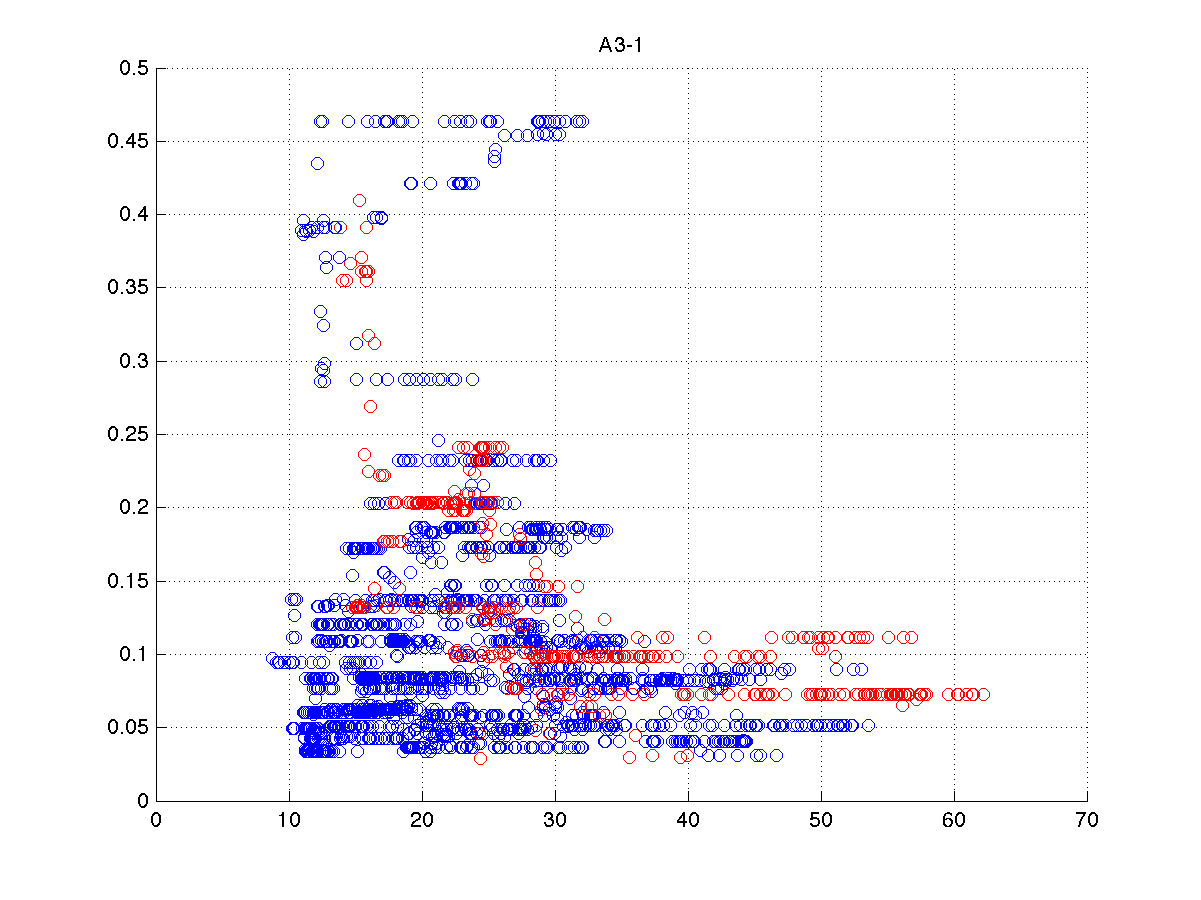}} &
\subfloat[$\mathscr{R}_{3}$correl]{\includegraphics[width = 1.7in]{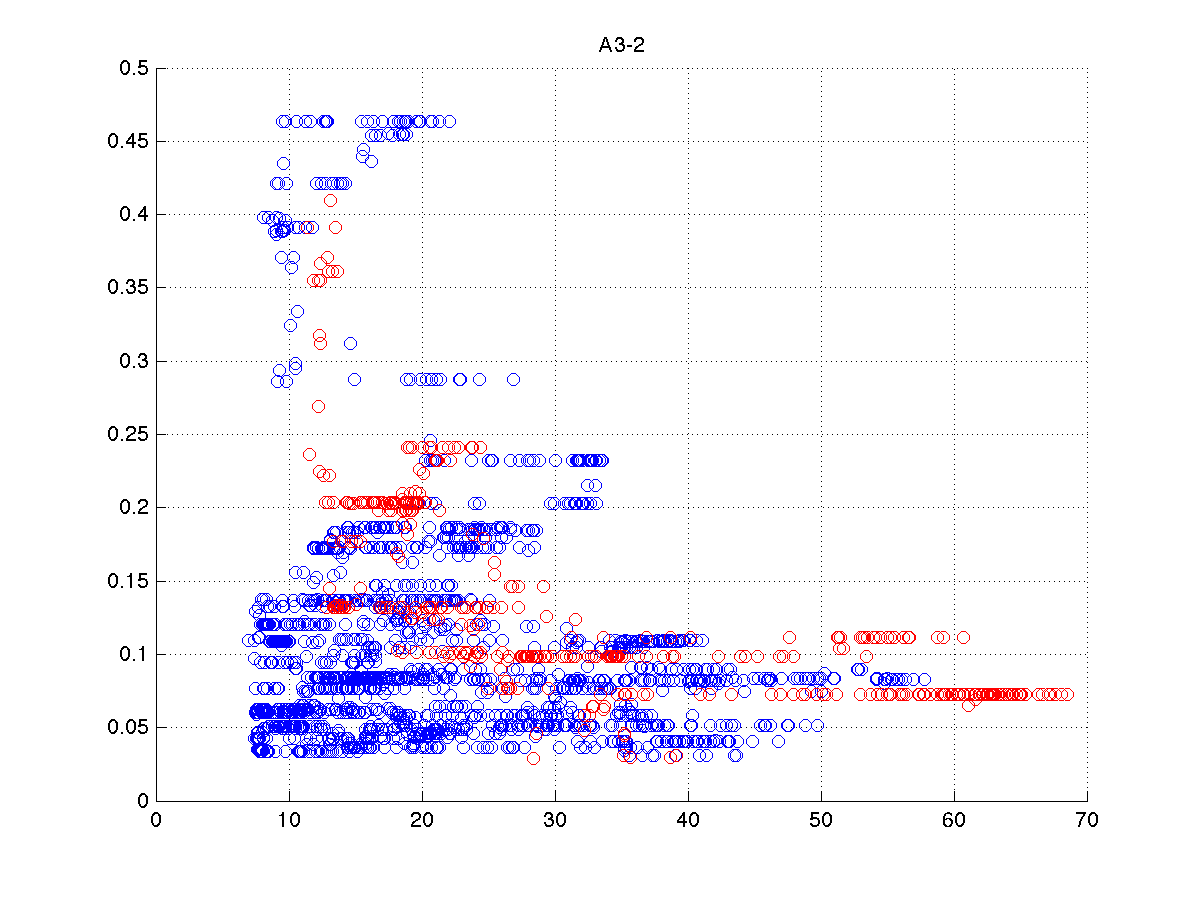}}\\
\subfloat[$\mathscr{R}_{3}$correl-volume]{\includegraphics[width = 1.7in]{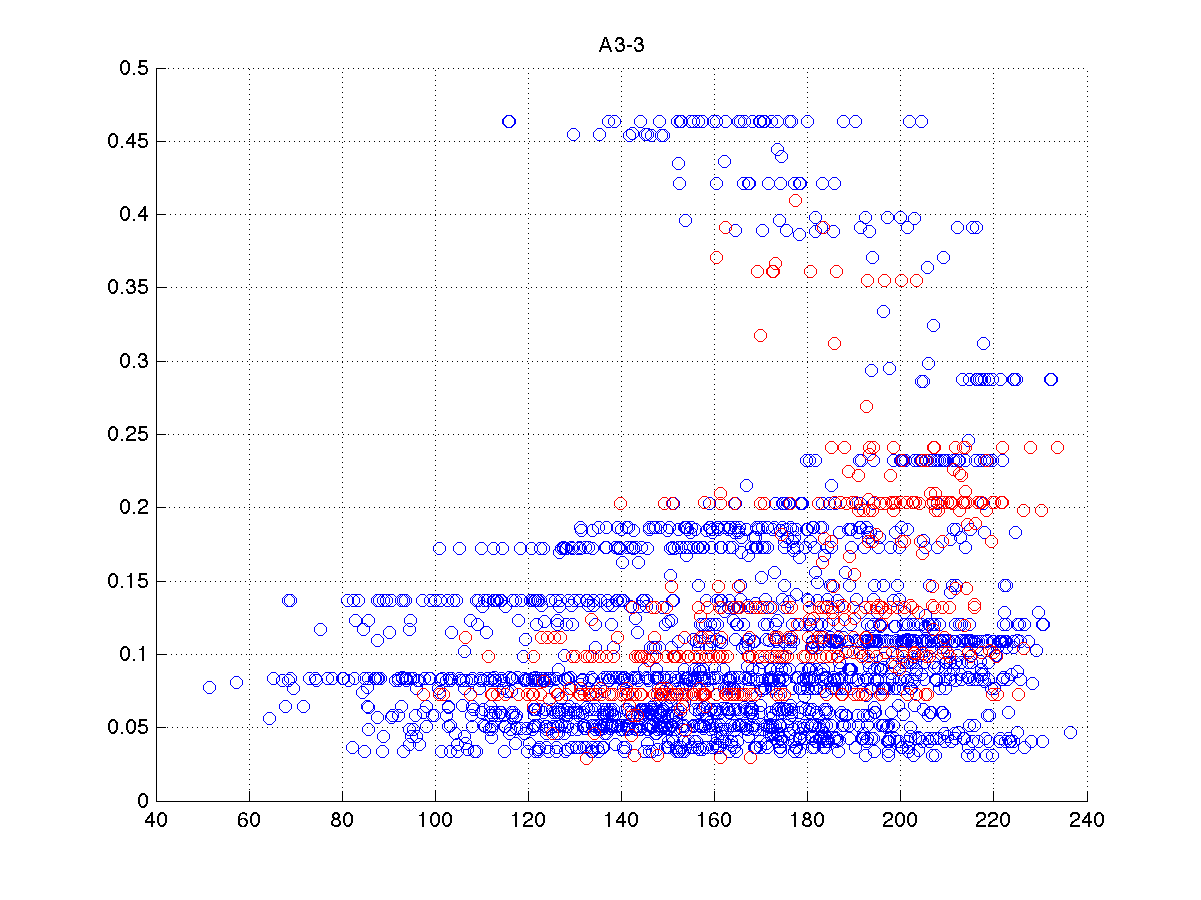}} &
\subfloat[$\mathscr{R}_{3}$correl-mcap]{\includegraphics[width = 1.7in]{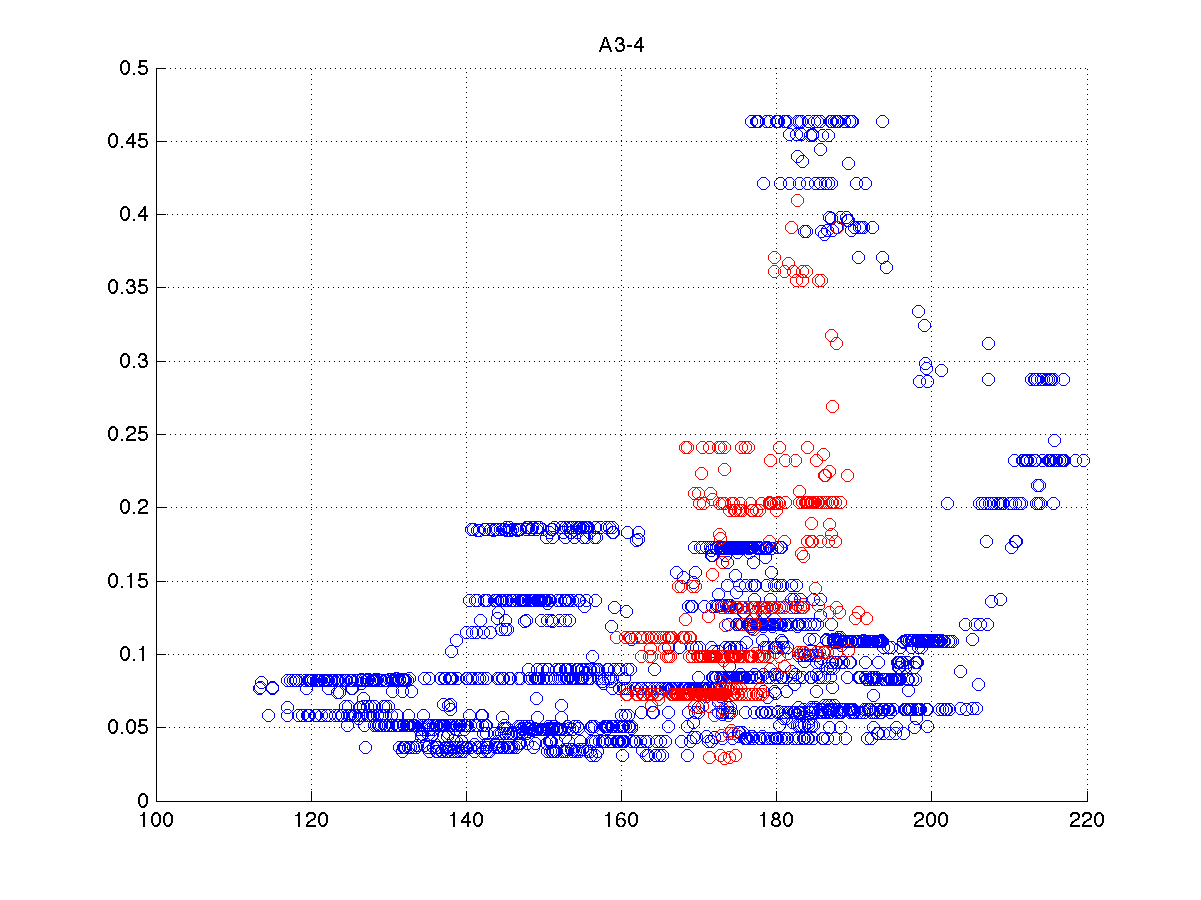}} &
\subfloat[$\mathscr{R}_{3}$correl-leverage]{\includegraphics[width = 1.7in]{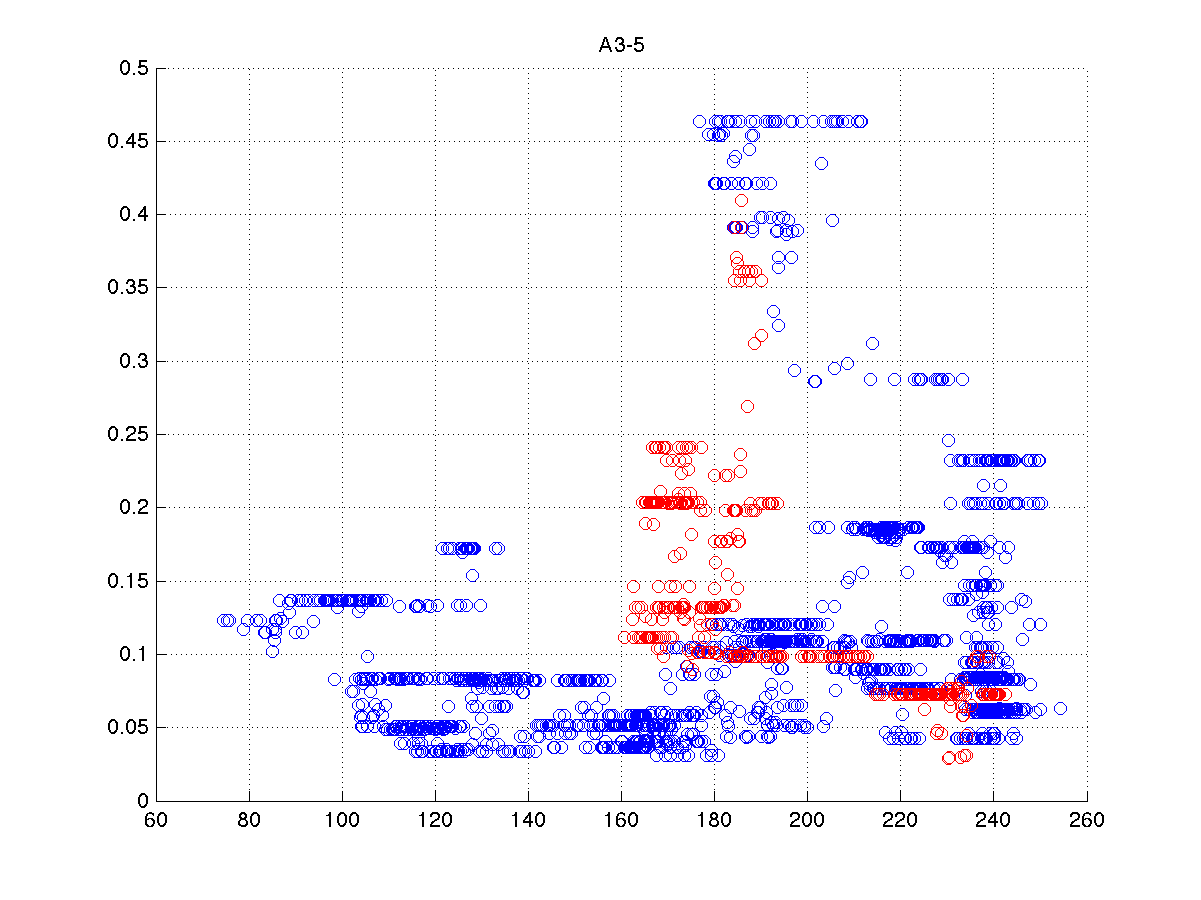}}\\
\end{tabular}
\captionsetup{labelformat=empty}
\caption{NASDAQ: Indicators of the $\alpha$-series. Red: in-sample ; Blue: out-of-sample}
\end{figure}

\begin{figure}[H]
\begin{tabular}{ccc}
\subfloat[rspec-covar]{\includegraphics[width = 1.7in]{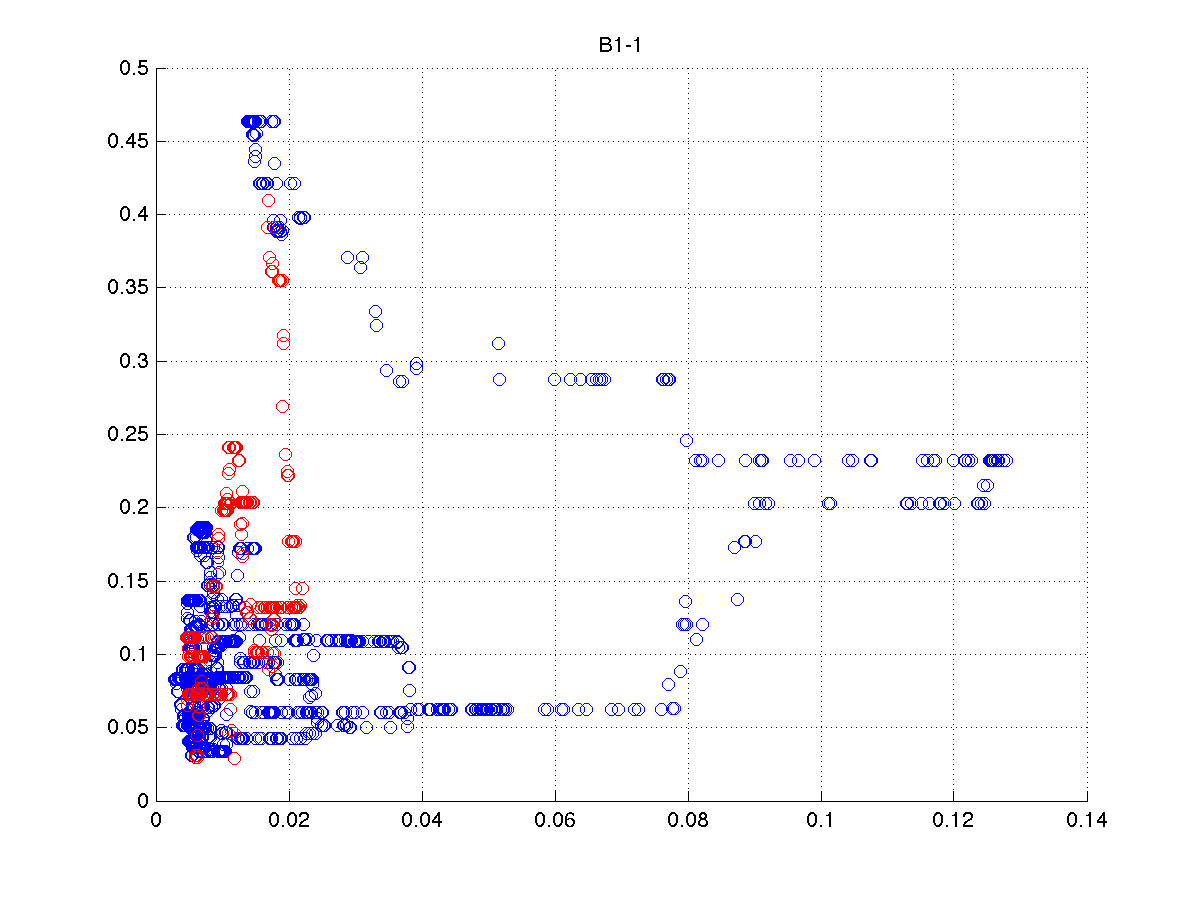}} &
\subfloat[rspec-correl]{\includegraphics[width = 1.7in]{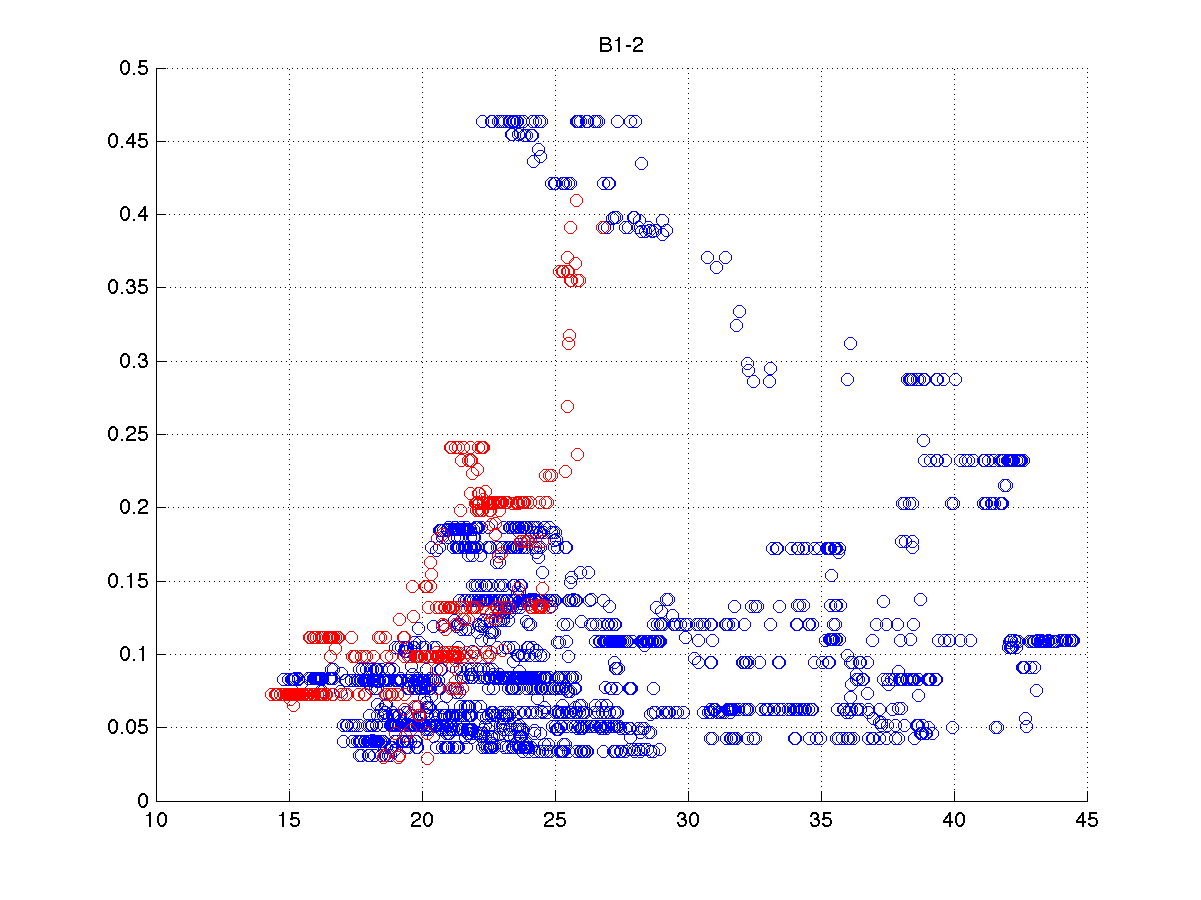}} &
\subfloat[rspec-correl-volume]{\includegraphics[width = 1.7in]{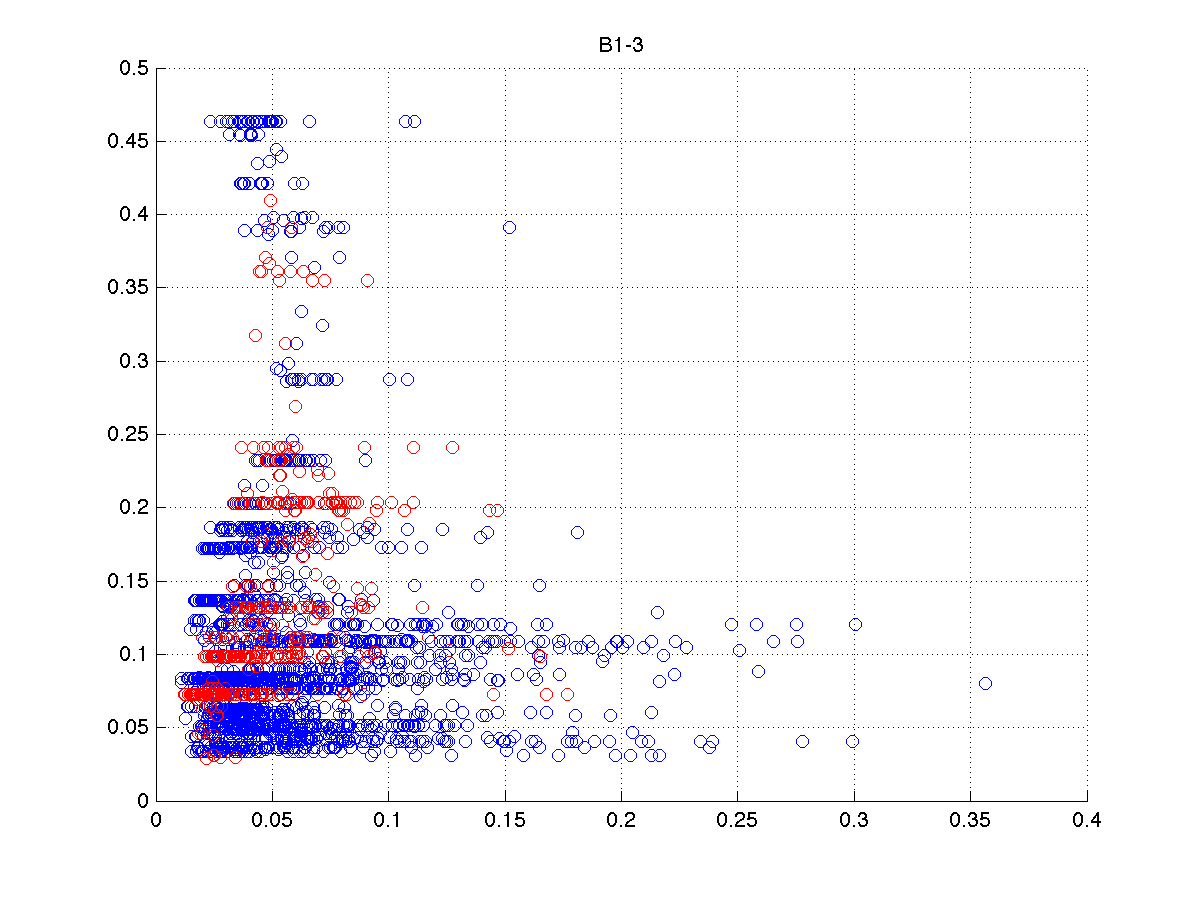}} \\
\subfloat[rspec-correl-mcap]{\includegraphics[width = 1.7in]{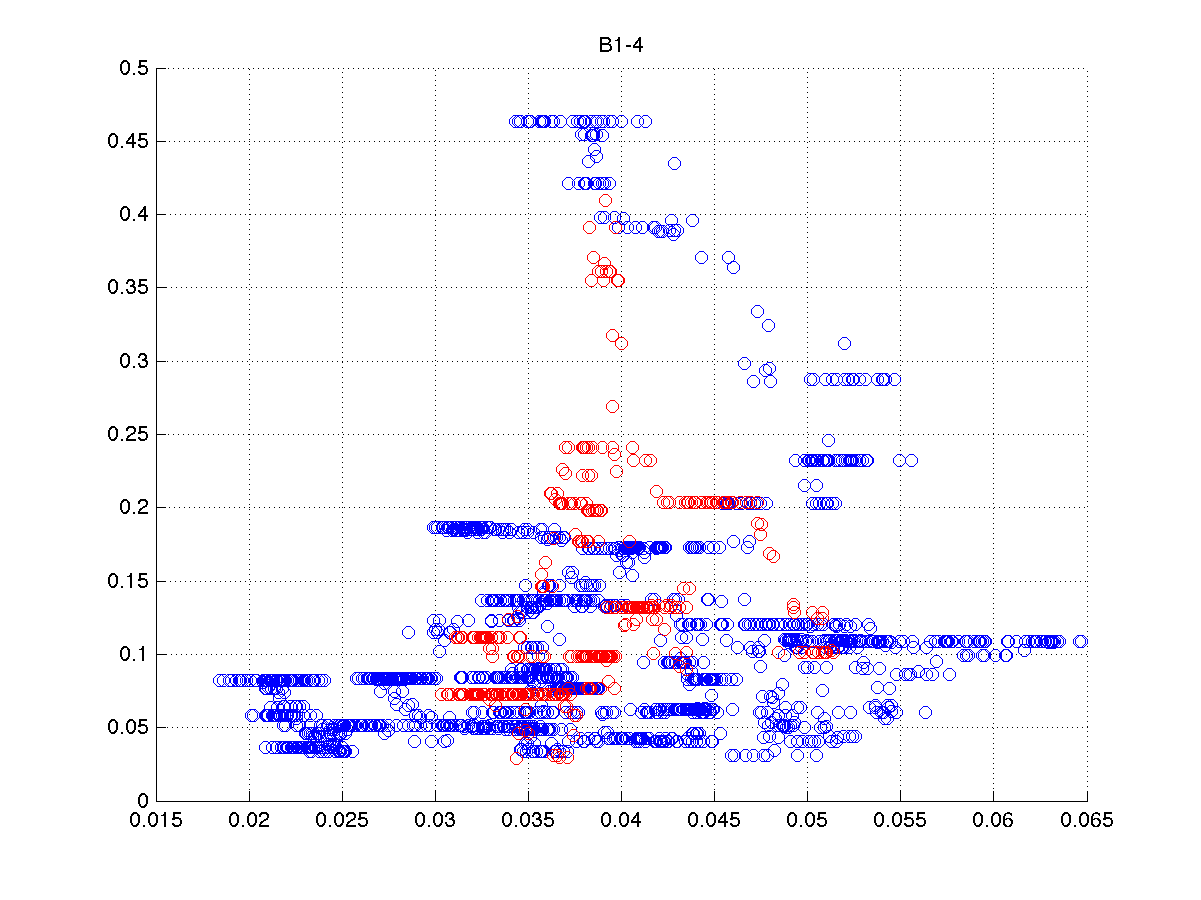}}&
\subfloat[rspec-correl-leverage]{\includegraphics[width = 1.7in]{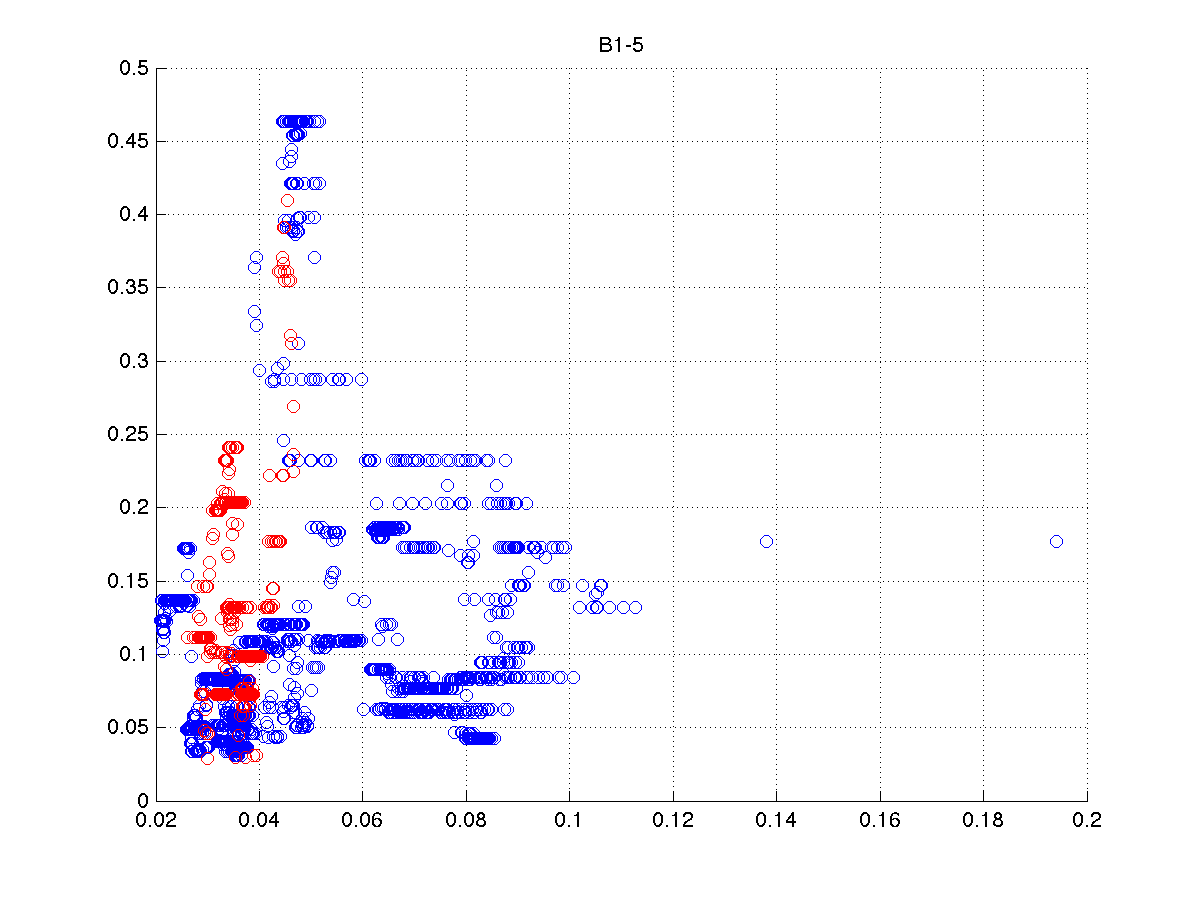}} &
\subfloat[trace-covar]{\includegraphics[width = 1.7in]{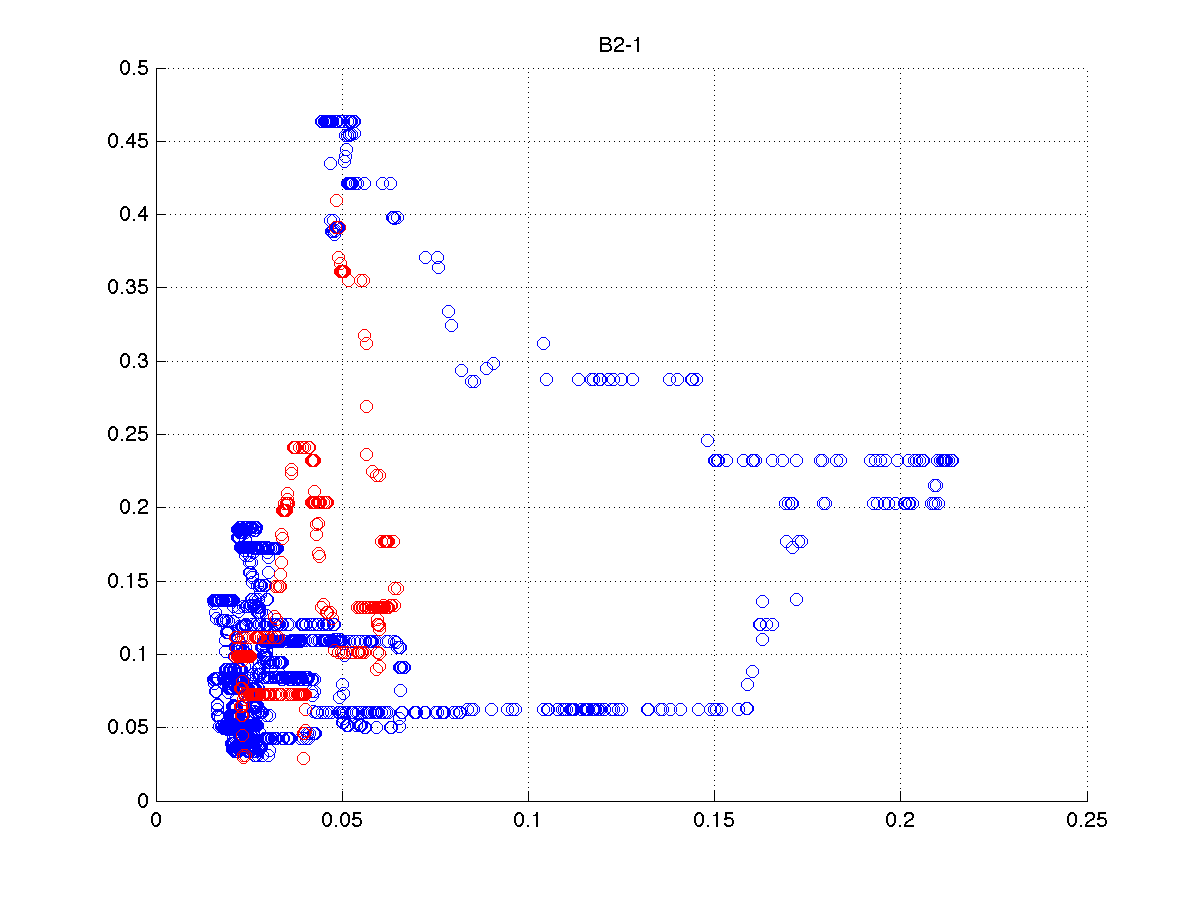}}\\
\subfloat[trace-correl-volume]{\includegraphics[width = 1.7in]{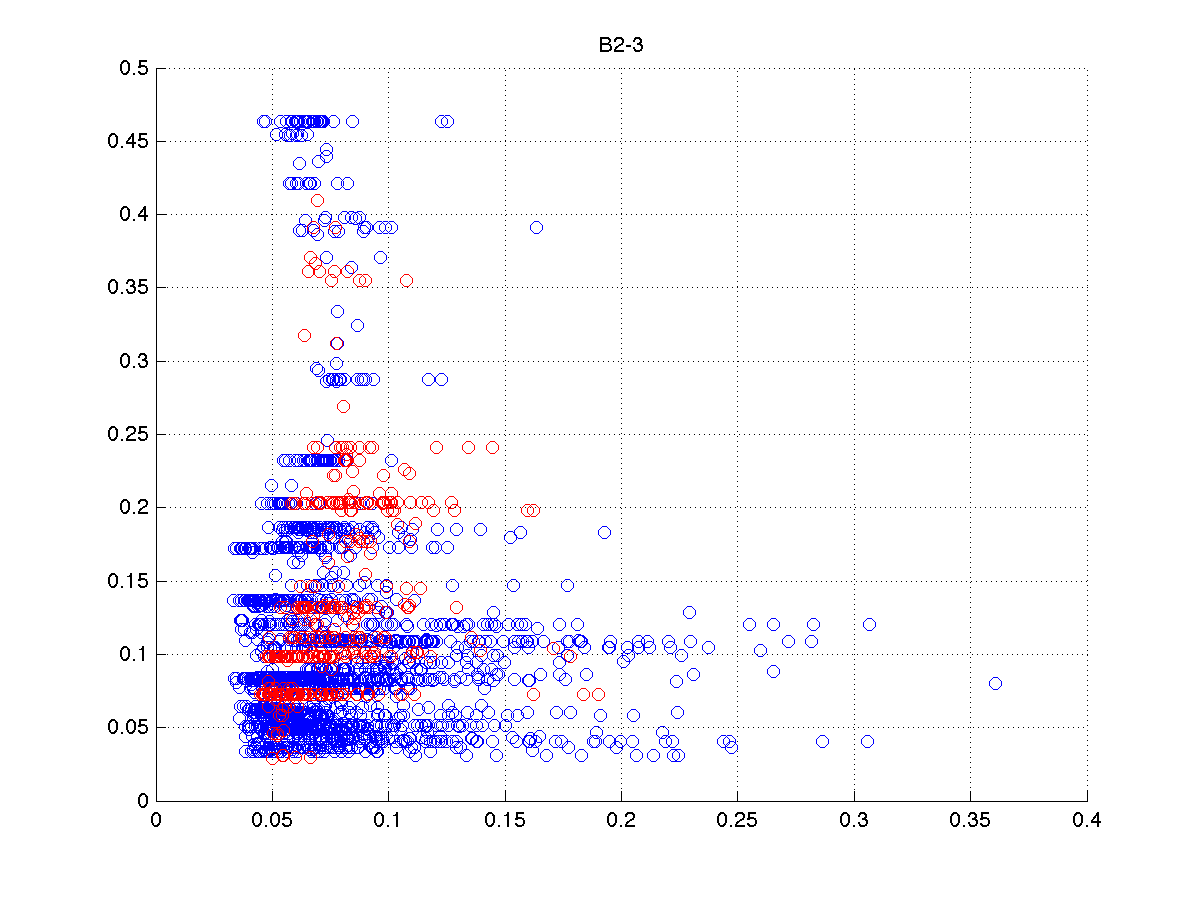}} &
\subfloat[trace-correl-mcap]{\includegraphics[width = 1.7in]{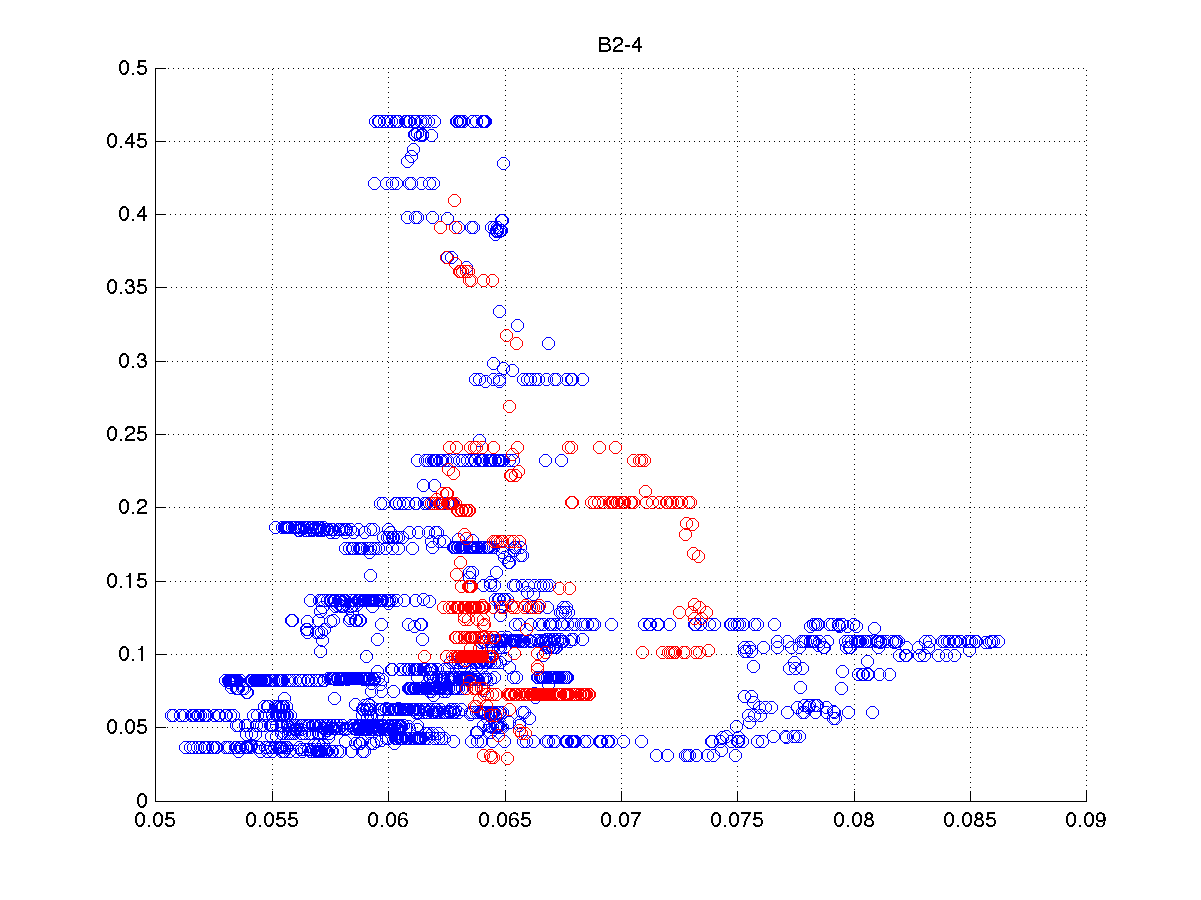}}&
\subfloat[trace-correl-leverage]{\includegraphics[width = 1.5in]{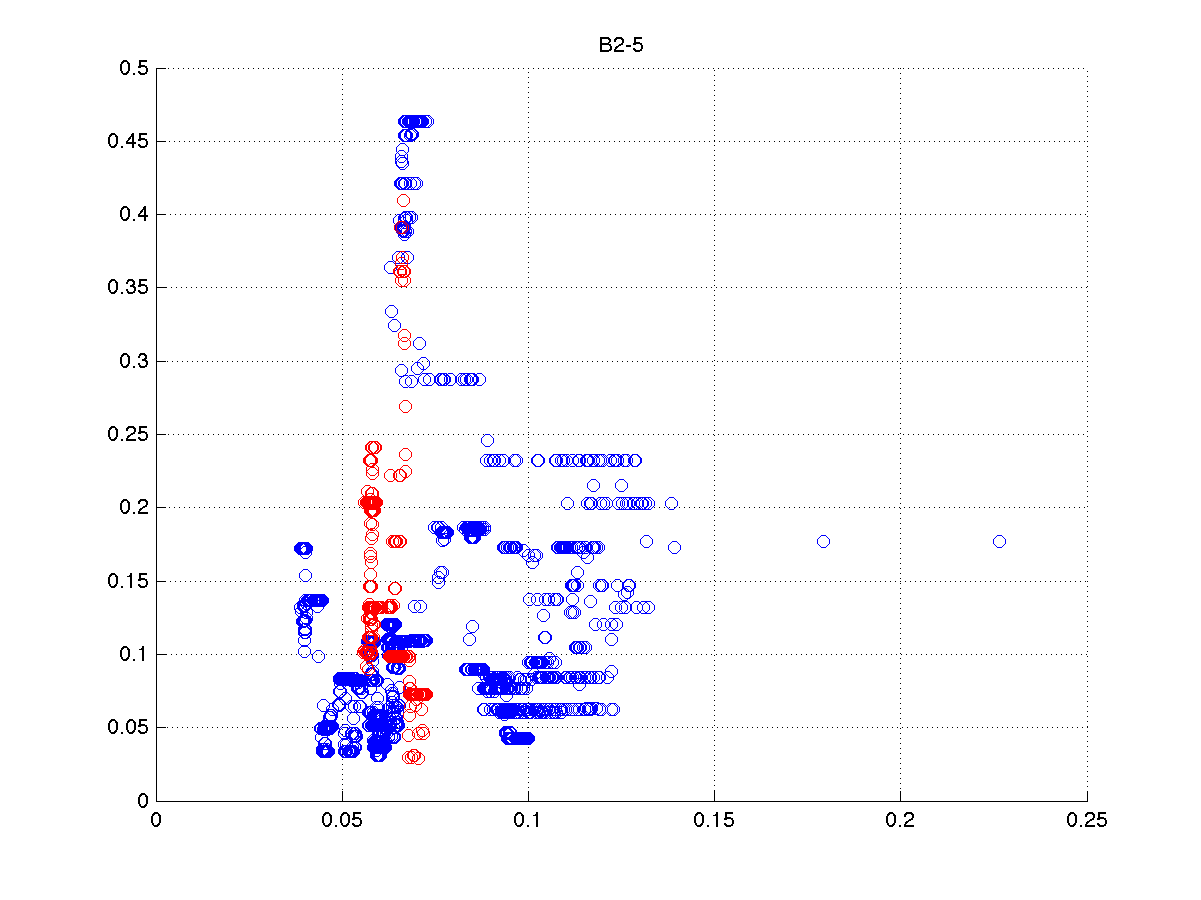}}\\
\subfloat[froben-covar]{\includegraphics[width = 1.7in]{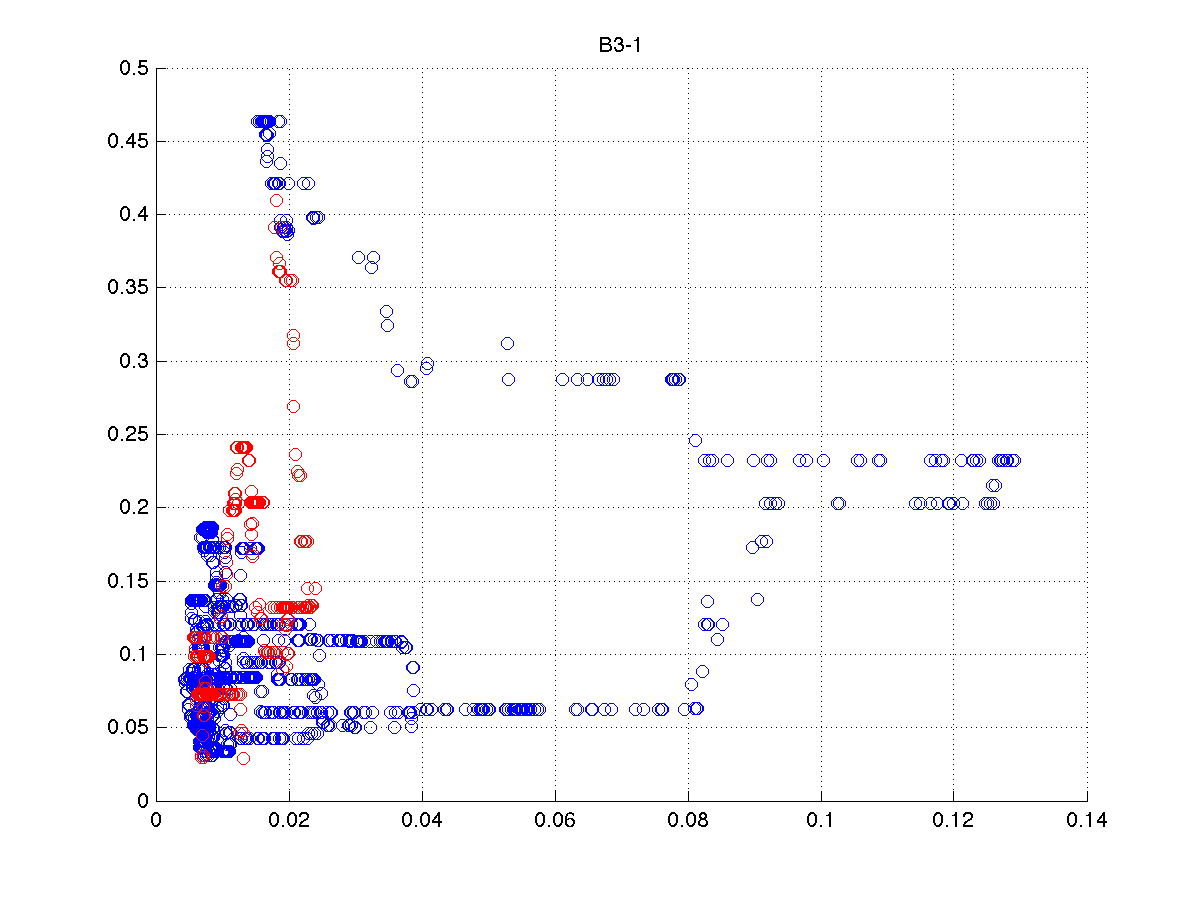}} &
\subfloat[froben-correl]{\includegraphics[width = 1.7in]{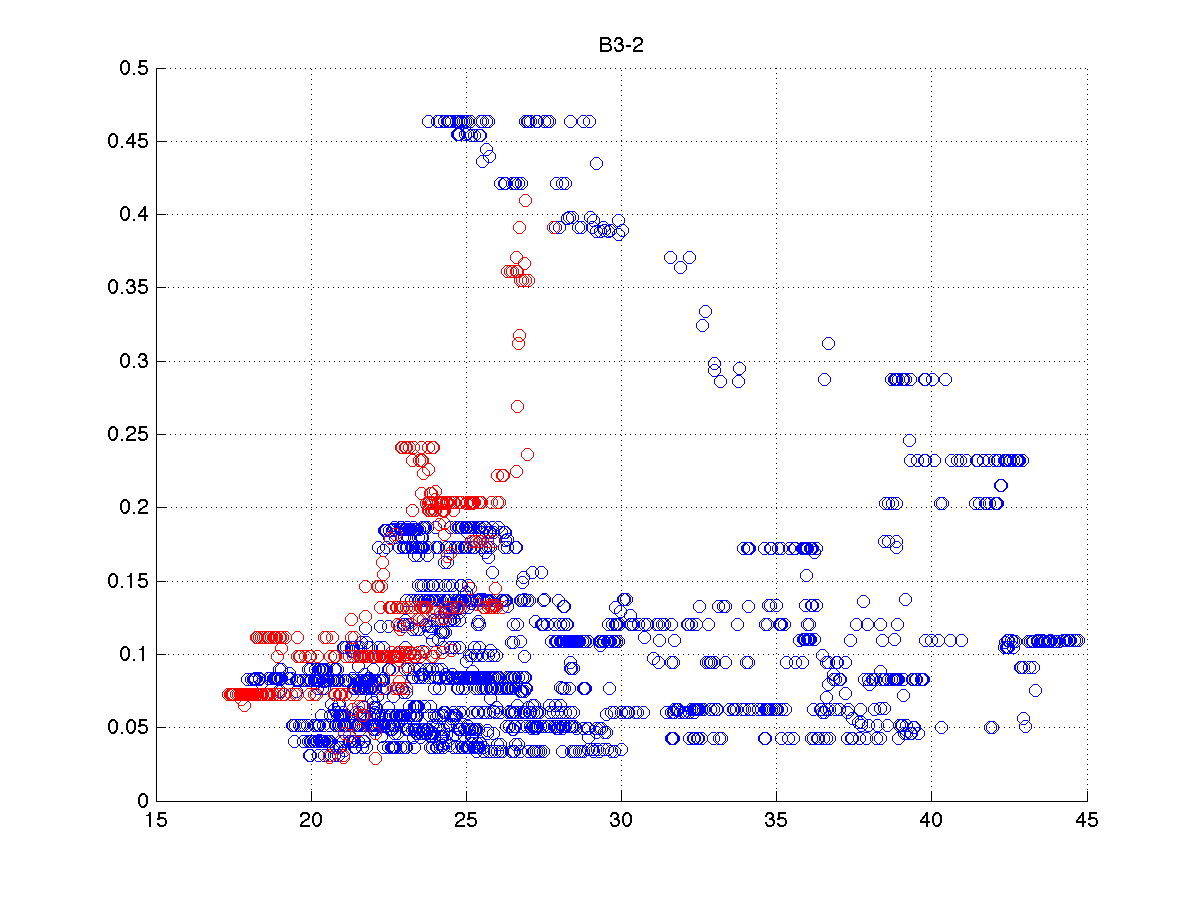}} &
\subfloat[froben-correl-volume]{\includegraphics[width = 1.7in]{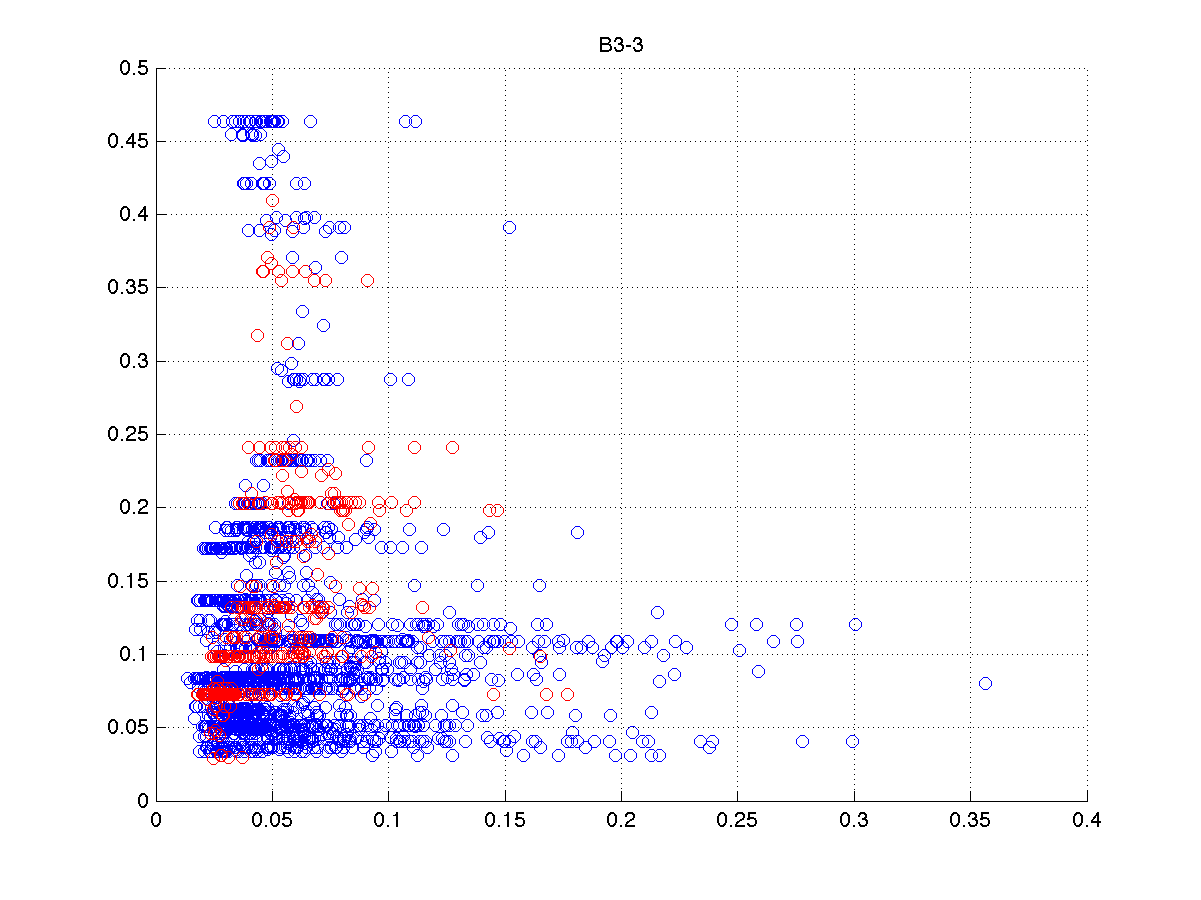}}\\
\subfloat[froben-correl-mcap]{\includegraphics[width = 1.7in]{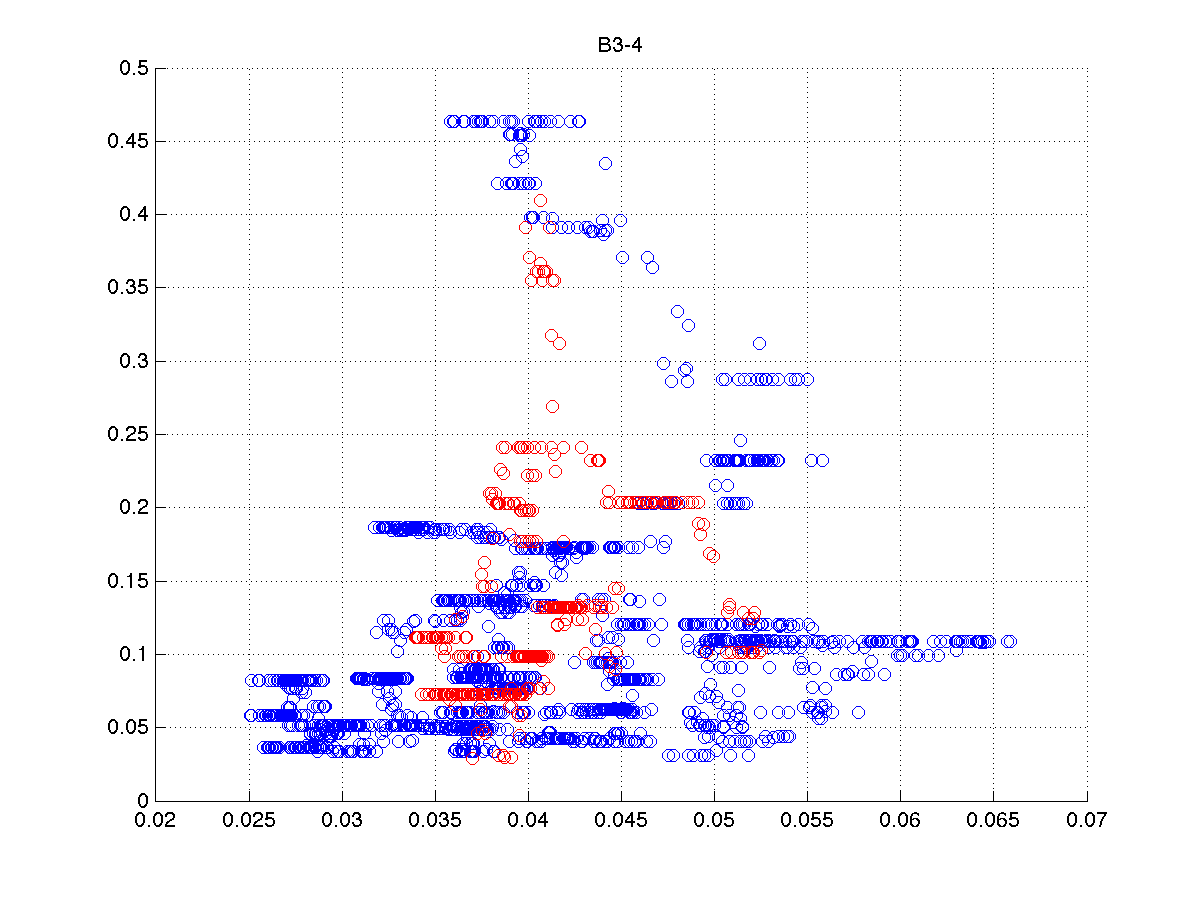}} &
\subfloat[froben-correl-leverage]{\includegraphics[width = 1.7in]{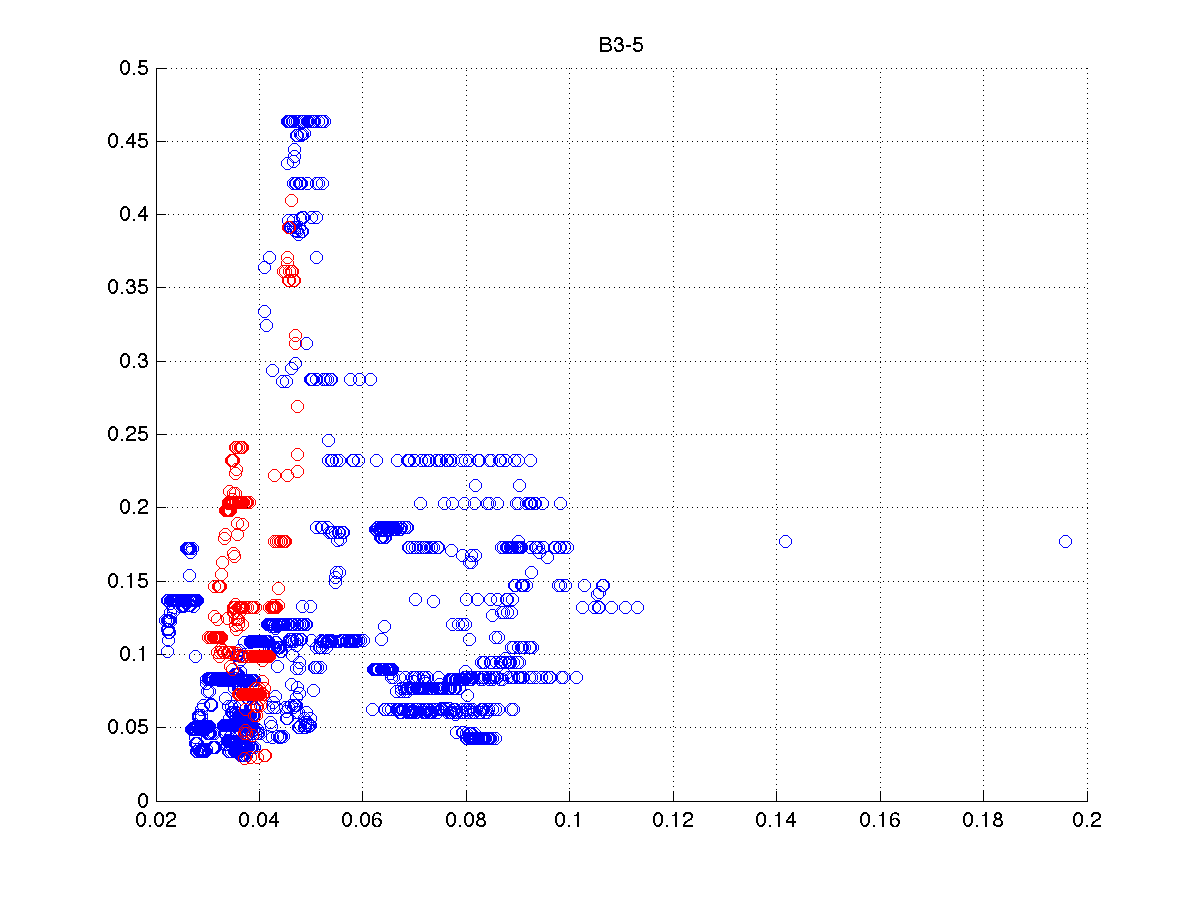}} &
\end{tabular}
\captionsetup{labelformat=empty}
\caption{NASDAQ: Indicators of the $\beta$-series. Red: in-sample ; Blue: out-of-sample}
\end{figure}

\begin{figure}[H]
\begin{tabular}{ccc}
\subfloat[$\mathscr{R}_{1}$covar]{\includegraphics[width = 1.7in]{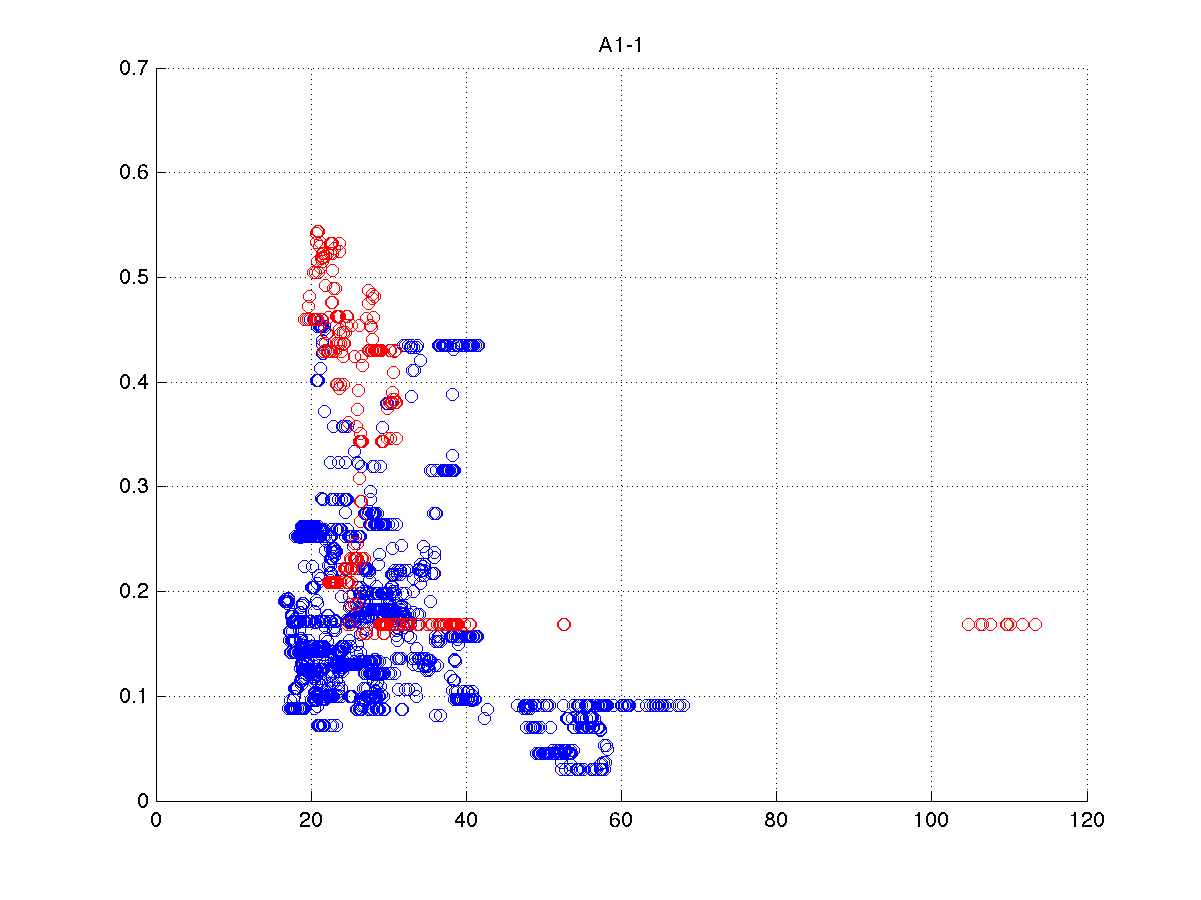}} &
\subfloat[$\mathscr{R}_{1}$correl]{\includegraphics[width = 1.7in]{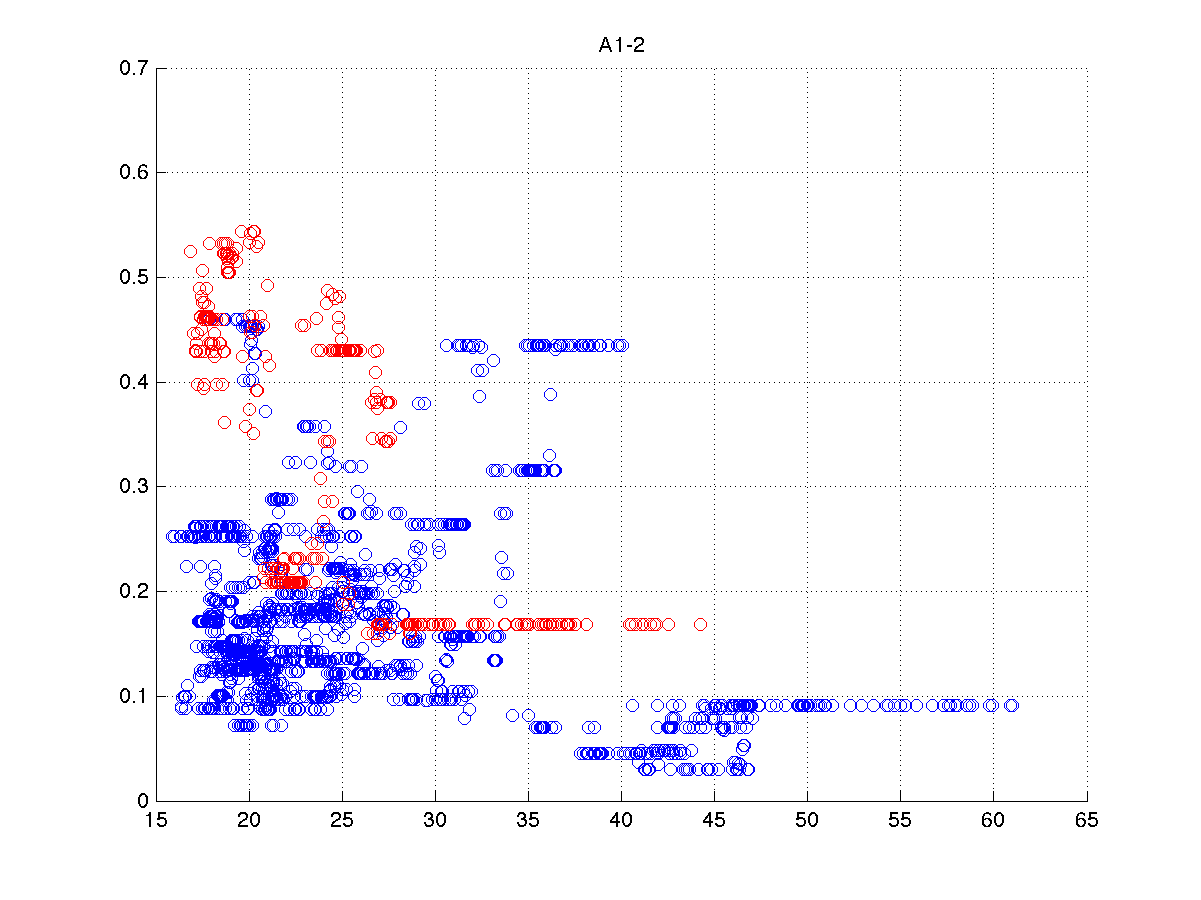}} &
\subfloat[$\mathscr{R}_{1}$correl-volume]{\includegraphics[width = 1.7in]{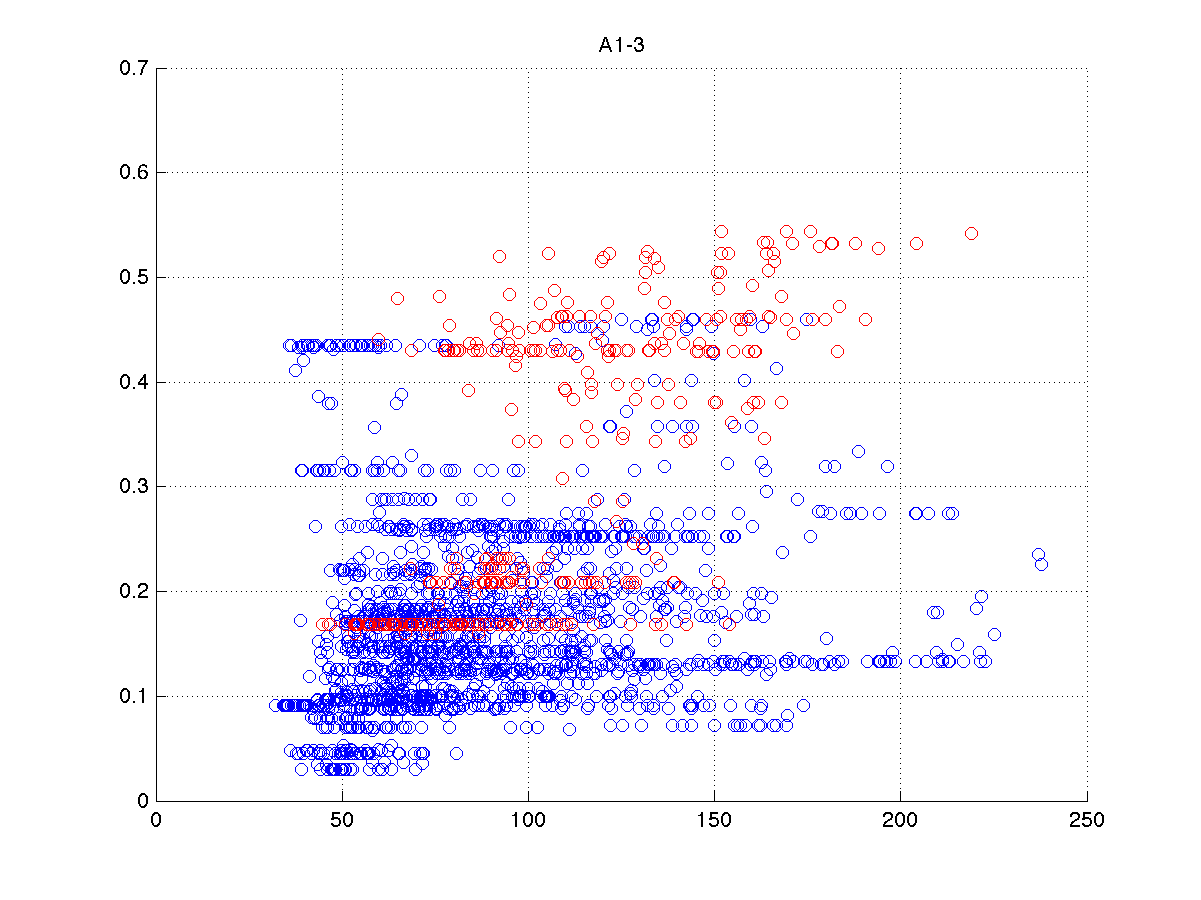}}\\
\subfloat[$\mathscr{R}_{1}$correl-mcap]{\includegraphics[width = 1.7in]{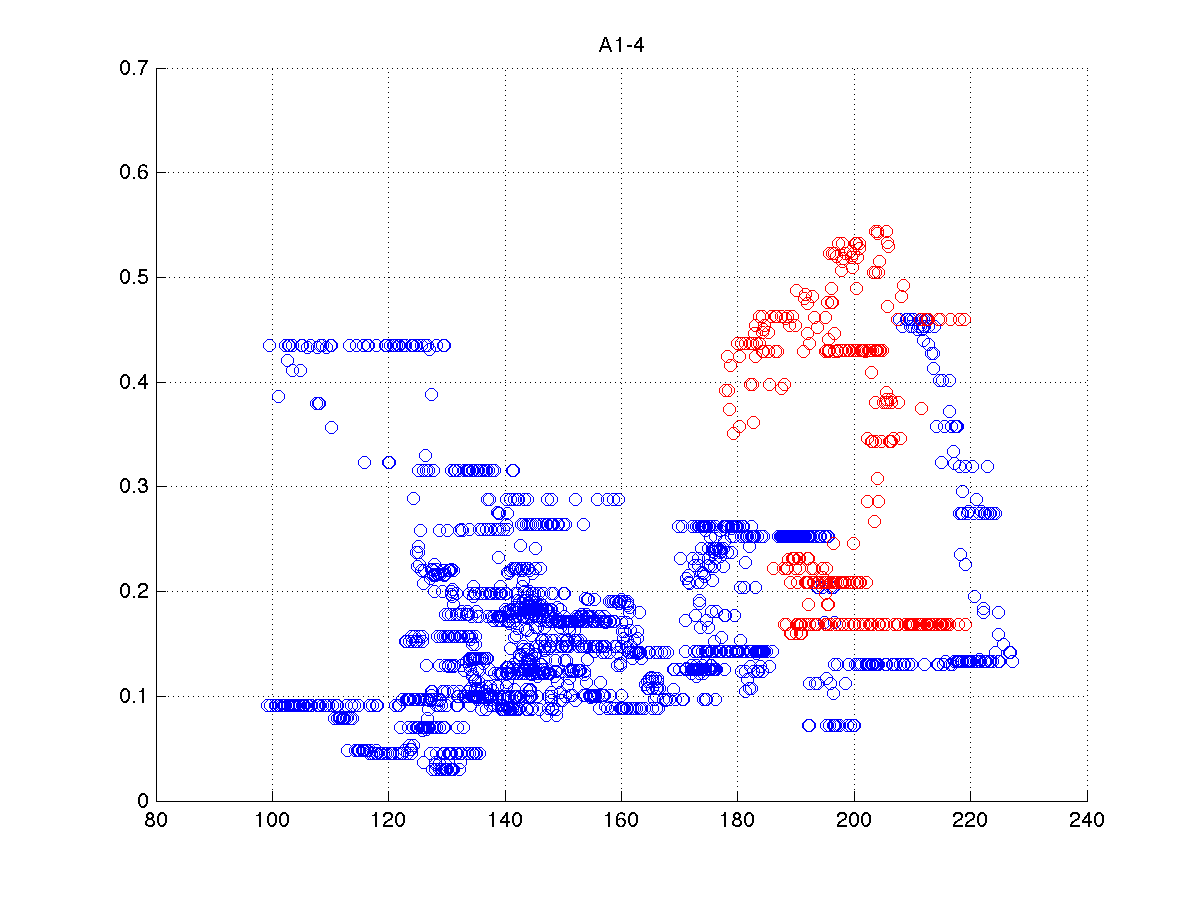}}&
\subfloat[$\mathscr{R}_{1}$correl-leverage]{\includegraphics[width = 1.7in]{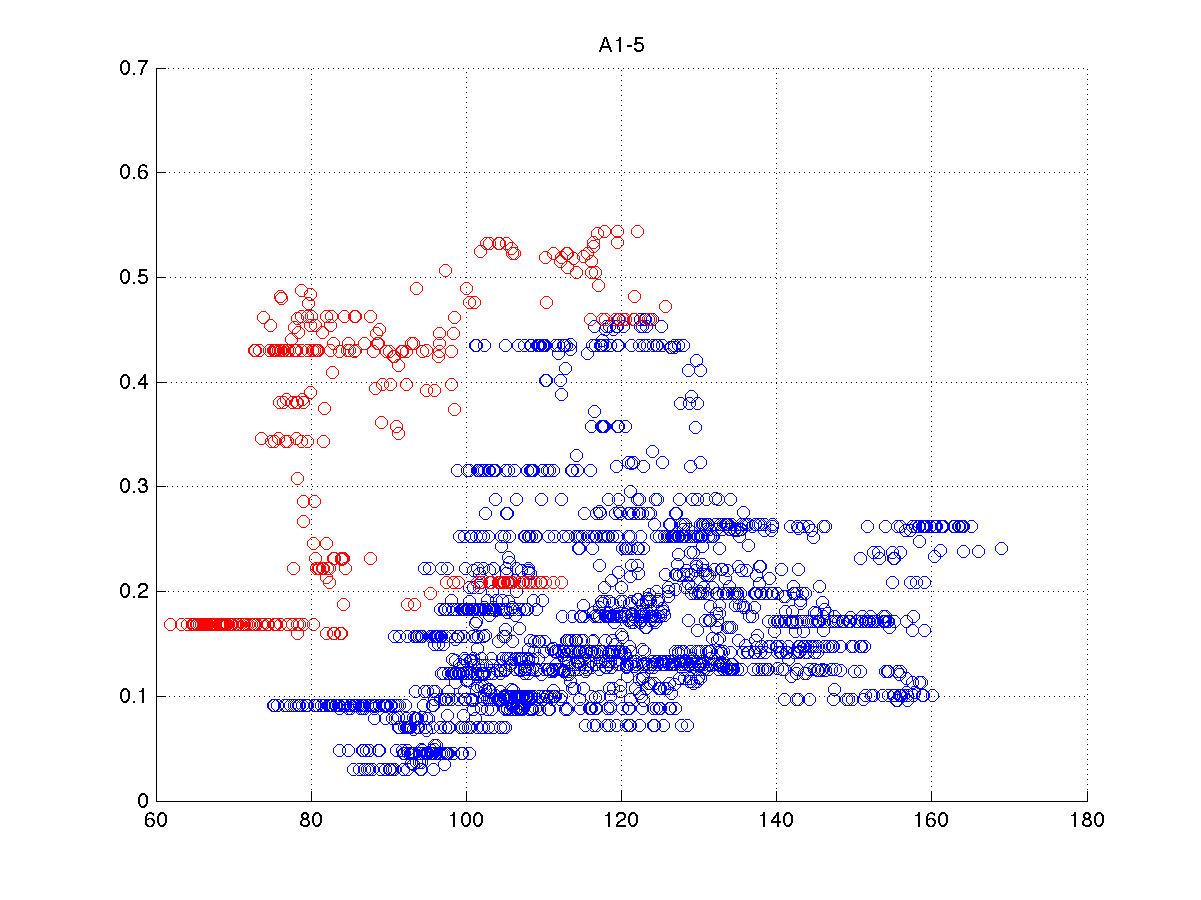}} &
\subfloat[$\mathscr{R}_{2}$covar]{\includegraphics[width = 1.7in]{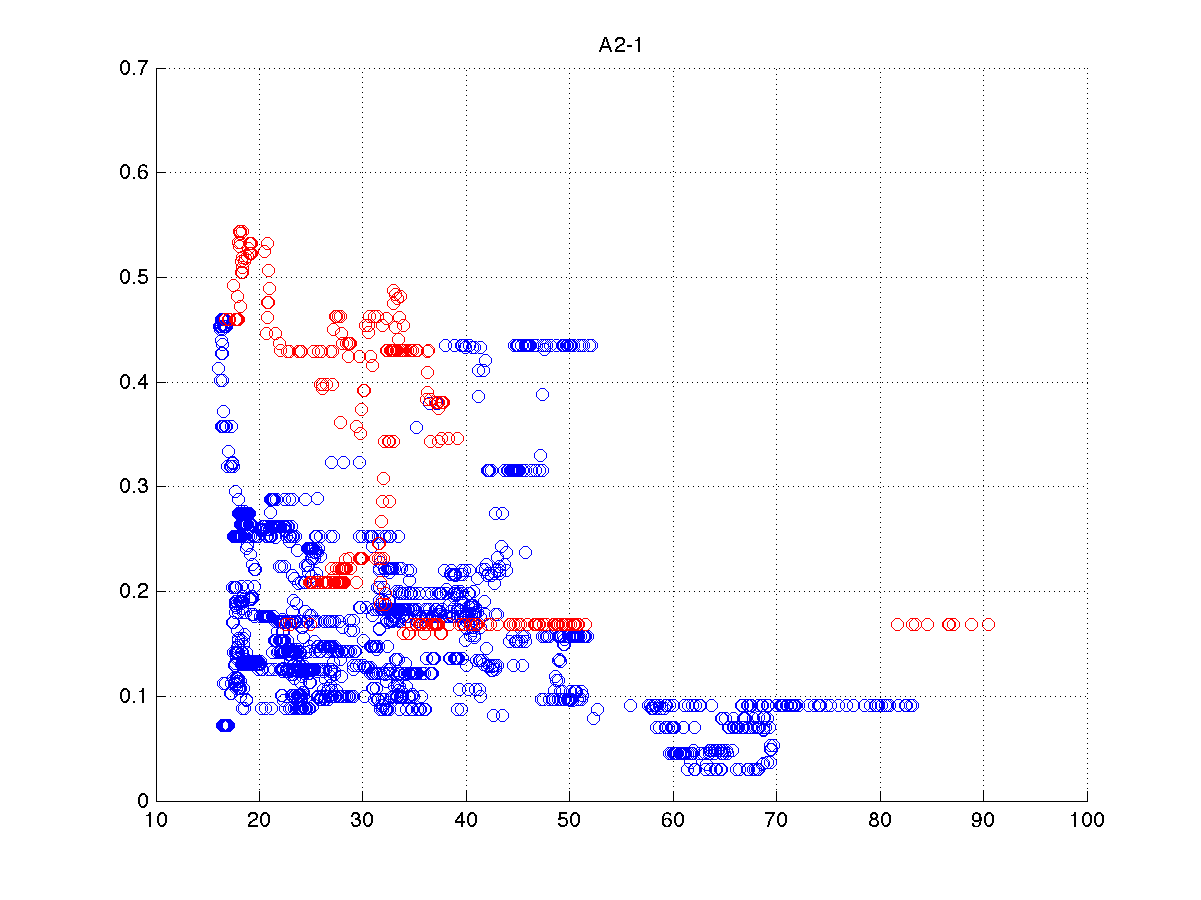}}\\
\subfloat[$\mathscr{R}_{2}$correl]{\includegraphics[width = 1.7in]{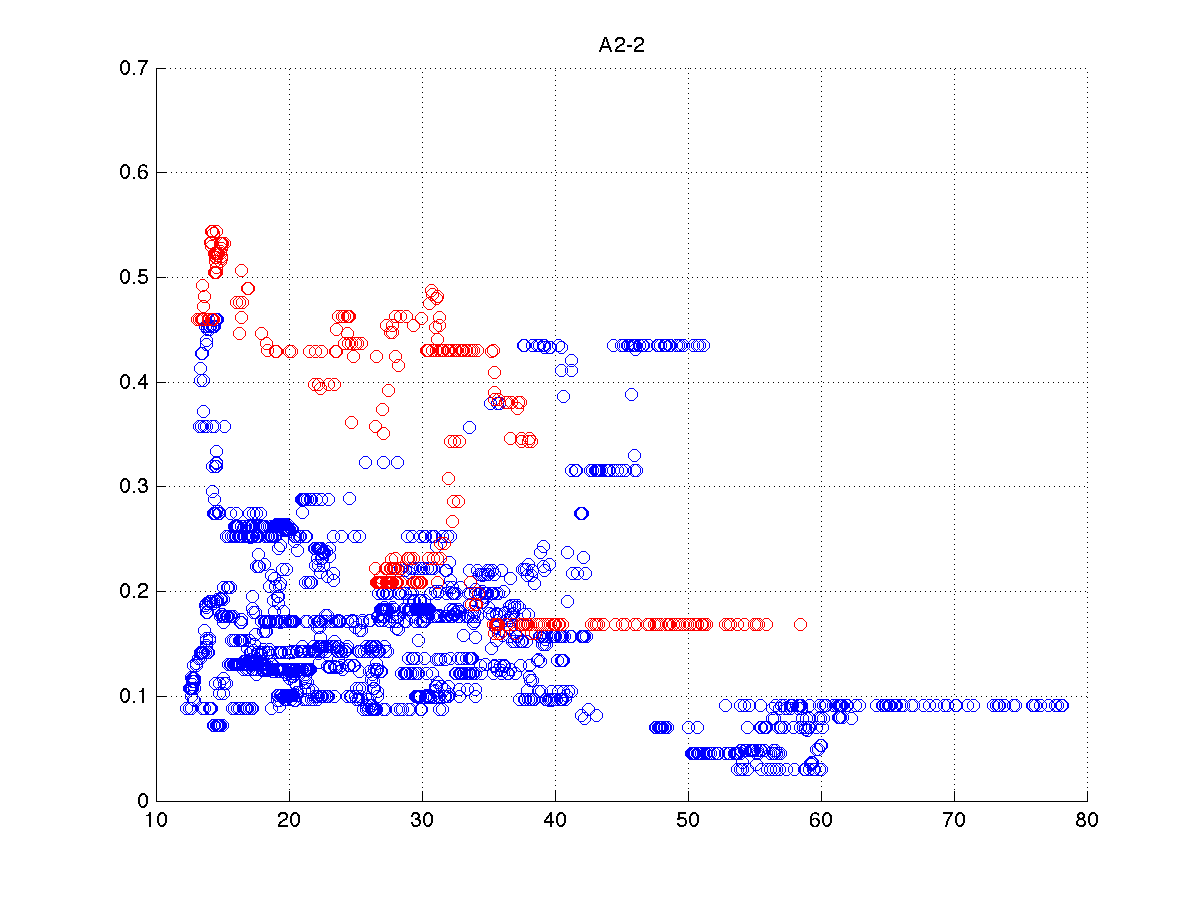}} &
\subfloat[$\mathscr{R}_{2}$correl-volume]{\includegraphics[width = 1.7in]{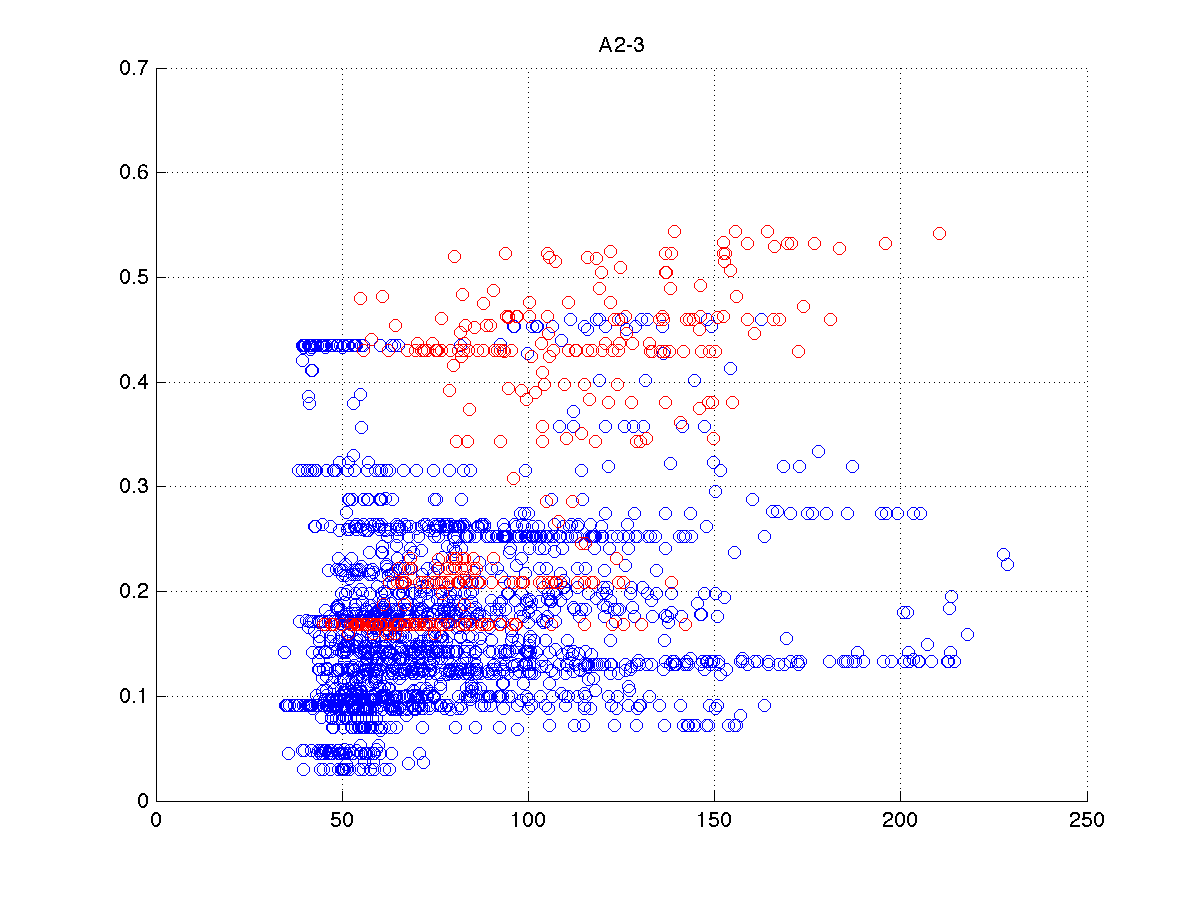}}&
\subfloat[$\mathscr{R}_{2}$correl-mcap]{\includegraphics[width = 1.7in]{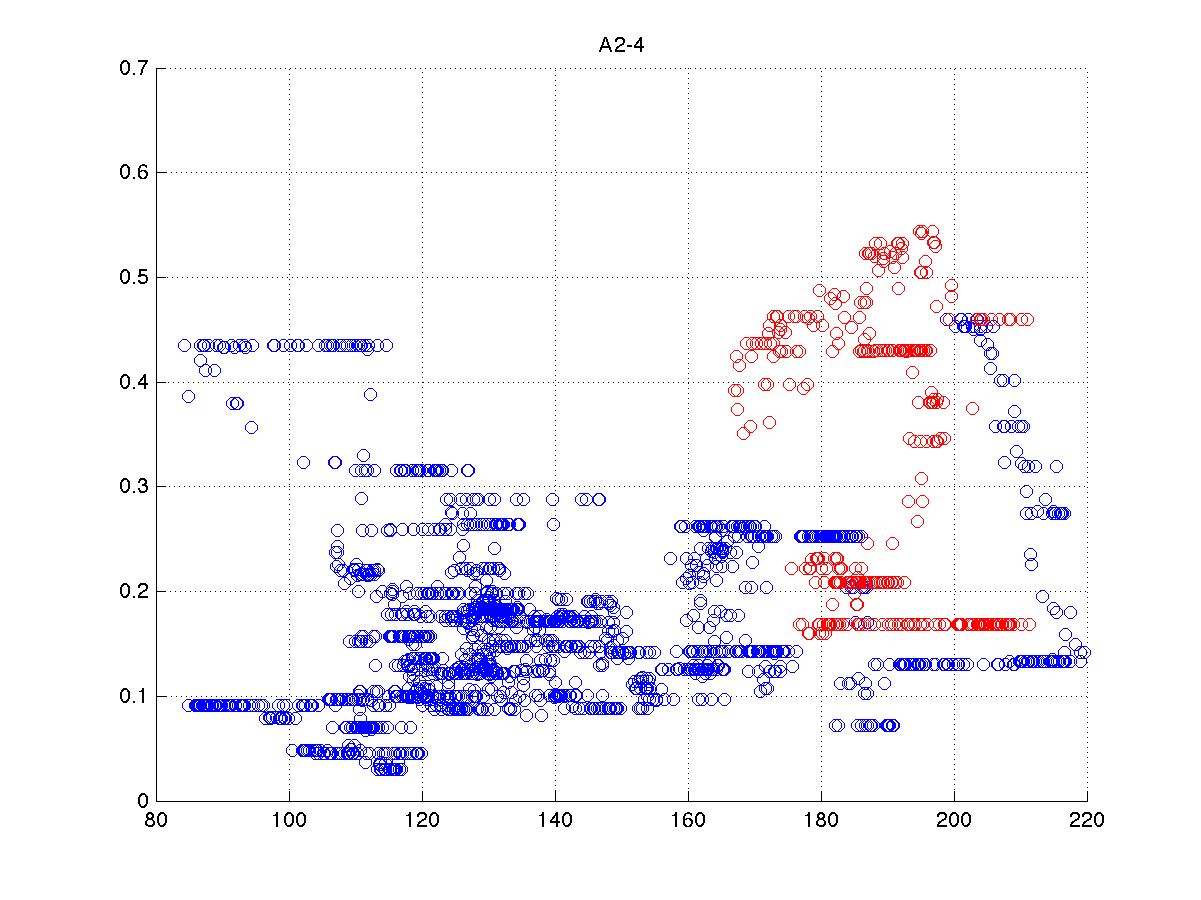}}\\
\subfloat[$\mathscr{R}_{2}$correl-leverage]{\includegraphics[width = 1.7in]{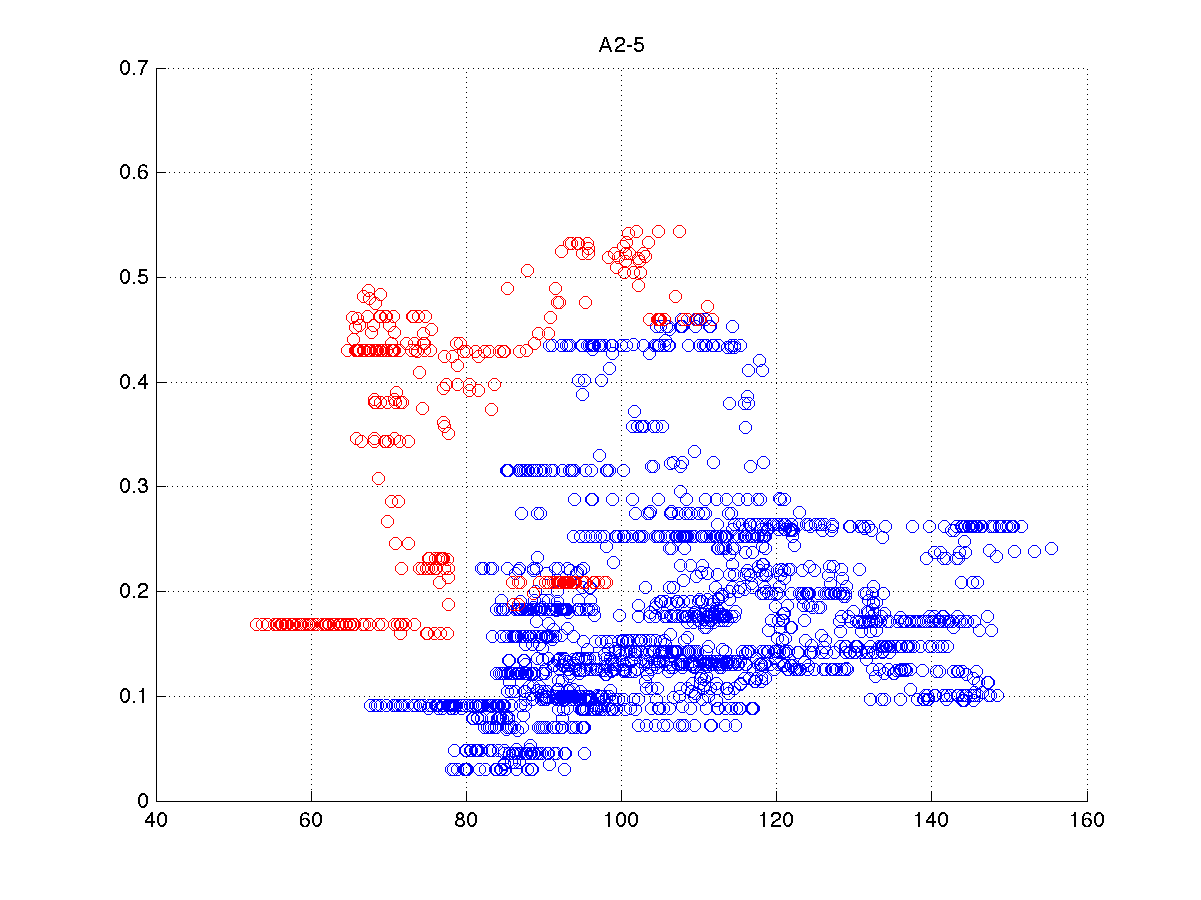}} &
\subfloat[$\mathscr{R}_{3}$covar]{\includegraphics[width = 1.7in]{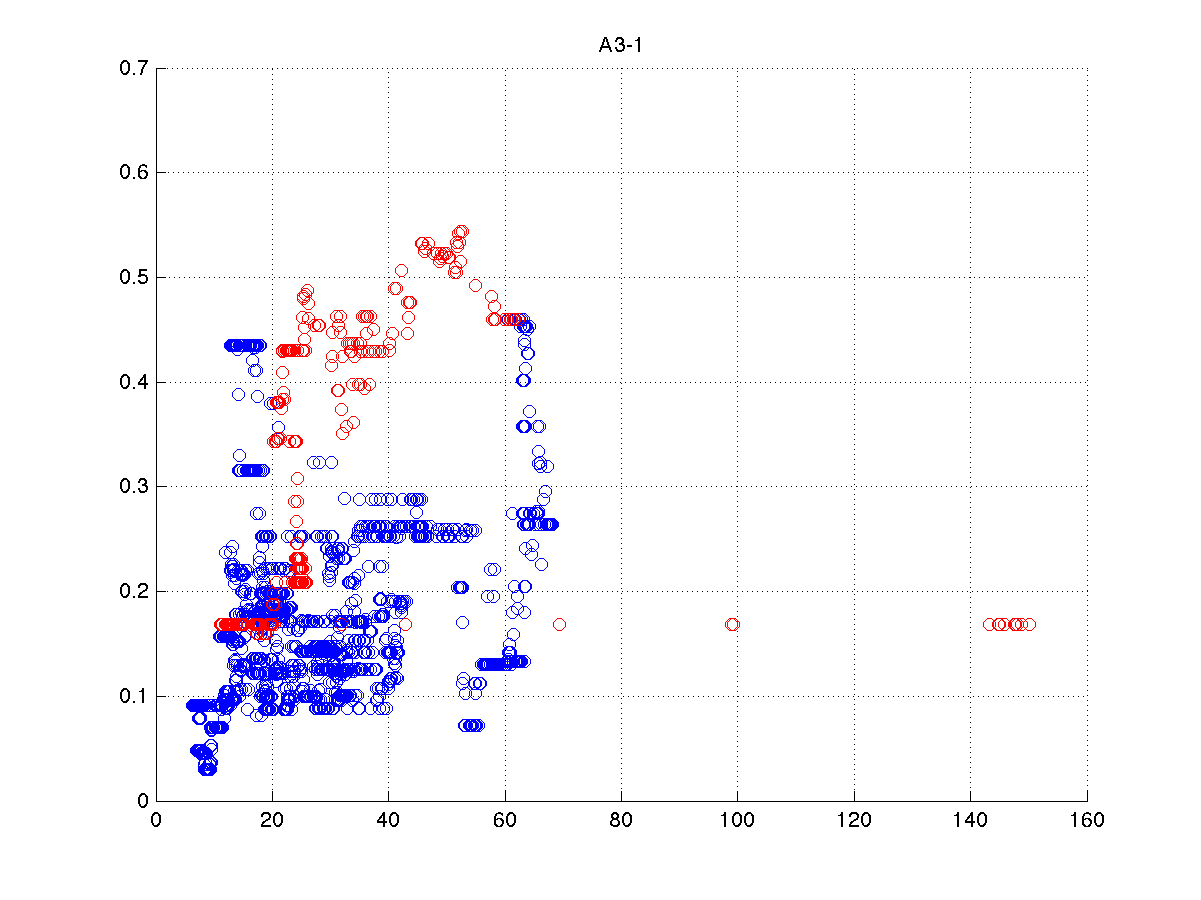}} &
\subfloat[$\mathscr{R}_{3}$correl]{\includegraphics[width = 1.7in]{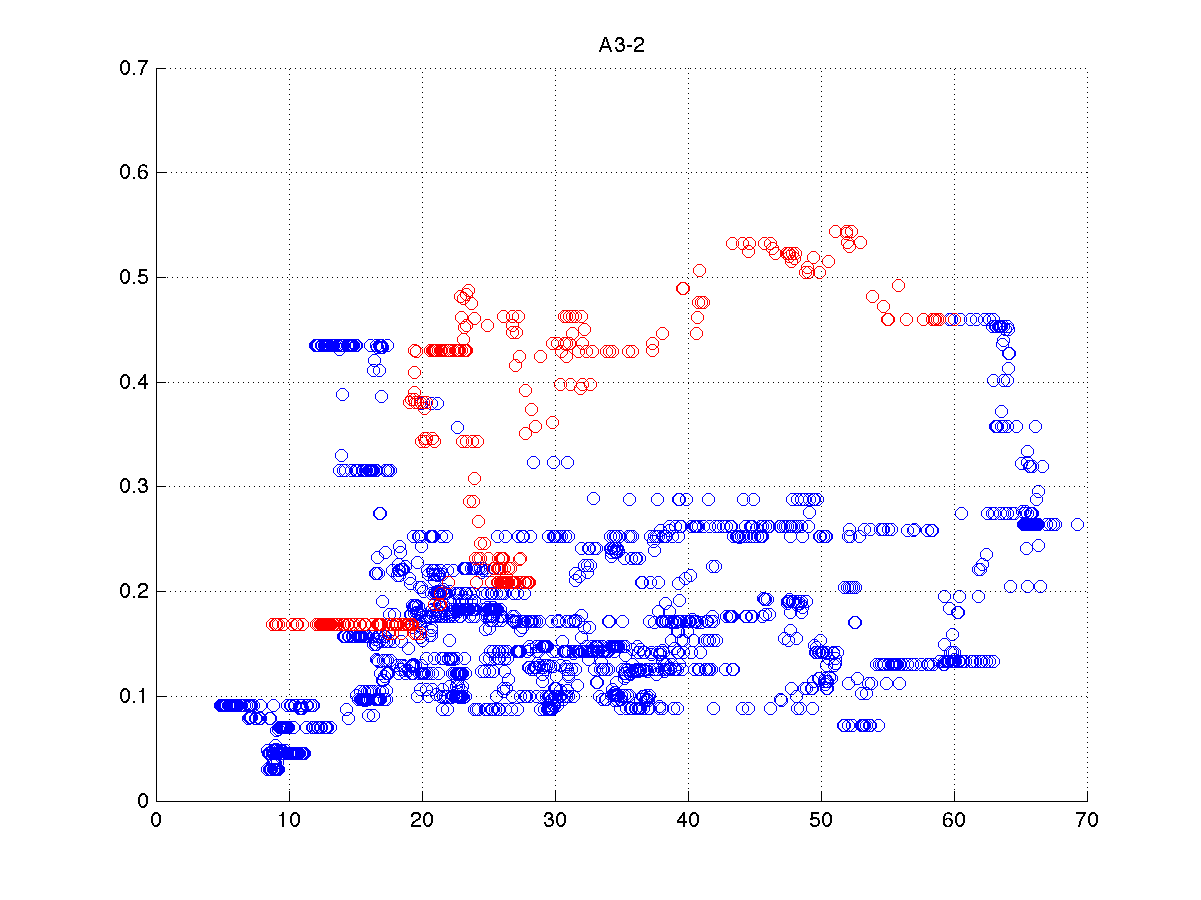}}\\
\subfloat[$\mathscr{R}_{3}$correl-volume]{\includegraphics[width = 1.7in]{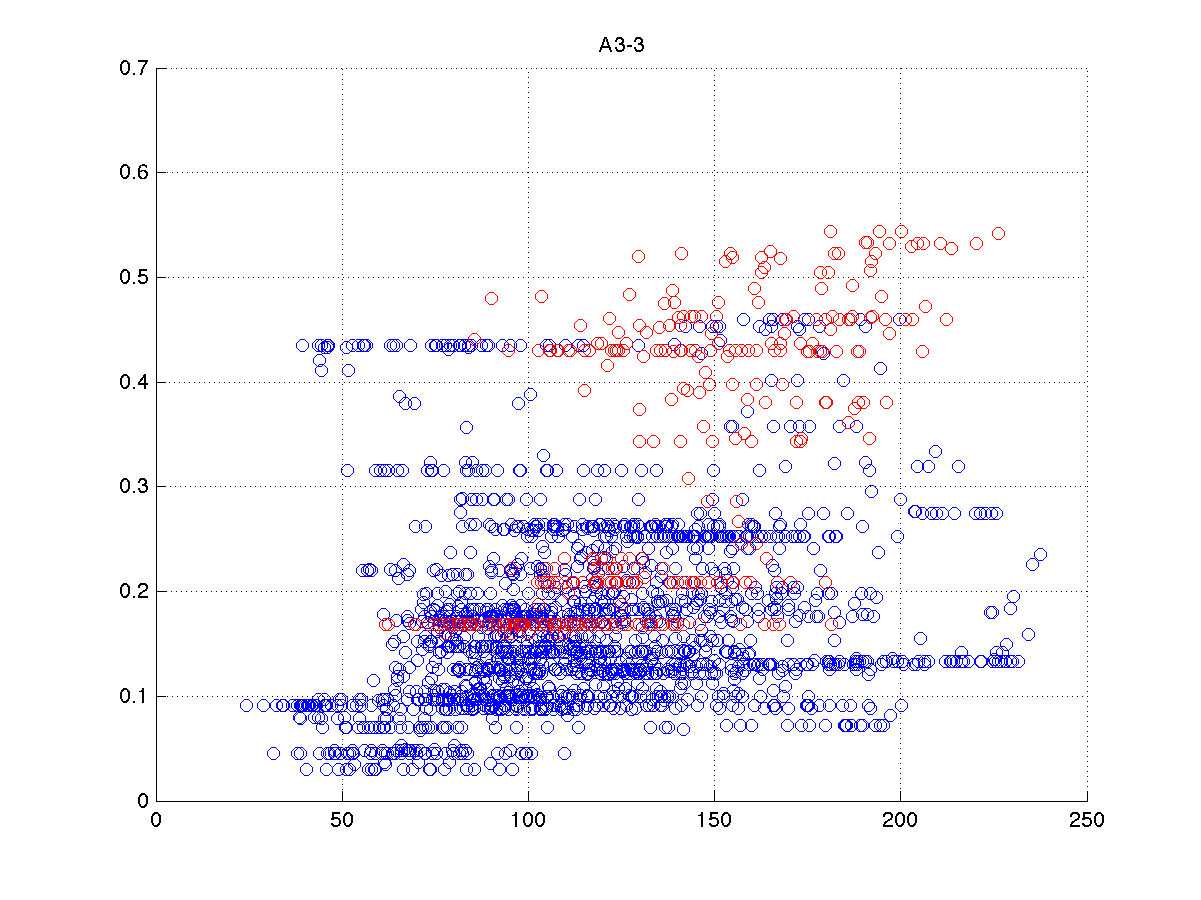}} &
\subfloat[$\mathscr{R}_{3}$correl-mcap]{\includegraphics[width = 1.7in]{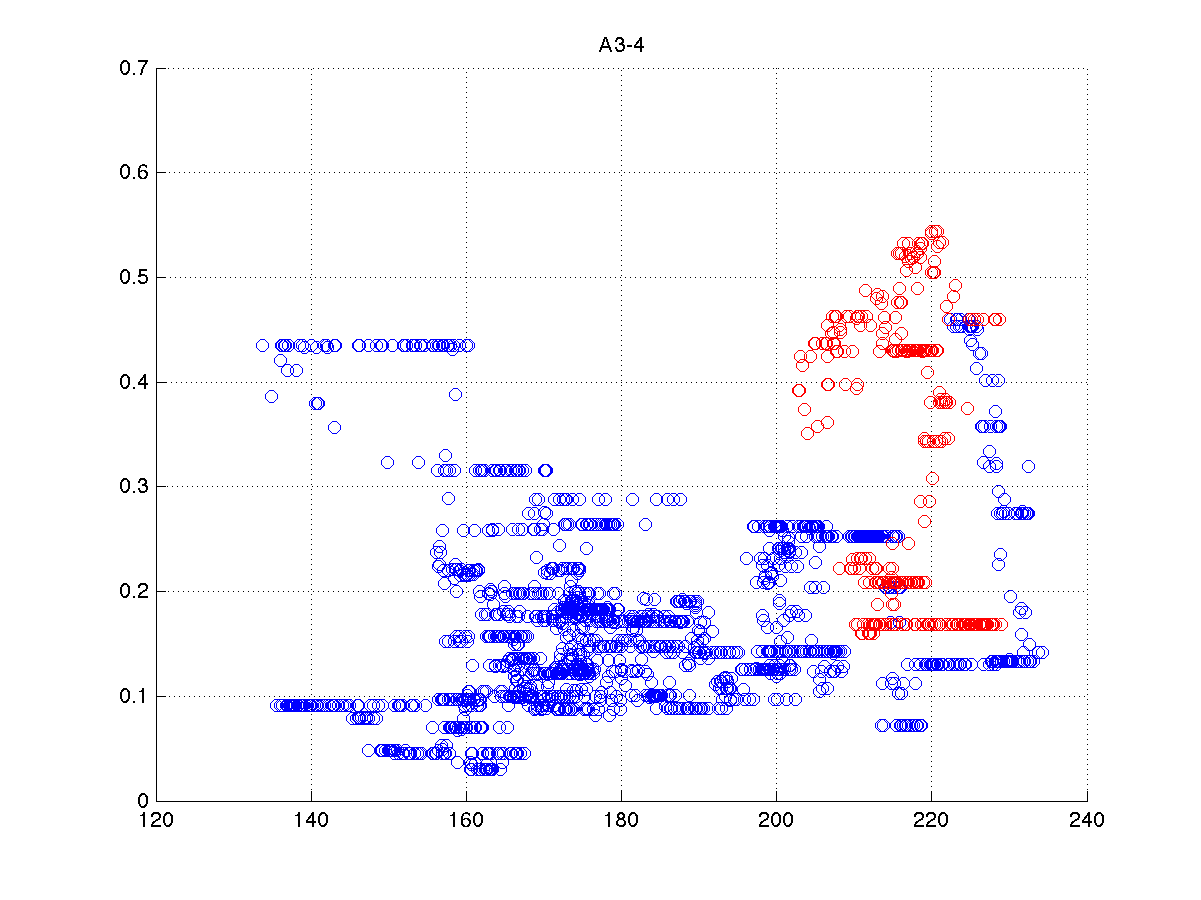}} &
\subfloat[$\mathscr{R}_{3}$correl-leverage]{\includegraphics[width = 1.7in]{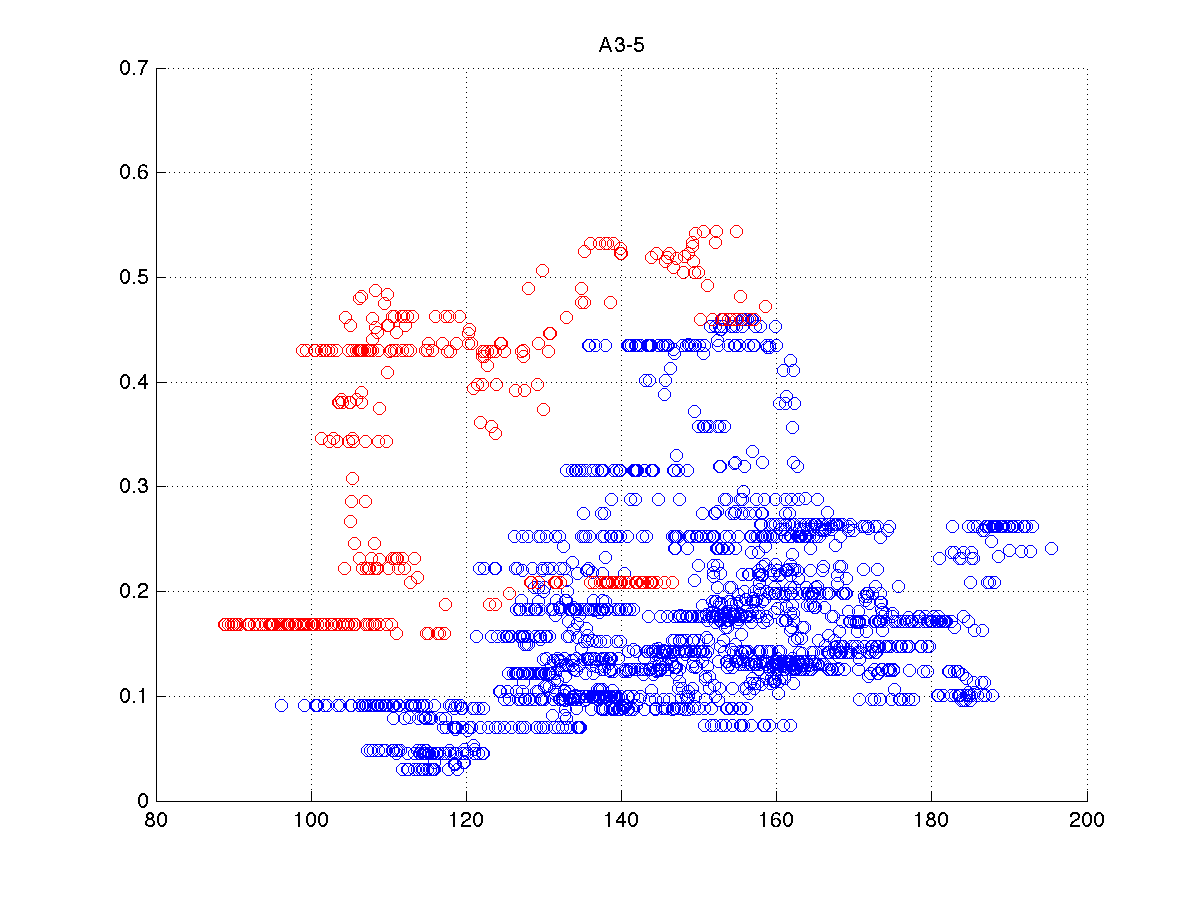}}\\
\end{tabular}
\captionsetup{labelformat=empty}
\caption{SHSZ300: Indicators of the $\alpha$-series. Red: in-sample ; Blue: out-of-sample}
\end{figure}

\begin{figure}[H]
\begin{tabular}{ccc}
\subfloat[rspec-covar]{\includegraphics[width = 1.7in]{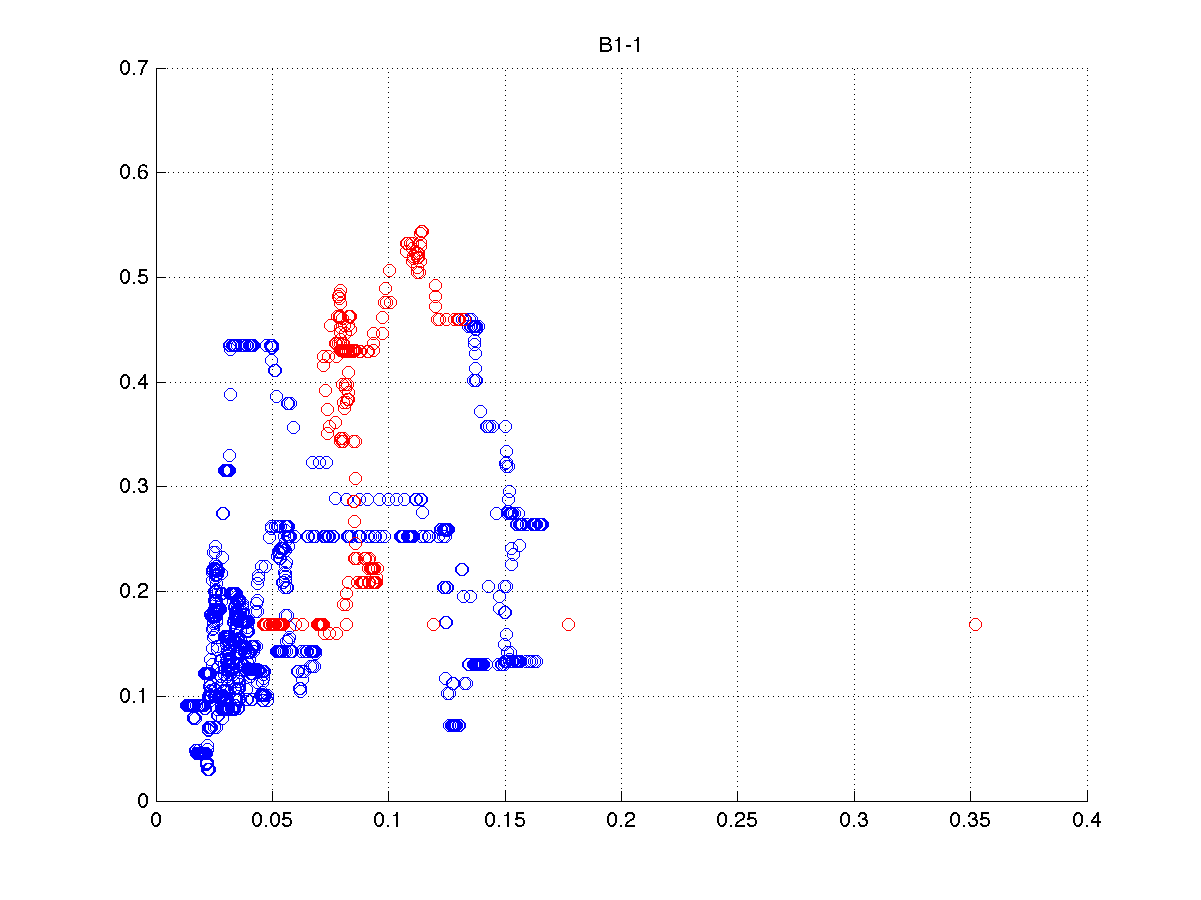}} &
\subfloat[rspec-correl]{\includegraphics[width = 1.7in]{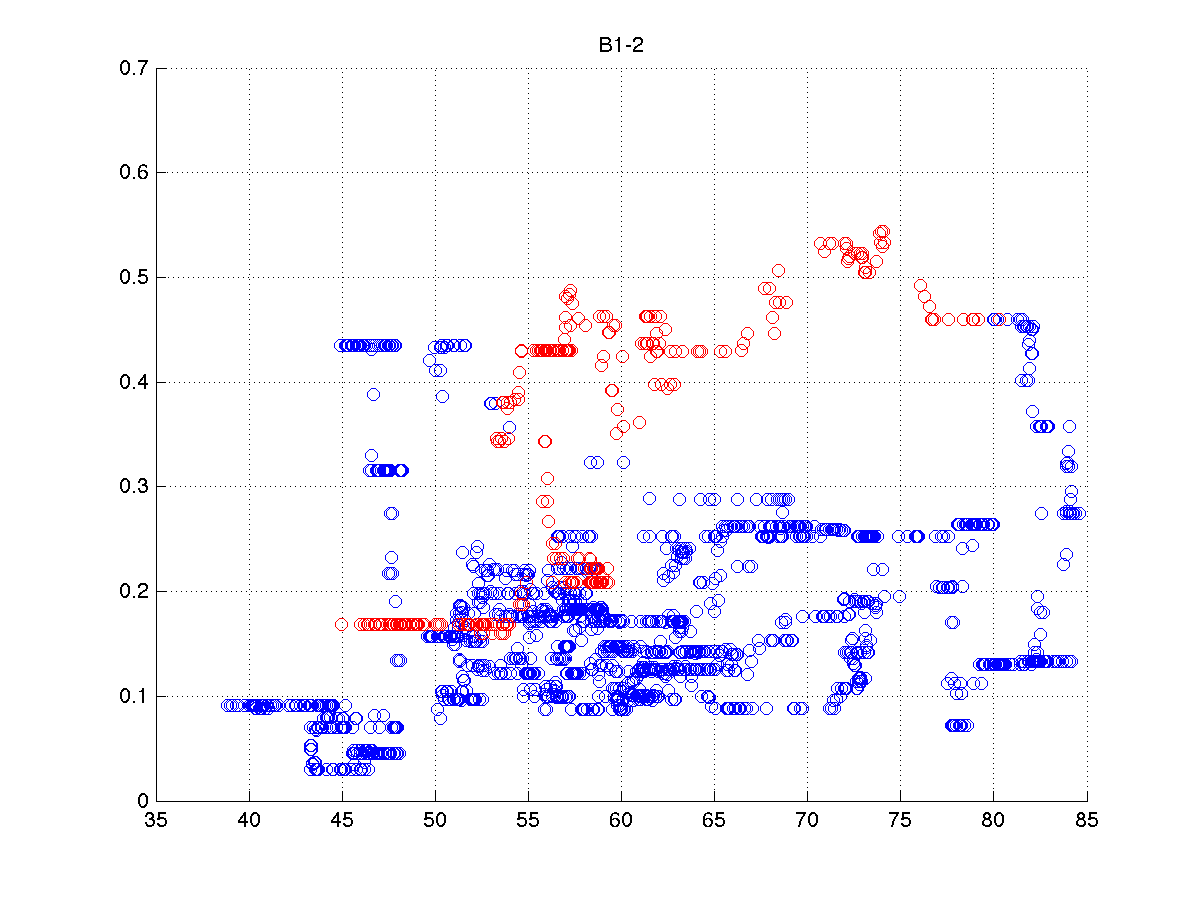}} &
\subfloat[rspec-correl-volume]{\includegraphics[width = 1.7in]{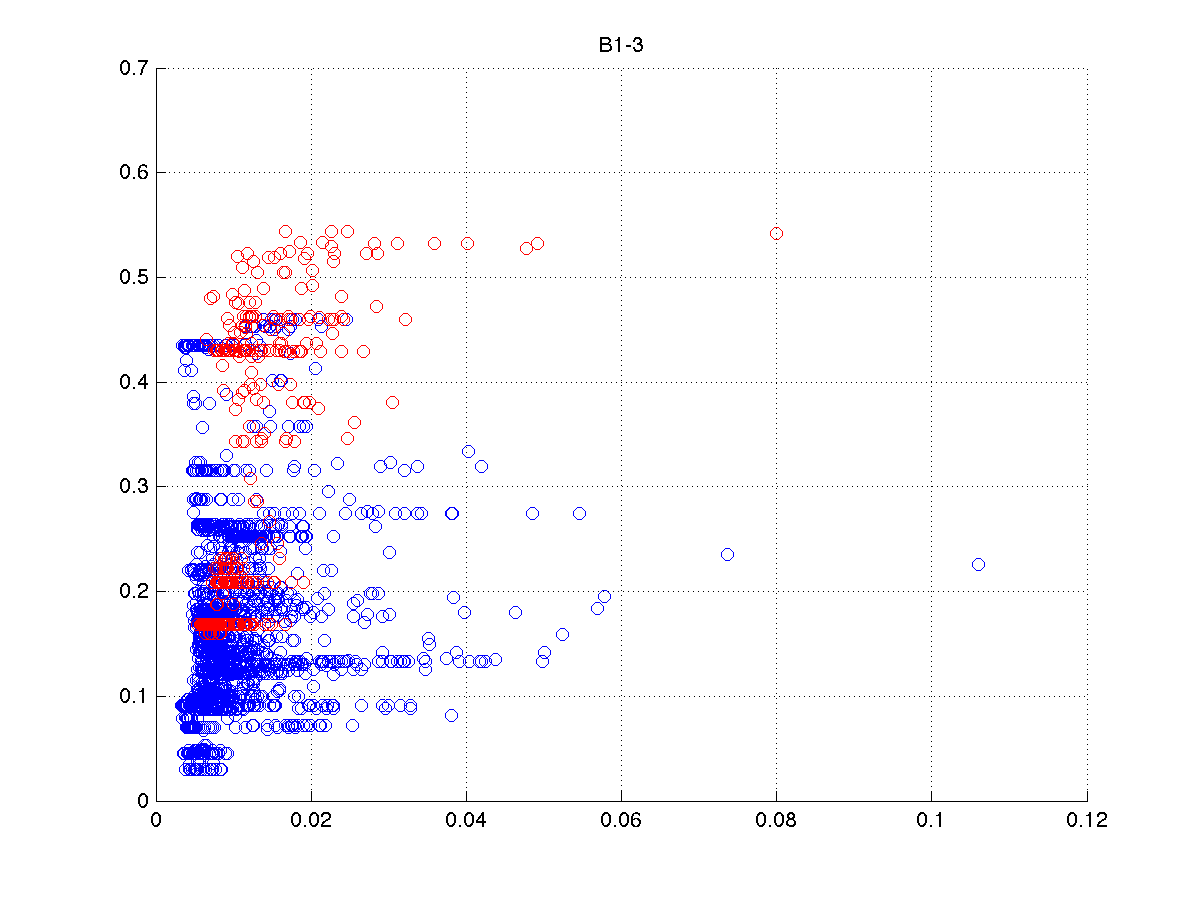}} \\
\subfloat[rspec-correl-mcap]{\includegraphics[width = 1.7in]{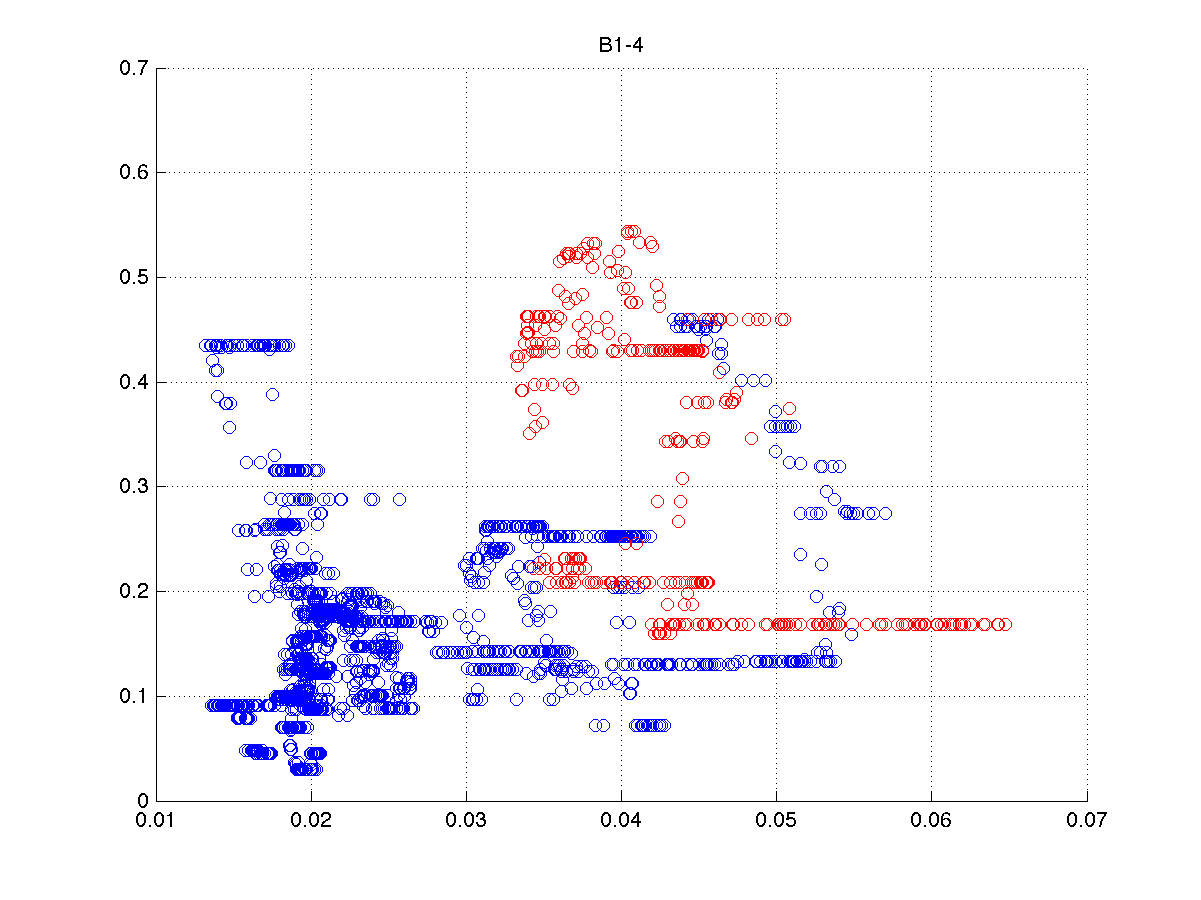}}&
\subfloat[rspec-correl-leverage]{\includegraphics[width = 1.7in]{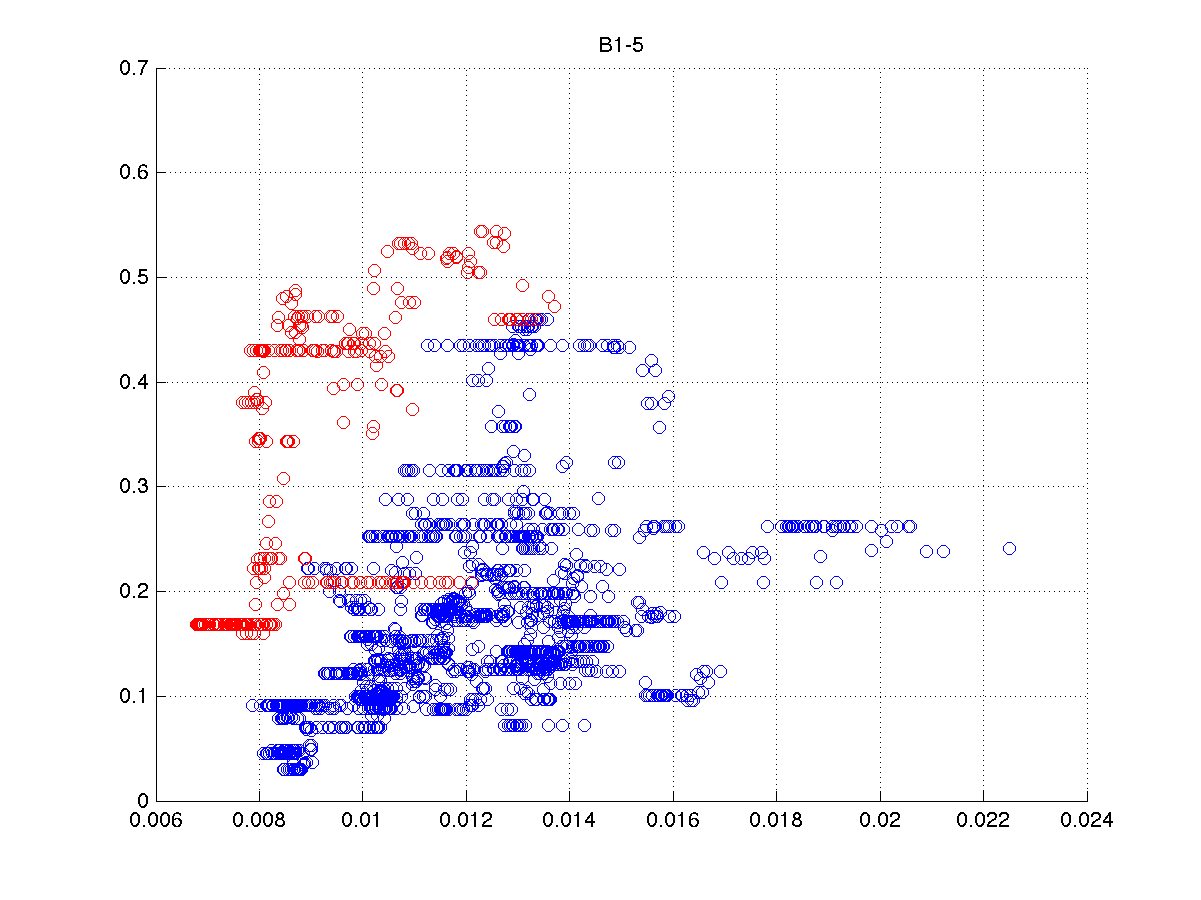}} &
\subfloat[trace-covar]{\includegraphics[width = 1.7in]{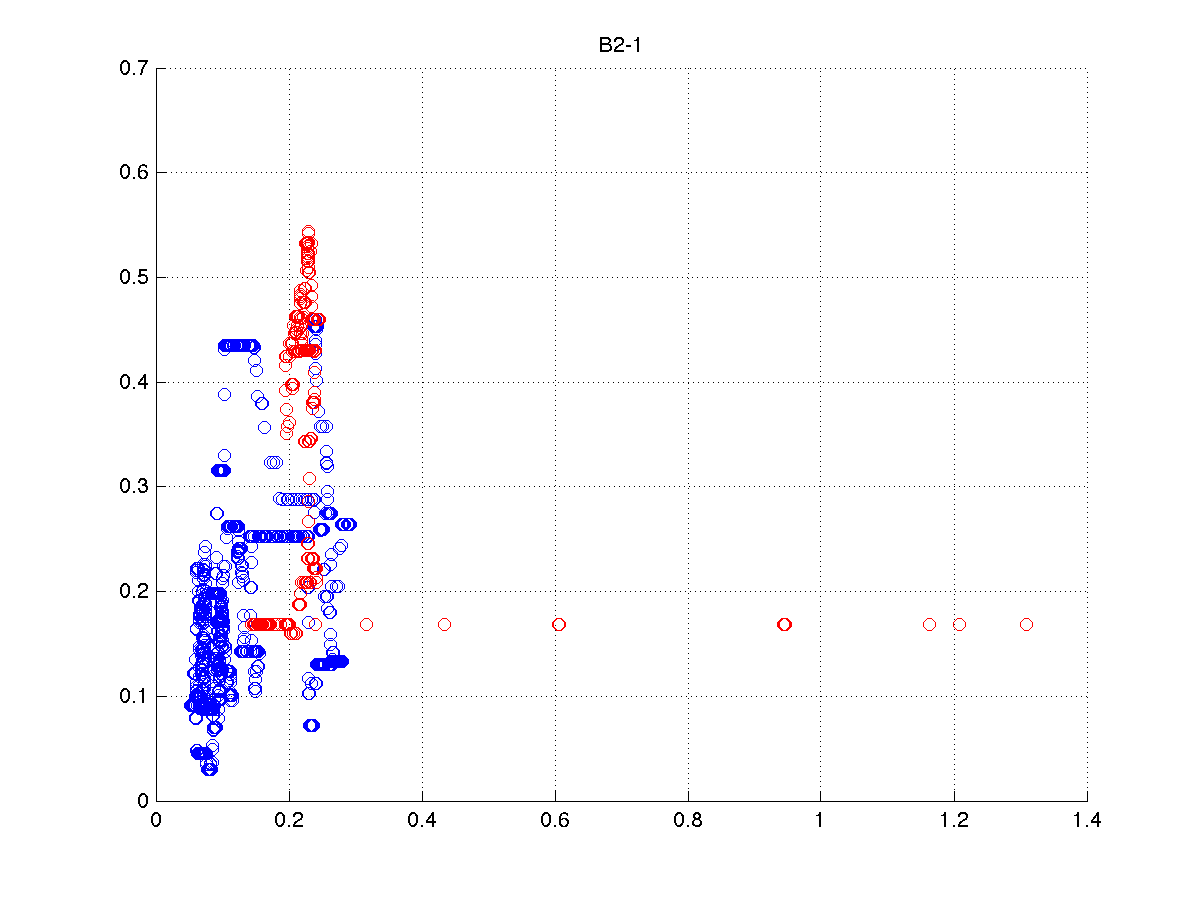}}\\
\subfloat[trace-correl-volume]{\includegraphics[width = 1.7in]{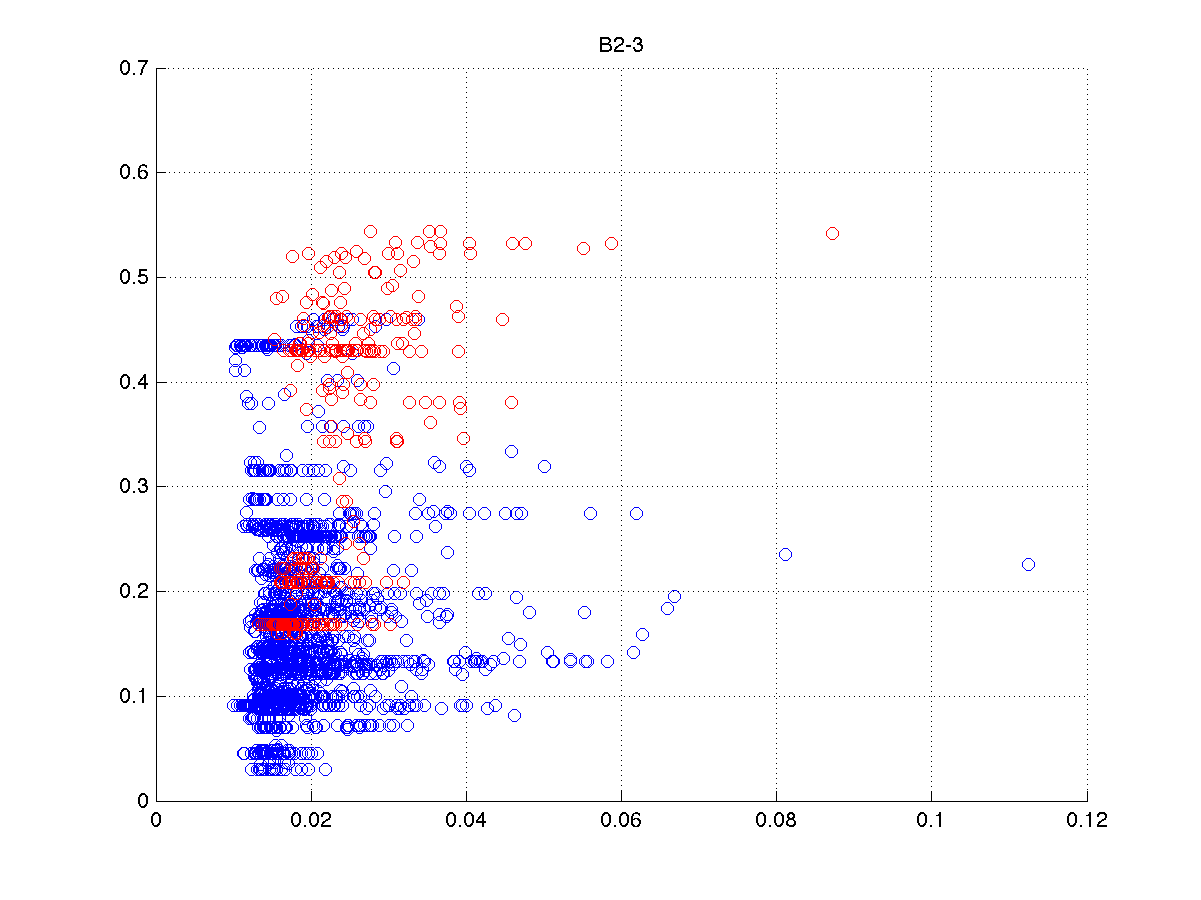}} &
\subfloat[trace-correl-mcap]{\includegraphics[width = 1.7in]{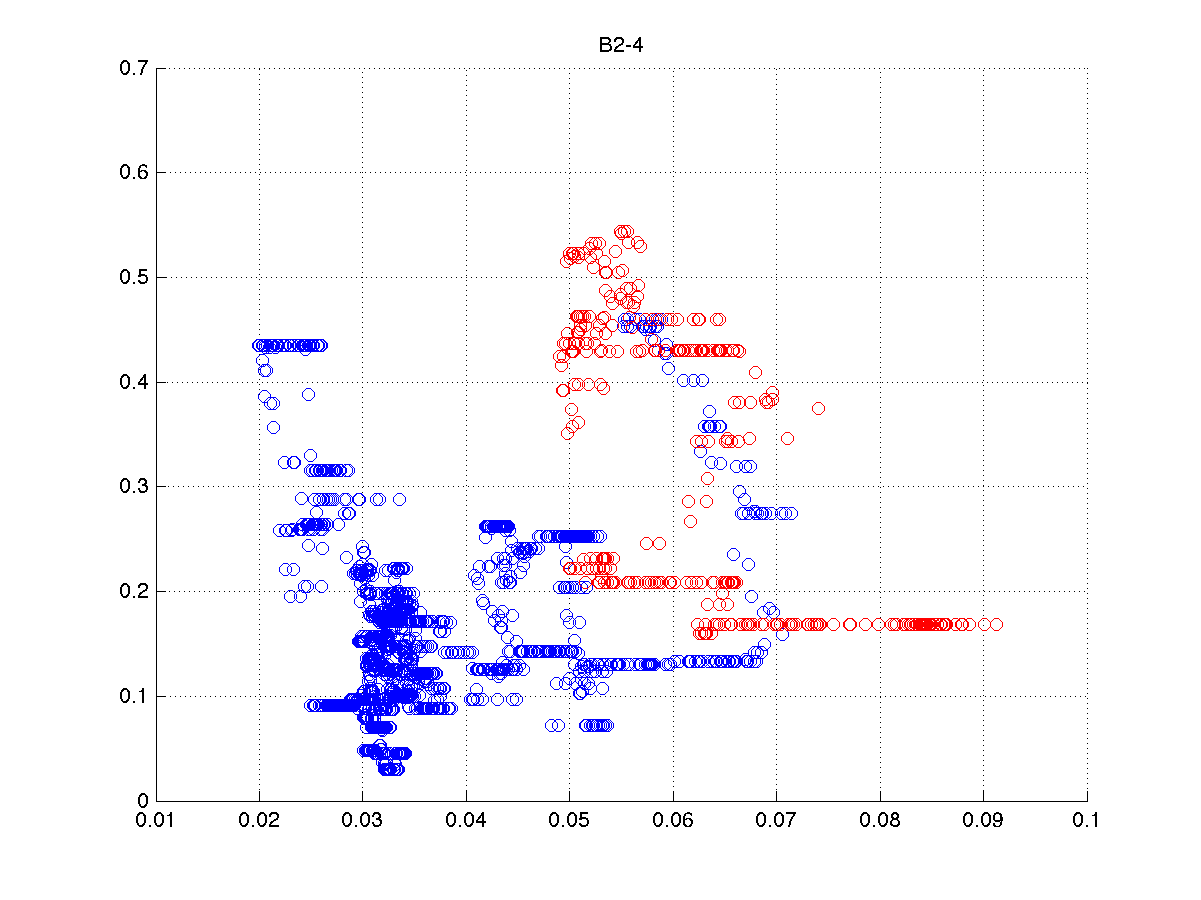}}&
\subfloat[trace-correl-leverage]{\includegraphics[width = 1.5in]{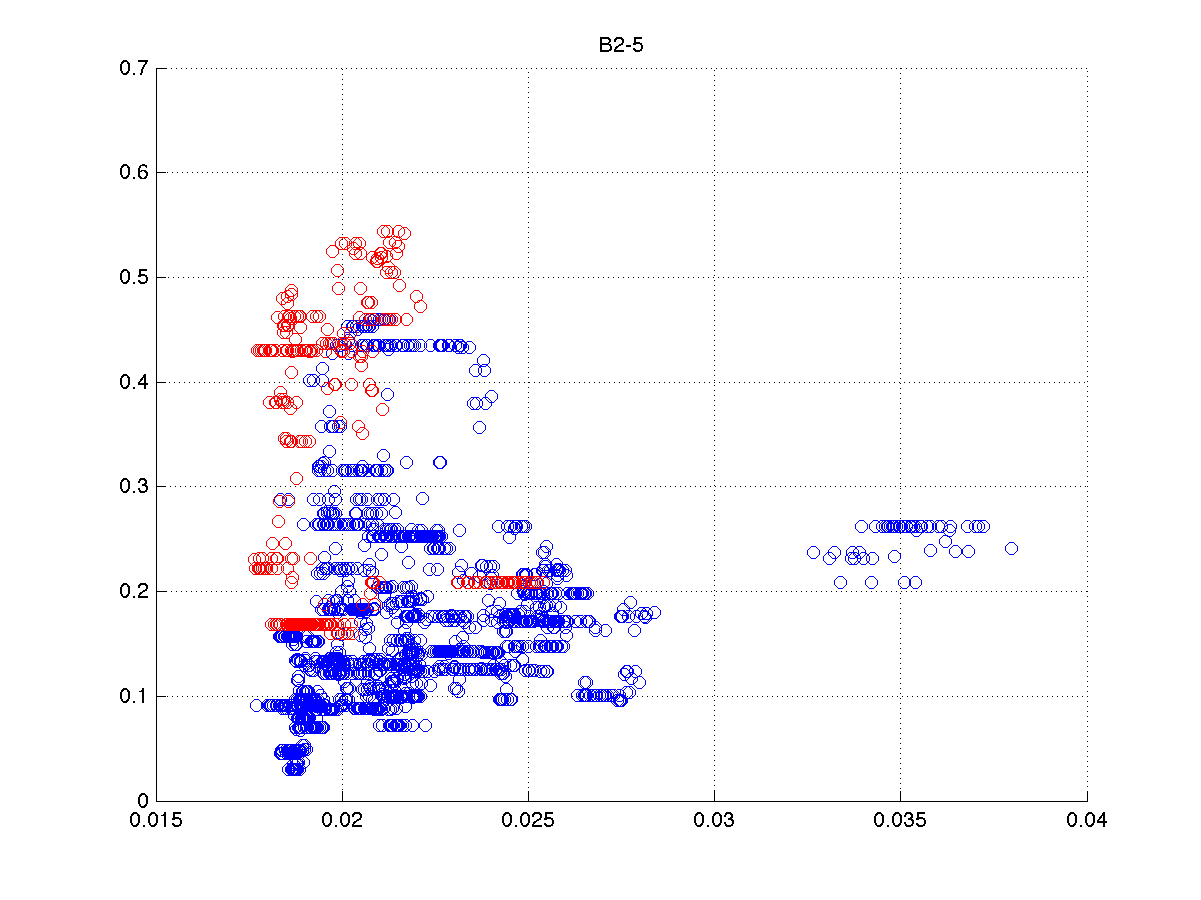}}\\
\subfloat[froben-covar]{\includegraphics[width = 1.7in]{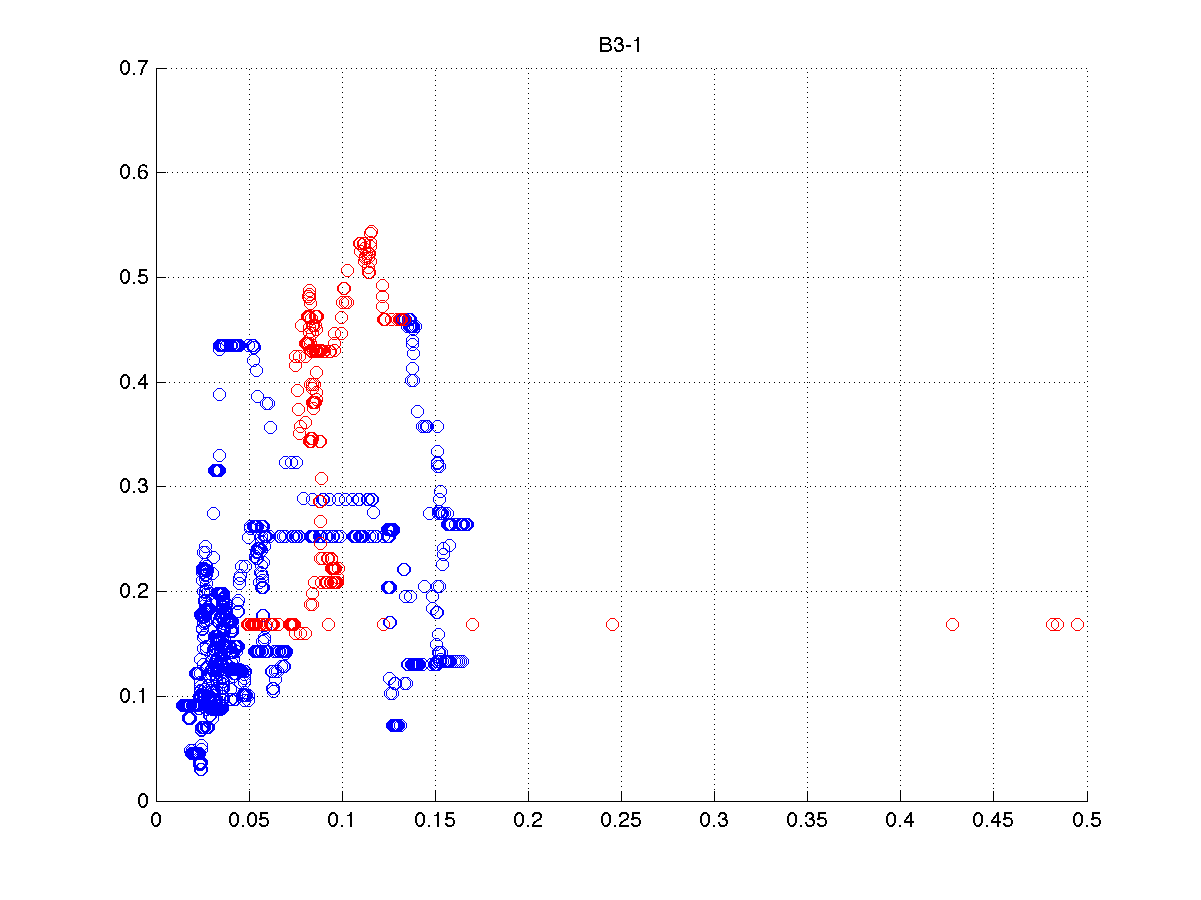}} &
\subfloat[froben-correl]{\includegraphics[width = 1.7in]{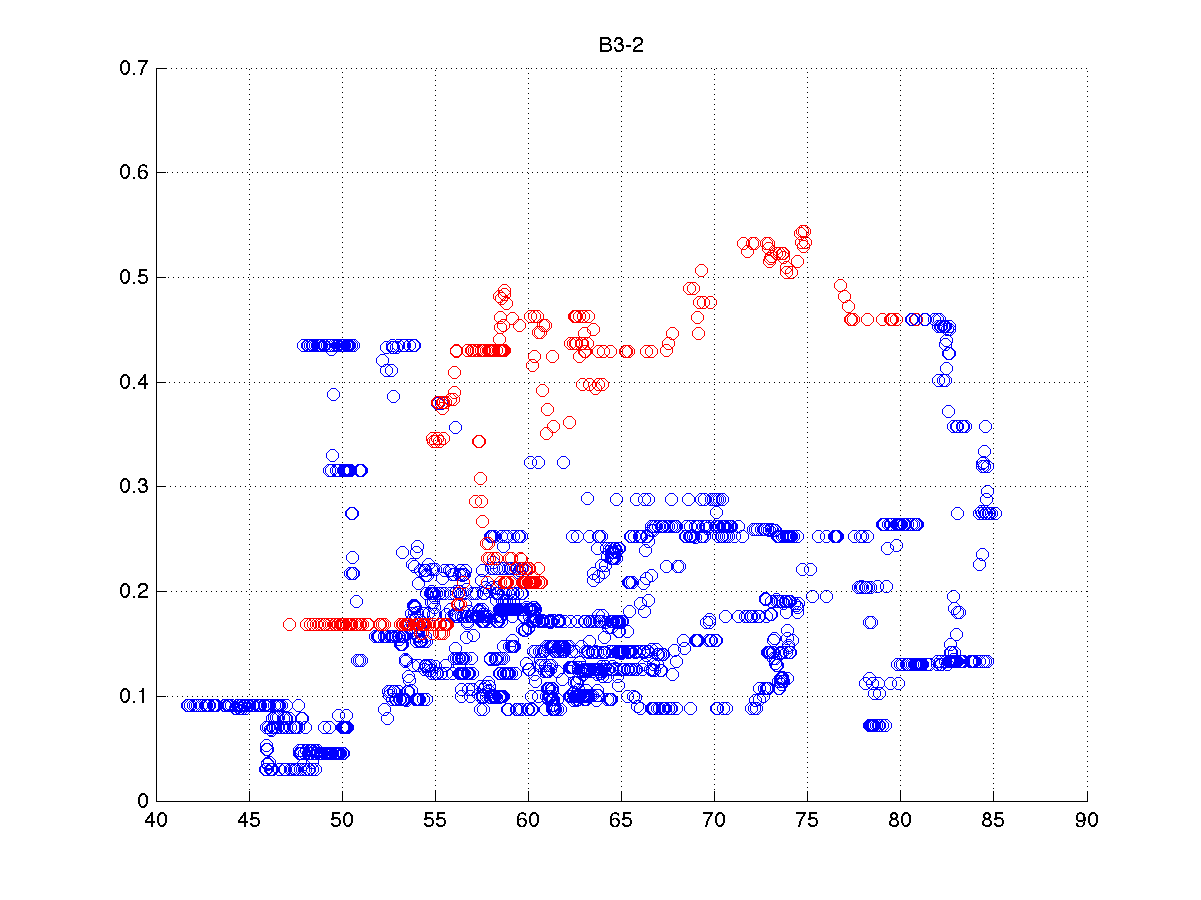}} &
\subfloat[froben-correl-volume]{\includegraphics[width = 1.7in]{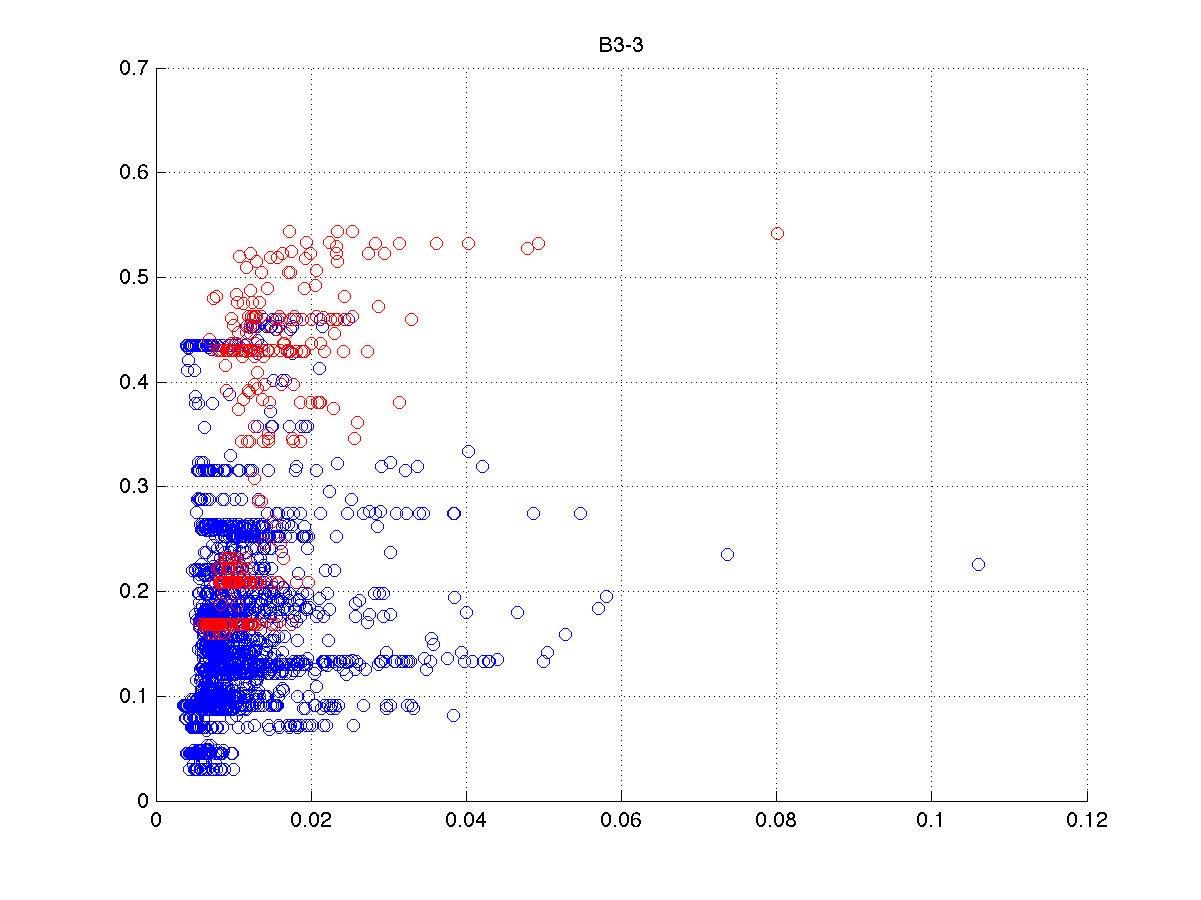}}\\
\subfloat[froben-correl-mcap]{\includegraphics[width = 1.7in]{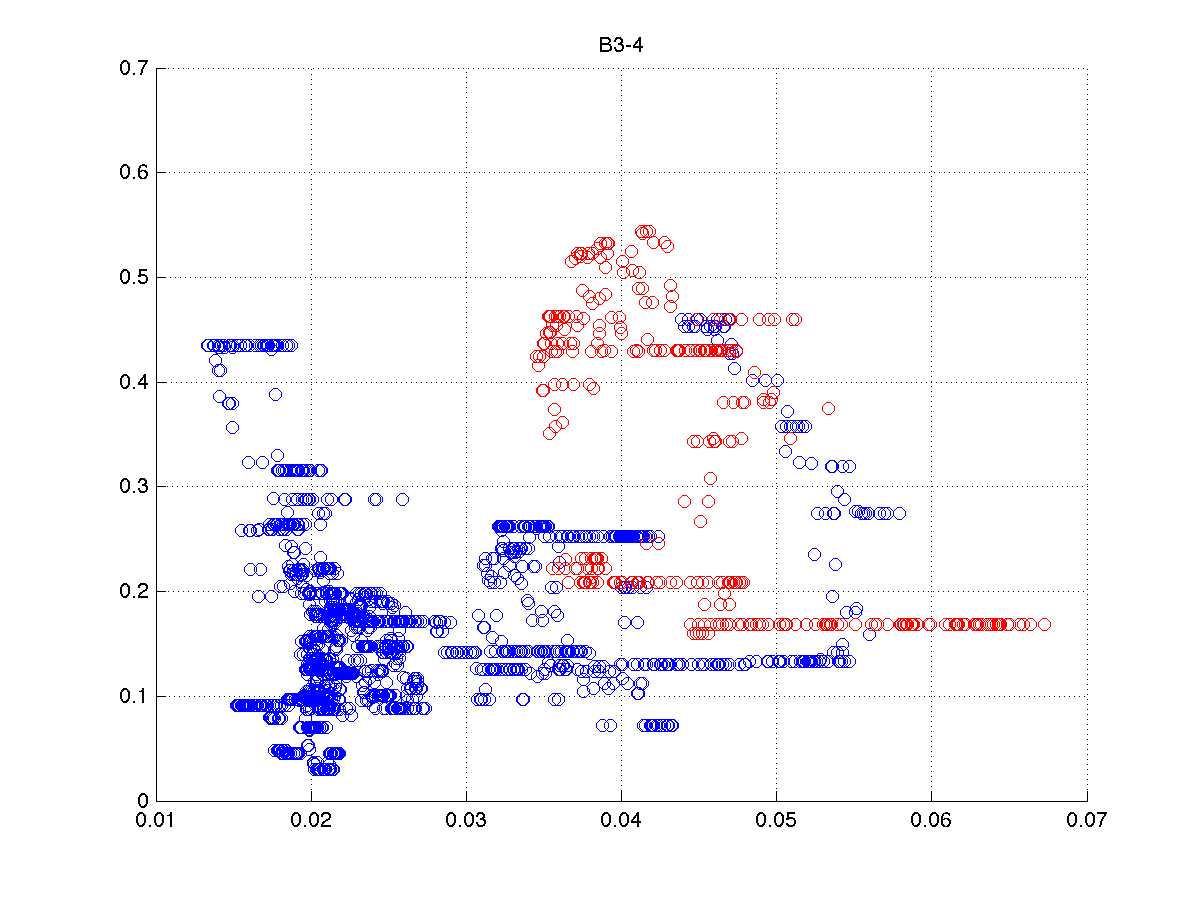}} &
\subfloat[froben-correl-leverage]{\includegraphics[width = 1.7in]{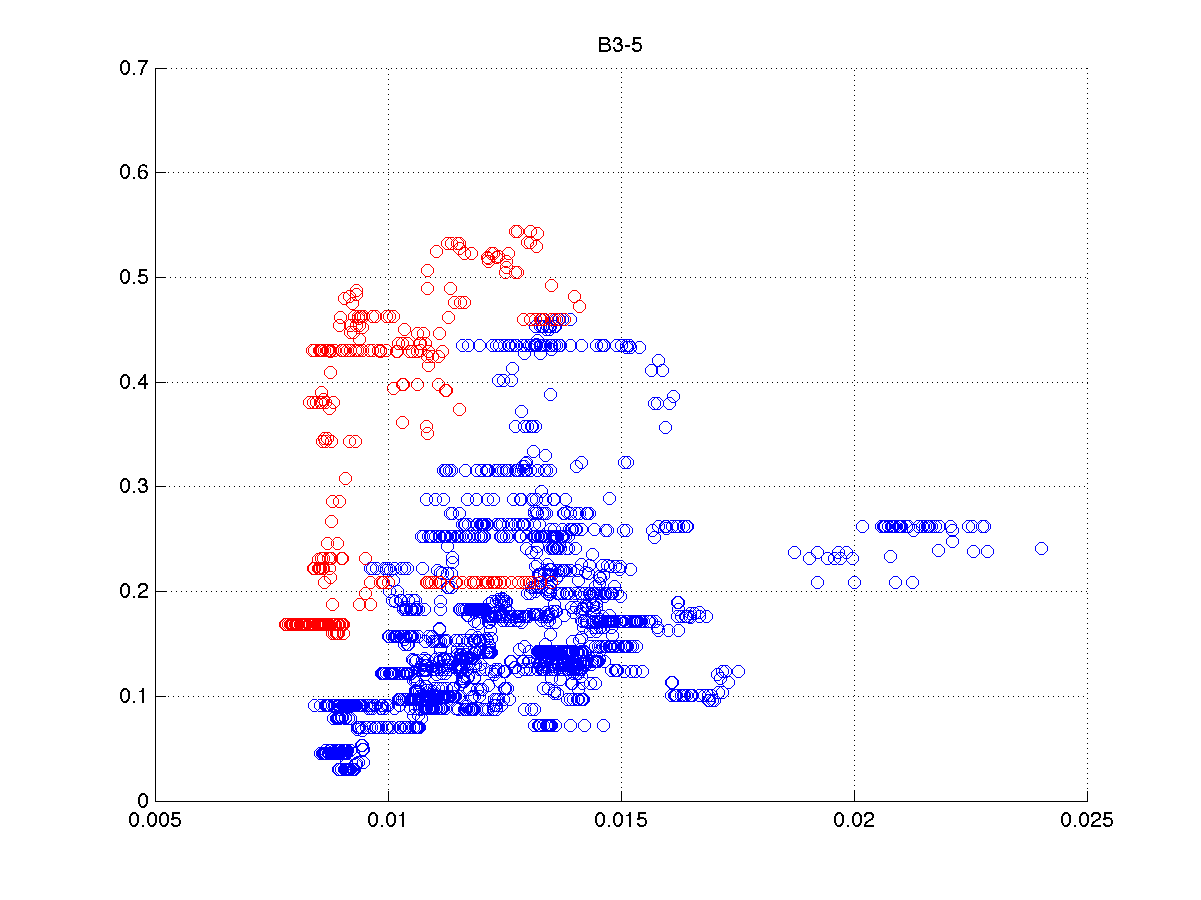}} &
\end{tabular}
\captionsetup{labelformat=empty}
\caption{SHSZ300: Indicators of the $\beta$-series. Red: in-sample ; Blue: out-of-sample}
\end{figure}

\section*{Funding}
This work was achieved through the Laboratory of Excellence on Financial Regulation (Labex RéFi) supported by PRES heSam under the reference ANR­10­LABX­0095. It benefited from a French government support managed by the National Research Agency (ANR) within the project Investissements d'Avenir Paris Nouveaux Mondes (Investments for the Future Paris ­New Worlds) under the reference ANR­11­IDEX­0006­02.
\end{document}